# The rotation and translation of non-spherical particles in homogeneous isotropic turbulence

-by-

Margaret Byron

A dissertation submitted in partial satisfaction of the requirements for the degree of

Doctor of Philosophy

in

Engineering—Civil and Environmental

in the

Graduate Division

of the

University of California, Berkeley

Committee in charge:

Professor Evan A. Variano, chair
Professor Mimi A.R. Koehl
Professor Ömer Savaş
Professor Mark T. Stacey

Summer 2015

# Abstract

The rotation and translation of non-spherical particles in homogeneous isotropic turbulence

by

Margaret Byron

Doctor of Philosophy in Engineering

University of California, Berkeley

Professor Evan A. Variano, chair

The motion of particles suspended in environmental turbulence is relevant to many scientific fields, from sediment transport to biological interactions to underwater robotics. At very small scales and simple shapes, we are able to completely mathematically describe the motion of inertial particles; however, the motion of large aspherical particles is significantly more complex, and current computational models are inadequate for large or highly-resolved domains. Therefore, we seek to experimentally investigate the coupling between freely suspended particles and ambient turbulence. A better understanding of this coupling will inform not only engineering and physics, but the interactions between small aquatic organisms and their environments. In the following pages, we explore the roles of shape and buoyancy on the motion of passive particles in turbulence, and allow these particles to serve as models for meso-scale aquatic organisms.

We fabricate cylindrical and spheroidal particles and suspend them in homogeneous, isotropic turbulence that is generated via randomly-actuated jet arrays. The particles are fabricated with agarose hydrogel, which is refractive-index-matched to the surrounding fluid (water). Both the fluid and the particle are seeded with passive tracers, allowing us to perform Particle Image Velocimetry (PIV) simultaneously on the particle and fluid phase. To investigate the effects of shape, particles are fabricated at varying aspect ratios; to investigate the effects of buoyancy, particles are fabricated at varying specific gravities. Each particle type is freely suspended at a volume fraction of $\Phi=0.1\%$, for which four-way coupling interactions are negligible. The suspended particles are imaged together with the surrounding fluid and analyzed using stereoscopic PIV, which yields three velocity components in a two-dimensional measurement plane. Using image thresholding, the results are separated into simultaneous fluid-phase and solid-phase velocity fields.

Using these simultaneous measurements, we examine particles' turbulent slip velocity and compare it to particles' quiescent settling velocity, which we measure directly. We observe that the slip velocity is strongly reduced relative to the quiescent case, and explore various mechanisms of particle loitering in turbulence. We further explore the relationship between the instantaneous particle velocity and the instantaneous fluid velocity, and develop a linear parametrization. By comparing our experimental data to a simple one-dimensional flow in the context of this parametrization, we elucidate aspects of slip velocity that are unique to turbulence.

We obtain the particles' angular velocity by applying the solid-body rotation equation to velocity measurements at points inside the particle. We find that the expected value of angular velocity magnitude does not vary significantly with particle aspect ratio, as long as particles are nearly neutrally buoyant. Stronger effects on rotation are found for more negatively-buoyant particles. We also investigate particles' inheritance of vorticity from turbulent velocity fields, and find that particle rotation can be predicted by applying a spatial filter to fluid-phase vorticity.

The results of this study will allow us to more accurately predict the motion of aspherical particles, giving new insights into oceanic carbon cycling, industrial processes, and other important topics. This analysis will also shed light onto biological questions of navigation, reproduction, and predator-prey interaction by quantifying the turbulence-driven behavior of meso-scale aquatic organisms, allowing researchers to sift out passive vs. active effects in a behaving organism. Lastly, processes that are directly dependent on particle dynamics (e.g., sediment transport, industrial processes) will be informed by our results.

*to Karen Rossman Styers, who started it all.*

ii

# Contents

# List of figures

## List of tables

## List of symbols used

| | |
|---|---|
| $\alpha$ | Particle aspect ratio $c/a$ |
| $A$ | Particle cross-sectional area |
| $A^*_{\mathrm{surf}}$ | Particle sphericity (normalized surface area) |
| $A^*_{\mathrm{proj}}$ | Particle normalized projected area |
| $C_D$ | Particle drag coefficient |
| $C_{\mathrm{shape}}$ | Newtonian correction factor for drag calculation |
| $\gamma$ | Particle specific gravity $\rho_p/\rho_f$ |
| $\delta_\nu$ | Viscous sublayer thickness |
| $\varepsilon$ | Turbulent dissipation rate |
| $\vec{\zeta} = \dfrac{1}{2}\vec{\omega}$ | Fluid rotation (one half of fluid vorticity) |
| $\eta$ | Kolmogorov microscale |
| $\theta$ | Slip direction |
| $f, f_\perp, f_\parallel, f_{\mathrm{shape}}$ | Stokes correction factors for drag calculation |
| $\lambda$ | Taylor microscale |
| $\Lambda$ | Particle shape factor $(\alpha^2 - 1)/(\alpha^2 + 1)$ |
| $L_f$ | Integral lengthscale |
| $L$ | Representative particle lengthscale |
| $m_p$ | Particle mass |
| $\mu$ | Fluid dynamic viscosity |
| $\hat{\boldsymbol{n}}$ | Unit vector parallel to particle's axis of symmetry |
| $\nu$ | Fluid kinematic viscosity |
| $q^2$ | Turbulent kinetic energy |
| $R$ | Particle radius (sphere) |
| $d_p$ | Particle diameter (sphere, or sphere of equivalent volume for nonspheres) |
| $Re_p$ | Particle Reynolds number $L \cdot U / \nu$ |
| $Re_\lambda$ | Taylor microscale-based Reynolds number |
| $Re_L$ | Integral lengthscale-based Reynolds number |
| $\rho_f$ | Density of fluid |
| $\rho_p$ | Density of particle |
| $T$ | Average eddy turnover time |
| $\tau_\eta$ | Kolmogorov timescale |
| $\tau_{st}$ | Stokesian or relaxation timescale |
| $\tau_w$ | Wall shear stress |
| $\vec{\boldsymbol{u}}_f$ | Fluid velocity |
| $\vec{\boldsymbol{u}}_p$ | Particle velocity |
| $\vec{\boldsymbol{u}}_s$ | Turbulent slip velocity |
| $u', u_{\mathrm{rms}}$ | Turbulent rms velocity |
| $u_T$ | Representative turbulent velocity scale |
| $v_q$ | Particle quiescent (still-water) settling velocity |
| $v_k$ | Kolmogorov velocity scale |
| $W$ | Stokes settling velocity (Wang and Maxey 1993) |
| $\langle \Delta V_1 \rangle$ | Altered particle settling velocity (Wang and Maxey 1993) |
| $\Phi$ | Particle volume fraction |
| $\vec{\boldsymbol{\omega}} = 2\vec{\boldsymbol{\zeta}}$ | Fluid vorticity |
| $\vec{\boldsymbol{\Omega}}$ | Particle angular velocity |
| $\vec{\boldsymbol{\Omega}}_\parallel$ | Particle spinning rate |
| $\vec{\boldsymbol{\Omega}}_\perp$ | Particle tumbling rate |

# Acknowledgements

It will come as no surprise to anyone that research is hard. Sleepless nights, experiments that don't work, deadlines met and missed, rejection, failure: all part and parcel of serving that ever-fickle mistress, Discovery. The process of earning a PhD adds hues of existential anxiety to this already volatile mix: what am I doing here? why have I chosen this path? what impact might I possibly have? and where might I end up afterwards?

That being said, this rocky road is substantially eased by those met along the way. I have been perhaps more fortunate than most in this regard. In this space, I can name only the tiniest fraction of these individuals: those who have loved me, listened to me, supported me, celebrated with me, mourned with me, laughed with me, and cried with me. I can only hope that the depth of my gratitude somehow seeps out of these pages and makes itself known to the rest of the community I have had the privilege of calling my own these past years.

First and foremost, my thanks go out to the members (past and present) of the Environmental Fluid Mechanics and Hydrology group at UC Berkeley: Andreas Brand, Rusty Holleman, Wayne Wagner, Maureen Downing-Kunz, Ian Tse, Cristina Poindexter, Rachel Pepper, Kevin "Fake Mary" Hsu, Nikola Marjanovic, Jason Simon, Bowen Zhou, Audric Collignon, David Wiersema, Rudi Schuech, Diane Taylor, Jingyi Bao, Olivia Hóang, Eli Goodfriend, Laura Mazzaro, Ruth Anne Lambert, Mike Dvorak, Alex Anderson-Connolly, Laurel Moll, David (and Darcy) Dralle, and Gabrielle Boisramé. You've all been unspeakably wonderful. Special thanks go out to Madeline Foster and Rachel Allen for their unending (and uncritical) support during particularly dire times. Other particular thanks go to Gabrielle Bellani for his mentorship, considerable expertise in particle-laden turbulence, and sense of humor; Colin Meyer for his friendship, positive attitude, depth of knowledge in fluid mechanics and other subjects, as well as his insatiable curiosity; and Ankur Bordoloi, for his help along the final stretch. And though they should also technically be included in this group, simple thanks fall woefully short of the debt I owe to Susan Willis and Megan Williams for their invaluable and much-cherished friendship.

My advisor, Evan Variano, has been an incredible resource for me. Anyone who works with him will quickly become aware of the considerable breadth of his scientific expertise; however, his true strength lies in his dedication to his students' mental and physical well-being, as well as their overall development as scholars. Additional thanks go out to the other members of my committee: Mark Stacey, Mimi Koehl, and Ömer Savaş. I'd also like to thank Tina Chow, whose mentoring I will carry with me when I someday step into a classroom of my own. Other thanks go to Bob Full, Robert Dudley, and all the students affiliated with CiBER (particularly Tom Libby and Evan Chang-Siu) for providing such a rich environment for exploration and collaboration. Lastly, I'd like to recognize all the undergraduates who helped me to produce the research contained in this thesis—most notably Yiheng Tao and Isabel Houghton. I wish you both the best of luck as you step out into your own graduate careers.

To all the current and former residents of the Hillegass-Parker cooperative house: I will treasure the time I spent at HiP, and all the things I learned there (how to sort trash and compost properly, how to cook for sixty people, how to make kick-ass granola, how to deal with other people's messes, how to learn from people whose lives are vastly different from my own, how to love unconditionally). Thanks to Jason Trager for his unapologetic passion and fierce loyalty; to Tim Ruckle for his music and his gentle spirit; to Erin Bergren for Thai food and knit nights; and to Chris Holdgraf for asking life's tough questions, like whether or not you can get cancer from eating a tumor. Special thanks to Katie Adamides for her matter-of-fact humor and inspiring dedication to helping others, and to Moe Turner for all those dinners and all those bottles of wine.

To the Veritas Graduate Christian Fellowship: thanks for a safe space to be my most authentic self. It's entirely because of you that I managed to (barely) hang onto my sanity throughout this wild ride. To all those who've passed through the Wednesday Small Group: thanks for all your prayers. To Jasmin Borja Miller—it has been such a joy to share my struggles and triumphs with you, and I'm looking forward to a lifetime of sisterhood (whether we are in the same town or not).

To the First Presbyterian Church of Berkeley: thanks for providing me with a third home (the first being my apartment, and the second being the lab). Thanks to Mitchell Covington and all the denizens of CAYAC for the beautiful music we made together, and for the loving and fun community you provided. Thanks to Carol Johnston, Courtney Grager, and the rest of FPCB's Stephen Ministry team for showing me how to B-L-E-S-S-E-D. Most of all, thank you to John Yoo for your friendship, mentoring, courage, music, and leadership.

To the Lake Merritt Rowing Club, and my beloved Women's team: if grad school didn't teach me the value of hard work and discipline, you certainly did. I admire each and every one of you for your commitment, tenacity, and exceptionally fine taste in spandex. Strong legs, quick arms, and smooth water to all of you.

And lastly, thanks to all those who don't fit into any of these categories. To Becca Friedman, for showing me so many new things and tolerating the fact that I only cooked spaghetti and tacos for a whole year. To the members of the Frye Lab at UCLA, for providing me with the desk, the jokes, and the coffee that produced most of this dissertation. To Paolo Domenici, John Steffensen, and the folks up at Friday Harbor Labs for letting me pretend to be a biologist for six weeks. To the American taxpayers and the National Science Foundation, who've paid my salary for five (soon to be seven) straight years. To Lex Smits, John Dabiri, and their affiliate labs, who adopted me at numerous conferences. To the organizers and attendees of the Microscale Ocean Biophysics symposium, for providing fertile soil in which to science. To Greg Somers and the State High nerds, for three straight years of quality math instruction (and, years later, quality beer). To Rachelle Hayes, for demonstrating the importance of ending well. To Joel Knopf, Eugene Joseph, Sus Long, Ian Walters, and the other brave souls who launch their music out into the world for our benefit. To the downtown Berkeley farmers' market vendors, staff, and patrons, and everyone who listened to my music and told me it was worth something—whether that something was a crumpled dollar bill, a bunch of dried lavender, a block of cheese, a bag of tiny oranges that tasted like sunshine, or even just a smile. To John Niehues, for teaching me so much about myself. To Michael, for his embodiment of gratitude in the face of adverse circumstances. To Lovejoy, for his unshakeable confidence in my ability to make my own way.

Of course, none of this would have been possible without the support of my family—my parents, Ron and Jeanne Byron; my sisters, Meredith LeRoy and Ronnie Byron; my brother-in-law Matt LeRoy and my beautiful niece and nephew, Adelynn Rae LeRoy and David Matthew LeRoy (as well as all the aunts, uncles, and cousins, both Byron and Watts). For twenty-six years they've tolerated my quirks and nerdiness—and even if they didn't quite understand why I got so excited about fluid mechanics, they nevertheless packed me up and sent me off to the Left Coast with their best wishes and their love. This one's for you, fambly.

Jean-Michel—you are so much more than a husband to me. You are my best friend, colleague, cheerleader, mentor, partner in crime, and fellow adventurer and truth-seeker. Your support and love has been my rock. I am so excited to share my forever with you.

Now, to Him who is able to do immeasurably more than all I could ever ask or imagine: *Soli Deo Gloria.*

# I.    Introduction

Interaction between suspended matter and fluid flow is ubiquitous in the natural and built environment. Sediment, aquatic organisms, and underwater vehicles are all subject to the whims of the complex flows in which they reside. Whether these suspended objects are as small as bacteria or as large as submarines, detailed knowledge of their motion in flow can improve our understanding of biology, physics, and engineering design.

Turbulence—with its wide range of length, time, and velocity scales—interacts in curious ways with suspended matter. Simultaneously chaotic and ordered, turbulence buffets, rotates, and aligns objects within its collection of tangled vortex tubes. For objects that are smaller than the largest turbulent scales, this buffeting can have a significant effect on the object's dynamics (and, in the case of living organisms, it can affect important biological processes like reproduction or feeding).

In this chapter, we will discuss the motivation for and consequences of an increased understanding of particle dynamics in turbulence. We begin by evaluating the importance of this topic in engineering, physics, biology, and robotics, and proceed to a survey of the current literature on these topics. Lastly, we outline our experiments and their general application to turbulent particle-laden flows.

## 1.1    Motivation

The study of suspended objects in flow is alternately called "particle-laden flow" or "multiphase flow", with the two terms serving as roughly interchangeable. Over the past several decades, this topic has evolved into a distinct subfield of fluid mechanics. Early studies focused on very small, spherical particles; these studies benefit from certain mathematical simplifications, such as the ability to neglect particle inertia and finite-size effects. The study of "large" particles (i.e., particles larger than the Kolmogorov scale of the surrounding turbulence) has blossomed in recent years, as has the study of nonspherical particles. This is very encouraging, since few naturally-occurring particles are perfectly spherical. A better understanding of the dynamics of nonspherical particles will inform countless models for particle transport and settling.

Of particular interest to us are particles whose sizes are comparable to the Taylor microscale. These particles are situated within the "inertial subrange" that is characteristic of turbulence, in which energy cascades losslessly from larger to smaller flow structures. At this size scale, particles are suspended in an extremely nonuniform field of vortices which are both larger and smaller than the particles. This heterogeneity in the surrounding flow should lead to strong size-dependence in particles' motion. Additionally, this is an interesting parameter space for both aquatic organisms and autonomous underwater vehicles. The large size range of the ambient flow structures means that both animals and robots at this scale will be subject to buffeting and fluctuating forces of a markedly different character than the forces experienced by very small or very large animals. This may have implications for navigation, locomotion, and control.

In this work, we examine statistical ensembles of Taylor-scale, non-spherical, near-neutrally-buoyant particles which are suspended in homogeneous isotropic turbulence. Particles of this size and density can serve as models for large flocs, small underwater vehicles, or the larger plankton.

### 1.1.1 Engineering and Physics

Though they are often treated as such, very few physically-relevant fluid flows are purely single-phase. From silt-laden estuaries to smoggy city skies, particles alter environmental and industrial flows. When researchers account for the presence of particles, it is usually an approximate solution only: an extra term in a numerical model, or a slight change of empirical constants. However, in many cases, scientists and engineers are interested in the dynamics of individual particles, and not just the bulk flow properties. A large body of work addresses the dispersion and diffusion of particles (B. Geurts, Clercx, and Uijttewaal 2007; Gill 1982; Elimelech, Gregory, and Jia 2013), and our understanding of this topic has led to great success in civil infrastructure development. The study of reactive and combusting flows—typically gaseous flows laden with droplets—has provided insight into the development of new technology in the automotive and energy industry (Kuo and Acharya 2012). These studies of particle dynamics feed back into larger-scale models, improving the accuracy with which we make predictions about particle-laden flows.

Applications in environmental engineering (e.g. sediment transport, waste management), mechanical engineering (e.g. combustion, particulate transport and deposition), or physics (e.g. turbulence modulation, particle clustering) require a mathematical understanding of particles in flow—or, at the very least, an accurate model of their dynamics. Often, particles are modeled as spheres despite the paucity of regular shapes in such processes. This is in part due to the complex and difficult nature of accounting for shape effects—even our most detailed equations of motion for nonspherical particles must make various assumptions about the particles' size, buoyancy, and/or their relationship to the surrounding flow.

Many industrial processes also involve non-spherical particles suspended in a fluid medium, as in papermaking or concrete mixing. In these industries, individual particle behavior is of interest, since particle or fiber alignment is a key determinant of the quality and strength of the manufactured material (Lundell, Söderberg, and Alfredsson 2011). In both industrial and environmental applications, particles are often suspended in complex flows (e.g. turbulence). It is therefore important to quantify the rotation and translation of particles in turbulence, and to measure how these vary with shape, size, and density.

### 1.1.2 Biology

The physics of individual particles suspended in turbulence are equally applicable to small aquatic organisms, many of which are near neutral buoyancy. Through our studies of particle-laden flows, we gain a better understanding of the passive buffeting experienced by organisms in complex flows. This contributes to our overall understanding of locomotion, navigation, reproduction, and predator-prey interactions. Of particular interest are the passive dynamics of meso-scale organisms, whose lengths are within the inertial subrange of ambient turbulence (in the ocean, the inertial subrange is from millimeters to centimeters (Thorpe 2005)). These organisms occupy a curious space in our "map" of turbulence: they are smaller than the largest turbulent eddies, but larger than the smallest eddies. Larger organisms, being larger than all turbulent scales, may safely ignore the subtleties of the surrounding flow. Smaller (sub-Kolmogorov) organisms, experiencing only linear shear, have little experience of the complexity that occurs at larger scales. In the classical view, large organisms are often called "nekton" (Greek: "swimmer") and small are called "plankton" (Greek: "drifter"), respectively (Aleev 1977). This classification implies that large organisms have complete control over their locomotion, whereas small organisms are at the mercy of the ambient flow. This is of course a gross oversimplification, as plankton have been shown to be capable of precise, controlled behaviors such as prey-capture maneuvers, escape responses, and directed swimming (Yamazaki and Squires 1996; Fields and Yen 1997; Kiørboe and Visser 1999). However, there is some utility in this view: the size scales of the ambient flow in relation to an organism's body size are certainly a large factor in the way it experiences its environment.

Organisms which exist at intermediate scales may be classified as nektoplankton (plankton tending towards nekton) or planktonekton (nekton tending towards plankton). In this view, meso-scale animals have only intermittent control over their locomotion because they are navigating through fields of turbulent flow structures which are both larger and smaller than themselves. Many studies (reviewed by (Liao 2007)) have shown that organisms which are slightly larger than the upper bound of the inertial subrange (e.g., many teleost fishes) can sense and exploit ordered, periodic vortex structures, which are usually about an order of magnitude smaller than themselves. However, smaller (meso-scale) organisms experience turbulent velocity fluctuations without the periodicity of these large structures. Exploration of the passive dynamics of these organisms, as modeled by large particles, may help to determine whether these smaller organisms have specialized methods of navigating in turbulence or managing aperiodic velocity fluctuations. This investigation will also help us to address the biological aspects of locomotion and control by sifting and separating passive behavior from active swimming.

### 1.1.3 Robotics

The above discussion of meso-scale aquatic organism navigation is equally applicable to small underwater robots. Autonomous underwater vehicles (AUVs) are used for scientific applications such as topographic surveying, biological sampling, and deep-sea explorations, as well as military applications including reconnaissance and weapons deployment (Griffiths 2002). AUVs are also useful for environmental applications, such as monitoring the spread of contaminants from underwater oil operations (Bingham et al. 2002). Of the many AUVs in common use, the smallest have lengths which are on the order of meters (Table 1.2). This is much larger than the largest size scales in oceanic turbulence, and thus these vehicles are relatively unaffected by turbulent velocity and vorticity structures (though they are still subjected to large-scale currents and eddies). Smaller, meso-scale vehicles, however, would encounter the same turbulent eddy field as a meso-scale organism of the same size.

Micro-autonomous underwater vehicles (μAUVs) have risen to prominence in recent years as a potential solution for many problems in the private and public sectors. Large underwater vehicles are limited in their maneuverability, and their conspicuous size may be a hindrance in military missions. μAUVs, with their potentially smaller turning radii and shorter stopping distances, are capable of navigating within complex terrain, such as wetland canopies and coral reefs. Smaller vehicles could potentially be built in larger numbers, opening up possibilities of sensor networks covering a large area for either scientific sampling or surveillance. However, such vehicles come with their own set of difficulties. In smaller vehicles, there is less space available for power supply, payload, and sensors. These severe space constraints set up propulsive efficiency as paramount.

Knowledge of particle dynamics provides *a priori* knowledge of the types of buffeting and rotation that meso-scale vehicles will experience in turbulence. This will help designers to maximize propulsive efficiency, and may also help with tasks such as station-holding or maintaining a preferred orientation in the midst of turbulence. Increased efficiency of propulsion, including an enhanced ability to deal with turbulent fluctuations, could either extend AUV range or free up payload so that the vehicle could include more or more powerful sensors.

## 1.2   Current Work

Having briefly outlined the importance of particle rotation and translation as applied to physics, biology, and engineering, along with the motivation for studying this topic, we will discuss the current state of the field as it is relevant to the research herein.

### A note on vocabulary

The following terms are used frequently when discussing particle-laden flow; they are defined here for ease of reading.

| Stokes flow/drag | In the Stokes flow regime, when particle Reynolds number is very low, the Navier-Stokes equation may be reduced to the viscous terms only. This condition is also referred to as "creeping flow". In the Stokesian regime, the drag on a particle is proportional to its velocity relative to the surrounding fluid. |
|---|---|
| Stokes number | The Stokes number is a nondimensional ratio: the numerator is the response time of a particle to turbulent fluctuations, and the denominator is a flow timescale (usually the Kolmogorov scale). So, if a particle is quick to respond to changes in flow, it will have a low Stokes number, and its behavior will be much like a fluid parcel. If the particle is slow to respond, it will have a high Stokes number. |
| Newtonian drag | When particle Reynolds number is very high, it is in the Newtonian drag regime; drag is proportional to the velocity squared. |
| One-way coupling | If a particle is very small and/or only weakly inertial, and are only sparsely suspended in the flow, it may be assumed that the particle has a negligible effect on the surrounding fluid. This is known as "one-way coupling." |
| Two-way coupling | The presence of particles in a flow may also alter the characteristics of the surrounding turbulence. This occurs if the particles are large or inertial enough to affect the surrounding flow. |
| Four-way coupling | If particles are inertial and densely suspended, there is four-way coupling: particles affect the fluid motion, fluid affects the particle motion, and particles collide with one another (and any solid boundaries which are present in the flow). |

Table 1.1: Definitions of key terms in particle-turbulence interaction.

### 1.2.1 Particle-laden turbulence

The first attempt to derive the analytic equation for a finite-sized particle in a non-uniform flow was completed by (Tchen 1947), building on the work of previous authors who had derived the equation of motion for an inertial particle in a fluid at rest (Basset 1888; Boussinesq 1903; Oseen 1927). Tchen's attempt, however, was soon criticized by (Corrsin and Lumley 1956) for failing to properly incorporate forces and torques due to the static pressure gradient and the spatial velocity gradient. Further augmentations to the equation were introduced to incorporate viscous shear, the stresses present in the undisturbed fluid, memory effects, and other subtleties (Buevich 1966; Riley 1971; Soo 1975; Gitterman and Steinberg 1980), moving the equation towards a true approximation of a finite-sized sphere in turbulence.

(Maxey and Riley 1983) incorporated the efforts of these authors, as well as the Faxén corrections for finite-size effects (Faxén 1922). It is their equation of motion which has entered the canon, and is often referenced as the Maxey-Riley equation:

$$
\begin{aligned}
m_p \frac{d\boldsymbol{u}_p}{dt} = & (m_p - m_f)\boldsymbol{g} + m_f \frac{D\boldsymbol{u}_f}{Dt} - \frac{1}{2} m_f \frac{d}{dt}\Big(\boldsymbol{u}_p - \boldsymbol{u}_f - \frac{1}{10}R^2\nabla^2\boldsymbol{u}_f\Big) \\
& - 6\pi R\mu\big(\boldsymbol{u}_p - \boldsymbol{u}_f\big) - \frac{1}{6}R^2\nabla^2\boldsymbol{u}_f \\
& - 6\pi R^2\mu \int_0^t \left( \frac{\frac{d}{d\tau}\big(\boldsymbol{u}_p(\tau) - \boldsymbol{u}_f(\tau) - \frac{1}{6}R^2\nabla^2\boldsymbol{u}_f\big)}{\big(\pi\nu(t-\tau)\big)^{\frac{1}{2}}} \right)
\end{aligned}
\tag{1.1}
$$

where $R$ equals the sphere's radius, $m_p$ is the particle's mass, $m_f$ is the mass of an equivalent volume of fluid, $\boldsymbol{u}_p$ equals the particle velocity and $\boldsymbol{u}_f = \boldsymbol{u}_f(t)$ is the fluid velocity, which in the equation above is always evaluated at the position of the particle center. This equation is discussed in detail in Chapter 3.

The Maxey-Riley equation accounts for the forces of buoyancy, pressure gradients, acceleration reaction, and drag, including both steady drag (using Stokes' drag law) and unsteady drag, which encompasses the effects of the particle's history. This last term is also called the "Basset history term," and it incorporates the particles' past interactions with the surrounding flow. However, the motion of particles outside the creeping flow regime is not well-described, even for spheres. For nonspherical particles at larger scales, there is as yet no theoretical or analytical approach (Mandø and Rosendahl 2010)—though we note that some empirical work has attempted to reconcile both Stokesian and Newtonian analytical drag models for nonspherical particles at transitional Re (discussed in detail in Appendix A). In the following sections, we survey those analytical approaches that have proven useful for non-spherical small particles, along with computational and experimental methods that provide some insight on the motion of non-spherical large particles.

**Approaches: Analytic, numerical, and experimental**

Our understanding of both spherical and nonspherical particles in turbulence has grown vastly in recent years (Toschi and Bodenschatz 2009; Balachandar and Eaton 2010). However, due to the mathematical considerations discussed above, research has largely focused on dynamics of very small particles. These small particles have sizes which are below the Kolmogorov scale $\eta$, and usually have Reynolds numbers Re which are much less than unity. In these types of problems, the fluid forcing can be well described by the linear, viscous terms of the Navier-Stokes equation. Analytical solutions for the motion of simple shapes are well-known in this low-Re regime (Stokes 1851; Jeffery 1922; Happel and Brenner 1983). A common approach is to perform a numerical simulation in which particles are treated as points, whose center-of-mass is advected with the flow and whose rotation is dictated by a variant of Jeffery's equations (Gustavsson, Einarsson, and Mehlig 2014; Zhao, Marchioli, and Andersson 2014; Ni et al. 2015).

For particles at intermediate Reynolds number, analytic solutions fail and researchers must turn to techniques in Computational Fluid Mechanics (CFD). These methods resolve the fluid-particle interaction at many locations along the particle surface, rather than computing the resultant force and torque on a particle from an analytical expression. In many cases, these methods also include the forcing which the particle applies to the surrounding fluid; these forces are not typically included in models which rely on Equation (1.1) or on Jeffery's equations (Jeffery 1922). For freely-suspended large particles (i.e., particles which are many times larger than the Kolmogorov scale, and subsequently several times larger than the simulation grid scale) in turbulence, the interfacial dynamics become very complex. Lagrangian CFD methods such as smoothed-particle hydrodynamics (SPH) and Lattice-Boltzmann methods (LBM) are often more efficient in this case; since they model the fluid as discrete elements rather than a continuum evaluated at points, the incorporation of local fluid-particle interactions is somewhat more elegant (Monaghan 2012; Aidun and Clausen 2010). However, Eulerian-Lagrangian techniques, in which turbulence is computed on an Eulerian grid and a particle is superimposed upon it, are the most widely-used. This approach has led to major insights on particle motion (Loth 2000; Subramaniam 2013).

Numerical simulations have given us a great deal of insight on shape-dependent variation in rotation, settling, and particle clustering (Andersson and Soldati 2013; Eaton and Fessler 1994; Gustavsson, Einarsson, and Mehlig 2014). The latest computational work is able to accurately simulate large, freely suspended particles in turbulence (Marchioli and Soldati 2013), some including two-way coupling effects (Lucci, Ferrante, and Elghobashi 2010; Zhao and Wachem 2013). However, the limits on available computing power restrict these experiments to computational domains that are still too small to simulate many interesting problems.

In addition to the analytical and numerical work referenced above, many researchers have pursued a greater understanding of particle motion through experiment. The most obvious limitation of the available analytical expressions is the failure to treat with large particles. Therefore, many researchers have kept their particles spherical, but increased the particles' size and/or inertia beyond the bounds of the Maxey-Riley equation. (Xu and Bodenschatz 2008) examined the acceleration statistics of particles larger than the Kolmogorov scale. They found that neutrally-buoyant large particles display different behavior than small heavy particles, even at the same Stokes number, and do not cluster in the same regions of turbulence. This work was continued by (Klein et al. 2013), who simultaneously measured the rotation and translation of large, near-neutrally-buoyant particles. (Zimmermann et al. 2011) also studied neutrally-buoyant large particles, exploring the effects of turbulent intermittency and showing the effects of lift forces on the particles. Several studies also examine the effects of finite size on the turbulent transport of particles, including the acceleration statistics of dispersed particles (Qureshi et al. 2008) and the relationship between particle size and turbulent intermittency (Qureshi et al. 2007).

Large nonspherical particles have also been studied in some detail. For example, a number of studies on flexible and rigid fibers in channel flow have explored flocculation/aggregation (Lundell, Söderberg, and Alfredsson 2011); rotation, as compared to spherical particles (Marchioli and Soldati 2013); and slip velocity (Zhao, Marchioli, and Andersson 2014). Other studies have focused on the orientation of nonspherical particles in turbulence, including studies of glass fibers in high-Reynolds number channel flow (Bernstein and Shapiro 1994) and rods in grid-generated turbulence (Parsa et al. 2012). Still more studies have examined at the modulation of turbulence due to the presence of nonspherical particles (Bellani et al. 2012).

There are many topics that are of interest when it comes to the motion of nonspherical particles in turbulence. We have briefly touched upon some specific studies to provide an impression of the very broad scope of this field. The subtopics that are most relevant to our work are particle settling and particle rotation; below, we give an in-depth overview of the analytical, numerical, and experimental work on these subtopics.

**Particle settling**

One of the most prominent justifications for the study of particles in turbulence is its applicability to particle settling. This topic has great application in environmental engineering, as it is crucial in the study of sediment transport. Another instance of particle settling in turbulence is the case of falling marine snow (i.e., aggregates of dead phytoplankton and other organic matter). Marine snow provides a significant source of carbon sequestration, since a nontrivial percentage of these particles settles all the way to the ocean floor. The settling velocity of marine snow is a key parameter in many models of the "biological pump", but it is only known approximately (De La Rocha and Passow 2007). Further applications of particle settling arise in industry, where settling particles play major roles in mining (Turian et al. 1992) and food production, such as winemaking (Robinson 2006), along with many other processes.

It is known that turbulence alters the settling velocity of particles (Murray 1970); however, the mechanisms by which this is accomplished are not well-defined, and the topic remains an area of much research. Most research to date has focused on varying the size and/or density of particles and measuring the subsequent changes in settling velocity. Few studies have explored the effects of particle shape on its settling velocity in turbulence, though some have attempted to quantify shape effects on quiescent settling (Christiansen and Barker 1965; Dietrich 1982). This is surprising, because the particles whose settling behavior is of applied importance are rarely spherical; naturally-occurring and biological particles are highly non-spherical, with the most common shapes being porous aggregates and long fibers (Mandø and Rosendahl 2010).

In Murray's experiments in 1970, particles at a Reynolds number (with $Re_p$ ranging from approximately $20 - 80$) showed a settling velocity in turbulence that was reduced by up to 30% relative to the same particles settling in still water (Murray 1970). He attributed this decrease, as did his predecessors who studied simple oscillating flow, to nonlinearity in the drag force (Field 1968; Brush, Ho, and Singamsetti 1962). (Tooby, Wick, and Isaacs 1977) elucidated another mechanism via elegant experiments involving stroboscopic images of heavy spheres in rotating flow, demonstrating that such particles tended to form long-lived circular orbits in regions of the flow which opposed their motion ("vortex-trapping"). Parallel investigations in the physics and applied mathematics community (as opposed to the environmental engineering and physical oceanography community) were also addressing this question, as applied to aerosol particles in the atmosphere (Yudine 1959; Csanady 1963; Meek and Jones 1973). (Maxey 1987) performed computational simulations which determined that for inertial particles in a Gaussian random flow, settling velocity was actually enhanced relative to the quiescent settling velocity. This was in contrast to the reduced-settling observed by Murray and others. Maxey's results were due to a trajectory-biasing effect that is known as "fast-tracking" (Maxey and Corrsin 1986; Maxey 1990). "Fast-tracking" was also observed in the biological community, when (Ruiz, Macías, and Peters 2004) observed the enhanced settling velocities of phytoplankton cells.

The discrepancy was reconciled in part by (Wang and Maxey 1993), who noted that for larger particles, nonlinear drag effects tempered the settling velocity enhancement. In other words, the settling velocity was still enhanced via fast-tracking, even in large particles, but the effect lessened when the simulation included nonlinear drag. An overall reduction in settling velocity was observed only at the uppermost bound of their parameter space. Wang and Maxey's parameter space was defined by $W/v_k$, the ratio of the Stokes settling velocity $W$ to the Kolmogorov velocity scale $v_k$. They then measured the particle settling velocity $\langle \Delta V_1 \rangle$, and normalized it with the turbulent rms velocity $u'$. Overall settling-velocity *reduction* occurred in computational particles with a Stokes number of 2.74, for which $W/v_k = 4$. The reduction was only observed when nonlinear drag was included in the model. This perhaps explains the substantial reduction observed by Murray, whose experimental parameters (when converted to Wang and Maxey's notation) denote much larger, heavier particles; Murray's particle Stokes numbers range from 2 to 7, and the ratio of settling velocity to turbulent rms velocity ranges from 2 to 11[1].

---

[1] It is difficult to compare these two cases, since one is a simulation that takes only dimensionless parameters (rather than physical quantities) and the other is an experiment in which the turbulence is not well-characterized. By using the data given in (Murray 1968), along with the empirical scaling relation $\varepsilon \approx v'^3/2L$ (where $v'$ is the turbulent fluctuating velocity and $L$ is the integral lengthscale), we are able to obtain approximations of the Kolmogorov timescale and velocity scale. This allows a relatively direct (if rough) comparison of the two studies.

The same mechanisms that reduce or enhance the settling of heavy particles also affect the rise rates of positively buoyant particles, including bubbles and droplets (though these two example cases are sometimes complicated by deformability and/or internal fluid motion). This question is particularly relevant for the oil and gas industries, which often encounter two-phase flows; for example, in the event of an oil spill, it is important to know how turbulence may impact the rise rate and dispersion of leaked oil droplets. (Friedman and Katz 2002) found that the turbulence-altered rise rate of positively-buoyant droplets is highly dependent on the ratio of the droplet's unaltered, original rise rate to the turbulent rms velocity. This is the same parameter used by (Wang and Maxey 1993) and (Murray 1970), except that they worked with negatively buoyant solid particles and therefore studied altered settling instead of altered rising. Friedman and Katz find that when the original rise rate is much larger than $u'$, rise rate is unaffected. When the original rise rate is small compared to $u'$, the altered rise rate is approximately equal to 25% of the turbulent rms velocity, regardless of the droplet's quiescent-flow rise rate or Stokes number—that is, the droplet's original rise rate does not matter, and the altered rise rate is determined wholly by the ambient flow. However, when the droplet's original rise rate is of the same order as $u'$, the altered rise rate is very sensitive to the Stokes number, with large-St particles experiencing a reduction in rise rate. (Poorte and Biesheuvel 2002) studied the rise rates of moderate-Reynolds number (Re~102) bubbles in turbulence, and found that their rise rate was reduced by up to 35%. This is in contrast to previous studies, nearly all of which (save (Murray 1970)) had found that settling or rising velocities were enhanced in turbulence. It is in this regime, where the particle settling velocity is comparable to the turbulent rms velocity, where we choose to conduct our experiments. In our experiments, we seek not only to match the particles' settling velocity to $u'$, but we also seek to match the particles' size to an intermediate eddy size.

Above, we have outlined three major mechanisms for settling-velocity alteration in turbulence: nonlinear drag (large particles), vortex-trapping (large particles), and fast-tracking (small particles). However, we do not have a firm grasp on when each of these mechanisms come into play, nor how they might depend on particle size, density, or relationship to the ambient turbulence. There may also be other mechanisms which have not yet been described.

The determination of particle settling velocity reduction or enhancement is an area of active research, and there is as yet no universal principle governing how settling velocity might be altered. Given the current state of the field, it appears that the three most important parameters are: the ratio of particles' quiescent settling velocity to $u'$; particle Stokes number; and particle Reynolds number. Little is known about the effects of particle shape on settling velocity alteration. For a collection of experiments on this topic, the reader is directed to the work of (Nielsen 2007) and (Good et al. 2014).

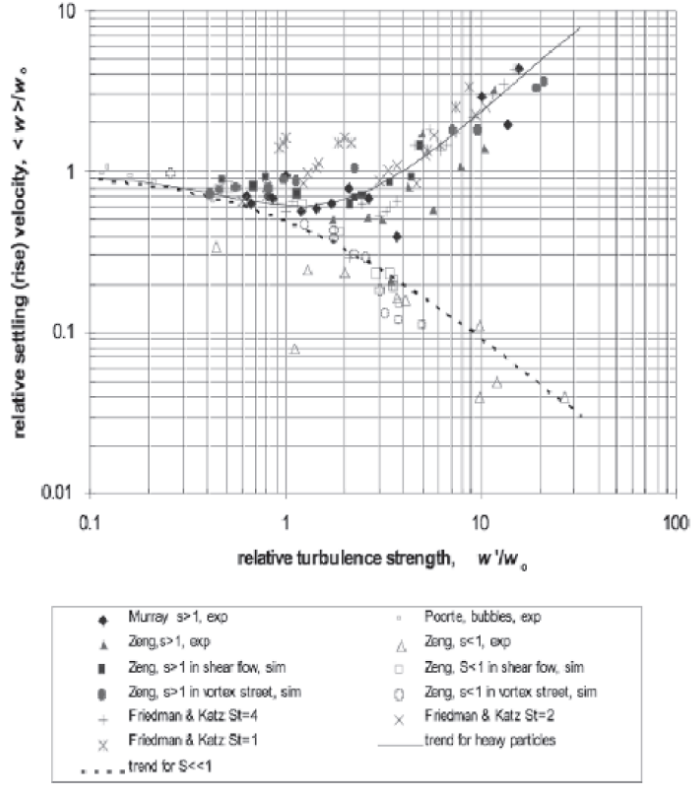

**Figure 1.1: A collection of data from experiments and simulations, reprinted from (Nielsen 2007), showing the regimes where the settling velocity is either reduced or enhanced.**

## Particle slip velocity in turbulence

The traditional definition of the settling velocity—that is, the time it takes for a particle to fall through a large distance divided by that distance—is not always practical or relevant in turbulence. This method requires the observer to track a particle over a long period of time, since the particle's vertical velocity may fluctuate drastically. Given a large enough ensemble of particles, one might assemble an ensemble-averaged vertical velocity, which we denote as the "group settling velocity". However, it is not the absolute particle velocity that determines the fluid drag on that particle. Rather, it is the relative velocity between the particle and the fluid—the "slip velocity." It is this parameter, defined as the particle velocity minus the fluid velocity, which appears several times in Equation (1.1), the Maxey-Riley equation. The slip velocity, not the absolute particle velocity, determines the fluid drag. The slip velocity is also the parameter that is most often used to define a particle Reynolds number, which is a crucial starting point for any study of particle dynamics. [2]

---

[2] The slip velocity should not be confused with the "slip length", a factor that is used to correct for non-ideal boundary layers in which the no-slip assumption is not entirely true (e.g., at very low pressure and/or in rarified atmosphere, when the continuum approximation still holds but very few molecules are present). Nor does the term "slip velocity" imply that the no-slip boundary condition does not hold at the particle surface—it does. The slip velocity refers to the relative velocity between a particle and a characteristic average velocity that represents the average flow in the immediate vicinity of the particle.

Slip is also a key parameter in studies of turbulence modulation due to the presence of particles. (Cisse, Homann, and Bec 2013) studied flow modification due to the slipping between large spherical particles and the surrounding homogeneous turbulence, including the viscous boundary layer around the particle. (Zhao, Marchioli, and Andersson 2014) also studied slip velocity in the context of turbulence modulation, investigating the effects of aspect ratio on the slip of long fibers in turbulent channel flow. A very detailed simulation was performed by (Zhao and Wachem 2013), in which experimenters used a full four-way coupling scheme (fluid affects particles, particles affect fluid, and particles can collide with one another). They simulated a turbulent channel flow that is sparsely laden with spherical and ellipsoidal particles, finding not only turbulence modulation but an anisotropic slip velocity (higher in the streamwise direction).

The slip velocity of a particle is defined as follows:

$$\vec{\boldsymbol{u}}_s = \vec{\boldsymbol{u}}_p - \vec{\boldsymbol{u}}_f \qquad (1.2)$$

where $\vec{\boldsymbol{u}}_p$ is the instantaneous particle center-of-mass velocity, and $\vec{\boldsymbol{u}}_f$ is a velocity which is representative of the local instantaneous fluid velocity field. Obtaining the particle velocity is relatively straightforward in both laboratory and numerical experiments. However, there are several choices for the fluid velocity $\vec{\boldsymbol{u}}_f$. The most intuitive is to use the fluid velocity that *would* exist at the location of the particle if the particle itself were not present—that is, the fluid velocity "seen" by the particle. In a numerical experiment, one may simply delete the particle, either by re-running the simulation with identical initial and boundary conditions, but no particles (Bagchi and Balachandar 2003), or—if the simulation is set up such that the particles have no impact on the fluid, but are merely advected in a one-way coupled fashion—one may use the predetermined fluid velocity at the particle center (Calzavarini et al. 2012). Neither approach is feasible for a laboratory experiment. If the particles are smaller than the scale at which turbulent velocity gradients become linear (the Kolmogorov scale), it is possible to interpolate a velocity to the particle center (though interpolation becomes complex for nonspherical particles).

In our experiments, we calculate an average fluid velocity $\vec{\boldsymbol{u}}_f$ in the neighborhood of the particle. Further details on how this neighborhood is bounded are given in Chapter 3.

**Particle rotation**

At small scales, when particle size is less than the Kolmogorov lengthscale, particles experience only local velocity gradients. If these small particles are neutrally-buoyant (i.e. without inertia), their center-of-mass is simply advected with the flow as a passive tracer, and their rotation can be calculated using Jeffery's equations (Jeffery 1922). This case has been studied in detail by a number of researchers, as discussed previously. Simulations by (Parsa et al. 2012) showed that tracer rods and disks rotate differently, finding that the tumbling rate of disks was substantially higher than that of rods. This was supported by the analytical work of (Gustavsson, Einarsson, and Mehlig 2014), who found the same pattern in simulated tracer rods and disks, and further validated by simulation and experiment in (Byron et al. 2015). Later work by (Parsa and Voth 2014) extended our knowledge well into the inertial range, studying the rotation rate of long rods in turbulence. Of particular interest to our work is Parsa and Voth's exploration of the effects of rod alignment with turbulent flow structures in the inertial range. When they compared their experiments to an analytic model predicting tumbling rate in randomly-oriented rods (i.e., not aligned with flow structures), they found substantial differences between experiment and model for short rods (less than $30\eta$). However, the effects of alignment were much weaker for particles above this lengthscale. In our experiments with large particles, the smallest dimension is $>40\eta$, and the aspect ratios of our particles are much smaller than Parsa and Voth's long rods (our longest rods are $60\eta$ with an aspect ratio of 4; in this study, the longest rods were $72.9\eta$ with an aspect ratio of 76). This is discussed further in Chapter 4.

## 1.2.2  Biological oceanography and animal locomotion

Small aquatic organisms inhabit a complex environment, and their abilities to feed and reproduce are strongly influenced by the flows in which they reside. For example, varying levels of turbulence in the water column may increase or decrease prey capture rates (MacKenzie et al. 1994; Rothschild and Osborn 1988), disrupt feeding currents (Sutherland et al. 2014), inhibit the ability to sense predators (Gilbert and Buskey 2005), affect habitat selection (Reidenbach, Koseff, and Koehl 2009; Sutherland et al. 2014), or obstruct chemical/odor cues (Weissburg et al. 2012; Visser 2001). Turbulence can also impact the development of plankton thin layers and large-scale distribution (Durham and Stocker 2012; Prairie et al. 2012). Alignment and rotation of individual organisms in turbulence may also influence boundary layer thickness and therefore nutrient uptake, influencing the chemistry and biological cycles of the environment at local and global scales (Karp-Boss, Boss, and Jumars 1996; Pahlow, Riebesell, and Wolf-Gladrow 1997; Nguyen et al. 2011).

Even at larger size scales, many organisms have developed physical adaptations (i.e. changes to the shapes of body shape and/or appendages) in order to better control and stabilize their position in the flow (Webb 1984b; Webb and Cotel 2010). Excessive buffeting and rotation can disorient fish and other organisms, potentially subjecting them to increased predation (Cada 1997). Fish swimming through eddies of various orientation demonstrated a loss of postural control, the magnitude of which was dependent on eddy orientation (Tritico and Cotel 2010).

For intermediately-sized organisms, such as copepods or some larvae, the ambient distribution of linear and angular velocities affects feeding, recruitment, and predator-avoidance (Peng and Dabiri 2009; Grünbaum and Strathmann 2003; Yamazaki and Squires 1996). Many organisms have developed adaptive behaviors to navigate and follow sensory cues in their noisy, fluctuating, turbulent environments (Grünbaum 1998; Reidenbach, George, and Koehl 2008). In addition to all of the noise in their environment, animals must also be able to adapt to rapid and often-unpredictable changes in their own orientation with respect to that environment. The orientation of an organism has been shown to be important for gravitaxis, settlement, and directed swimming; turbulent shear and mixing interferes with these processes (Machemer and Braucker 1991; Roberts and Deacon 2002).

We are curious about the effects of body shape on organism-environment interaction; our experiments focus on how model organisms (i.e., passive particles) of various aspect ratios and densities rotate and translate in turbulence. Shape has been shown to be an important factor for swimming efficiency, nutrient uptake, gravitaxis, rheotaxis, and many other behaviors, across a wide range of organism sizes and taxa (Marcos et al. 2012; Roberts and Deacon 2002; Pahlow, Riebesell, and Wolf-Gladrow 1997; Peng and Alben 2012). Aquatic organisms tend to be very close to neutrally-buoyant, with most animals' densities falling within 2-3% of the density of water (Aleev 1977). However, very small changes in density may account for strong differences in the passive response of an organism to the ambient turbulence. We will explore this parameter space by slightly changing the density of our experimental models, within the same range of specific gravities which occur in aquatic organisms.

**Biological navigation and body shape**

Animals' styles of navigation are frequently correlated with their body shape and size (Domenici 2001; Tytell et al. 2010). For example, aquatic organisms navigating at low Re have substantially different body plans than high-Re swimmers, which tend to be streamlined. At very high Reynolds numbers, streamlined body plans converge to a fineness ratio (body length divided by body diameter) of 4.5—the same ratio which was independently discovered to be optimal for aircraft fuselage (Ohlberger, Staaks, and Holker 2006; von Mises 1959). Further specialization occurs based on the animal's preferred prey and predation method, as is illustrated by the "Webb Triangle" (Webb 1984a). In this schematic illustration, "cruising" fish such as tuna have streamlined, relatively stiff bodies, whereas "acceleration specialists" such as pike have long, slender, flexible bodies for quick-start strikes. A third category brings in "maneuverers" such as the butterflyfish, which have disc-like bodies with well-distributed fins that are capable of very precise applications of thrust. This diversity illustrates that hydrodynamic efficiency is not always paramount, and that morphology may change depending on other aspects of the animals' lifestyle.

Even taking into account the morphological variations mentioned above, high-Re swimmers (e.g., fishes and cetaceans) have generally similar-looking, streamlined body plans. However, diversity reigns supreme at low Reynolds number. There is no single representative body plan for a low-Re swimmer; even relatively non-locomotory organisms, such as diatoms, occur as a variety of disks, stars, chains, and other fanciful shapes. For these animals, hydrodynamic issues such as drag reduction are not necessarily of great concern. Hydrodynamic efficiency may be secondary to other evolutionary pressures, which produce the diversity of shapes we see in the smallest of the plankton. In very small swimmers, the laminarity of the surrounding flow dictates that the predominant component of drag will be proportional to the total surface area of the organism. This is in contrast to high-Re swimmers, for whom the flow-normal area is a large contributor to the drag force. However, some hydrodynamic adaptations are evident even in the smallest swimmers— for example, a bacterium's shape influences its sinking speed, nutrient-absorbing capacity, and its ability to form biofilms (Dusenbery 2009; Dusenbery 1998). Copepods use their long, slender appendages to circumvent the surrounding viscous boundary layer and find prey (Kjellerup and Kiørboe 2012; Andrews 1983).

At transitional Reynolds numbers and mid-range lengthscales, where our experimental particles lie, it is far from clear whether there is an "optimal" body shape. Organisms at these intermediate scales occupy a unique position in the turbulent lengthscale spectrum, falling in between the viscous and energy-containing limits. Therefore, the hydrodynamic forces on these organisms will be extremely variable, alternately placing them into the high-Re and low-Re regimes. Previous work has shown that in very small particles, shape changes do affect particle kinematics, with the strongest relative effects occurring at aspect ratios near unity (Byron et al. 2015). We investigate larger particles, of transitional Reynolds number, and the effects of shape change on particle rotation and translation.

A group of organisms that is directly comparable to our experimental particles is the cydippid ctenophores, particularly *Pleurobrachia* and *Hormiphora* (Hutchins and Olendorf 2004). These animals' body plans are roughly spheroidal, with the exception of a pair of long thin tentacles and eight ciliated ridges that longitudinally circumscribe the body. *Pleurobrachia* have a body aspect ratio of approximately unity, whereas *Hormiphora* have a body aspect ratio of between one and two (Figure 1.2). These animals locomote via a combination of passive drifting and active swimming, and are approximately the same size scale as the particles with which we conduct our experiments (≈1 cm). Our study will shed light on the passive behavior of these and other transitional-Reynolds-number organisms.

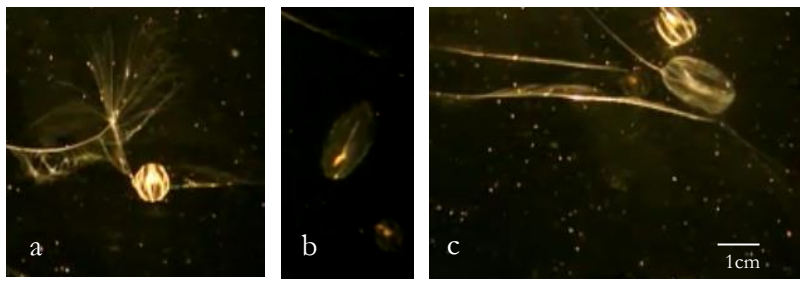

**Figure 1.2: (a)** *Pleurobrachia bachei*, **with an aspect ratio of approximately unity; (b)** *Hormiphora californiensis*, **with an aspect ratio of approximately two; (c) the two species side-by-side. All images captured at the Monterey Bay Aquarium in March 2014.**

### Ocean particulates and the global distribution of planktonic organisms

Further biological questions arise when considering the bulk transport of particulates in aquatic environments. The settling of "marine snow" particles, which are on the order of one centimeter in size (Alldredge and Gotschalk 1989), is a crucial link in the global carbon cycle (Shanks and Trent 1980; Alldredge and Silver 1988). Marine snow settling rates are dependent on both size and shape (Asper 1987; Logan and Wilkinson 1990). However, the effects of turbulence on particle settling are still far from quantified, both in the biological literature (Kiørboe 1997; Piazza 2014) and in the physical sciences and engineering (see section 1.2.1).

The ocean can be influenced both physically and chemically by the presence of small objects and organisms. Particle transport and aggregation are important in studies of ecology, oceanography, and climate (Katija and Dabiri 2009; Alldredge, Passow, and Logan 1993; Thornton 2002).

Additionally, the global transport and distribution of planktonic organisms plays an important role in the aquatic food web. These plankton prey upon the primary producers who play a critical role in carbon sequestration (Largier et al. 2006; Hirche et al. 1991). The degree to which individual-organism dynamics impact global populations is an open question; however, studies have suggested that behaviors such as turbulence avoidance can play a large role in global distribution (Pringle 2007).

### 1.2.3 Underwater robotics

Autonomous underwater vehicles (AUVs) have rapidly developed over the past 50 years, but major constraints still apply to vehicle design. As vehicles become smaller, limited space for an onboard power supply requires vehicles to have effective maneuvering and efficient propulsion. AUV navigation has also advanced over the past several years. Since GPS signals rapidly attenuate in water, a typical AUV will navigate with one of three systems: A) dead-reckoning/inertial, B) acoustic/beacon-based, and C) geophysical (Paull et al. 2014). Approach B requires prior placement of underwater beacons or a mothership to help the vehicle triangulate its position; approach C) requires a detailed map of the terrain so that the vehicle can check its position against known landscape features. However, these approaches are typically unfeasible for exploratory missions, where prior knowledge of the terrain is limited. This leaves approach A, which is undesirable because errors in position increase over time. Though advances in inertial guidance systems have provided accurate navigation to within 0.01% of the plotted destination (Leonard et al. 1998), this still may not provide enough accuracy for missions of long duration or those requiring more precision. Dead-reckoning may also fail when vehicles are exposed to extreme turbulent velocity fluctuations that knock them off course (since errors are integrated over time).

Additional challenges arise for smaller AUVs as vehicles begin to deal with ambient turbulence, including velocity fluctuations and intermittent gusts (Watson, Crutchley, and Green 2011; Watson and Green 2010). A fine-level control system is required to respond to these challenges. Such a control system would also further enhance AUV maneuverability at small scales for obstacle avoidance or rapid directional change. Engineers have begun to consider how animals address these challenges, reflected in the biologically-inspired design of robots that move like fish (Barrett, Grosenbaugh, and Triantafyllou 1996), jellyfish (Villanueva, Smith, and Priya 2011), and other aquatic animals ((Roper et al. 2011; Seo, Chung, and Slotine 2010; Yu et al. 2012; Moored et al. 2011)).

A significant portion of today's cutting-edge research on ocean chemistry, biology, and physics comes from data collected by Remotely Operated Vehicles, or ROVs (Hudson, Jones, and Wigham 2005). In contrast to AUVs, ROVs are tethered (allowing more space for on-board sensors, since power is typically transmitted through the tether) and require continuous human input (giving them more flexibility in operations). However, ROVs come with high operating costs due to the need for a mothership, and may carry safety risks for human operators (Yuh, Marani, and Blidberg 2011).

**Underwater vehicle design**

The largest design constraint for AUVs has been and remains the issue of power supply. We frame our discussion of AUV design acknowledging that the ultimate goals of improved navigation, locomotion, and other capabilities are directly served by increasing efficiency and decreasing power consumption. Increased efficiency has the potential to prolong mission duration and/or increase available space for sensors and other payload.

For a non-tethered vehicle, the power supply must either be carried on board or somehow harvested from the environment. Battery technology has advanced significantly since the early days of AUV development, but the longest mission durations for battery-powered AUVs are on the order of days (Parry 2013). Some AUVs use fuel cells, increasing mission range (Yamamoto et al. 2004). A specific class of vehicle called a glider—so called because of its navigation path of repeated rise, then free-falling glide, thus harnessing gravity for propulsion—has had great success in converting buoyant forces to propulsion (Eriksen et al. 2001; Davis, Eriksen, and Jones 2002). Earlier models harnessed the temperature difference between the surface water and the deep ocean (relatively large, at 10-25 degrees), converting the steep thermal gradient into power supply (Jenkins et al. 2003). These innovations are able to extend mission duration from hours to weeks or months (Hines 2005).

Most AUVs in common usage today are very large with respect to turbulent velocity fluctuations in the ocean (often on the order of 10 meters in length). These vehicles are used most often for seafloor mapping, geologic surveys and general ocean research. However, other missions (e.g. mine countermeasures, hull inspection, et al) may require more agility and maneuverability. A great diversity of small (≤1m) AUVs has arisen to meet these challenges (Table 1.2). However, no commonly-used vehicle yet exists on the size-scale of centimeters, though some prototypes have been developed (Watson, Crutchley, and Green 2012; Kodati et al. 2008; Walker 2006; Hobson et al. 2001).

Many aquatic organisms in the 0.5 – 15cm size range (which we will here denote as the meso-scale) have served as inspiration for underwater vehicles (Roper et al. 2011), though the vehicles themselves are often much larger than the inspiration organism. However, as discussed previously, most AUVs are still larger than this scale, constraining mission possibilities. Smaller vehicles could negotiate complex environments such as wetland canopies or reef crevices, and could also potentially take measurements of critical parameters such as ocean acidity, dissolved gases, and nutrient levels in these ecologically sensitive and often-threatened habitats. For such a vehicle to be feasible, it would need very finely tuned position and orientation control, with the ability to dexterously maneuver through complex terrain and variable flows, avoid unexpected obstacles, and quickly change direction.

Since oceanographers can predict only large-scale currents and some meteorological changes, designers do not have an a priori knowledge of the exact flow environment that will be encountered by an AUV. This is especially important for small AUVs, which are more susceptible to turbulent fluctuations. In high levels of turbulence, AUVs could potentially lose their heading or sensor calibrations. Several researchers have proposed course-correcting control algorithms for path-planning in heavy turbulence (Garau, Alvarez, and Oliver 2006; Yang and Zhang 2009; Yoerger and Slotine 1985). However, few researchers have yet investigated the concept of hydrodynamic stability through mechanical design. Mainstream AUV designers' focus on shape has thus far been limited to their attempts to reduce drag via streamlining. Often, vehicle shape is not a priority at all—and if it is, the focus is on efficiency, and not stability. However, the natural world provides countless examples of body plans which are adapted for stability (and not necessarily streamlined). For example, the boxy shape of ostraciiform fish has a unique ability to maintain its trajectory despite flow disturbances (Weihs 2002; Kodati et al. 2008). We hypothesize that the shape of a small underwater vehicle may be a significant factor in its ability to reject perturbation due to turbulence.

| **Name** | **Institution** | **Max length** | **Weight** | **Max speed** | **Uses/Notes** |
|---|---|---|---|---|---|
| Nessie IV | Heriot-Watt University | 0.70 m | 41 kg | 2.6 m/s | Hull inspection, general ocean research. Hover-capable, battery-powered. |
| Robopike (Wanda) | MIT | 0.82 m | 3 kg | 0.09 m/s | Science demonstrator. Bioinspired (chain-pickerel). Further development by iRobot. |
| Aqua2 | Independent Robotics | 0.64 m | 16.5 kg | 0.51 m/s | General ocean research. Bioinspired (insect). Maneuverable, portable. |
| Aqua Penguin | Festo Corp. | 0.77 m | 9.6kg | 1.39 m/s | Highly maneuverable; Bio-inspired (penguin) |
| NARO – Tartaruga | ETH Zurich | 1.00 m | 75 kg | 2 m/s | Bio-inspired (turtle) |
| HAUV | Bluefin Robotics | 0.98m | 82 kg | 0.26 m/s | Hull-inspection, hovering |
| Sea Scout | QinetiQ | 0.91 m | not available | 7.7 m/s | Hull-inspection, mine countermeasures |
| Ranger | iRobot | 0.86 m | 9.07 kg | 7.7 m/s | Mine countermeasures and defense, ocean research |
| HSAUV | Virginia Tech | 1.00 m | 3.6 kg | 7.7 m/s | General ocean research |
| 475 AUV | Virginia Tech | 0.86 m | 8.3 kg | 1.54 m/s | Environmental monitoring, sensor networks |
| Yellowfin | Georgia Tech | 0.89m | 7.71kg | 1.02 m/s | Environmental monitoring, sensor networks |

**Table 1.2: Technical specifications of a selection of commonly-used or commercially-available AUVs whose maximum lengthscale is ≤ 1m. Data were taken from the Autonomous Undersea Vehicle Applications center at www.auvac.org and, in some cases, the home websites of each robot. This list does not include prototype robots that are in development, university-based robots that are primarily intended for concept demonstrations rather than actual deployment, or biomimetic robots whose primary purpose is entertainment (toys, animatronics, etc).**

Though we have discussed several aspects of AUV development which are relevant to our work, we have only scratched the surface of the breadth and depth of research in this field. The reader is directed to several excellent reviews of the history of AUV development (Blidberg 2001), the basics of AUV design (Wang et al. 2009), recent advances (Yuh, Marani, and Blidberg 2011), and a summary of bio-inspired technology as is applicable to AUV development (Roper et al. 2011; Bandyopadhyay 2005).

*A note on biology as inspiration*

One might assume that evolution, given millions of years, would converge on a so-called "best" design, and that engineers should therefore look to nature as the paragon of efficiency and optimization. However, this is almost never true. Unlike engineering design, the process of evolution is gradual and constrained by its past trajectory, so that major discrete changes in design are usually not selected. Additionally, natural selection's criteria are merely that the organism survives—not fastest, smartest, or strongest, but simply fast, smart, or strong *enough*. Lastly, an organism is multi-purposed: it must find food, locomote, reproduce, escape from predators, and otherwise survive in a harsh world. Optimization for a single, specialized trait (e.g. hydrodynamic efficiency) could be detrimental to other traits, so it is not likely that any organism is completely optimized for any single trait. Though engineers are increasingly building multi-purposed robots, vehicles, and machines, we have not yet approached the complexity of animal life. This is in part because engineering systems are often designed to perform a single task (in contrast to the intricately interwoven networks of simultaneous requirements which are characteristic of biological systems).

Both human ingenuity and the natural world have provided inspiration for engineering design. Wheels, gears, and uniquely human inventions are as integral to today's cutting-edge technology as airfoils, sonar, or other innovations based on observations of biological phenomena. Increasingly, interdisciplinary work at this intersection has shown that strict biomimicry is often not the best solution to engineering design problems; rather, a nuanced understanding of the underlying biology is required. In a similar vein, biologists who attempt to understand locomotion without knowledge of the physics involved will have, at best, an incomplete picture of whole-organism behavior. As biology and physics continue their cautious overtures toward one another, we would do well to keep in mind that neither supersedes the other, and both can provide useful tools for our understanding of the natural world *and* for our development of new technologies.

## 1.3   Dissertation outline

In this dissertation, we present a novel method for simultaneous measurements of freely-suspended solid particles and the fluid flow around them (Chapter 2). We fabricate hydrogel particles in a range of shapes, including cylinders and spheroids. These particles, which are refractive-index-matched to water, are suspended in homogeneous isotropic turbulence. We use stereoscopic particle image velocimetry to simultaneously calculate the in-particle velocity field and the surrounding fluid flow; from the in-particle measurements, we can calculate the particle's translational and angular velocities.

By comparing the particle's translational velocity to the average velocity of the surrounding fluid, we calculate the turbulent slip velocity (Chapter 3). We compare the turbulent slip velocity to the quiescent settling velocity, and explain mechanisms by which particle settling velocity may be altered in turbulence. The relationship between the particle and fluid velocity is explored in detail, both in the experimental data and in a simple one-dimensional model flow.

We assemble a large number of independent measurements of particle rotation, and calculate the mean rotation rate across different shapes (Chapter 4). We combine this experimental work with a numerical simulation of small (sub-Kolmogorov-scale) nonspherical particles in turbulence; for these small particles, we examine how their rotation rate is distributed about the symmetry axis, and how they preferentially orient with respect to the fluid strain eigensystem. We propose a qualitative framework for the angular momentum transfer or "vorticity inheritance" between the ambient turbulence and the suspended particles.

Lastly, we summarize our conclusions and present a number of future avenues for analysis in both particle-laden turbulence and aquatic biomechanics (Chapter 5). We point out specific ways in which our data will be useful in the study of boundary layers, particle alignment, and turbulence modulation. Furthermore, with our information-rich measurement technique, we may be able to shed light on the mechanisms by which turbulence is attenuated by the presence of particles. We also present a plan of study for the next step in this investigation: the application of our results to biological systems.

# II.   Experimental Methods

The goal of this research is to study particle dynamics within the inertial subrange of turbulence, and investigate how those dynamics depend on particle shape and density. Due to the limitations of numerical methods, as outlined in Chapter 1, we pursue this question experimentally. The turbulent flow in our experiments is homogeneous (i.e., statistically uniform in space) and isotropic (i.e., statistically independent of coordinate system orientation). Experiments were conducted by suspending refractive-index-matched hydrogel particles in a 600-gallon (2.3 m³) stirred-tank facility, capable of generating an approximately 30 cm by 30cm by 30cm volume of homogeneous isotropic turbulence via two facing jet arrays. We will refer to this apparatus as the "turbulence tank," illustrated in Figure 2.1. Suspended particles are imaged via stereoscopic Particle Image Velocimetry (PIV), which gives velocity measurements in a planar slice of the homogeneous isotropic volume. This method yields all three velocity components (u, v, and w) in a two-dimensional measurement volume (this is sometimes called 2D3C velocimetry). In this section, we will describe the configuration of the tank, along with its measured turbulent statistics; the specifications and set-up of the PIV system; and the use of the refractive-index-matched hydrogel particle method, which was developed specifically for this project.

## 2.1   Turbulence generation facility

The generation of turbulence in the laboratory has a long and storied past, beginning with Osborne Reynolds' 1883 experiments showing the turbulent mixing of dye in a long pipe. Grids have been used to great effect to generate turbulence, beginning with the experiments of G.I. Taylor and his contemporaries (Simmons and Salter 1934; Taylor 1935). Further refinements on this method were provided by S. Corrsin, who eventually showed that the decaying turbulence behind passive grids was not quite isotropic (Corrsin 1942). Other studies using grid-generated turbulence gave insight on eddy diffusivity (Towle, Sherwood, and Seder 1939), thermal mixing (LaRue and Libby 1981), and the effects of mean shear (Burgers and Mitchner 1953). (Makita 1991) greatly advanced the field with the introduction of active grids for turbulence generation, allowing for much higher Reynolds numbers in smaller facilities. The use of a single grid (active or passive) may produce turbulence that achieves 2D homogeneity and isotropy within planar slices downstream of the grid; however, the turbulence overall is necessarily decaying, and this approach therefore cannot produce 3D homogeneity or isotropy.

Many recent experiments have used Makita's active grids as a starting point to accomplish the goal of generating homogeneous and isotropic (H.I.) turbulence in a 3D volume. To achieve this, experimenters have injected energy into the flow from multiple locations, allowing the signatures from the forcing elements to merge in some central location. This results in approximately H.I. turbulence. For forcing elements, researchers have used oscillating grids (Villermaux, Sixou, and Gagne 1995), speakers (Hwang and Eaton 2004), fans (Birouk, Sarh, and Gökalp 2003), or jets (Krawczynski, Renou, and Danaila 2010). Earlier experiments, usually conducted with passive grids in wind tunnels, generated turbulence which was superimposed on a (sometimes large) mean flow. In contrast, the "turbulence box" approach has been effective in reducing or eliminating the underlying mean flow (Hwang and Eaton 2004; Jong et al. 2009). Our turbulence-generating facility is based on these active-grid, multiple-injection-point approaches, and produces a relatively large volume of H.I. turbulence with a high Reynolds number.

H.I. turbulence is a curious phenomenon; it is highly sought-after in the laboratory, yet rarely exists in nature. Since we purport to apply our research to environmental flows, it is worthwhile to question our purpose in generating H.I. turbulence. However, it is precisely *because* the applications of this work are so broad that we choose to measure particles in H.I. turbulence, rather than simulated riverine, estuarine, or oceanic flows. Performing the study in H.I. turbulence will allow us to uncover some of the fundamental physics of particle-flow interaction, common across all turbulent flows. The community may then use this work as a springboard to examine particle-laden flows in a variety of specialized environments.

### 2.1.1 Experimental setup

To generate H.I. turbulence, we use two facing randomly-actuated synthetic jet arrays or RASJAs (Variano, Bodenschatz, and Cowen 2004). Each jet array contains 64 pumps mounted on an equally-spaced 8x8 Cartesian grid, so that each outflow jet (diameter 2.19cm) is 10 cm from its neighbors. Since the intake of the jet is only 7cm away from the outflow, they may be considered roughly as co-located; therefore, the jets inject only momentum, not mass, into the test section and can be considered as zero net mass flux (ZNMF) jets. The two facing arrays of pulsating jets are driven stochastically: both the on- and off-time for each jet is normally distributed with a known mean and standard deviation, based on the algorithms in (Variano and Cowen 2008). In the center region of the tank, the signatures of the individual jets are no longer discernable, and the turbulence is homogeneous and isotropic. The stochasticity of the firing jets keeps tank-scale mean flow weak, and the symmetry of the two arrays contributes to isotropy. Though some tank-scale circulation is present, it is minimized via this method (Bellani et al. 2012). The tank's working fluid is tap water that has been filtered to 5 microns, and is continuously UV-purified to prevent microbial growth.

The region of homogeneous isotropic (H.I.) turbulence extends over several integral length scales in each direction (for complete characterization of the turbulence within the tank and a quantification of the homogeneity and isotropy in the working section, please see (Bellani and Variano 2014)). The large H.I. region allows the particle to pass through a large region of turbulence before entering the measurement plane, so that the motion which we measure is determined primarily by turbulence that is statistically uniform. This is critical to studies of particles in flow due to the presence of a memory term in most formulations of the equation of motion (Reeks and McKee 1984; Armenio and Fiorotto 2001).

The tank cross-section is 80 cm x 80 cm, and the working section of the tank is 75cm long. It is bounded on either end by a solid divider to which the jets are attached, leaving an "endcap" on either side that is 96.5 cm long. This endcap allows experimenters to re-position the jet panels if desired in order to expand or contract the working section of the tank. The endcaps also serve as a recirculation zone, since water can flow freely around the edges of the solid divider. In our experiments, the jets were positioned 45.5 cm away from the mesh screens which form the tank's working section, and the distance between the jets was 166 cm.

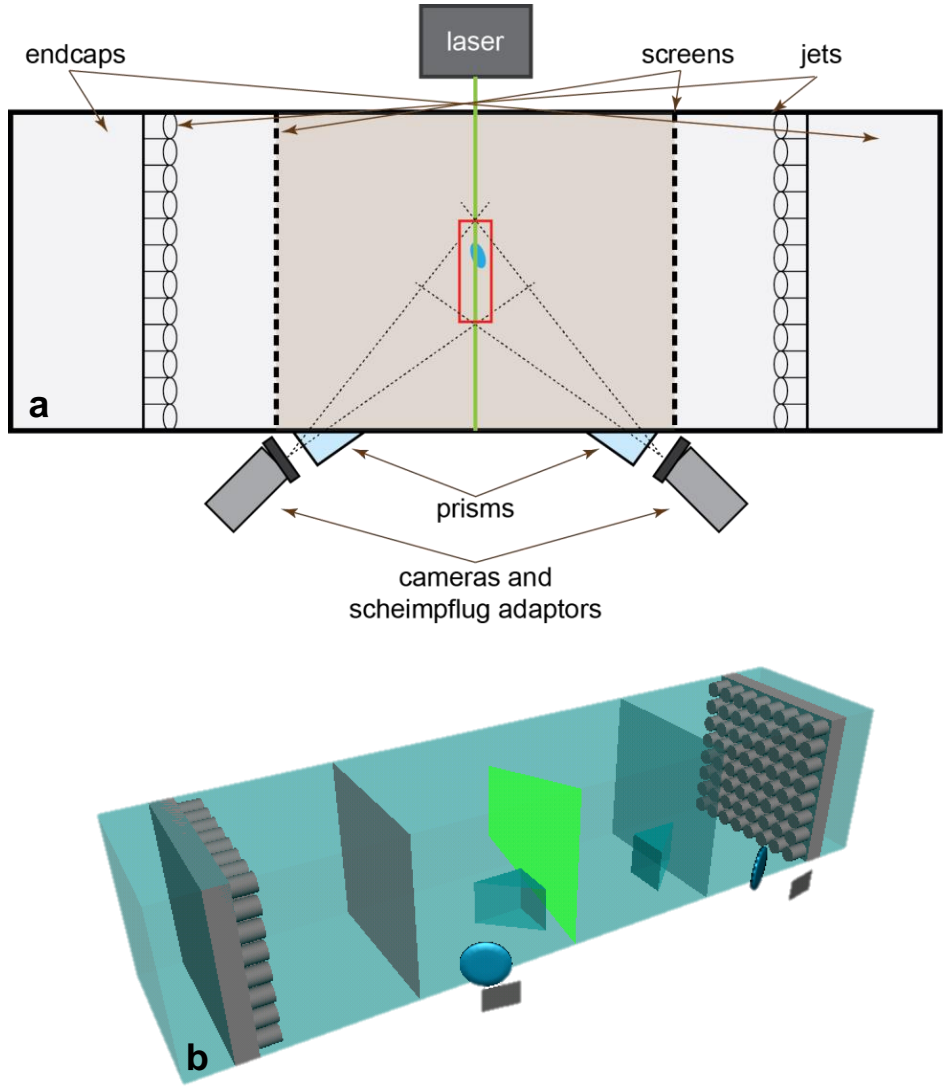

**Figure 2.1 (Schematic): (a) Turbulence-generating tank, top view. Turbulence is generated via two facing jet arrays, whose signatures pass through screens to form a central volume of homogeneous isotropic turbulence. Two cameras focus stereoscopically on a two-dimensional central image window (further described in Section 2.3). Not to scale. (b) Isometric view of turbulence tank, showing jets, screen, laser sheet, prisms, and cameras (stylized as lens and CCD chip).**

The particles we study are refractive-index-matched hydrogels, which are relatively fragile. When these hydrogel particles collide with a jet intake, or with the divider separating the tank endcaps from the working section, they are susceptible to breakage and deformation. This necessitates the addition of two screens (5mm square-opening mesh[1]) on either side of the working section, which protects the particles from the destructive jet intakes. These screens change the properties of the turbulence, lowering the Reynolds number and thus decreasing the separation between the largest and smallest scales. However, the screens do not affect the isotropy or homogeneity of the flow, nor do they introduce a mean flow. All experiments described here were conducted with the screens installed, and so we report the turbulent properties with the screens in the tank. This configuration is identical to that which is reported in (Bellani, Nole, and Variano 2013).

## 2.1.2 Tank properties

The tank's turbulent properties are fully described in a number of published papers (Bellani and Variano 2014; Bellani and Variano 2012; Bellani et al. 2012), and summarized here. In our tank-based coordinate system, the x-direction is the lateral (cross-tank, horizontal) direction; the y-direction is the vertical (gravity-coupled) direction; and the z-direction is the axial direction, aligned with the jets and out-of-plane in the stereoscopic imaging (Figure 2.1). The corresponding velocities are denoted as u, v, and w, as is conventional.

| | | | |
|---|---|---|---|
| Lateral velocity scale, $\sqrt{\langle \overline{u^2} \rangle}$ | [x $10^{-2}$ m s$^{-1}$] | 1.65 | *[1.61, 1.69]* |
| Vertical velocity scale, $\sqrt{\langle \overline{v^2} \rangle}$ | [x $10^{-2}$ m s$^{-1}$] | 2.36 | *[2.34, 2.38]* |
| Axial velocity scale, $\sqrt{\langle \overline{w^2} \rangle}$ | [x $10^{-2}$ m s$^{-1}$] | 2.07 | *[2.04, 2.09]* |
| Turbulent velocity scale, $u_T = \sqrt{\frac{1}{3}\left(\langle \overline{u^2} \rangle + \langle \overline{v^2} \rangle + \langle \overline{w^2} \rangle\right)}$ | [x $10^{-2}$ m s$^{-1}$] | 2.05 | *[2.02, 2.07]* |
| Kinematic viscosity, $\nu$ | [x $10^{-7}$ m$^2$ s] | 8.93 | - |
| Turbulent kinetic energy, $q^2 \equiv \frac{3}{2}u_T{}^2$ | [x $10^{-4}$ m$^2$ s$^{-2}$] | 6.30 | *[6.12, 6.43]* |
| Taylor microscale, $\lambda_f$ | [x $10^{-2}$ m] | 1.60 | *[1.49, 1.80]* |
| [2]Integral lengthscale, $L_f$ | [x $10^{-2}$ m] | 8.3 | *[8.1, 8.9]* |
| Turbulent dissipation rate, $\varepsilon \equiv 30\nu u_T^2 \cdot \lambda_f{}^{-2}$ | [x $10^{-5}$ m$^2$ s$^{-3}$] | 4.40 | *[3.37, 5.17]* |
| Kolmogorov lengthscale, $\eta \equiv (\nu^3/\varepsilon)^{1/4}$ | [x $10^{-2}$ m] | 0.036 | *[0.034, 0.038]* |
| Kolmogorov timescale, $\tau_\eta \equiv (\nu/\varepsilon)^{1/2}$ | [s] | 0.14 | *[0.13, 0.16]* |
| Taylor microscale-based Reynolds number, $Re_\lambda \equiv u_T \cdot \lambda \cdot \nu^{-1}$ | - | 367 | *[337, 417]* |
| Eddy turnover time, $T \equiv L_f \cdot u_T{}^{-1}$ | [s] | 4.05 | *[4.01, 4.11]* |
| Integral lengthscale-based Reynolds number $Re_L \equiv u_T \cdot L \cdot \nu^{-1}$ | - | 1905 | *[1876, 1924]* |

**Table 2.1: Turbulent statistics of experimental facility, measured in the image plane (excepting the integral lengthscale $L_f$—see footnote). Overbars here denote time-averaging, and brackets denote a spatial average over the image window. Details of calculations are given in the text or in (Pope 2000).**

The longitudinal Taylor microscale $\lambda_f$ is estimated by calculating the two-point longitudinal spatial autocorrelations, and fitting an osculating parabola to the first three (discrete) points of the autocorrelation curve, which is resolved at $dx = 0.66$mm. The Taylor microscale is the location on the x-axis at which this osculating parabola crosses the y=0 mark. The Taylor scale, by virtue of its construction, is larger than the Kolmogorov scale and smaller than the integral lengthscale. It may be thought of as the average distance between stagnation points within a turbulent flow (Sreenivasan, Prabhu, and Narasimha 1983). In isotropic turbulence, the transverse Taylor microscale is given by $\lambda_g = \lambda_f / \sqrt{2}$ , and the dissipation rate is given by $\varepsilon = 15 \nu u_T{}^2 \cdot \lambda_g{}^{-2} = 30 \nu u_T{}^2 \cdot \lambda_f{}^{-2}$ (Pope 2000). We note that the turbulence generated in this facility is not strictly isotropic, as the lateral, vertical, and axial velocity scales differ slightly from one another; however, this variation is small (less than 20%) and may be considered isotropic for our purposes.

---

[1] The mesh wire diameter is 1.67mm, giving an open area of 77.4%; therefore, we do not expect the screens to generate significant turbulence or jets from the mesh openings.

[2] Note: this parameter is taken from (Bellani, Nole, and Variano 2013), which was conducted under identical experimental conditions but with a larger image area (via shorter focal-length lenses on the same cameras used in our experiments). This larger image area allowed for the computation of the integral lengthscale.

The properties of the turbulence in our facility are comparable to small-scale ocean turbulence (Jimenez 1997), save the largest lengthscales. This is understandable, as it is impossible to replicate the ocean's largest eddies—where energy is injected on the scale of wind and tides—in a laboratory setting. Additionally, since our focus is at the smaller end of the turbulent cascade, this failure to match the large scales is not of great concern. The Taylor-scale Reynolds number, along with the dissipation rate, turbulent velocity scale, turbulent kinetic energy, and Kolmogorov microscales are well within the range that has been measured in the ocean. This shows that our facility generates turbulence that is very similar to that which might be experienced by aquatic animals.

| Turbulent velocity scale, $u_T$ | [x $10^{-2}$ m s$^{-1}$] | 1 - 3 |
|---|---|---|
| Turbulent kinetic energy (*TKE*) | [x $10^{-4}$ m$^2$ s$^{-2}$] | 2 - 14 |
| Taylor microscale-based Reynolds number, $Re_\lambda \equiv u_T \cdot \lambda \cdot \nu^{-1}$ | - | $200 - 10^4$ |
| Turbulent dissipation rate, $\varepsilon$ | [x $10^{-5}$ m$^2$ s$^{-3}$] | $10^{-2}$ - 10 |
| Kolmogorov lengthscale, $\eta$ | [x $10^{-2}$ m] | $0.03 - 0.2$ |
| Large-eddy lengthscale | [m] | 2 - 100 |

Table 2.2: Properties of small-scale ocean turbulence, as listed in (Jimenez 1997).

## 2.2 Refractive-index-matched hydrogel method[1]

Particle image velocimetry (PIV) has been an important tool for measurements in fluid mechanics (Raffel et al. 2007). However, this technique is not easily extended to the specific case of flow around solid objects. This is because PIV requires the use of a laser light sheet to illuminate the region of interest. When measuring flow around an opaque or translucent object, the object itself will interfere with the illumination by casting shadows and/or scattering light. Of specific interest to this work is the case of large particles (regular and irregular three-dimensional shapes, of lengthscale ≈1 cm) suspended in a flow. When these particles cast shadows, we often cannot see their wakes, nor can we see into the interior of a dense suspension. These problems are also encountered when using PIV around models (e.g., of organisms or turbomachinery): shadows occlude large portions of the image area, preventing a complete analysis of the flow field (Sciacchitano, Dwight, and Scarano 2012).

To resolve this issue, refractive index matching (RIM) has been employed with great success (Budwig 1994; Wiederseiner et al. 2010; Dijksman et al. 2012). This permits the use of optical techniques such as laser Doppler velocimetry and PIV (Hassan and Dominguez-Ontiveros 2008). In RIM-PIV, the test objects (in our case, large particles) are made of a material that is transparent and refractively matched to the surrounding fluid. This avoids blockage or distortion of the laser light sheet and grants optical access to the entire flow field. One commonly used refractive-index-matched pair is mineral oil and glass or fused quartz (Stoots et al. 2001; Thompson, Vafidis, and Whitelaw 1987; Ezzein and Bathurst 2011). This pair has been used by biologists to conduct dynamically matched experiments of low-Reynolds-number phenomena such as lobster antennule flicking (Reidenbach, George, and Koehl 2008). It has also been used by hydrologists to study flow through porous media (Lachhab, Zhang, and Muste 2008). The high viscosity of mineral oil, however, precludes examination of high-Reynolds-number phenomena, including turbulent flows. By using aqueous sodium iodide solution matched with glass or acrylic, one can reach much higher Reynolds numbers, but at significant economic expense (Uzol et al. 2002). (Butscher et al. 2012) used resin paired with anisole to investigate flow through porous structures. Neither glass, acrylic, nor resin, however, allows for the study of flexible or deformable materials, which are common in biology.

---

[1] Note: this section is in large part reprinted from (Byron and Variano 2013), with permission.

We explore the use of two hydrogels, which by virtue of their chemistry are nearly refractive-index-matched to water. These hydrogels can be easily manufactured in the laboratory using injection molding (as discussed in following sections); more complex shapes can be obtained through stereolithography (Arcaute, Mann, and Wicker 2010). The material cost is small, and the use of water as the working fluid greatly expands experimental options. Furthermore, because hydrogels have adjustable density and flexibility, they can be used to model myriad objects that are of interest in biological fluid dynamics.

## 2.2.1 Materials and methods: hydrogel particle fabrication

One candidate for our investigation is polyacrylamide (PAC) hydrogel. PAC is an organic polymer with subunit formula $-CH_2CHCONH_2-$ (acrylamide). Aqueous acrylamide at low concentrations can be chemically or photochemically polymerized to form a highly water-retentive hydrogel. This gel is commonly used as a matrix for DNA gel electrophoresis, a staple technique in molecular biology (Shapiro, Viñuela, and Maizel Jr. 1967). Additionally, PAC is used in agricultural soil treatment (Sojka, Lentz, and Westerman 1998; Boatright et al. 1997); aesthetic surgery (Pallua and Wolter 2010; S. T. Christensen et al. 2007); and soft contact lenses (Steffen, Turner, and Vanderlaan 2005).

Polyacrylamide gel also has a number of qualities which are desirable for RIM-PIV: it is straightforward to manufacture, it can be cast easily into a variety of shapes, it is nearly transparent, and it has a refractive index close to that of water (Stein 2010). Its optical properties, density, and elastic modulus can be controlled by chemical composition and manufacturing method (Franklin and Wang 2003; Borzacchiello and Ambrosio 2009; Grattoni et al. 2001; Ferruzzi, Pan, and Casey 2000). Commercially produced PAC spheres have been used for refraction-matched quantitative imaging (Mukhopadhyay and Peixinho 2011; Klein et al. 2013). Via injection molding, we are able to fabricate PAC (and other hydrogels) in custom shapes. This process allows us to seed the gel with internal tracers, facilitating tracking during imaging studies. Using this last feature, we will demonstrate that images of large PAC particles can yield detailed vector fields describing the solid-body motion of the particles as well as the surrounding flow field, using a standard commercial PIV system (Figure 2.3).

A second option for RIM-PIV with hydrogels is agarose. Agarose is a polymer with subunit formula $-C_{24}H_{28}O_{10}$ $(OH)_8$— (agarobiose). It is most commonly used as a laboratory growth medium for microorganisms. Agarose has been used successfully in RIM-PIV in previous experiments in our facility (Bellani et al. 2012). Agarose is more fragile than PAC and is less transparent; PAC has superior optical clarity, robustness, and longevity. However, agarose particles can be made at a significantly lower density while retaining coherent shapes, can be made more cheaply and more quickly, and carry less health risk during production. We are interested in the motion of particles of roughly neutral buoyancy and will primarily use agarose hydrogel in our experiments.

**Procedure for fabricating polyacrylamide models**

We discuss two types of particle shapes in this work: spheroids and cylinders. Spheroids (and many other complex shapes, if desired) may be made via injection molding. Plastic molds are 3D printed (purchased from Protocafe Inc. or printed in-house using ProJet 3000 HD printer and proprietary acrylic polymer) to create desired shapes. The molds are lightly coated with mineral oil, clamped into their assembled form, and set aside before mixing the acrylamide solution. Cylindrical particles are made by casting the hydrogel into a sheet of uniform thickness, equivalent to the desired cylinder height. Hydrogel cylinders may then be cut from the sheet, using sections of thin-walled pipe at a specified diameter.

De-ionized water is seeded with 11-μm glass spheres (Sphericel, manufactured by Potters Industries) at a concentration of 0.05 % by mass, to act as embedded optical tracers for PIV. The initial concentration of the acrylamide solution (30% acrylamide, manufactured by Bio-Rad Laboratories) is diluted such that acrylamide composes 8% of the total solution volume (see Table 2.3). The ratio of acrylamide isomers (in this case, 37.5 parts acrylamide to 1 part bis-acrylamide) is responsible for the cross-link density and therefore the consistency of the gel. Note that acrylamide in its unpolymerized form is a potent neurotoxin; to avoid adverse health effects, latex gloves are worn at all times when fabricating particles out of polyacrylamide.

| 30% Acrylamide/bis solution | 26.7 mL |
|---|---|
| 10% Ammonium persulfate solution | 0.5 mL |
| TEMED | 0.1 mL |
| Sphericel glass microspheres | 0.05 g |
| Deionized $H_2O$ | 72.7 mL |
| **Total solution volume** | **100 mL** |

Table 2.3: Solute masses/volumes for 8% PAC hydrogel.

A 10% aqueous ammonium persulfate (APS) solution is mixed using de-ionized water and ammonium persulfate crystals (manufactured by Bio-Rad Laboratories). The 10% APS solution is added to the acrylamide-seed solution, constituting 0.5 % of the total solution volume. The catalyst tetramethylethylenediamine (TEMED, manufactured by Bio-Rad Laboratories) is then added to the solution in a quantity constituting 0.1 % of the total solution volume. The total solution is mixed well to ensure even distribution of the two polymerizing agents (APS and TEMED) and to improve suspension of tracer particles. Excessive stirring is not desirable, since oxygen will absorb the free radicals necessary for polymerization. The solution is then injected into the custom-shaped molds using a hypodermic syringe. The solution begins to polymerize into a gel in approximately 10 min, and polymerization is completed in 2-3 h. For experimenter safety, the full length of time must elapse before particles are removed from the molds (or, if fabricating cylinders, the sheet of uniform thickness must be completely polymerized before cutting cylinders).

After complete polymerization has occurred, the particles are removed from the molds and placed in containers of de-ionized water. After 24 h in water, the particles expand by approximately 10% in length dimensions. Particles have been observed during 6 months of aging with no visible signs of degradation in shape or optical clarity. We store the particles in a screw-top container, refrigerated, and submerged in a water bath. To prevent microbial growth in the water bath, a small amount of detergent is added. There is no microbial growth within the PAC particles themselves.

As previously discussed, a second option for RIM–PIV is agarose hydrogel. Our materials and procedure for manufacturing agarose particles are as follows.

**Procedure for fabricating agarose particles**

A beaker of de-ionized water is heated in a hot water bath to a temperature of approximately 50C. Agarose powder (Apex BioResearch Products general purpose agarose, low electroendosmosis value) is added at 0.4 % by mass (Table 2.4). 13–44-μm glass spheres are added at 0.2 % by mass, again to provide embedded optical tracers (larger glass spheres are used for agarose than for PAC to accommodate differences in background light scattering between the two gels; this is explained in detail in the following section). The mixture is stirred until agarose powder dissolves. This mixture is injected into clamped, oiled molds as in the PAC procedure outlined above and refrigerated for 10 min in order to set the gel. If cylinders are desired, the gel is cast into a sheet and allowed to set at room temperature. After polymerization, particles are removed from the mold or sheet and placed in water, as above. Agarose does not visibly expand or further hydrate after casting. Agarose particles are significantly more fragile than PAC particles and must be handled with care to avoid breakage.

| Agarose powder | 2.0 g |
| Deionized H$_2$O | 500 mL |
| 13 – 44 μm glass microspheres | 1.0 g |

Table 2.4: Solute masses/volumes used in 0.4% agarose hydrogel formulation.

## 2.2.2 Assessment of method

**Physical properties**

Physical properties for the two hydrogel materials are given in Table 2.5. All refractive indices and densities are measured at 23C. Index of refraction is measured without embedded tracers using a refractometer (Atago Inc., PAL-RI model) using 589 nm light. For the de-ionized water with which the particles are manufactured, the refractive index is 1.332 at 23 C. Elastic properties are measured using an Instron tensometer; slices of hydrogel (with known cross-sectional area) are placed under compressive strain, and the resultant compressive stress is measured by the instrument. Elastic modulus for high- and low-strain regions is estimated by averaging stress–strain curves from N=7 (0.4% agarose) or N=8 (8% PAC) different samples (Figure 2.2).

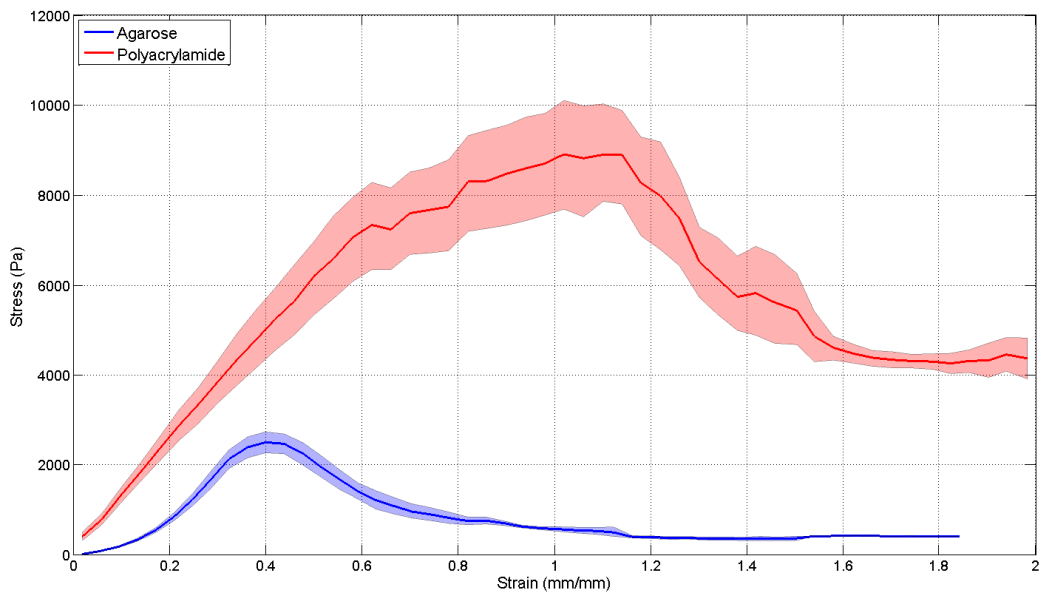

**Figure 2.2: Stress-strain curve for agarose (blue lower curve) and polyacrylamide (red upper curve) hydrogels. Shaded areas represent 95% confidence interval, calculated via bootstrapping with 1000 replicates.**

Note that higher-percentage gel formulations for both PAC and agarose yield a less flexible and more dense gel. These formulations may be useful for other researchers, but for our application (free-floating, nearly neutrally buoyant large particles), we prefer the agarose gel (up to 0.8% concentration).

| Property | 8% PAC | 0.4% Agarose |
|---|---|---|
| Index of refraction (IoR) | 1.349 | 1.3329 |
| Compressive breaking stress (kPa) | $36.6 \pm 2.5$ | $5.1 \pm 0.2$ |
| Compressive breaking strain (mm/mm) | $0.61 \pm 0.03$ | $0.36 \pm 0.03$ |
| Elastic modulus at low strains (near zero) (kPa) | $10.9 \pm 2.2$ | $1.9 \pm 0.2$ |
| Elastic modulus at high strains (near breaking) (kPa) | $134.0 \pm 1.2$ | $22.4 \pm 1.6$ |
| Density (g cm$^{-3}$)[1] | 1003.3 kg/m$^3$ | 999.6 kg/m$^3$ |
| Specific gravity | $1.006 \pm 0.001$ | $1.003 \pm 0.001$ |

**Table 2.5: Properties of PAC and agarose hydrogels. All error ranges shown are the standard error.**

---

[1] Since particles are very close to neutrally buoyant, we must measure density very precisely; a simple volume displacement method is not sufficient due to the high error range. Details of the density measurement method are given in Chapter 3, since the technique used to measure density is similar to the analyses described there.

Hydrogels are much less rigid than other common refractive-matched materials, and this makes them potentially useful in modeling the interaction of fluid flow and flexible structures such as biological tissue. In such studies, the formulation of the hydrogel should be tuned to give the desired material and mechanical properties. To provide a starting point for these modifications, we measure the maximum compressive stress and strain for 8% PAC and agarose gels. Both materials exhibit non-Hookean behavior and neither exhibit complete elastic recovery. Thus, a single elastic modulus cannot be defined, but the ranges shown in Table 2.5 may be used when rough comparisons are needed. The maximum compressive stresses and strains of 8% PAC and of 0.4% agarose are listed in Table 2.5; these values are in agreement with our qualitative observations of the particles' relative fragility.

**Performance in PIV**

The example images in Figure 2.3 show an agarose prolate spheroid (Figure 2.3a-b) and a PAC prolate spheroid (Figure 2.3c-d) passing through the PIV measurement plane. The surrounding flow is homogeneous and isotropic turbulence, as generated by the turbulence tank described in Section 2.1. The 2D3C vector fields shown were computed by tracking tracer particles either within the moving particle (Figure 2.3b,d) or suspended in the surrounding flow (Figure 2.3a, c). The laser used is a frequency-doubled pulsed Nd:YAG laser at a wavelength of 532 nm as described in Section 2.1. These data show that it is possible to track hydrogel particle translation using common PIV technology (in our case, commercial software from LaVision GmBH). Gradients in the within-particle velocity field allow us to calculate particle rotation, which is useful in its own right, and to improve the precision of particle translation measurements (Bellani and Variano 2012). This allows us to calculate the particle's translational velocity by taking the average over the in-plane slice, rather than select a single point at the slice's center-of-area, improving the robustness of the measurement.

The particle phase scatters slightly more light than the water phase, allowing the two to be separated via basic image processing. In the images above, we implement an intensity threshold followed by an erosion-dilation sequence to isolate the particles, using the built-in masking algorithm in DaVis 7.2 from LaVision GmBH. This method easily and robustly separated the two vector fields for the purposes of data analysis. This step is not necessary for vector computation, but merely separates the in-particle vectors from the fluid-phase vectors so that we can distinguish between particle and fluid dynamics.

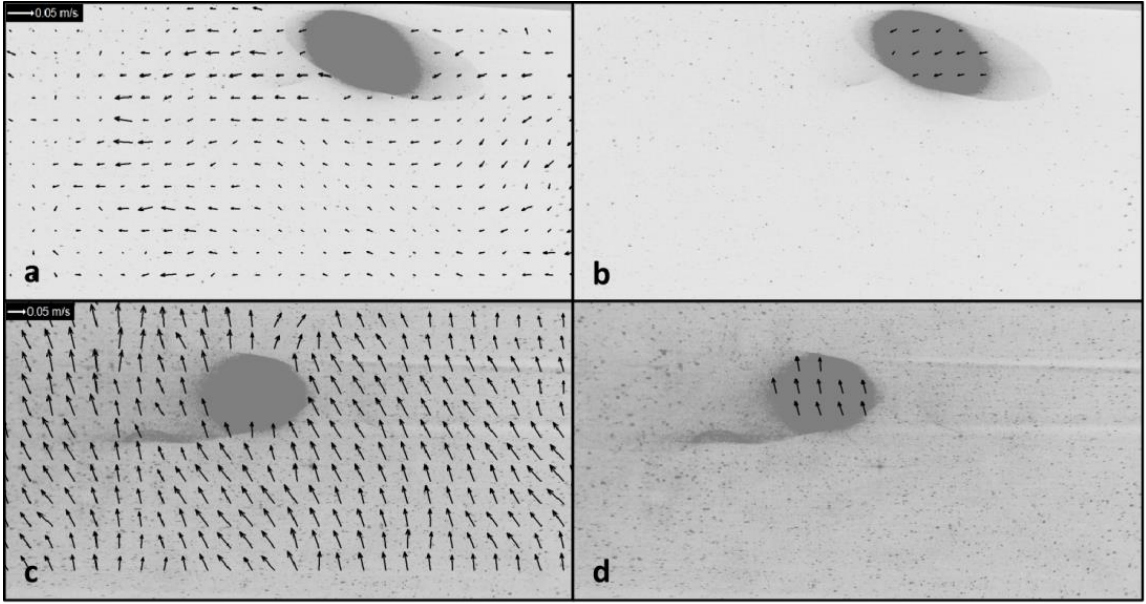

Figure 2.3: example vector fields for agarose hydrogel particles (panels a-b) and polyacrylamide hydrogel particles (panels b-c), demonstrating the results of the refractive-index-matched hydrogel method. For clarity, only every other vector has been retained in panels a-b, and every fourth vector has been retained in panels b-c. Background image shows hydrogel ellipsoid and surrounding tracer-laden flow; note that to achieve best display, the color balance has been adjusted such that no tracers appear within the gel. In reality, the gel is laden with glass microspheres which serve as tracers for the PIV algorithm.

**Limitations**

Two major limitations of this technique are that (1) injection molding becomes challenging for features smaller than approximately 1 mm and that (2) hydrogels are inherently flexible, thus attempts to formulate extremely rigid particles may interfere with their optical properties. However, in cases where rigidity is required, there are many choices for refractive-matched pairs. Currently, model size is limited by manufacturing method (not by the material itself); future manufacturing methods may improve upon those stated here.

The PIV and image thresholding method above requires a large number of tracers that are all internal to a single particle with relatively unambiguous borders (i.e., borders which can be detected by traditional image-processing methods). For very small particles, which cannot contain as many tracers, analysis can be improved by switching from PIV to particle tracking velocimetry (PTV). Such an approach has been demonstrated by (Klein et al. 2013) using fluorescently labeled tracers. More exact refractive index matching may be required for experiments involving dense suspensions, since small discrepancies will compound when light is directed through a large number of hydrogel particles. Additionally, using image intensity thresholds to separate the particle phase from the fluid phase will only work when the particle scatters a significant amount of light. In cases where this is not the case, the circular Hough transform may be useful to pick out (spherical) particles from cluttered backgrounds (Rizon et al. 2005).

As previously mentioned, hydrogel particles (as we have fabricated them) are not completely transparent and scatter a small amount of light. Some degree of light scattering is useful in most applications (such as the one illustrated by Figure 2.3), otherwise the gel becomes essentially invisible and the object or particle boundaries cannot be detected. The light-scattering properties are sensitive to illumination (laser power level, wavelength, and scattering angle relative to the cameras). Thus, fine-tuning will be needed in any new application. Additionally, if the application requires imaging through multiple layers of material, a more exact refractive index match is required; the reader is directed to (Wiederseiner et al. 2010) and (Dijksman et al. 2012) for a collection of useful techniques.

### 2.2.3  Discussion of method and its potential uses

Both PAC and agarose, as we have produced them, are near neutral buoyancy (within 1% of the density of water), though agarose is the less dense of the two. Both also have indices of refraction that are within 2% of water. PAC is robust to mechanical stresses, with a high maximum compressive strain (approximately 7 times that of agarose). This makes it useful for flows in which particles experience a high degree of stress. Agarose is less resilient than PAC but can still be used in RIM–PIV experiments (Bellani et al. 2012).  In its favor, agarose is closer to neutral buoyancy, has less health risk during production, is less expensive, and can be more quickly mass-produced.

The potential applications of this technique are wideranging— relevant to biology, engineering, medicine, and many other fields. In biomechanical studies, hydrogel models of anemones, fish, or plankton could provide new insight into functional morphology and biohydrodynamics. Environmental engineers could also investigate sedimentation processes with this technique. The adjustable flexibility and deformability of hydrogels points toward the modeling of tissues, including flow through blood vessels or porous structures. We have illustrated only one application here, but we are confident that this method can be tuned and used for a wide variety of functions.  Some have already been pursued, building on our work with PAC, by (Weitzman et al. 2014).

### 2.2.4  Particle specifications

For the experiments described in subsequent chapters, we have fabricated both spheroids and cylinders from agarose.  The primary results that we present are derived from the agarose cylinders. We present secondary data from agarose spheroids as a point of reference, to support our conclusions about the agarose cylinders and to justify our assumption that cylinders and spheroids behave similarly. In all experiments, particles were added (by type) at a volume fraction $\Phi = 0.1\%$. This volume fraction ensures that particles regularly pass through the image window, but that particle-particle collision is not significant (Elghobashi 1994).  PAC was not used for any of the experiments contained herein; the method for fabricating it is described here for completeness, as well as to illustrate the novelty and utility of the method.

**Primary dataset: cylinders**

To measure the effects of both shape and density on particle rotation and translation, we fabricate agarose cylinders at four different aspect ratios and two different gel concentrations, for a total of eight individual particle types.  The cylinders were cut from a uniform sheet using the method described above, and submerged in water for storage. Initially, particles were cut to be volume-matched to one-another. However, when the particles were submerged in water, they experienced slight changes in length dimensions (and therefore in volume). These changes resulted in the cylinder dimensions found in Table 2.6. After a long time submerged in water, particle volume varied by no more than 21% around a mean of 0.415 cm$^3$. Interestingly, the surface area of each particle varies by no more than 5% around a mean of 3.30 cm$^2$ (though the original mold dimensions were designed to be volume-matched). This suggests that the water-absorption equilibrium is controlled by particle surface area, rather than particle volume.

| Height 2c (mm) *(Mold dimensions)* | | Diameter 2a = 2b (mm) *(Mold dimensions)* | | α (=H/D) *(Mold dimensions)* | | Volume (cm$^3$) | Surface Area (cm$^2$) |
|---|---|---|---|---|---|---|---|
| 4.77 ± 0.11 | *(5.5)* | 10.60 ± 0.13 | *(11)* | 0.45 | *(0.5)* | 0.421 | 3.35 |
| 8.24 ± 0.18 | *(8.8)* | 8.72 ± 0.06 | *(8.8)* | 0.94 | *(1)* | 0.492 | 3.45 |
| 12.99 ± 0.14 | *(13.6)* | 6.41 ± 0.11 | *(6.8)* | 2.02 | *(2)* | 0.419 | 3.26 |
| 18.91 ± 0.06 | *(21.2)* | 4.70 ± 0.04 | *(5.3)* | 4.02 | *(4)* | 0.328 | 3.14 |

**Table 2.6: Dimensions of hydrogel cylinders. Volume varies by no more than 21% around a mean of 0.415 cm²; surface area varies by no more than 5% around a mean of 3.30 cm². Errors marked are the standard error. Dimensions immediately after manufacturing ("mold dimensions") are shown in italicized parentheses.**

Particles were manufactured at two concentrations of agarose: 0.4% and 0.8%. This corresponds to specific gravities of $\gamma_1$=1.003 and $\gamma_2$=1.006. For a detailed description of the method used to measure particle specific gravity, see Section 3.1.3.

### Secondary dataset: spheroids
Agarose spheroids are fabricated via injection molding as outlined in Section 2.2.1. We choose two shapes, a sphere of diameter 8mm and a prolate ellipsoid of major axis 16mm, minor axes 8mm. Over time, the hydrophilic behavior of the gel led to the dimensions shown in Table 2.7. Agarose spheroids were made at 0.4% concentration (specific gravity $\gamma_1$=1.003).

| | 2a (mm) *(Mold dimensions)* | | 2b (mm) *(Mold dimensions)* | | 2c (mm) *(Mold dimensions)* | |
|---|---|---|---|---|---|---|
| Sphere | 7.70 ±0.05 | *(8)* | 7.70 ±0.05 | *(8)* | 7.70 ±0.05 | *(8)* |
| Prolate ellipsoid | 7.57 ±0.05 | *(8)* | 7.57 ±0.05 | *(8)* | 15.25 ±0.12 | *(16)* |

**Table 2.7: Dimensions of agarose spheroids. Errors shown are standard error. Dimensions immediately after manufacturing ("mold dimensions") are shown in italicized parentheses.**

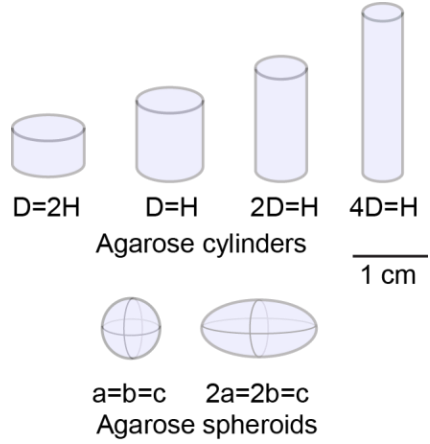

**Figure 2.4: schematic of particle dimensions, to scale. Cylinders are matched by their volume, and spheroids are matched by their minor axes.**

**Particle kinematic properties**

We are interested in the motion of particles at and around the Taylor microscale, which places them within the inertial subrange of the surrounding turbulence. At this size, particles are well above the Kolmogorov microscale and experience locally non-linear flow. However, as a point of reference, we may compute a Stokesian or relaxation timescale, though the particle Reynolds number is on the order of $10^2$ and thus particles are far outside of the creeping flow regime (see Section 3.1.2). This will guide our discussion of particle length and timescales, and how they may be related to corresponding scales within the turbulent flow. The relaxation timescale for a sphere is given by:

$$\tau_{st} = \frac{\rho_p d_p^{\;2}}{18\rho_f \nu} \tag{2.1}$$

where $\rho_p$ is the particle density, $d_p$ is the particle diameter, $\rho_f$ is the fluid density, and $\nu$ is the fluid kinematic viscosity. This equation was derived for a very small sphere in creeping flow and thus is not valid, except as a point of reference, for large shapes such as those we study in this work. To adapt this equation for ellipsoids, we look to the work of (Zhang et al. 2001) for the following expression:

$$\tau_{st,ell} = \tau_{st,sph} \cdot \alpha \frac{\ln(\alpha + \sqrt{\alpha^2 - 1})}{\sqrt{\alpha^2 - 1}} \tag{2.2}$$

where $\tau_{st,sph}$ is the relaxation timescale of a sphere whose diameter is equivalent to the ellipsoid's minor axes, and $\alpha$ is the particle aspect ratio. This expression is based on the average drag over possible ellipsoid orientations, assuming that every orientation is equally probable. Note that both of these expressions give the translational relaxation timescale, rather than the rotational relaxation timescale (i.e., the timescale over which a particle reaches rotational equilibrium with the ambient flow). In general, the rotational relaxation timescale is smaller than the translational relaxation timescale (Zhao et al. 2015), and is dependent on the particle's moments of inertia. We use this formula for spheroids and cylinders, even though it is not precisely true for the latter. However, cylinders may be expected to behave similarly to spheroids of similar axial dimensions, albeit with increased drag due to the increase in surface area (Loth 2008). For cylinders with an aspect ratio that is close to unity, the approximation becomes less accurate, but is still useful as an order-of-magnitude comparison.

We use Equation (2.2) to calculate the timescale for prolate cylinders ($\lambda \geq 1$), and the following equation (from (Challabotla, Zhao, and Andersson 2015)) to calculate the timescale for oblate cylinders or discs ($\lambda < 1$):

$$\tau_{disc} = \frac{2\rho_p a^2}{9\rho_f \nu} \frac{\alpha(\pi - C)}{2(1 - \alpha^2)^{1/2}} \tag{2.3}$$

where $a$ is the semimajor axis of the disk, $\alpha$ is the aspect ratio as before, and $C = 2\tan^{-1}\left(\alpha(1 - \alpha^2)^{1/2}\right)$.

| | $\tau_{st}$ [s] | |
| --- | --- | --- |
| | $\gamma_1 = 1.003$ | $\gamma_2 = 1.006$ |
| Cylinder, $\alpha$=0.5 | 4.70 | 4.72 |
| Cylinder, $\alpha$=1 | 2.60 | 2.61 |
| Cylinder, $\alpha$=2 | 3.58 | 3.59 |
| Cylinder, $\alpha$=4 | 4.24 | 4.26 |
| | $\gamma_1 = 1.003$ | |
| Agarose sphere D (=2a=2b=2c) = 7.7mm | 3.70 | |
| Agarose ellipsoid 2c=15.25mm, 2a=2b=5.7mm | 3.78 | |

Table 2.8: Stokes timescales of all particles, calculated from Equations (2.1), (2.2), and (2.3).

All particle timescales are on the order of 3-4 seconds, which is comparable to the large-eddy turnover time of 4.05 seconds, and much larger than the Kolmogorov timescale of 0.14 seconds. This indicates that particles will not respond to the smallest flow fluctuations, but that their motion will be on the scales of the larger eddies.

## 2.3   Stereoscopic PIV

Suspended particles were stereoscopically imaged, simultaneously with the surrounding fluid, using two CCD cameras (Imager PRO-X, 1600 x 1200 pixels, both fitted with a 105mm Nikkor lens and a Scheimpflug/tilt adapter). Each camera is equipped with a polarizing filter. These filters increase the contrast between the embedded tracers within the hydrogel particles and the hydrogel itself. Polarizer alignment is optimized to maximize this contrast, and a misalignment of the polarizers can cause the PIV algorithm to yield fewer or no vectors. Cameras have a maximum frame rate of 29.5 fps, and a minimum inter-frame time of 10 μs. This translates to an effective maximum frame rate of 14.773 fps for image pairs, using the frame-straddling technique illustrated in Figure 2.5. To minimize interfacial distortion, the cameras focused through 35° water-filled acrylic prisms on a 75x35mm window in the center of the homogeneous and isotropic region of the tank. This window aligned with a 1mm laser sheet (Quantel/Big Sky Lasers, 532nm) which bisected the tank. Flow was seeded using 11μm neutrally buoyant glass spheres (Sphericel, Potters Industries). Image pairs were collected at an effective framerate of 14.773Hz, with a 4ms separation between adjacent images.

Using image-intensity thresholding, particle and fluid phase data were separated before vector processing. Subsequent application of the PIV algorithm yielded exterior and interior velocity fields (Figure 2.3). For the interior vector fields, additional image preprocessing tools (sliding background subtraction and particle normalization) were used to identify the tracers within the hydrogel cylinders. Multipass PIV computation was performed using a commercial software package (DaVis, Lavision Inc; Goettingen, Germany), using 128x128 to 64x64 pixel interrogation windows with 50% overlap. Vector fields were resolved at 1.3mm, approximately 4 times the Kolmogorov scale. This grid spacing is small compared both to our particles and to the Taylor microscale.

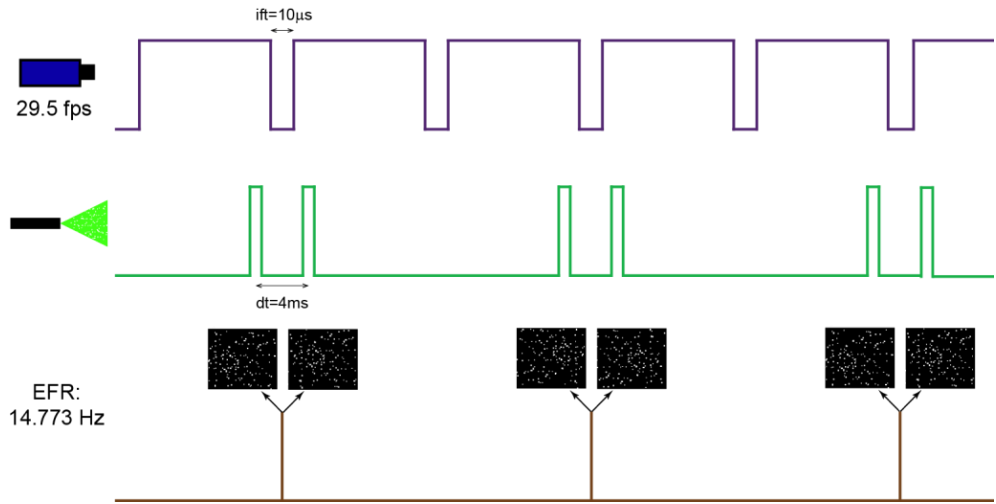

**Figure 2.5: Schematic of data collection using frame-straddling to decrease separation time. CCD cameras have a maximum framerate of 29.5 frames per second, and an interframe time (ift) of 10μs, as enabled by an onboard interline transfer chip. Laser pulses are separated by a time interval dt of 4ms, which translates to an effective frame rate (EFR) of 14.773 Hz.**

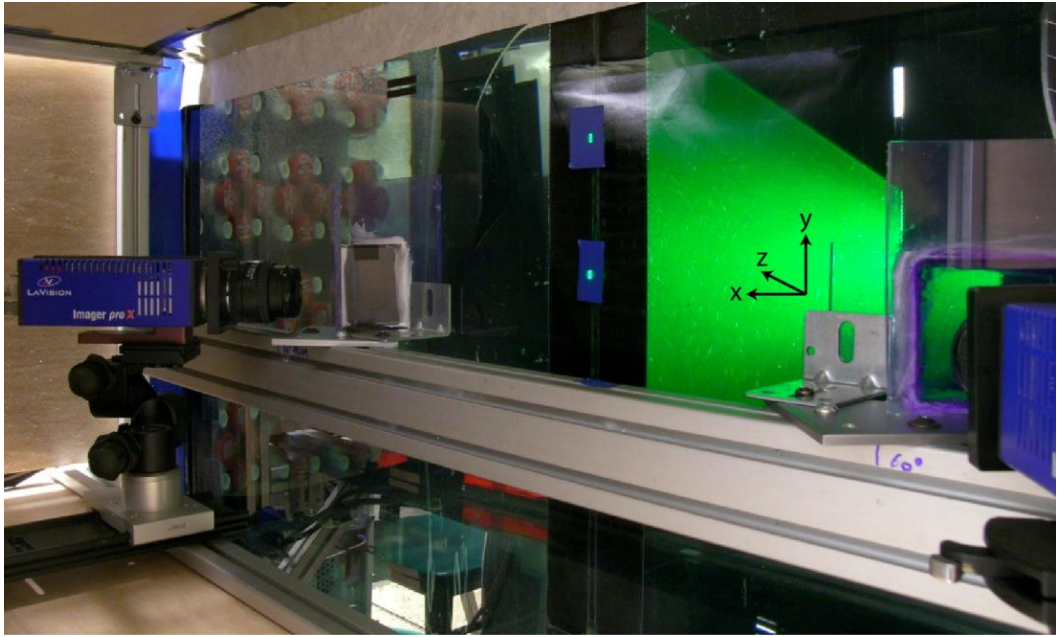

**Figure 2.6:** Photograph of imaging setup, showing cameras, Scheimpflug adapters, water-filled prisms, jet array, and laser sheet. A right-handed coordinate system is used as shown. Photo courtesy of Gabriele Bellani.

As mentioned in the previous section, the cameras focus on a small window that is entirely within the central volume of homogeneous, isotropic turbulence. Our coordinate system originates at the center of this window, with the in-plane dimensions (x-, cross-tank, and y-, vertical) and the out-of-plane dimension (z-, axial) as shown.

Note that the two CCD cameras are focused on opposite sides of the laser light sheet, which is an unconventional choice for stereoscopic PIV; for 2D3C PIV, most cameras view the same side of the laser light sheet but are separated by a small angle in order to measure the out-of-plane velocity component. Due to the presence of hydrogel particles in our flow, and the accompanying difficulties involving differential light-scattering between the particle and fluid phase, the camera configuration is an important variable in our experimental setup. Our configuration was empirically found to maximize the light-scattering from interior tracers (i.e., those inside the hydrogel particles) as compared to the surrounding gel matrix, while still maintaining strong light-scattering from the exterior tracers (i.e., those suspended in the fluid). The cameras are separated by a large angle, which minimizes the image-stretching that is caused by the non-orthogonality of the cameras to the light-sheet.

# III. Particle Settling

The effect of turbulence on particle settling velocity is well-studied (see Section 1.2.1), but much remains unknown. The physical parameter space is large, and includes particle density, size, and shape. Particle settling velocity also depends on the properties of ambient turbulence—more specifically, it depends on the relationship of the turbulent scales to the particle size, time, and velocity scales. Some nondimensional parameters governing the settling velocity include (e.g.) the ratio of particles' quiescent settling velocity to the rms velocity of the surrounding turbulence ($v_q/u_{rms}$), the ratio of particle size to the turbulent lengthscales ($L/\eta$), and the particle specific gravity ($\gamma$). The full set of parameters is discussed below, and in detail in Chapter 1.

In general, particles are classified as "large" when their characteristic lengthscale is larger than the turbulent Kolmogorov microscale. In this regime, particles experience locally-nonlinear velocity gradients, and the drag and lift on a particle are not predicted by the Stokesian solution (Happel and Brenner 1983). The history forces for large inertial particles are complex, and likely to have a strong effect on instantaneous particle drag. It is therefore difficult to predict *a priori* how the settling of large particles will be altered in the presence of turbulence. In this chapter, we investigate particles of varying shape and density, whose lengthscales are comparable to the Taylor scale of ambient turbulence. Particles' quiescent settling velocity, which we measure directly, is on the order of the turbulent rms velocity. Therefore, our particles' length and velocity scales are comparable to the intermediate length and velocity scales of the surrounding flow; that is, the turbulence is of moderate strength with respect to the particles. We will call this the "moderate-intensity" regime.

In addition to the quiescent settling velocity, we also measure the turbulent slip velocity of these large particles. By calculating this parameter, we find that turbulence leads to a strong reduction in the settling velocity (relative to the quiescent case). The reduction that we find is stronger than the reductions observed in comparable previous work which is also in the moderate-intensity regime. We also investigate the relationship of the slip velocity to the instantaneous surrounding flow, and explore how the slip velocity may trend with the particle and fluid velocities.

## 3.1 Quiescent (still-water) settling velocity

### 3.1.1 Experimental setup

To measure quiescent settling velocity $v_q$, particles of varying aspect ratio $\alpha$ and specific gravity $\gamma$ (properties listed in Section 2.2.4) are released at the top of a hexagonal water-filled tank of 500 mm vertical extent (Figure 3.1). One face of the tank is overlaid with an opaque plate, with 2 mm-wide slits laser-cut at 50 mm intervals. Two handheld lasers (Wicked Lasers, Hong Kong), both fitted with screw-in sheet optics, are mounted on a ring stand and positioned such that the laser planes are parallel, horizontal and shining directly through two of the precision-cut slits. Under the tank, a digital camera (Nikon J1) is positioned so that it focuses upward through the transparent tank floor, filming at 60 frames per second. When the particle passes through one of the two parallel laser planes, it is brightly illuminated for a short time. This clearly-delineated short flash of light allows the digital camera to measure the time at which the particle passes through the plane. By positioning the laser planes at a known vertical interval $dz$, and observing the time interval $dt$ between the laser flashes, the average settling velocity over the interval may be calculated as $dz/dt = v_q$.

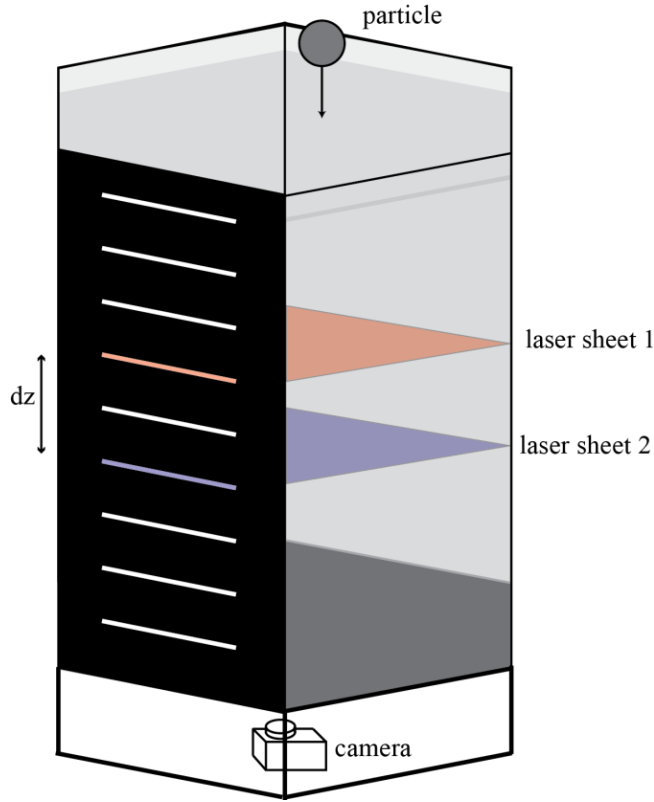

**Figure 3.1: Hexagonal tank for measuring quiescent settling velocity of large particles.**

To determine the terminal settling velocity, the experiment is repeated over several vertical intervals, starting near the top of the tank and progressing downwards. Initially, a 5cm $dz$ is used to calculate particles' average vertical velocity over the interval. This permits the determination of the location $z_t$ at which each particle type reaches its terminal velocity. The experiment is then repeated for $z > z_t$, with the interval expanded to $dz = 10$cm to improve accuracy. For each iteration of the experiment, including the preliminary 5cm $dz$ intervals, a sample size larger than N = 20 is used. Through the remaining transparent walls of the hexagonal tank, the particles' fall orientations (with respect to their axes of symmetry) are qualitatively observed. Terminal quiescent settling velocities ($v_q$) are recorded in Table 3.1 and plotted in Figure 3.2. Particle properties and dimensions are recorded in Table 2.5 through Table 2.8

### 3.1.2 Results

In general, particles with aspect ratio $\alpha = 1$ fell faster than other shapes at the same density (with the exception of $\alpha$=0.5, $\gamma_1$=1.003 cylinders). We attribute this in part to the particle orientations, which show that particles tended to fall in a drag-maximizing configuration: disks fall with their axis of symmetry vertical, and rods fall with their axis of symmetry horizontal. This is a well-known effect of the recirculation in the particle wake and subsequent pressure distribution (Mandø and Rosendahl 2010). Since $\alpha = 1$ particles tended to fall in a vertical orientation, the presenting face had a flow-normal area of only 0.60cm², while the longer $\alpha = 4$ rods fell broadside with a flow-normal area of 0.89 cm². This increased area resulted in higher drag force on the falling particle, and therefore lower terminal velocity $v_q$. At lower densities, close to neutral buoyancy, particle fall orientation was consistently either horizontal ($\alpha > 1$) or vertical ($\alpha \leq 1$). However, deviations from this norm increased as particles became more negatively buoyant; particles tended to "wobble" as they fell. This provides a visual illustration of the development of a stronger vortex wake, a function of the particle Reynolds number (Kundu and Cohen 2007). The data in Table 3.1 are plotted in Figure 3.2.

| Aspect ratio $\alpha$ | Density $\gamma$ | Quiescent settling velocity $v_q$ | Reynolds number $Re_p = \frac{v_q \cdot L_p}{\nu}$ | Predicted settling velocity (Loth 2008) | Fall orientation (direction of symmetry axis) |
|---|---|---|---|---|---|
| 0.5 | $\gamma_1$=1.003 | -1.47 ± 0.14 cm s⁻¹ | {79, 175}; 153 | -1.20 cm s⁻¹ | vertical |
| 1 | $\gamma_1$ | -1.34 ± 0.10 | {123, 130}; 147 | -1.80 | vertical (95%) |
| 2 | $\gamma_1$ | -1.10 ± 0.10 | {78, 160}; 114 | -1.49 | horizontal |
| 4 | $\gamma_1$ | -0.98 ± 0.09 | {52, 207}; 94 | -1.13 | horizontal |
| 0.5 | $\gamma_2$=1.006 | -1.59 ± 0.16 | {85, 189}; 166 | -1.83 | vertical |
| 1 | $\gamma_2$ | -2.48 ± 0.17 | {229, 242}; 272 | -2.73 | vertical (80%) |
| 2 | $\gamma_2$ | -1.65 ± 0.16 | {118, 240}; 172 | -2.23 | horizontal (80%) |
| 4 | $\gamma_2$ | -1.64 ± 0.13 | {86, 347}; 157 | -1.74 | horizontal (80%) |
| 1 (sphere) | $\gamma_1$ | -1.91 ± 0.25 | 165 | -1.83 | n/a |
| 2 (spheroid) | $\gamma_1$ | -1.51 ± 0.17 | {128, 258}; 162 | -1.67 | horizontal |

**Table 3.1: Quiescent settling velocities, Reynolds number, and fall orientations of all particle types (8 cylinders and 2 spheroids). Reynolds number range represents the range of $Re_p$ calculated using the particles' shortest and longest lengthscales (bracketed), followed by the $Re_p$ based on the diameter of the sphere of equivalent volume. Error indicated is the standard deviation of the measurements. Parentheses in last column indicate the proportion of tested particles that fell in the specified orientation; the remainder fell at some angle close to the predominant orientation.**

Since the particle Reynolds numbers are on the order of $10^2$, they fall into neither the Stokesian regime (in which drag is proportional to velocity) nor the Newtonian regime (in which drag is proportional to velocity squared). Though no analytic drag model exists for particles in this Reynolds number range, we calculate a theoretically predicted settling velocity based on models developed by Ganser, Cheng, Clift and Gauvin, as reviewed in (Loth 2008). To approximate particle drag coefficients, we use the normalized expression

$$C_D^* = \frac{24}{Re_p^*} + \left[1 + 0.15\left(Re_p^*\right)^{0.687}\right] - \frac{0.42}{1 + \frac{42{,}500}{\left(Re_p^*\right)^{1.16}}} \qquad (3.1)$$

where $C_D^* = C_D/C_{shape}$ and $Re_p^* = C_{shape} Re_p / f_{shape}$, and $C_{shape}$ and $f_{shape}$ vary according to particle aspect ratio (a full explanation of this model is provided in Appendix A). We see that in general, the measured settling velocities agree with the predicted trend, though our measured values are lower on average than the predicted values. We attribute this to the cylindrical shapes of our particles. The empirical models (dotted lines in Figure 3.2) were derived for oblate and prolate spheroids and then corrected to describe cylinders. Therefore, the model is not expected to perfectly describe cylinders. This is supported by the fact that the model describes the spheroid particles well, to within experimental error.

The particle Reynolds number $Re_p$, when based on the quiescent settling velocity and the largest particle dimension, varies from 130 to 347. This is a regime of dramatic transition in particle wakes and drag forces. A transitional threshold has been observed at approximately $Re_p = 210$ in both numerics and experiments (Magarvey and Bishop 1961; Reddy et al. 2010; Ormières and Provansal 1999; Natarajan and Acrivos 1993; Jenny, Bouchet, and Dušek 2003), where the wake behind a falling sphere transitions from "single-threaded" to "double-threaded." At lower Reynolds numbers, the wake is axisymmetric, but at $Re_p = 210$ (or thereabouts), the axisymmetry is broken. After the transition to a non-axisymmetric, double-threaded wake type, lift on the object dramatically increases as instabilities cause it to rotate. This Reynolds-number transition point for a given particle type is very sharp, though it can vary somewhat around the given value of $Re_p = 210$ (Magarvey and Bishop 1961). Wake instabilities could affect the boundary layer separation point, potentially altering the settling velocity and allowing a particular particle type to fall faster or slower than would otherwise be predicted.

Since many of our particle types fall at or around this threshold of $Re_p$, we expect wake instabilities that are not accounted for in the empirical model (dotted lines in Figure 3.2). This could be another explanation for the discrepancy between the model and the experimental data. In general, the $\gamma_2$ cylinders are closer to this transition point; this may explain the differing trends seen in the $\gamma_1$ and $\gamma_2$ cylinders. We note that the relationship between wake structure and boundary layer separation point is still unclear (Veldhuis et al. 2005), and unlikely to provide a full explanation for our observations.

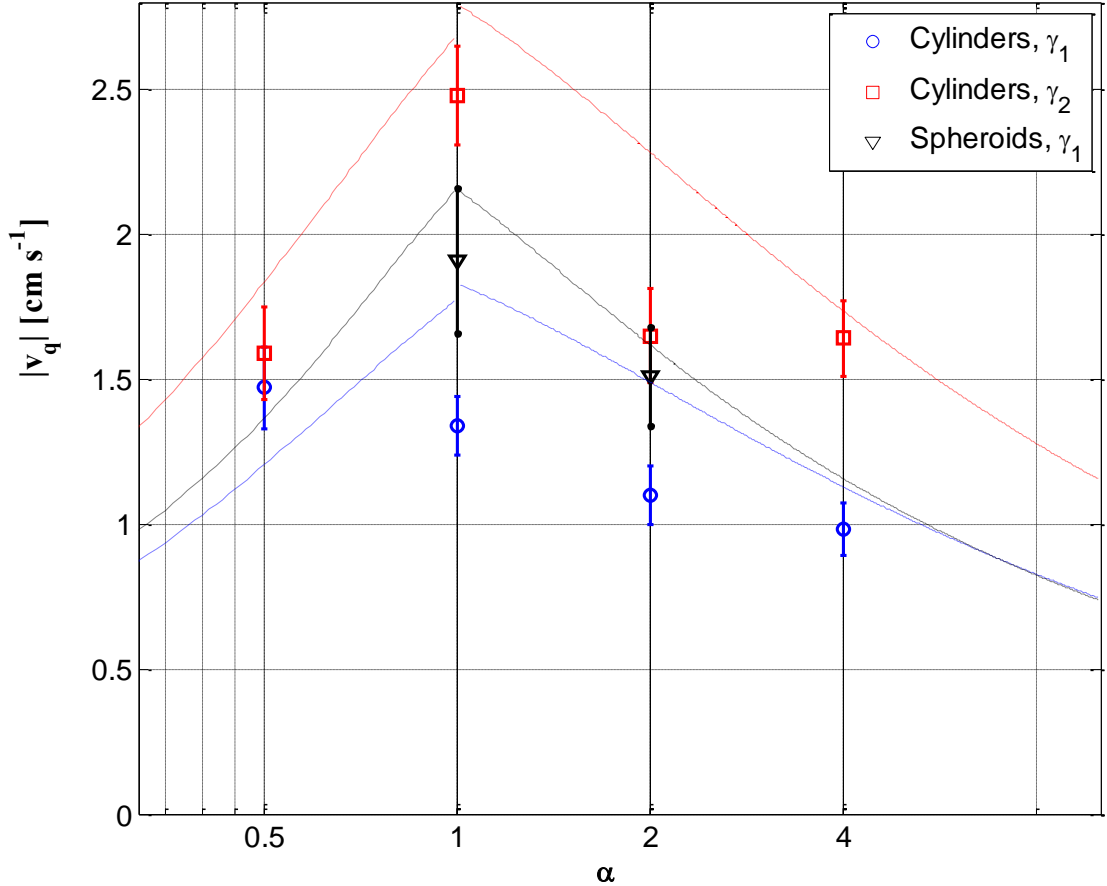

Figure 3.2: Measured quiescent settling velocities $v_q$ for cylindrical particles at four different aspect ratios $\alpha$ and two different specific gravities $\gamma_1=1.003$ and $\gamma_2=1.006$, as well as spheroidal particles at $\alpha=1$ (spheres) and $\alpha=2$ (prolate ellipsoids). Dotted lines indicate predictions from the particle drag model based on (Loth 2008) for preferentially-oriented cylinders at $\gamma_1=1.003$ (red dotted line) and $\gamma_2=1.006$ (blue dotted line), as well as preferentially-oriented spheroids at $\gamma_1=1.003$. Note that absolute value of settling velocity is shown, so that more positive values represent faster settling. Error bars show standard deviation of measurements.

In general, we see that shape is less important than density when it comes to determining the quiescent settling velocity of large particles. The small differences between shapes in our case may be explained by the difference in effective cross-sectional area caused by preferential fall orientation, as described previously and as shown in Table 3..

### 3.1.3 Particle density measurement method

Because our goal is to explore the physics of near-neutrally buoyant particles, any density measurement method must be extraordinarily precise. The primary set of particles (agarose cylinders) was fabricated using 0.4% and 0.8% agarose hydrogel. At both concentrations, particles are >99% water, but their still-water settling velocities $v_q$ are non-negligible (on the order of 1-2 cm s$^{-1}$). This settling velocity is approximately the same as the turbulent rms velocity we have measured in our tank. This "moderate-intensity" regime is exactly the region of the parameter space which we seek to explore (see Figure 1.1 and accompanying discussion). The rms velocity of the turbulence we are able to generate is approximately 2 cm s$^{-1}$. In our desired size range (millimeters to centimeters), a still-water settling velocity of 2 cm s$^{-1}$ corresponds to particles whose specific gravity is approximately $\gamma = 1.001 - 1.01$ (according to empirical models such as those described in Appendix A). Therefore, any method we use to measure density must be precise to at least four significant figures.

We attempted a simple method using volume displacement (i.e., Archimedes' principle) to measure particle density. Particles were stored in water, then removed from their storage and submerged in a known volume of water in a graduated cylinder which rested on a digital mass balance. The resultant increase in volume was taken to be the particle's volume, and the increase in mass was taken to be the particle's mass. However, this method was insufficiently precise. Because the particles are hydrophilic, they retained an outer layer of water when being transferred from the storage container to the graduated cylinder. This caused a non-proportional overestimation of both volume and mass. Furthermore, though the digital balance was capable of great precision, the available graduated cylinders did not provide a precise enough volume measurement, even when very large numbers of hydrogel particles were tested simultaneously. A related method, in which a large rectangular volume of hydrogel was manufactured and weighed, suffered the same lack of precision. The most challenging aspect of density measurement for hydrogels is the shifting volume of any given hydrogel object. Even if a known volume of hydrogel was measured immediately after manufacturing, the volume changes when the hydrogel is placed in water. This method would overestimate particle density, since our particles are submerged in and free to absorb water.

When volume displacement methods failed, we sought to quantify hydrogel density using the still-water settling velocity, $v_q$, and the existing drag models for intermediate-Reynolds number particles (see Appendix A). We tested this drag model using borosilicate glass spheres of known diameter and density (McMaster-Carr), with a specific gravity of $\gamma=2.2$ and a diameter D=4.76mm (giving a $v_q$-based Reynolds number on the order of $10^3$). The model precisely predicted the particles' still-water settling velocity of 45 cm s$^{-1}$, verifying this method as an accurate way to measure the density of intermediate-Reynolds-number spheres. We then measured the still-water settling velocity of agarose spheres at both 0.4% and 0.8% concentration, and used the model to calculate their specific gravities (which we denote as $\gamma_1$ and $\gamma_2$). This level of precision would have been difficult and perhaps impossible to achieve using standard volume-displacement methods. This method also has the added benefit of being directly tied to our experimental process and one of our desired investigation topics: particle settling.

Using this method, we measured the specific gravity of the 0.4% agarose spheres to be 1.0033 ± 0.0007, and the specific gravity of the 0.8% agarose spheres to be 1.0063 ± 0.0005. That is, 0.4% agarose is 0.3% denser than water, and 0.8% agarose is 0.6% denser than water. The factor-of-two difference between both the concentration and the density difference provides further support for this method's accuracy.

## 3.2 Turbulent slip velocity

### 3.2.1 Analysis method

Inertial particles in flow do not behave as passive fluid tracers—the effects of gravity and finite size will cause the particle's path to deviate from the corresponding Lagrangian fluid trajectory. The relative magnitude of this deviation may be quantified by the particle Stokes number $= \tau_p/\tau_f$, where $\tau_p$ is the relaxation time of the particle (i.e., the characteristic timescale of velocity change due to fluid drag) and $\tau_f$ is a fluid timescale, usually either the Kolmogorov timescale or the turbulent eddy turnover time at the lengthscale of the particle. A low Stokes number indicates that a particle behaves as a "tracer", which responds quickly and advects with the fluid flow. Inertial particles in strong turbulence typically have a long response time compared to the timescales of ambient velocity fluctuations, and therefore have a high Stokes number. This disparity results in an often-substantial difference between the particle velocity and the local fluid velocity, known as the "slip velocity." This name is slightly misleading, because the no-slip condition does hold at the particle-fluid boundary. The "slip" referred to in "slip velocity" is defined relative to a characteristic velocity describing fluid motion near the particle.

Though the concept of a turbulent slip velocity is qualitatively easy to grasp, it may be quantitatively defined in many ways (Bellani and Variano 2012). In numerical models, the local fluid velocity is defined as the fluid velocity that would be measured at the location of the particle's center-of-mass if the particle itself were not present. This can be thought of as the velocity which is "seen" by the particle. However, in an experimental context, this information is not available. If the particles are smaller than the Kolmogorov scale, linear interpolation of the velocity to the particle center would perhaps yield a good approximation of the local fluid velocity. However, when the particles are larger than the Kolmogorov scale, linear interpolation of the velocity field to the desired point (the particle center-of-mass) will have little relevance to the particle dynamics. In our case, we also do not necessarily know the particle's center-of-mass velocity, or even the location of the particle center-of-mass. We have velocity data only for the slice of the particle that lies within the laser plane. Therefore, we define slip velocity as follows:

$$\overline{\boldsymbol{u}}_s \equiv \overline{\boldsymbol{u}}_p - \overline{\boldsymbol{u}}_f \tag{3.2}$$

where $\overline{\boldsymbol{u}}_p$ is the average in-plane particle velocity, and $\overline{\boldsymbol{u}}_f$ is the average fluid velocity in a neighborhood surrounding the particle. We prefer this fluid-averaging approach to interpolation methods. Even setting aside the difficulty of accurate interpolation for nonspherical particles, large particles sample the flow at a scale many times larger than the Kolmogorov scale; we therefore consider the spatially-averaged fluid velocity to be more relevant than any interpolated point velocity at the particle center.

Raw PIV images, obtained using the experimental setup described in Chapter 2, were masked using image intensity thresholding and filtering. This provided a boundary between the fluid and solid phases and enabled the computation of separate interior and exterior vector fields. To calculate the spatially-averaged fluid velocity, we dilate the particle border to produce a shape-preserving annulus around the particle (Figure 3.3). The inner bound of the fluid-averaging annulus is constructed to exclude the immediate particle boundary layer, which would unduly bias the slip velocity towards zero due to the no-slip condition. The boundary layer on a large, freely-suspended, nonspherical particle in turbulence is not yet well-understood, and so any attempt to predict the boundary layer thickness must be somewhat imprecise. However, in the interest of finding the most accurate slip velocity possible, we will define an approximate particle boundary layer thickness. Following (Cisse, Homann, and Bec 2013), we define a viscous sublayer thickness $\delta_\nu = \nu/u_\tau$ where $u_\tau$ is the friction velocity, $u_\tau = (\tau_w)^{1/2}$. Since our data do not allow for the calculation of the kinematic "wall" shear stress $\tau_w$, we take $u_\tau$ to be approximately equal to the turbulent velocity scale $u_T = 0.02$ m s$^{-1}$. This gives a viscous sublayer thickness of approximately $5 \cdot 10^{-5}$ m. We then define a wall unit $r^+ = r/\delta_\nu$. The outer layer of the (turbulent) logarithmic boundary layer is traditionally understood to begin at $r^+ \approx 50$ (Kundu and Cohen 2007; Pope 2000), which for our particles corresponds to an absolute distance of 2.5mm. It is this distance that we take to be the inner bound for our fluid-averaging annulus. This excludes the fluid that is most directly influenced by the no-slip boundary layer.

We would like our fluid-averaging annulus to contain the fluid that has the strongest influence on the particle's motion. In many studies involving particle dynamics, the diameter of the sphere of equivalent volume, $D_{sph}$, is taken as a representative lengthscale (Mandø and Rosendahl 2010). We therefore set the outer bound of the annulus so that it contains all fluid that is within $D_{sph}$ of the fluid-solid boundary (Figure 3.3)—that is, the outer bound is located at $r = D_{sph}$ from the particle surface. This approach gives consistency across all particle types, and across the various cylindric sections that appear in the laser plane. Changes in either the inner or outer annulus bounds yield qualitatively similar results (see Appendix B). Decreasing the inner radius decreases the calculated slip velocity, as the annulus begins to include the no-slip boundary layer. Increasing the outer radius increases the calculated slip velocity, as this biases the fluid-averaged velocity to zero (due to the zero-mean properties of the tank).

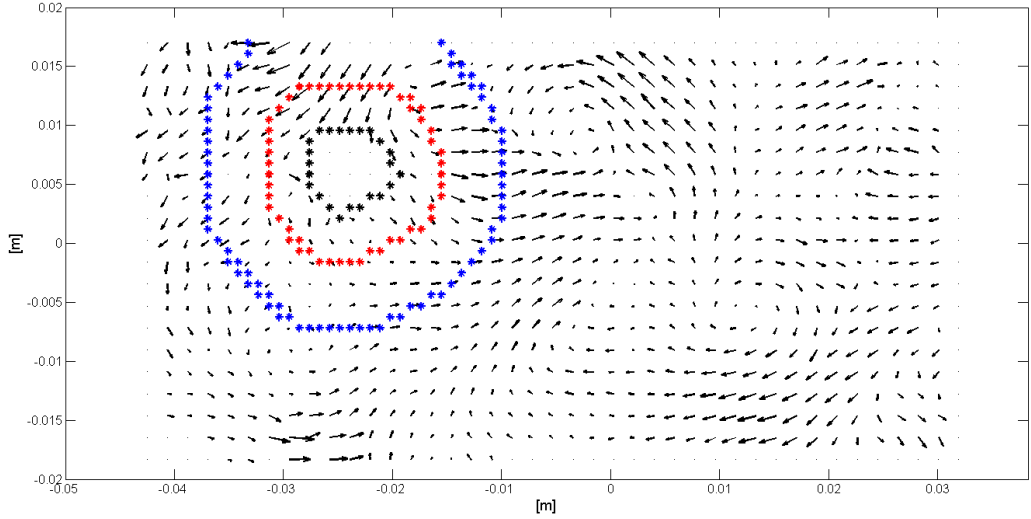

**Figure 3.3: shape-preserving annulus for calculating average fluid velocity. Black outline is the detected particle border, red outline is the inner bound of the annulus, and blue outline is outer bound of the annulus. The vector field is downsampled for clarity, showing only every fourth velocity vector.**

Since our measurement method includes only a two-dimensional slice of the velocity field in both the fluid and solid phases, we must consider potential bias in our results. If the particle is rotating, the in-plane slice may be moving in a different direction than the particle's actual center of mass—in fact, it is unlikely that the particle center of mass is in the laser plane at all. However, recall that the fluid velocity is also measured in the same plane. We compare the fluid flow in this plane with the particle cross-section which it immediately influences, which is the slice that is visible to us. (Bellani and Variano 2012) showed that the error in particle velocity due to randomly located measurement planes is negligible. Additionally, since the ambient turbulence is homogeneous and isotropic, we expect that any error introduced via this method will be random, not bias error, and thus cancel to zero when particle rotation and slip velocity statistics are calculated.

### 3.2.2   Results and discussion

Figure 3.4 shows the components of $\overline{\boldsymbol{u}}_p$ and $\overline{\boldsymbol{u}}_f$ for a large number of independent particle-fluid measurements (for clarity, the figure shows only one representative particle type, namely α=4, γ=1.006). There is a clear offset between the vertical particle velocity, $v_p$, and the two lateral velocity components $u_p$ and $w_p$. In general, the particles—which are slightly negatively buoyant—are moving downward with respect to the flow. The theoretical tracer limit, where $\overline{\boldsymbol{u}}_p = \overline{\boldsymbol{u}}_f$, is line with a slope of one and an intercept of zero (shown as the magenta dotted line in Figure 3.4). We note that the lateral velocities cluster isotropically about this line, whereas the vertical velocities seem to be linearly offset below. The offset is indicative of particles' vertical settling, which is discussed further in the following section. The relationship between vertical fluid velocity and vertical particle velocity (shown in red in Figure 3.4) can be reasonably described with a linear fit (cyan dashed line in Figure 3.4, with $R^2$ values given in Table 3.2). The intercept of this fitted line can be considered as the "turbulence-altered settling velocity", and agrees very well with the ensemble-averaged vertical slip, $\langle v_s \rangle$ (see Table 3.2).

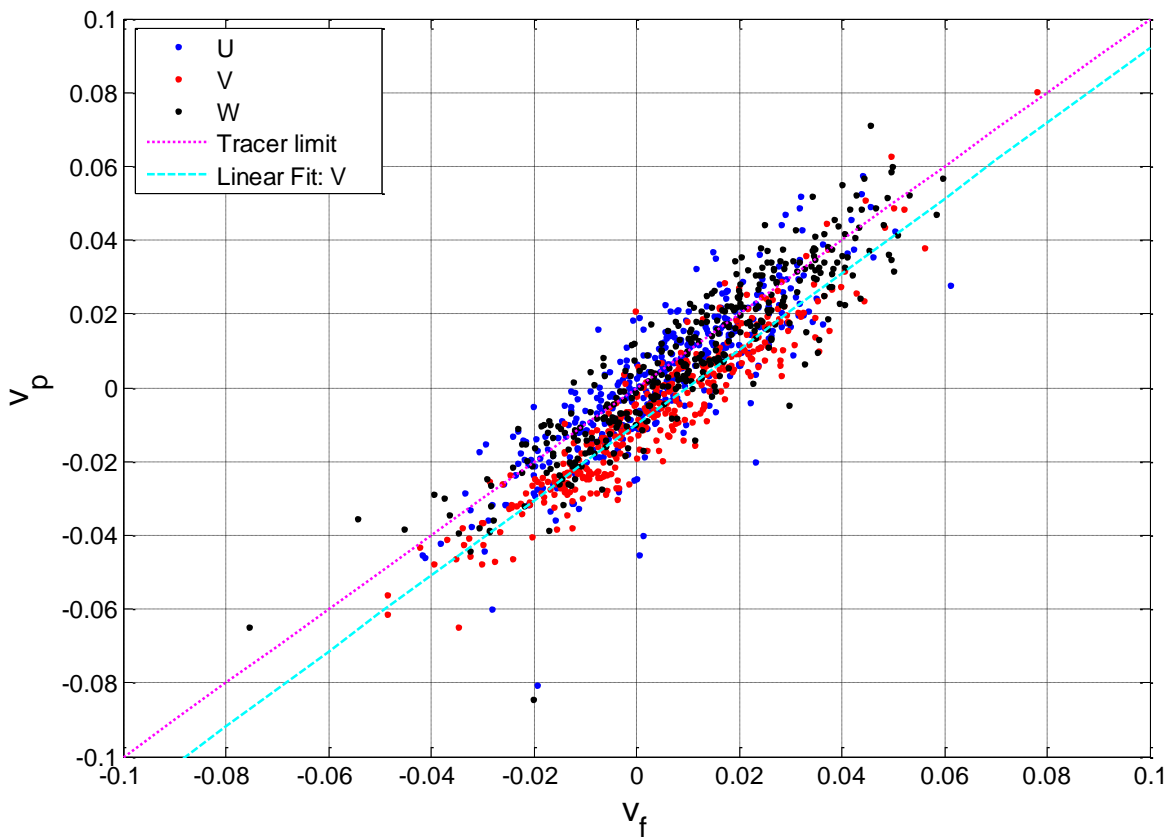

Figure 3.4: fluid velocity $\vec{u}_f = \{u_f, v_f, w_f\}$ vs. particle velocity $\vec{u}_p = \{u_p, v_p, w_p\}$, shown here for one particle type ($\alpha$=4, $\gamma$=1.006). Magenta line represents 1:1 relationship between $\vec{u}_f$ and $\vec{u}_p$ that would occur for the theoretical perfect fluid tracer. Cyan line is the linear best-fit to the vertical velocity, showing substantial offset between the lateral (U, W) and vertical (V) velocities.

| **C Y L I N D E R S** | | | | | | |
|---|---|---|---|---|---|---|
| | | $\gamma = 1.003$ | | | $\gamma = 1.006$ | |
| | $R^2$ value | fit intercept (mm s$^{-1}$) | mean slip $\langle v_s \rangle$ (mm s$^{-1}$) | $R^2$ value | fit intercept (mm s$^{-1}$) | mean slip $\langle v_s \rangle$ (mm s$^{-1}$) |
| $\alpha = 0.5$ | 0.89 | -5.7 | -5.6 | 0.90 | -7.5 | -7.5 |
| $\alpha = 1$ | 0.93 | -5.9 | -5.8 | 0.90 | -11.2 | -11.5 |
| $\alpha = 2$ | 0.88 | -5.6 | -5.7 | 0.89 | -8.2 | -8.3 |
| $\alpha = 4$ | 0.94 | -4.1 | -4.0 | 0.88 | -10.0 | -9.9 |
| **S P H E R O I D S** | | | | | | |
| $\alpha = 1$ | 0.90 | -5.8 | -5.9 | | | |
| $\alpha = 2$ | 0.88 | -5.6 | -5.6 | | | |

Table 3.2: goodness of fit for vertical-velocity linear regression (as shown in Figure 3.4) and comparison between the actual mean slip velocity and the y-intercept of the best-fit model, representing the average offset between the vertical fluid and particle velocities. This offset averages to zero in the lateral (non-gravity-coupled) velocities.

Two things here are particularly remarkable: first, that $\vec{u}_p$ is well-described as a linear function of $\vec{u}_f$, and second, that the slope of this linear dependence is approximately one. This is not necessarily the case for all flows, but is unique to turbulence; this will be discussed further in Section 3.3.2.

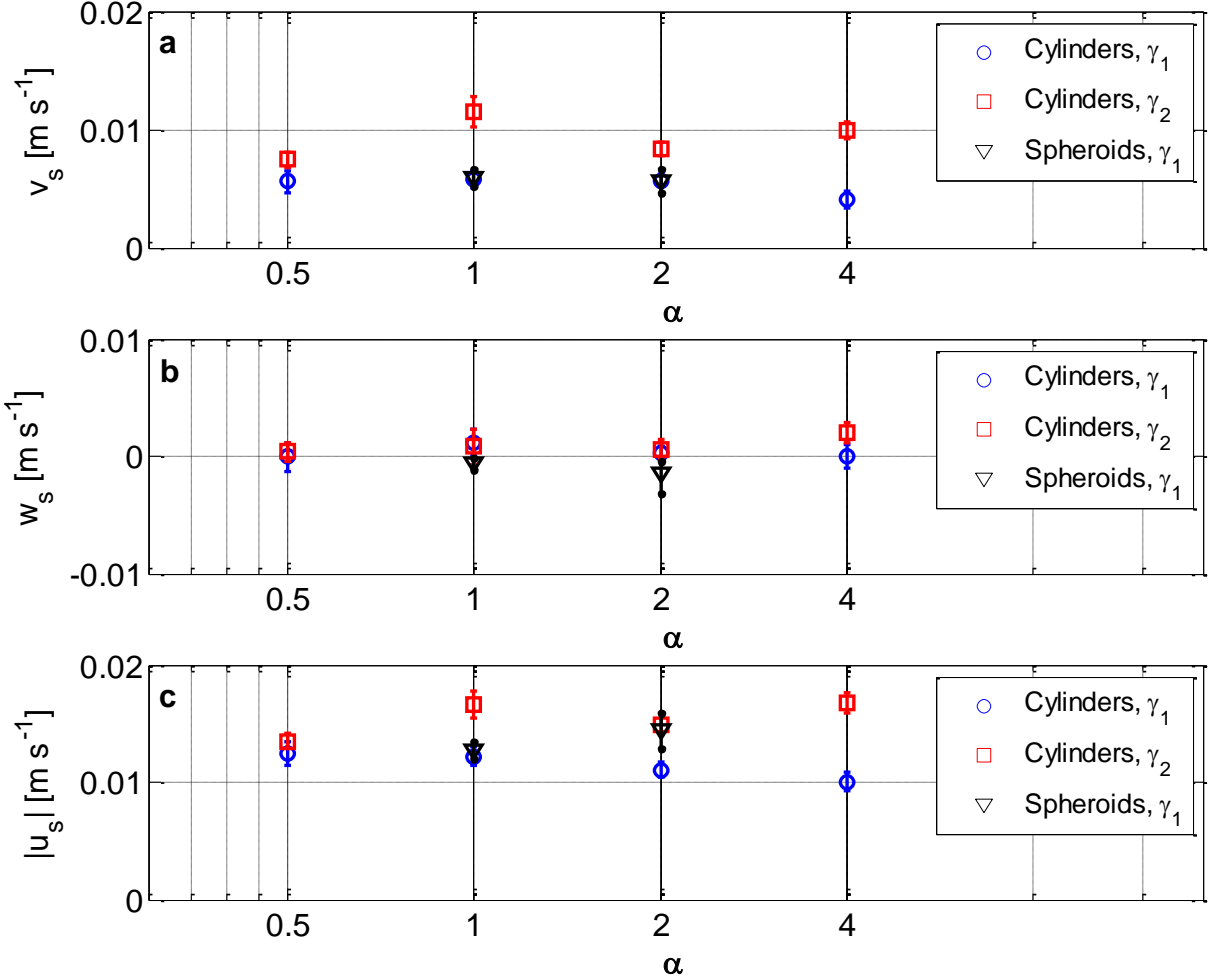

Figure 3.5: Ensemble average slip velocity $\langle \vec{u}_s \rangle$ of large aspherical particles in homogeneous isotropic turbulence. a) vertical component of slip $v_s$, b) horizontal component of slip $w_s$ (out-of-plane dimension), c) magnitude of the slip velocity vector $|\vec{u}_s|$ (i.e., including all dimensional components). Errorbars represent 95% confidence intervals obtained via bootstrapping with 1000 replicates.

In Figure 3.5, we illustrate several views of the ensemble-averaged slip velocity $\langle \vec{u}_s \rangle$, including the vertical slip $\langle v_s \rangle$ (Figure 3.5a), a horizontal component of the slip $\langle w_s \rangle$ (Figure 3.5b), and the overall magnitude of the slip velocity vector, $\langle |\vec{u}_s| \rangle$ (Figure 3.5c). The mean horizontal (out-of-plane) slip velocity $\langle w_s \rangle$ is an order of magnitude lower than the mean vertical slip velocity $\langle v_s \rangle$, and is very close to zero (within 95% confidence intervals in all but one particle type). In both Figure 3.5a and Figure 3.5c, particles with $\gamma_2$=1.006 displayed higher values of $\langle v_s \rangle$ and $\langle |\overline{\boldsymbol{u}}_s| \rangle$ than did particles with $\gamma_1$=1.003. For $\langle v_s \rangle$, this is intuitively obvious: denser particles will fall faster. The reason for the increase in $\langle |\overline{\boldsymbol{u}}_s| \rangle$ is slightly less apparent, but no less intuitive. The increased density of the $\gamma_2$=1.006 particles raises their inertia, causing them to respond more slowly to fluctuations in the ambient flow. This leads to more pronounced differences between the fluid velocity and the particle velocity, which we see manifested in the slip velocity.

In Figure 3.5a, the vertical slip $\langle v_s \rangle$ is highest for particles with $\alpha$=1, $\gamma_2$=1.006. Recall that in our measurements of the quiescent settling velocity, this particle type settled the fastest. The fact that this trend persists in the turbulent case suggests that some features present in the quiescent case are robust to strong turbulent perturbation.

Another high value of slip is observed for $\gamma_2 = 1.006$, $\alpha = 4$. We did not observe this pattern in the quiescent settling velocity (Figure 3.2). This increase may be related to the longer lengthscale present in this particle type; at this lengthscale, particles are sampling the velocity gradients at a larger scale, and therefore may experience more rotation (this is discussed in detail in Section 4.3). Our measurement method measures the particle velocity within the in-plane slice, which may be any cylindric section of the particle. For the $\alpha$=4, rod-like particles, the outer edges of the rod are likely to be overrepresented—the total ensemble of available slices for a rod will include more cylindric sections of the rod "arms" than of the rod center. If the rods are tumbling and rotating within the flow, the "arms" will be experiencing a higher velocity than the center. Therefore, the slip velocity measured with our method may be higher than the slip velocity measured by a fully three-dimensional method. We attribute the observed higher slip velocity for this particle type to this two-dimensional sampling bias, in which the rod "arms" are overrepresented compared to the center.

In Figure 3.4, we see that scatter in the two lateral directions (x- and w-) is roughly symmetric about the one-to-one tracer limit. This symmetry can also be observed in the angle histograms of the slip direction, projected into two dimensions (Figure 3.6). The slip direction is defined as $\theta = \tan^{-1}(u_s/v_s)$, and measured from the horizontal in-plane axis (i.e., the x-axis). Figure 3.6 shows that particles tend to go downwards relative to the surrounding fluid (i.e. towards 270°) and not upwards relative to the surrounding fluid (towards 90°). The distribution is symmetric about 90°. The most neutrally buoyant set of particles (row 1 in Figure 3.6, where $\gamma_1$=1.003) shows a small but finite percentage of positive vertical slip, but this percentage decreases in the second row ($\gamma_2$=1.006). The scatter about the one-to-one limit is discussed further in Section 3.3.1.

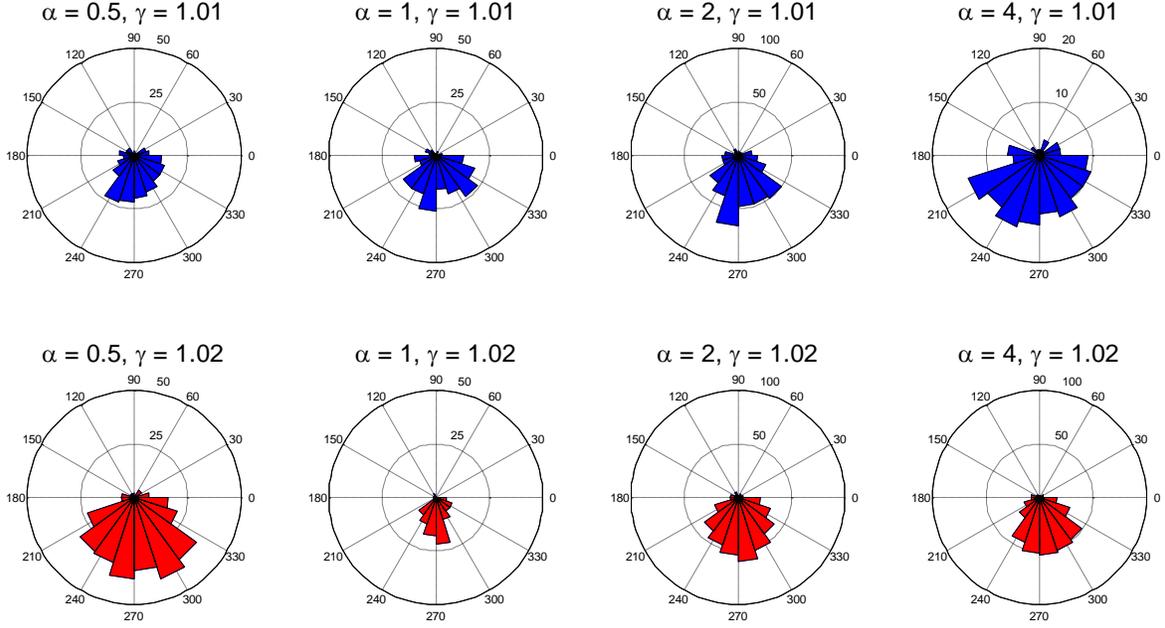

**Figure 3.6:** angle histogram plots of two-dimensional particle slip direction $\theta = \tan^{-1}(u_s/v_s)$. This quantity is the overall slip direction, which is projected into two dimensions X and Y to show symmetric behavior. Wedge area corresponds to the number of particles whose slip direction falls into that bin; total area is not normalized between plots.

### Discussion of particle and fluid timescales

In contrast to the quiescent settling velocity experiments, particles in turbulence may not display a preferential orientation relative to their slip direction. We hypothesized in Section 3.1.2 that the U-shaped curve observed in Figure 3.2 is caused by changes in the flow-normal area, which controls the bulk of the particle drag. In turbulence, this preferential orientation may or may not occur, and is dependent on the particle Stokes number[1], which we define as the ratio between the particle's Stokesian response timescale and the turbulent inertial-range timescale at the same size. If the particle response timescale is much longer than the fluid timescale (large $St$), particles should not display a preference for the drag-maximizing orientation (as they do in the quiescent case), because particle re-orientation time would exceed the lifetime of the immediately-surrounding eddy or flow structure. By contrast, if the particle response timescale is much smaller than the fluid timescale (small $St$), particles may display alignment and preferential orientation behavior.

The particles we examine are far outside of the creeping-flow regime, with Reynolds numbers on the order of $10^2$ (Table 3.). Therefore, we do not expect them to experience linear drag as in the Stokesian model (see discussion in Section 1.2.1, and calculations in Appendix A). However, as a point of reference, we can compute the Stokesian timescale for a sphere of equivalent volume to the cylindrical particles:

$$\tau_{St} = \frac{\rho_p d_p{}^2}{18\mu} = 5.3 \; s \tag{3.3}$$

where $\rho_p$ is the particle density, $d_p$ is the diameter of the sphere of equivalent volume, and $\mu$ is the fluid's dynamic viscosity. In the surrounding turbulent flow, an inertial-range timescale calculated at the same lengthscale gives a characteristic timescale of:

$$\tau_c = \tau_\eta \left(\frac{r_p}{\eta}\right)^{\frac{2}{3}} = 0.77s \tag{3.4}$$

where $\tau_\eta$ is the Kolmogorov timescale, $r_p$ is the radius of the sphere of equivalent volume, and $\eta$ is the Kolmogorov lengthscale. This gives us a particle Stokes number of $St = \tau_{St}/\tau_c = 6.88$.

---

[1] Note that the particle stokes number $St$ is often calculated as the ratio of the particle's Stokesian response time $\tau_{St}$ to the Kolmogorov timescale $\tau_\eta$, regardless of particle size. A more appropriate fluid timescale for large particles is the inertial-range timescale at the scale of the particle, $\tau_c$. We use this definition here and throughout this work, and caution the reader to be cognizant of the other definition.

The high ($St$ >1) particle Stokes number suggests that particle motion is occurring on a much longer timescale than the fluid motions at the same scale; particles therefore cannot adjust their orientation to present a different flow-normal cross-section, and no effects of preferential orientation should be observed. This implies that the shape effects we observed in Figure 3.5 are not based on the particle preferentially presenting a certain cross-section to the flow. Particle response time is slow compared to fluid fluctuation at the same scale, and so preferential orientation is unlikely.

The high Stokes number calculated above is a representative value of $St$ based on an inertial range timescale $\tau_c$ at the scale of the particle, and implies that particles cannot instantaneously align with local flow structures of comparable size to the particle. This does not preclude the possibility of particles aligning with very large vortex structures, with correspondingly larger values of $\tau_c$. In these vortex structures, particle $St$ would be smaller and alignment and/or preferential orientation could potentially occur.

### 3.2.3 Comparison between quiescent settling and vertical slip

It is clear that the average particle slip velocity in turbulence (that is, $\langle v_s \rangle$) is reduced with respect to $v_q$ by 40-60% (Table 3.3). This is interesting, given that maximum reductions in settling velocity for inertial particles have been previously observed to be on the order of 30-40% (Nielsen 2007). This reduction was observed to be strongest for "moderate-intensity" turbulence, in which particles whose quiescent settling velocities $v_q$ were on the order of the turbulent rms fluid velocity $u_{rms}$. This is the case for our particles. However, the data presented in (Nielsen 2007) were primarily drawn from the experiments of Murray (Murray 1970) and Zeng (Zeng 2001). Murray's particles were 2mm in diameter, much smaller than our particles (though their particles' settling velocities were comparable to our particles' settling velocities). This suggests that a simple ratio of $v_q$ to $u_{rms}$ (in Nielsen's notation, a ratio of $w_0$ to $w'$) is insufficient to predict particle behavior. The inertial properties of our particles are drawn from both gravity and finite-size effects, which may not be adequately captured in the $v_q/u_{rms}$ ratio. We note also that for our particles, the overall slip magnitude $|\overline{\boldsymbol{u}}_s|$ is comparable to the quiescent settling velocity magnitude $|\boldsymbol{v}_q|$.

| γ = 1.003 | | | | | | |
|---|---|---|---|---|---|---|
| | $v_q$ [mm s⁻¹] | | $\langle v_s \rangle$ [mm s⁻¹] | | $\langle|\overline{\boldsymbol{u}}_s|\rangle$ [mm s⁻¹] | |
| α = 0.5 | 14.7 | *[14.4, 15.0]* | 5.6 | *[4.6, 6.5]* | 12.4 | *[11.4, 13.6]* |
| α = 1 | 13.4 | *[13.2, 13.6]* | 5.8 | *[5.0, 6.6]* | 12.2 | *[11.4, 13.0]* |
| α = 2 | 11.0 | *[10.8, 11.2]* | 5.7 | *[5.2, 6.2]* | 11.0 | *[10.5, 11.7]* |
| α = 4 | 9.8 | *[9.6, 10.0]* | 4.0 | *[3.3, 4.8]* | 10.0 | *[9.3, 10.8]* |
| γ = 1.006 | | | | | | |
| | $v_q$ [mm s⁻¹] | | $\langle v_s \rangle$ [mm s⁻¹] | | $\langle|\overline{\boldsymbol{u}}_s|\rangle$ [mm s⁻¹] | |
| α = 0.5 | 15.9 | *[15.5, 16.3]* | 7.5 | *[6.8, 8.1]* | 13.5 | *[12.8, 14.3]* |
| α = 1 | 24.8 | *[24.5, 25.1]* | 11.5 | *[10.2, 12.8]* | 16.6 | *[15.4, 17.8]* |
| α = 2 | 16.5 | *[16.2, 16.8]* | 8.3 | *[7.7, 9.0]* | 14.9 | *[14.2, 15.9]* |
| α = 4 | 16.4 | *[16.1, 16.7]* | 9.9 | *[9.2, 10.6]* | 16.7 | *[15.9, 17.6]* |

**Table 3.3: Comparison of quiescent settling velocity $v_q$, average mean slip velocity $\langle v_s \rangle$, and average overall slip magnitude $\langle|\overline{\boldsymbol{u}}_s|\rangle$. $\langle v_s \rangle$ is reduced by approximately $40-60\%$ relative to $v_q$.**

Mechanisms for the alteration of settling velocity in turbulence remain far from clear. In some cases, e.g. small particle "fast-tracking", settling velocity may be enhanced by turbulence. This occurs in very strong turbulence, for which $v_q \ll u_{rms}$ (Nielsen 2007). In moderate-intensity turbulence ($v_q \approx u_{rms}$), particles may experience vortex trapping, in which heavy particles preferentially sample upgoing flows (Tooby, Wick, and Isaacs 1977). However, this mechanism has been demonstrated only for particles which are small compared to the size of the vortex; finite-size effects, which are present in our case, may alter vortex-trapping for large particles.

Our particles are one to two orders of magnitude larger than the Kolmogorov scale of the ambient turbulence, so the velocity gradients they experience are nonlinear. Furthermore, their measured slip velocities yield a Reynolds number which is on the order of $10^2$. In this transitional regime, particles experience neither Stokesian drag (linearly proportional to velocity) nor Newtonian drag (proportional to the velocity squared), but something in between, as illustrated by Equation (3.1). Nonlinear drag has been shown to reduce the settling velocity (Stout, Arya, and Genikhovich 1995), but is estimated to have a very weak effect compared to vortex trapping (Nielsen 2007).

A third mechanism for settling reduction is "shear-lift loitering," in which particles falling through shear will invariably experience a lift force that pulls them into flow that more strongly opposes their motion (see schematic in Figure 3.7). However, this mechanism is an oversimplification—large particles in turbulence do not experience linear shear, and their response time is long enough that this mechanism may not play a major role in the settling velocity reduction that we observe. However, this is an interesting topic that has not yet been explored to the extent of either vortex-trapping or nonlinear drag, and could be a major factor in reduced-settling in small or medium-sized particles.

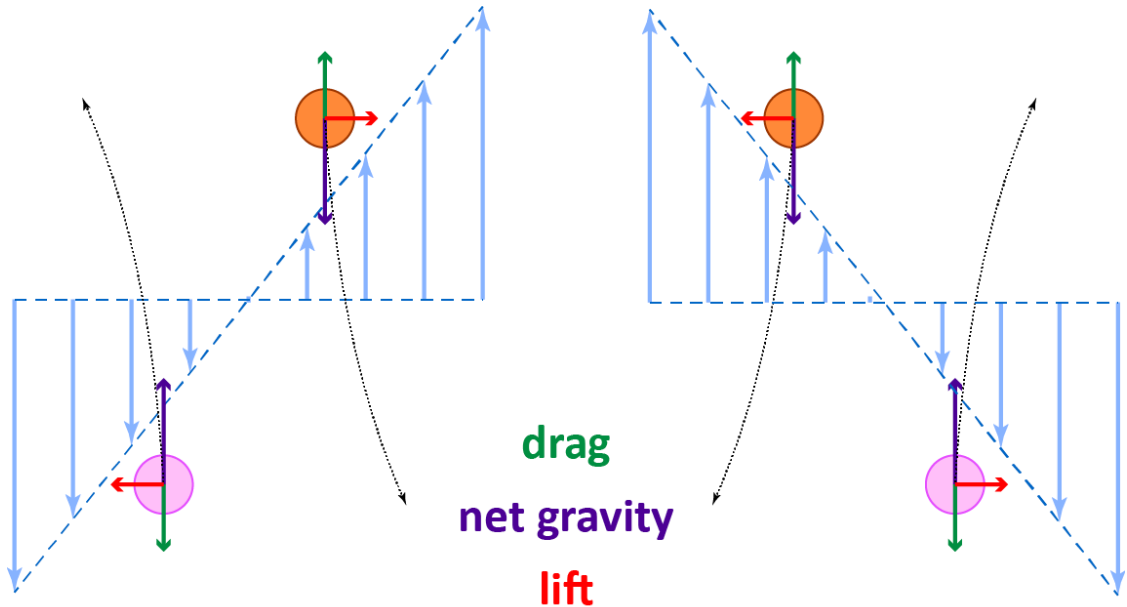

**Figure 3.7:** Bubbles, drops, or particles falling (orange) or rising (pink) through a region of linear shear will experience a net buoyancy/gravity force (purple), a drag force opposing their motion (green), and a lift force (red). This lift force will cause the trajectory to deviate from a straight path through the sheared region, and will always draw the particle towards the flow that is more strongly opposed to its motion.

## 3.3 Dependence of slip on local fluid velocity: instantaneous vs. bulk metrics

The slip velocity is often reported as a bulk statistic: particles at a given size, shape and density in a given turbulent flow will have only a single characteristic slip velocity. This is how we calculate the slip velocity in the previous section, and for our flow this certainly seems to be true. The linear model we describe in Figure 3.4 and the accompanying discussion show that $v_p$ is, on average, equal to $v_f$ plus some constant offset, which we assume to be $v_s$. However, this approach ignores subtleties in how the slip velocity may vary according to the instantaneous flow: for example, do particles exhibit systematically higher or lower instantaneous slip when subjected to the infrequent strong velocity fluctuations that are the hallmark of turbulent flows? Do negatively buoyant particles experience higher instantaneous slip when they are embedded in downdrafts vs. updrafts? Our data are ideal for investigating these and related questions.

### 3.3.1 Experimental data: effects of history on slip velocity measurements

For a qualitative framework, we return to Figure 3.4, which shows the particle vertical velocity $v_p$ vs. the fluid vertical velocity $v_f$. The upper half of this plot, where $v_p$ is positive, indicates upgoing particles, whereas the lower half indicates downgoing particles. The left half, where $v_f$ is negative, indicates particles which are experiencing downdrafts; in the right half, particles are experiencing updrafts. There are therefore four behavioral regimes for particles in flow: upgoing particles in updrafts, downgoing particles in updrafts, downgoing particles in downdrafts, and upgoing particles in downdrafts (we will use this terminology from this point on). This is schematically illustrated in Figure 3.8. It is not immediately apparent whether all four of these situations will result in statistically similar slip.

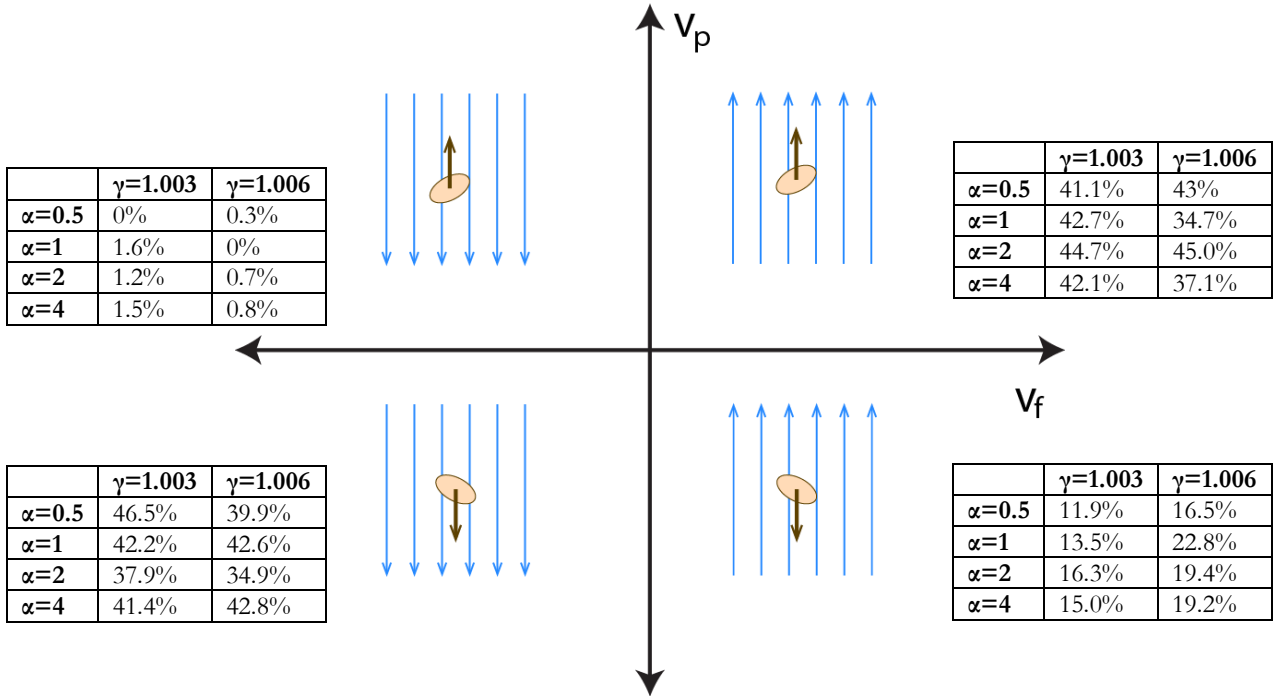

| | $\gamma$=1.003 | $\gamma$=1.006 |
|---|---|---|
| $\alpha$=0.5 | 0% | 0.3% |
| $\alpha$=1 | 1.6% | 0% |
| $\alpha$=2 | 1.2% | 0.7% |
| $\alpha$=4 | 1.5% | 0.8% |

| | $\gamma$=1.003 | $\gamma$=1.006 |
|---|---|---|
| $\alpha$=0.5 | 41.1% | 43% |
| $\alpha$=1 | 42.7% | 34.7% |
| $\alpha$=2 | 44.7% | 45.0% |
| $\alpha$=4 | 42.1% | 37.1% |

| | $\gamma$=1.003 | $\gamma$=1.006 |
|---|---|---|
| $\alpha$=0.5 | 46.5% | 39.9% |
| $\alpha$=1 | 42.2% | 42.6% |
| $\alpha$=2 | 37.9% | 34.9% |
| $\alpha$=4 | 41.4% | 42.8% |

| | $\gamma$=1.003 | $\gamma$=1.006 |
|---|---|---|
| $\alpha$=0.5 | 11.9% | 16.5% |
| $\alpha$=1 | 13.5% | 22.8% |
| $\alpha$=2 | 16.3% | 19.4% |
| $\alpha$=4 | 15.0% | 19.2% |

**Figure 3.8: Qualitative depiction of the four "behavioral quadrants" (clockwise from upper right): I) upgoing particles in updrafts, II) downgoing particles in updrafts, III) downgoing particles in downdrafts, and IV) upgoing particles in downdrafts. Tables show percentage of data lying in each behavioral quadrant for each type of particle (two different densities, four different aspect ratios).**

Because our particles are slightly negatively buoyant, particles which pass through our measurement window are more likely to be travelling downward than upward. This is borne out by the tables shown in Figure 3.8, which show that approximately 80% of measurements are relatively evenly distributed between quadrants I and III, with approximately 20% of measurements in quadrant II, and virtually no measurements in quadrant IV. The two updraft quadrants therefore represent approximately 60% of the data: we measure more particles in updrafts than particles in downdrafts. Though the vertical fluid velocities in the measurement window average to zero, the vertical velocities we *record* are—on average—positive.

Because we only record measurements for which a particle is present in the image window, we are more likely to observe updrafts than downdrafts. This is not reflective of the actual velocity distribution within the tank, which is Gaussian and symmetric about zero. This could indicate a preferential concentration effect, in which negatively-buoyant particles are more likely to be found in updrafts than in downdrafts. This is consistent with vortex-trapping, discussed in Section 1.2.1 as a potential mechanism for re-suspending particles in turbulence: large, negatively-buoyant particles are routed into the part of the flow that most opposes their motion, which leads to a reduction of the settling velocity.

However, the data do not allow us to conclusively determine that vortex-trapping or preferential concentration is responsible for the asymmetric fluid distribution that we observe. To definitively establish that vortex-trapping is occurring, we would need a "zoomed-out" view in which we could observe a single particle for a long period of time. We must therefore consider the possibility that the asymmetric distribution of $\overline{\boldsymbol{u}}_f$ which we observe is a result of some experimental or sampling bias. Important questions arise: does this bias affect our overall measurements of slip? If we were truly sampling particles at random (i.e., if our measurements of $\overline{\boldsymbol{u}}_f$ followed a Gaussian distribution), would we calculate the same average slip?

Our data take the form of a set of simultaneous measurements of $\overline{\boldsymbol{u}}_f$ and $\overline{\boldsymbol{u}}_p$, which may be broken down into their components $\{u_f, v_f, w_f\}$ and $\{u_p, v_p, w_p\}$. We know that $\overline{\boldsymbol{u}}_p$ is a function of $\overline{\boldsymbol{u}}_f$, and that the dependence appears to be linear (Figure 3.4). When $v_p$ is linearly regressed to $v_f$, we observe high values of $R^2$—above 0.88 in every case. However, our observations of $v_f$ are weighted towards upgoing flow. This may be a sampling bias (i.e., particles are preferentially concentrated in updrafts, therefore we observe more updrafts), or it may be some kind of experimental limitation (e.g., particles passing through our image window *are* more likely to occur in updrafts, but perhaps this is because our tank has finite size). Whatever the cause, it is illustrative to construct a linear fit in which our observations of $v_f$ are weighted according to the normal distribution that we expect in non-particle-laden flow (Figure 3.9).

We see that the linear model still appears to be a good description for $v_p = v_p(v_f)$, even when the data are weighted such that $v_f$ follows a normal distribution. The $R^2$ values do not change appreciably, as seen in Table 3.4. This indicates that even though our measurements skew toward positive values of $v_f$, our calculation of the average slip velocity is not affected by this bias. This also provides further support for the model of $v_p$ as a linear function of $v_f$: measurements taken in any region of the $v_f$-$v_p$ parameter space would yield the same one-to-one linear fit with the same y-intercept (and therefore the same average slip $\langle v_s \rangle$).

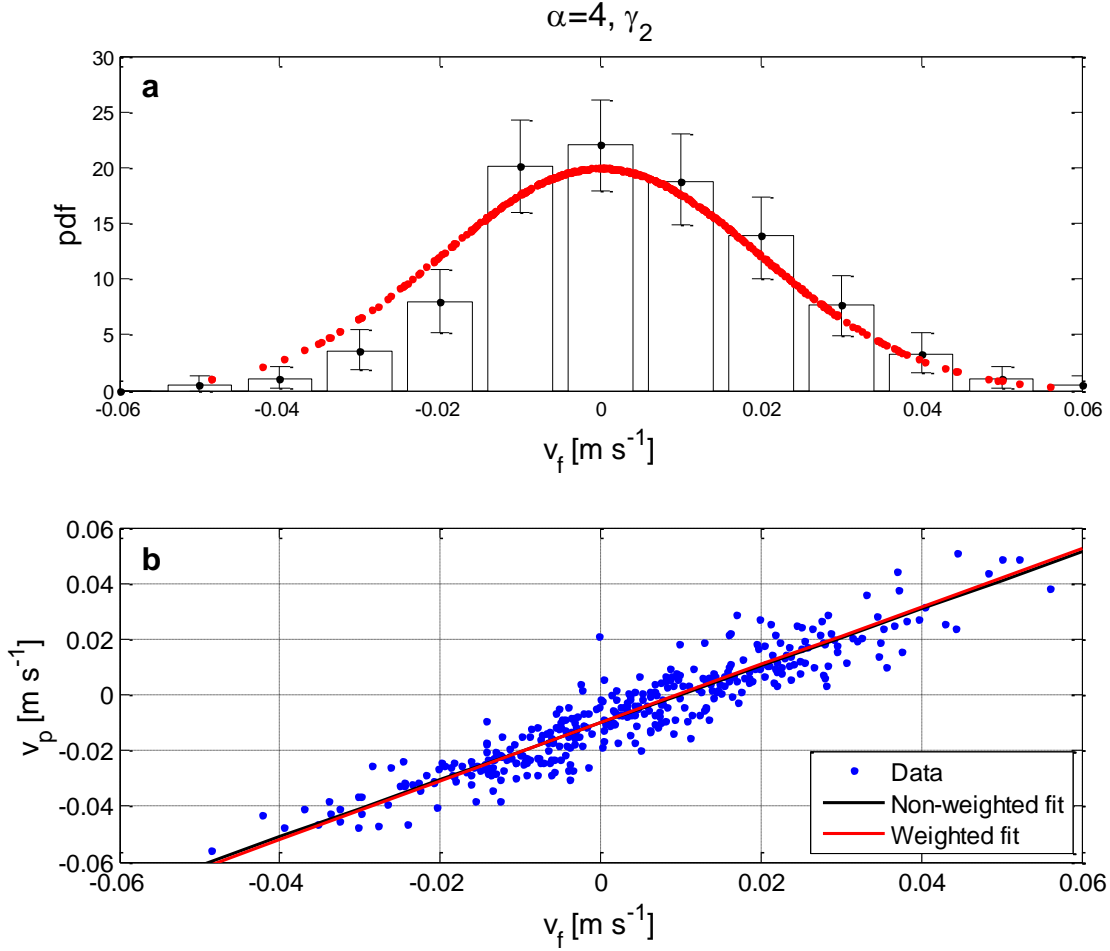

**Figure 3.9:** (a) Measured, asymmetric distribution of $v_f$ (bars) together with the Gaussian weighting function assigned to each measurement of $v_f$ (red dots), and (b) weighted vs. non-weighted linear fit of $v_p$ regressed to $v_f$. Shown is a single representative case, for which $\alpha$=4 and $\gamma$=2 (the same data are displayed in Figure 3.4).

| | $\gamma$=1.003 | | | | $\gamma$=1.006 | | | |
|---|---|---|---|---|---|---|---|---|
| | Fit intercept $\approx \langle v_s \rangle$ | | R$^2$ | | Fit intercept $\approx \langle v_s \rangle$ | | R$^2$ | |
| $\alpha$ | *Weighted [mm s$^{-1}$]* | *Unweighted [mm s$^{-1}$]* | *Weighted* | *Unweighted* | *Weighted [mm s$^{-1}$]* | *Unweighted [mm s$^{-1}$]* | *Weighted* | *Unweighted* |
| **0.5** | -5.8 | -5.7 | 0.89 | 0.89 | -7.7 | -7.5 | 0.90 | 0.90 |
| **1** | -6.0 | -5.9 | 0.93 | 0.93 | -11.6 | -11.2 | 0.88 | 0.90 |
| **2** | -5.7 | -5.6 | 0.88 | 0.88 | -8.3 | -8.2 | 0.89 | 0.89 |
| **4** | -4.4 | -4.1 | 0.94 | 0.94 | -10.0 | -10.0 | 0.88 | 0.88 |

**Table 3.4: Fit parameters of weighted vs. nonweighted linear fit, as seen in Figure 3.9.**

To qualitatively illustrate this point, we divide the vertical fluid velocities $v_f$ into four bins, centered at -0.03 m s$^{-1}$, -0.01 m s$^{-1}$, 0.01 m s$^{-1}$, and 0.03 m s$^{-1}$. The bins have equal width, so that e.g. the third bin goes from 0 to 0.02 m/s. For $v_f$, these bins may be qualitatively considered as "strong downdrafts", "weak downdrafts", "weak updrafts," and "strong updrafts," respectively. The results, across particle shape and particle density, are displayed in Figure 3.10.

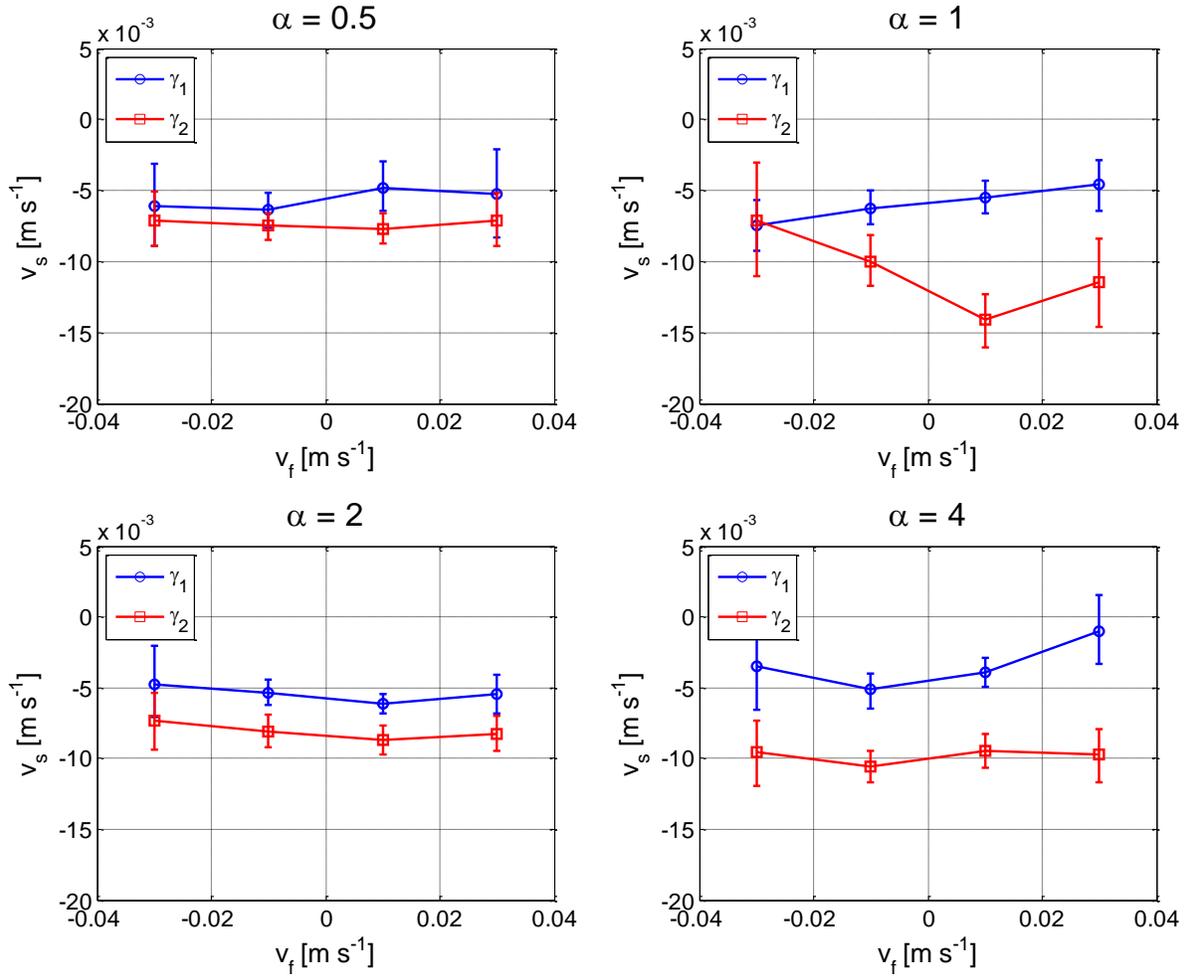

**Figure 3.10:** Fluid velocity $v_f$, binned into strong and weak updrafts and downdrafts, vs. vertical slip velocity $v_s$. Slip velocity becomes more positive when fluid velocity becomes more extreme. Errorbars represent 95% confidence intervals obtained via bootstrapping with 1000 replicates.

The slip velocity as it is defined in Equation (3.2) is dependent on both the fluid velocity $\vec{\boldsymbol{u}}_f$ and the particle velocity $\vec{\boldsymbol{u}}_p$. In Figure 3.10, we observe only no significant dependence of $v_s$ on the specific value of $v_f$: slip is statistically constant across the entire spectrum of fluid velocities. This supports the one-to-one linear model, which shows $v_s$ as the offset from the tracer limit which is constant for all $v_f$. However, in some datasets there is a slight suggestion of a U-shaped curve, in which slip is more positive (in most cases) at more extreme fluid velocities—that is, in strong updrafts or downdrafts, $\vec{\boldsymbol{u}}_p$ is closer to $\vec{\boldsymbol{u}}_f$.

This makes sense due to the nature of turbulence. In strong updrafts or downdrafts, the particle is likely to have been in this flow structure for some length of time (since higher-velocity flow structures in turbulence are longer-lived). This means that on average, the flow has had time to overcome the inertia of the particle, and the particle is traveling with the flow (save for some small constant negative offset due to buoyancy forces). In other words: we know that in turbulent flow, the velocity, size, and duration of fluid motions vary together. Therefore, when particles experience an extreme fluid velocity, they are likely in a larger and longer-lived vortex structure. The particle is then smaller compared to the scale of the ambient flow, giving it a lower Stokes number and more tracer-like properties. We therefore expect to see lower slip velocities at more extreme fluid velocities. And indeed, this is what we see in Figure 3.10.

A similar approach, in which $v_s$ is regressed to $v_p$, shows an apparent correlation of $v_s$ with $v_p$ (Figure 3.11). This apparent correlation is the result of two separate processes: first, the presence of random noise; second, and much more importantly, the effects of the particle's history—that is, the fluid flow along the particle's trajectory before it entered the measurement plane. The apparent correlation between $v_s$ and $v_p$ may be understood as a simple consequence of our analysis method. We understand $v_p$ to be a linear function of $v_f$, with a coefficient of approximately one, such that $v_p = v_f + b$ where $b$ is some constant offset. However, in reality, our measurements of $v_p$ contain some noise such that $v_p = v_f + b + \varepsilon$ (where we assume that $\varepsilon$ is random and normally distributed). The slip velocity $v_s$ is defined as $v_s \equiv v_p - v_f = b + \varepsilon$. So, when we plot $v_p$ versus $v_s$, we are actually plotting $(v_f + b + \varepsilon)$ versus $(b + \varepsilon)$. We assume $v_f$ to be independent of both $b$ (a constant) and $\varepsilon$ (random noise), but obviously $\varepsilon$ is perfectly correlated with itself. This noise results in the small positive slope of the regression line we see in Figure 3.11.

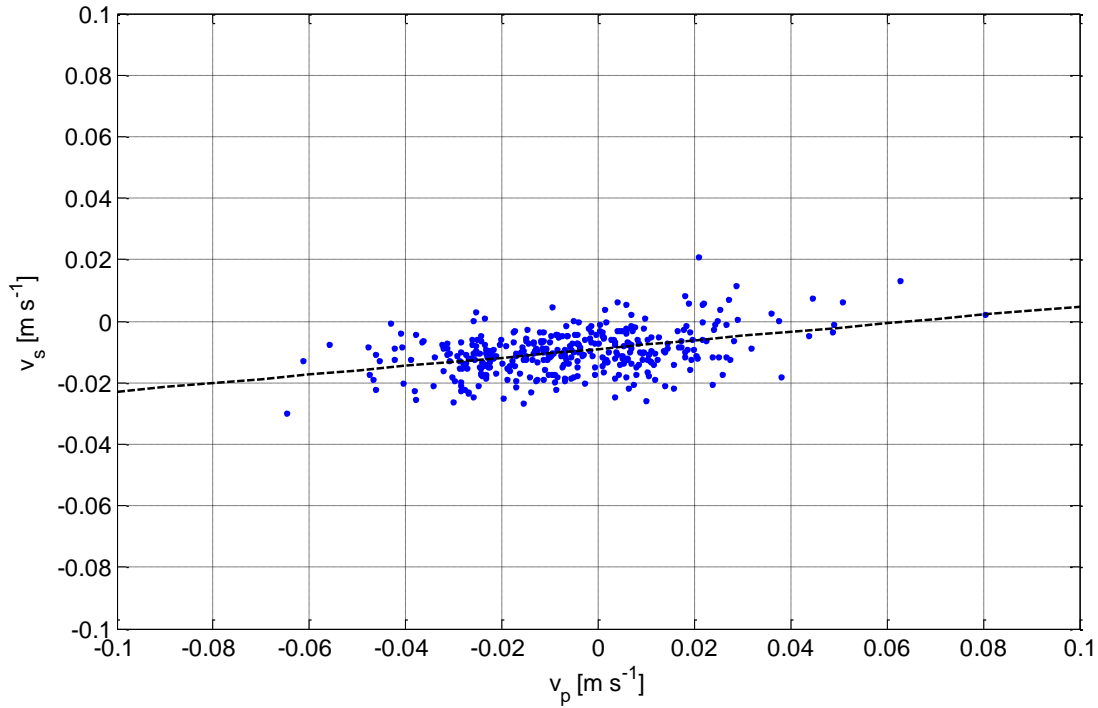

**Figure 3.11:** When we regress $v_s$ to $v_p$, an apparent trend emerges as a result of both random noise in the data and the effect of the history term on the particle's instantaneous slip velocity. Shown are data for particles with $\alpha=4$, $\gamma=1.006$ (identical to Figure 3.4).

In Figure 3.4, we observe a nontrivial amount of scatter around the linear-fit lines for all three components of $\vec{u}_f$ and $\vec{u}_p$. As discussed above, some part of this scatter may be attributed to random noise. However, even if our measurements were completely noise-free, there should still be some scatter around the linear fit due to the influence of the particle's history. An example is pictured in Figure 3.12. In Figure 3.12a, a particle is embedded within a very strong updraft, and has some positive vertical velocity $v_p$. It has been in this updraft for a very long time, and so its velocity is very close to the fluid velocity $v_f$ and there is very little slip. In the next instant, the particle is ejected from the small updraft into a weak downdraft. Its velocity $v_p$ is unchanged, since the particle's inertia keeps it from immediately adjusting to the change in $v_f$. This particle would have an extremely large and positive instantaneous slip velocity $v_s$. However, if a particle is embedded in a very strong downdraft, then enters the same weak downdraft (Figure 3.12b), the instantaneous value of $v_s$ would be slightly negative. Therefore, two identical measurements of $v_f$ would result in two very different measurements of $v_s$. This results in some, if not most, of the scatter about the regression line.

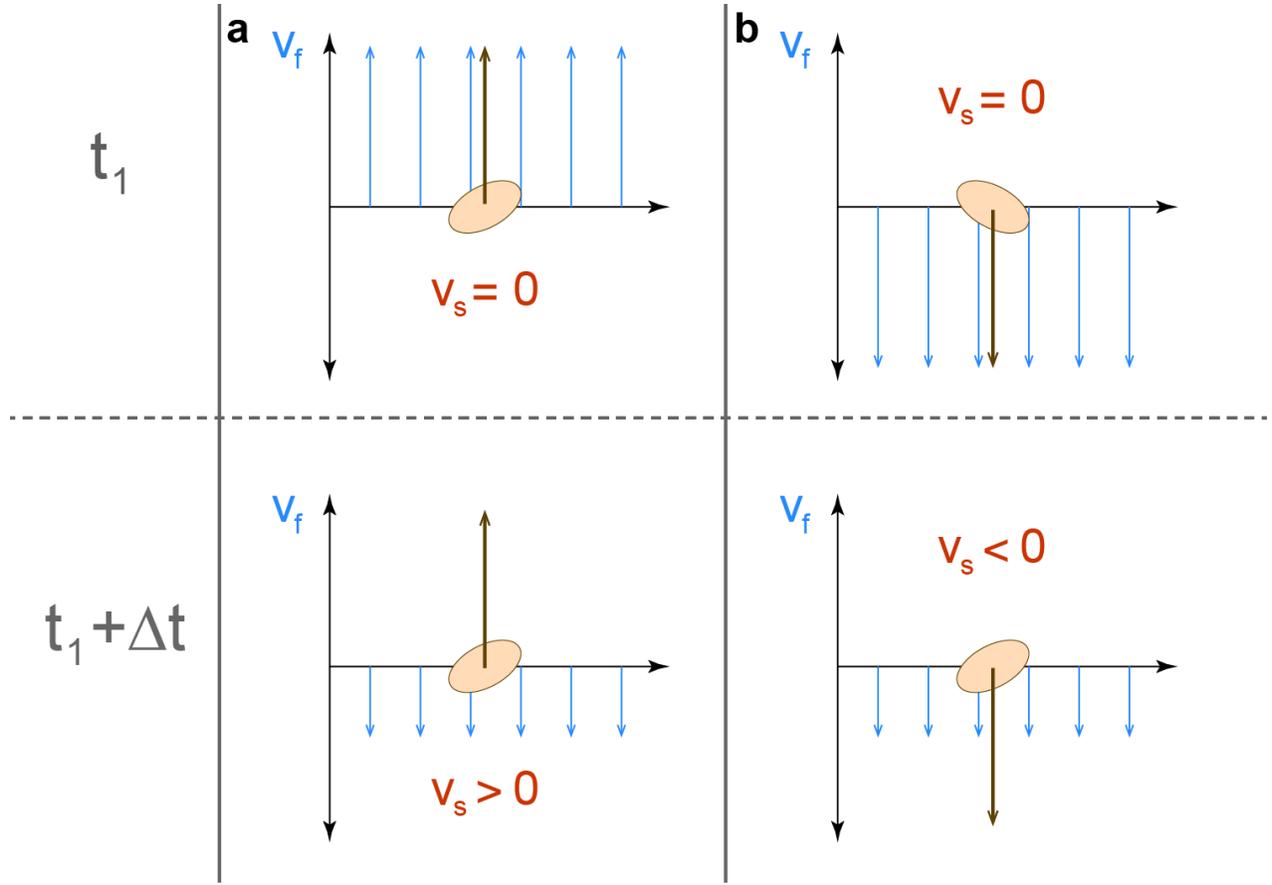

**Figure 3.12:** Schematic illustrating the importance of particle history in determining the instantaneous slip. (a) shows a particle which transitions from a strong updraft at time $t_1$ to a weak downdraft at time $t_1+\Delta t$. (b) shows a particle which transitions from a strong downdraft at time $t_1$ to a weak downdraft at time $t_1+\Delta t$. At the later time, the two particles are immersed in identical local flow conditions, but have very different slip velocities.

In our illustration, the noise term $\varepsilon$ is not only accounting for measurement error, but all the effects of the particle's history. Using our experimental method, it is impossible to separate the two processes. This is especially true because we expect both the measurement noise and the history effect "noise" to follow a normal distribution. Since our flow is homogeneous and isotropic turbulence, all three components of $\vec{u}_f$ should have identical normal distributions, and particles should encounter all possible values of $u_f$, $v_f$, and $w_f$ according to these distributions. Scenarios such as the ones depicted in Figure 3.12a and Figure 3.12b must therefore occur with equal probability, resulting in symmetric scatter about the one-to-one line.

To quantify the scatter, $\varepsilon$, we can examine the residuals about the best-fit line. The "spread" of the data about the best-fit line gives us a qualitative estimate of the influence of the history term, with more spread signifying a greater dependence of $\vec{u}_s$ on the particle history. We define $\sigma$ as the variance of the residuals about the best fit line. The slope $m$, intercept $b$, and residual-variance $\sigma$ are recorded in Table 3.5. We note that for all particle types, the slope of the best-fit line is approximately one (to within 95% confidence), and that the intercept of this line is approximately equal to the ensemble-averaged slip velocity $\langle v_s \rangle$ in every case (Table 3.2). The residual-variance $\sigma$ is generally larger for particles with $\gamma = 1.006$ than for particles with $\gamma = 1.003$. This tentatively points to the importance of inertia in the history term for large particles. However, in order to adequately capture the full distribution of possible particle histories, we would need a much larger sample size. This is an interesting topic that deserves further study.

| | $\gamma = 1.003$ | | | $\gamma = 1.006$ | | |
|---|---|---|---|---|---|---|
| $\alpha$ | m | b $\approx \langle v_s \rangle$ [m s$^{-1}$] $\cdot$ 10$^{-2}$ | $\sigma$ [m s$^{-1}$] $\cdot$ 10$^{-4}$ | m | b $\approx \langle v_s \rangle$ [m s$^{-1}$] $\cdot$ 10$^{-2}$ | $\sigma$ [m s$^{-1}$] $\cdot$ 10$^{-4}$ |
| 0.5 | 1.03 [0.96, 1.09] | -0.57 [-0.66, -0.47] | 0.45 [0.34, 0.57] | 1.00 [0.97, 1.04] | -0.75 [-0.81, -0.69] | 0.37 [0.29, 0.45] |
| 1 | 1.04 [1.00, 1.08] | -0.59 [-0.66, -0.52] | 0.27 [0.21, 0.33] | 0.94 [0.85, 1.01] | -1.12 [-1.25, -1.00] | 0.40 [0.26, 0.55] |
| 2 | 0.99 [0.95, 1.02] | -0.56 [-0.62, -0.51] | 0.27 [0.23, 0.32] | 0.98 [0.95, 1.01] | -0.82 [-0.88, -0.76] | 0.42 [0.34, 0.49] |
| 4 | 1.05 [1.00, 1.10] | -0.41 [-0.49, -0.34] | 0.18 [0.13, 0.24] | 1.02 [0.98, 1.06] | -1.0 [-1.07, -0.93] | 0.49 [0.41, 0.58] |

**Table 3.5: Fit parameters for linear fits: $v_p = m \cdot v_f + b + \varepsilon(\sigma)$, where $\varepsilon$ is the error arising from measurement noise and history effects, and $\sigma$ is the variance of the residual (representing the degree of scatter about the best-fit line).**

In summary, we have observed the following functional dependence for the particle velocity $\vec{u}_p$:

$$\vec{u}_p = \vec{u}_f + \vec{u}_s + \varepsilon, \quad (3.5)$$

where $\vec{u}_s \approx \langle v_s \rangle \hat{\jmath}$, a constant vertical offset. Two features of this model are remarkable: first, that $\vec{u}_f$ and $\vec{u}_p$ follow a one-to-one relationship, and second, that the slip velocity $\vec{u}_s$ is not a function of the instantaneous fluid flow (Figure 3.10). For the latter, it is clear that more investigation is needed to understand the physics of the slip velocity—this study has investigated only one aspect of many. $\vec{u}_s$ is likely to depend on geometric/physical variables (e.g. particle shape, size, and specific gravity), bulk-flow variables and their relation to particle properties (e.g. $v_0/u_T$ and $L_p/L_f$, where $L_p$ is the particle lengthscale and $L_f$ is a representative fluid lengthscale). The slip velocity will also depend heavily on the particle history.

The first feature of the linear parametrization (the one-to-one relationship between $\vec{u}_f$ and $\vec{u}_p$) deserves further investigation. Is this dependence unique to turbulence, or does it occur in other unsteady flows? To further explore this question, it is illustrative to construct a one-dimensional model of a freely-suspended particle in a simple time-varying flow. This will allow us to compute the slip velocity of a particle in a simple non-turbulent flow, to see if it behaves in qualitatively similar ways as our experimentally-measured slip velocities.

### 3.3.2 One-dimensional model of a sphere in a simple unsteady flow

**Equation of motion**

As outlined in Chapter 2, the analytic equation for particle motion in unsteady flow is known in full only for small spherical particles (Maxey and Riley 1983):

$$
\begin{aligned}
m_p \frac{d\boldsymbol{u}_p}{dt} = (m_p - m_f)\boldsymbol{g} &+ m_f \frac{D\boldsymbol{u}_f}{Dt} - \frac{1}{2}m_f \frac{d}{dt}\left(\boldsymbol{u}_p - \boldsymbol{u}_f - \frac{1}{10}R^2\nabla^2\boldsymbol{u}_f\right) \\
&- 6\pi R\mu\left(\boldsymbol{u}_p - \boldsymbol{u}_f\right) - \frac{1}{6}R^2\nabla^2\boldsymbol{u}_f \\
&- 6\pi R^2\mu \int_0^t \left(\frac{\frac{d}{d\tau}\left(\boldsymbol{u}_p(\tau) - \boldsymbol{u}_f(\tau) - \frac{1}{6}R^2\nabla^2\boldsymbol{u}_f\right)}{\left(\pi\nu(t-\tau)\right)^{\frac{1}{2}}}\right) d\tau
\end{aligned}
\tag{3.6}
$$

where $R$ equals the sphere's radius, $m_p$ is the particle's mass, $m_f$ is the mass of an equivalent volume of fluid, $\boldsymbol{u}_p = \boldsymbol{u}_p(t)$ equals the particle velocity and $\boldsymbol{u}_f = \boldsymbol{u}_f(t)$ is the fluid velocity, which in the equation above is always evaluated at the position of the particle center. The surface and body forces on the sphere are accounted for on the right-hand side of the equation, where the terms are (1) buoyancy, (2) pressure gradients, (3) acceleration reaction (added mass), (4) steady (Stokes) drag, and (5) the history (Basset) term, which includes the effects of unsteadiness (note that terms (3), (4), and (5) also include the Faxén corrections for finite-size effects). However, the underlying assumptions of this equation—that the particles are spherical, and they experience only Stokes flow—are limiting.

For large inertial particles, the analytic equation of motion is not known. However, following the Maxey-Riley equation above, we can qualitatively define forces on the particle and examine the dependencies of those forces on particle velocity and acceleration (along with environmental variables such as fluid density and viscosity):

$$
\begin{aligned}
\sum \boldsymbol{F} = m_p \frac{d\boldsymbol{u}_p}{dt} \\
= \boldsymbol{F}_B(V, \boldsymbol{g}, \rho_f, \rho_p) &+ \boldsymbol{F}_{PG}(\rho_f, \boldsymbol{u}_f, \dot{\boldsymbol{u}}_f) + \boldsymbol{F}_{AR}\left(V, \rho_f, \frac{D\boldsymbol{u}_s}{Dt}\right) \\
&+ \boldsymbol{F}_D(\boldsymbol{u}_s, L, \mu, C_D, \rho_f) + \boldsymbol{F}_H(t, \boldsymbol{u}_s, \dots)
\end{aligned}
\tag{3.7}
$$

$\boldsymbol{F}_B$ is the (constant) buoyancy force depending on the particle volume $V$, fluid density $\rho_f$, particle density $\rho_p$, and the acceleration of gravity $\boldsymbol{g}$. $\boldsymbol{F}_{PG}$ is the force due to fluid pressure gradients and is dependent on $\rho_f$, along with the fluid velocity $\boldsymbol{u}_f$ and its accelerations. $\boldsymbol{F}_{AR}$ is the acceleration reaction force, also known as the "added mass" force, which is dependent on $V$, $\rho_f$, and the material derivative of the relative velocity $\boldsymbol{u}_s$. $\boldsymbol{F}_D$ is the steady drag force, dependent on the relative particle velocity $\boldsymbol{u}_s$, a characteristic lengthscale $L$, a shape-dependent drag coefficient $C_D$, fluid viscosity $\mu$, and $\rho_f$. $\boldsymbol{F}_H$ is the history force, including the effects of unsteadiness; it is dependent on a variety of complex factors, including the elapsed time t and the slip velocity $\boldsymbol{u}_s$, integrated along the previous particle trajectory.

The functional dependency of these terms, especially the history term, is not necessarily known for large inertial particles. This is particularly true for particles at intermediate Reynolds number, which fall between the Newton and Stokes drag regimes (see Appendix A). Further complications arise for nonspherical particles, since the dependence of particle motion on particle shape is not known, nor is it likely to be easily captured analytically. Without an analytical equation of motion, computations of particle motion are possible but difficult. To solve for particle motion using an Eulerian-Lagrangian framework, such as those discussed in Chapter 1, one must solve both Newton's equations and the Navier-Stokes equations at each grid point on the surface of the particle. Because particle boundaries may not fall exactly on the grid, an immersed boundary method is needed (alternatively, a Lagrangian method such as Lattice-Bolzmann may be used for both the fluid and particle phase). This arduous process must be repeated for each time step for both the particle and the fluid, taking into account the intricate coupling between the fluid forcing on the particle and the particle forcing on the fluid.

For complex flows, such as turbulence, numerical simulation becomes very challenging and demands vast computing resources. However, we can construct an analytical, reduced-order model of a large sphere in a prescribed flow, including only the steady drag, buoyancy, and acceleration reaction forces. This simplified case yields an ordinary differential equation that can be solved numerically. We can then use this model to gain insight about the behavior of the slip velocity $v_s$ and its functional dependence on the fluid velocity $v_f$.

**Model parameters**

We consider a sphere of radius $R$ and density $\rho_p$ falling through a quiescent fluid of density $\rho_f$. To initialize the model, we calculate the sphere's quiescent settling velocity $v_q$, which is determined by the steady-state balance between the vertical steady drag $F_D$ and gravitational/buoyancy forces $F_B$:

$$\sum F = F_D + F_B = -\frac{1}{2} C_D \rho_f \cdot \pi R^2 \cdot |v_q| v_q + \frac{4}{3} \pi R^3 (\rho_f - \rho_p) g = 0 \qquad (3.8)$$

where $C_D$ is the drag coefficient on a sphere ($C_D$=0.47 for a rough sphere at high Reynolds number) and $g$ is the acceleration of gravity ($\boldsymbol{g}$ = 9.81 m s⁻¹). Here we use the Newtonian formulation of steady drag, which is not strictly accurate for our intermediate-Reynolds-number particles, but is adequate for a simple model (note also the shift from boldface $\vec{\boldsymbol{u}}$, denoting three-dimensional vector quantities, to plain text $v$, which is a scalar since the model is one-dimensional). The balance between buoyancy and steady drag holds as long as the falling (or rising) sphere is in equilibrium. At $t = 0$, we "turn on" a fluid forcing $v_f(t)$ such that the sphere is no longer in equilibrium but is accelerating, with a velocity $v_p(t = 0) = v_q$. This introduces the acceleration reaction force, $F_{AR}$:

$$F_{AR} = -\frac{1}{2}\left(\frac{4}{3}\pi R^3\right)\rho_f \dot{v}_s \qquad (3.9)$$

where $v_s$ is the slip velocity, defined as $v_s(t) \equiv v_p(t) - v_f(t)$. The steady drag term will also depend on the slip velocity, leaving an overall force balance of

$$-\frac{1}{2}|v_s|v_s A\rho_f C_D + V(\rho_f - \rho_p)g - \frac{1}{2}V\rho_f \dot{v}_s = \rho_p V \dot{v}_p \qquad (3.10)$$

where $V$ and $A$ have been used to denote the volume and cross-sectional area of the sphere. The overall particle acceleration is equal to $\dot{v}_p = \dot{v}_f + \dot{v}_s$; therefore, if $v_f$ is prescribed, the equation is a nonlinear first-order differential equation in $v_s$.

In our experiments, conducted in homogeneous, isotropic flow, $v_p$ was well-described as a linear function of $v_f$:

$$v_p(\vec{x},t) = v_f(\vec{x},t) + v_s + \varepsilon\left(\int v_p(\vec{x},t),\ldots\right) \qquad (3.11)$$

where $v_s$ is a constant, and $\varepsilon$ contains all the effects of particle history. It is an open question whether Equation (3.11) adequately describes other unsteady (time-varying) flows. The assumption that $v_s$ is statistically constant across all values of $v_f$ is a strong one, and may not hold for all unsteady flows.

With this simple model, based on Equation (3.10), we will explore the functional dependence of $v_s$ upon $v_f$. We will then compare the simplified one-dimensional case to our experimental data to see which effects are inherent results of unsteady flow, and which effects may be related to the phenomenological properties of turbulence (e.g. vortex stretching and/or the alignment of particles with coherent structures).

**Results and comparison to experimental data**

We set the fluid forcing to a simple oscillatory flow, described by:

$$v_f = u_T \sin\left(\frac{2\pi}{\tau_L}t\right) \qquad (3.12)$$

where $u_T = 0.02$ m s$^{-1}$ is the turbulent velocity scale in the experimental tank, and $\tau_L = 5$ s is the approximate eddy turnover time. In this way, our oscillatory model flow is somewhat representative of typical flow structures experienced by the particles in the experimental tank (though in the experiment, the flow structures are three-dimensional. We also match other environmental variables $(\rho_p, \rho_f, \text{etc})$ such that the quiescent settling velocity of the model sphere is the same as the quiescent settling velocity of the comparable experimental cylinder. However, we are unable to incorporate the effects of shape, or the more complex terms in the equation of motion as summarized in the previous sections. These terms, along with phenomenological effects such as alignment, may have large roles to play in determining the slip velocity of large aspherical particles in turbulence.

We solve Equation (3.10), using MATLAB's ode45 command, for the oscillating flow prescribed by Equation (3.12). The model is initialized with $v_f\big|_{t=0} = 0$ and $v_p\big|_{t=0} = v_q$, the still-water settling velocity. The model yields simultaneous measurements of $v_p(t)$, $v_f(t)$, and $v_s(t)$, along with a vector of time measurements $t$. We obtain a long time measurement (t = 500 s) and randomly sample this measurement (neglecting the initial transients, so as to mimic our experimental data). For comparison with the experimental data, we again choose cylinders with α = 4 and γ = 1.006, which have a quiescent settling velocity $v_q$ =-1.64 cm s$^{-1}$. We then plot the particle velocity $v_p$ versus fluid velocity $v_f$ (Figure 3.13a-b, analogous with Figure 3.4); the slip velocity $v_s$ versus the fluid velocity $v_f$ (Figure 3.13c-d, analogous to Figure 3.10); and the slip velocity $v_s$ versus the particle velocity $v_p$ (Figure 3.13e-f, analogous to Figure 3.11). Note that all three plots display exactly the same data, but this method of display is useful to illustrate the relationships between $v_s$, $v_f$, and $v_p$.

In Figure 3.13a, both the model (left) and the experimental data (right) show a positive correlation between the fluid and solid velocities, as they should. The oval shape of the model results is due to the model's periodicity, which sets up only one one-dimensional "eddy". A greater distribution of eddy sizes would result in a more scattered distribution. Interestingly, the linear fit seems to appropriately describe the oscillatory model as well as the experimental data.

We note that in the experiment, $v_p$ and $v_f$ experience an approximately one-to-one relationship (Figure 3.13b), but in the model, the slope of the same linear-fit line is around 1.5 (Figure 3.13a). This is our first important insight: a one-to-one linear relationship between $v_f$ and $v_p$ does not occur in a simple oscillatory flow. Our experimental observations are likely unique to turbulence and may arise from particle interactions with coherent structures. This is also a further indicator of the importance of the history term in the particle equation of motion; this term is missing in the model but present in the experiment.

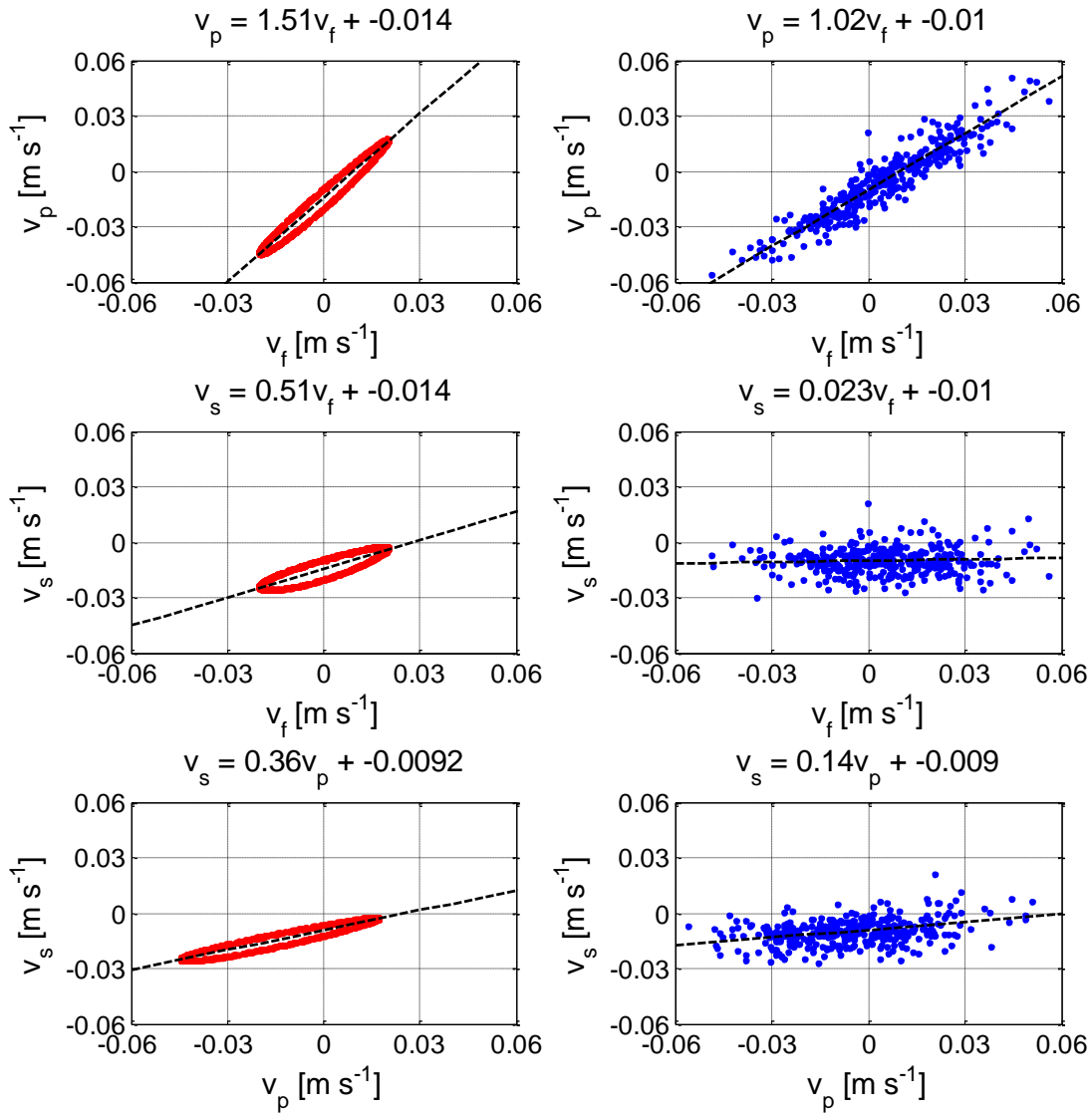

**Figure 3.13:** Model output (left column) compared with experimental data (right column). Oval shape of model data is due to oscillatory flow. (a-b) shows the fluid velocity $v_f$ plotted against the particle velocity $v_p$. (c-d) show slip velocity $v_s$ regressed to fluid velocity $v_f$. (e-f) show $v_s$ regressed to $v_p$.

When $v_p$ is regressed to $v_f$, a slope of exactly unity would indicate that there is no difference in slip, on average, between upgoing and downgoing particles and/or particles traveling in updrafts vs. downdrafts. This is true for our experimental data, and can also be seen in Figure 3.13d (as well as in Figure 3.10). In the oscillatory flow, this is not the case. In Figure 3.13c, we see more clearly that the slip velocity $v_s$ is strongly dependent upon $v_f$.

The correlation between $v_s$ and $v_p$ is small in our experiment (Figure 3.13f), and can be explained by the self-correlation of history effects and random noise (see Figure 3.11 and accompanying discussion). These self-correlation effects are also present in the model (Figure 3.13e), but much of the nonzero slope of the linear-fit line in this plot may be explained algebraically. If $v_p = m \cdot v_f + b$, and $v_s = v_p - v_f$, we can construct two linear models of $v_s$:

$$v_s = (1 - m)v_f + b \tag{3.13}$$

$$v_s = \left(\frac{1 - m}{m}\right)v_p + \frac{b}{m} \tag{3.14}$$

For Figure 3.13e-f, this gives the following expected best-fit parameters:

$$v_s = 0.34v_p + 0.01 \text{ (model)} \tag{3.15}$$

$$v_s = 0.02v_p - 0.01 \text{ (experiment)} \tag{3.16}$$

Though the intercept values of the expected best-fit parameters agree well with the actual best-fit parameters in Figure 3.13e-f, the slopes are slightly different due to the scatter about the line, which has self-correlated. The experimental data are more scattered than the model data, and so the self-correlation effect is stronger.

For both the experiment and the model, the quiescent settling velocity $v_q$ is equal to -1.64 cm s$^{-1}$. The reduced settling velocity, as measured by the intercept of the best-fit line of $v_p$ regressed to $v_f$, is -1.4 cm s$^{-1}$ for the model, and -1.0 cm s$^{-1}$ for the experiment. Therefore model is indeed experiencing reduced settling, which is well-known to occur in simple oscillatory flows (Field 1968); the model displays a weaker reduction than the experiment. In the oscillatory (model) case, $v_s$ is strongly dependent on $v_f$, because $v_f$ and $v_p$ do not have a one-to-one dependence. Across the entire range of sampled fluid velocities, the average slip velocity $\langle v_s \rangle$ is the same as the intercept of the best-fit line. However, if some experimental limitation had caused us to sample only one region of the $v_f$ parameter space, $\langle v_s \rangle$ would not be equal to the intercept value. For example, if only positive values of $v_f$ were recorded, the intercept of the best-fit line would remain relatively unchanged (at $b = -1.5$ cm s$^{-1}$), but the average slip velocity would be drastically different (at $\langle v_s \rangle = -0.76$ cm s$^{-1}$).

## Discussion of model and experimental results

We have compared two cases: 1) the slip velocity in several simple, one-dimensional unsteady flows around a large model sphere and 2) the slip velocity calculated from experiments with large inertial particles in isotropic turbulence. From the model, we see that we should not necessarily expect $v_s$ to be statistically uniform across $v_f$ and $v_p$. In the experimental case, $v_s$ is statistically uniform across $v_f$—not so in the model. In future studies, we should be careful not to take this relationship as the default state. For example, convective flows are often cellular, with regular patterns similar to our one-dimensional oscillating model. In such cases, the measured slip velocity may depend strongly on the ambient fluid flow, and so the measurement region must be chosen with care.

The one-dimensional model also shows us that the functional dependence of $v_p$ on $v_f$ is not necessarily a one-to-one linear relationship with a constant offset, even when most of the forces on the particle are accounted for. In the simple model case, the inclusion of buoyancy, steady drag, and acceleration reaction resulted in a roughly linear correlation between $v_p$ and $v_f$, but with a slope of approximately 1.5 (in contrast to the experimental data, which have a regressed slope of approximately 1). This is perhaps due to the presence of many different wavenumbers in turbulence, as opposed to just one in the model.

Our model is very simple and necessarily incomplete. As discussed earlier, large inertial particles are also influenced by the history term and pressure gradient term. These terms are of course included in the experimental data, but are absent from the model. Since velocity fluctuations on the scale of the particle are not very large, we assume that the pressure gradient term is less important than the history term, which also includes the effects of unsteadiness from the turbulent flow. We also note that in our experiments, particles are approximately thirty to sixty times the Kolmogorov lengthscale. The experimental particles are within the inertial subrange of the ambient turbulence, with a size comparable to (though slightly less than) intermediate eddies and vortex structures. This size allows for the possibility of alignment with flow vorticity, potentially impacting the effects of unsteadiness. These effects, in addition to those discussed above, are responsible for the slip velocities observed.

## 3.4   Settling: Conclusions

In this chapter, we have explored the effects of turbulence on large-particle settling velocity, and how the settling may be altered between quiescent and turbulent flow. We directly measured the still-water (quiescent) settling velocity $v_q$ for near-neutrally-buoyant cylindrical particles of four different aspect ratios α and two different specific gravities γ (as well as two spheroid shapes). Particles tended to settle in a drag-maximizing configuration, with rods falling broad-side and disks falling with their circular face downward; because of differences in the presenting cross-sectional area, particles with an aspect ratio of unity tended to have higher settling velocities than other shapes. Particle Reynolds numbers based on $v_q$ were on the order of $10^2$; therefore, the drag forces on the particles lie between the Stokesian and Newtonian regimes, and both the total surface area and presenting cross-sectional area are important in determining particle drag. In this Reynolds number regime, wake instabilities begin to form and may affect the forces on the falling particles.

Using refractive-index-matched stereoscopic particle image velocimetry (RIM-SPIV), we measure the velocities of freely-suspended particles simultaneously with the surrounding turbulent flow.

Using these data, we calculate the turbulent slip velocity $\vec{u}_s$, based on the average in-plane particle velocity $\vec{u}_p$ and the average fluid velocity in the neighborhood of the particle, $\vec{u}_f$. We find that the particle velocity seems to be linearly related to the fluid velocity, with a slip velocity that is constant across the spectrum of fluid velocities observed. In the vertical direction, this can be expressed as follows:

$$v_p = v_f + v_s + \varepsilon \tag{3.17}$$

where $v_s$ is constant with respect to $v_f$, and $\varepsilon$ represents both measurement noise and the effects of the particle's history. The history/noise term $\varepsilon$ is apparent in the scatter that surrounds an otherwise one-to-one relationship between $v_p$ and $v_f$. This one-to-one relationship does *not* necessarily hold for all unsteady flows, as shown by a simple one-dimensional model of a small sphere in a simple oscillating flow.

The close coupling between $v_p$ and $v_f$ signifies that though particle history plays a major role in determining slip, the instantaneous fluid velocity also has a strong influence. This is not obvious *a priori*; the effects of the particle's previous trajectory might have been so strong that $v_p$ would have little to no correlation with $v_f$. This may be the case for other particle-turbulence regimes, e.g. high-Stokes-number particles in very strong turbulence. However, in our case, $v_p$ is well-described by Equation (3.17).

Finally, we compare the average vertical turbulent slip velocity $\langle v_s \rangle$ and compare it to the measured still-water settling velocity $v_q$. We find that particle settling velocity is reduced by 40-60% in turbulence (relative to the quiescent case). This is a greater reduction than that which is previously observed in the literature. In our experiments, we are slightly more likely to measure particles which are situated in upgoing regions of the flow—our measurements of $v_f$ (conditioned on the presence of a particle in the measurement window) are 60% "updrafts" and 40% "downdrafts." This is evidence of vortex trapping, a phenomenon in which large particles preferentially concentrate in areas of the flow which oppose their gravity-driven motion. Other mechanisms which may also reduce the settling velocity include nonlinear drag effects and shear-lift loitering. Our slight bias towards updrafts does not affect the value of $\langle v_s \rangle$, providing further support that $v_p$ is a one-to-one linear function of $v_f$ and that $v_s$ is constant with respect to $v_f$.

# IV. Particle Rotation

Freely suspended particles in turbulence have six kinematic degrees of freedom. In the previous chapter, we considered the first three by examining particle translation via settling and slip. Here we investigate the latter three degrees of freedom by measuring particle rotation. We first consider the rotation of very small particles in turbulence, whose analytic equation of motion is known. We then compare these results to experimental measurements of large particles in turbulence, and explore how particle rotation may vary with shape and density. Lastly, we consider how particle rotation is inherited from fluid vorticity, and how knowledge of the ambient flow may help to predict particle rotation.

## 4.1   Rotation of large and small nonspherical particles

We consider axisymmetric particles, both cylinders and spheroids, with an axis of symmetry (length $2c$) and two other axes of equal length ($2a = 2b$). We define the aspect ratio $\alpha \equiv c/a$, which is greater than one for rod-like particles and less than one for disk-like particles ($\alpha = 1$ for spheres, and for cylinders whose height is equal to their diameter). We also define a unit vector $\hat{\boldsymbol{n}}$ which is parallel to the axis of symmetry of the particle.

When freely suspended in turbulence, an axisymmetric particle will take on both a translational velocity $\vec{\boldsymbol{u_p}}$ and an angular velocity $\vec{\boldsymbol{\Omega}}$. We may then define two different rotation rates, one which is parallel and one which is perpendicular to the particle's axis of symmetry:

$$\vec{\boldsymbol{\Omega}}_\parallel = \hat{\boldsymbol{n}} \cdot \vec{\boldsymbol{\Omega}} \tag{4.1}$$

$$\vec{\boldsymbol{\Omega}}_\perp = \vec{\boldsymbol{\Omega}} \times \hat{\boldsymbol{n}} = \dot{\boldsymbol{n}} \tag{4.2}$$

We will refer to $\vec{\boldsymbol{\Omega}}_\parallel$ as the "spinning rate" and $\vec{\boldsymbol{\Omega}}_\perp$ as the "tumbling rate."[1] Note that the tumbling rate, $\vec{\boldsymbol{\Omega}}_\perp$, is the sum of the rotation rates about both of the two equatorial particle axes. If rotation was isotropically distributed about all particle axes, the tumbling rate would be twice the spinning rate. The tumbling and spinning rates are analogous to the terms "roll", "pitch", and "yaw" in aerodynamics and other fields. Our spinning rate $\vec{\boldsymbol{\Omega}}_\parallel$ (i.e., the component of $\vec{\boldsymbol{\Omega}}$ which is parallel to the axis of symmetry) is directly comparable to "roll". However, since our particles are axisymmetric, we cannot define two meaningfully different axes which are perpendicular to the axis of symmetry. Therefore, the tumbling rate takes into account both "pitch" and "yaw." In Section 4.1, we will explore the dynamics of $\vec{\boldsymbol{\Omega}}$, $\vec{\boldsymbol{\Omega}}_\parallel$ and $\vec{\boldsymbol{\Omega}}_\perp$, and their dependence on particle shape, buoyancy, and inertia.

---

[1] We note that in our construction, $\vec{\boldsymbol{\Omega}}_\parallel$ and $\vec{\boldsymbol{\Omega}}_\perp$ do not add up to $\vec{\boldsymbol{\Omega}}$. Our defined tumbling rate, $\vec{\boldsymbol{\Omega}}_\perp$, is mutually perpendicular to both $\vec{\boldsymbol{\Omega}}$ and $\hat{\boldsymbol{n}}$ (and therefore also perpendicular to $\vec{\boldsymbol{\Omega}}_\parallel$). However, the magnitude of $\vec{\boldsymbol{\Omega}}_\perp$ is equal to the magnitude of the "true" tumbling rate, $\vec{\boldsymbol{\Omega}}_\top = \vec{\boldsymbol{\Omega}} - \vec{\boldsymbol{\Omega}}_\parallel$ (this vector lies in the plane spanned by $\vec{\boldsymbol{\Omega}}$ and $\vec{\boldsymbol{\Omega}}_\parallel$). This is easily seen with a trigonometric identity: $\left|\vec{\boldsymbol{\Omega}}_\parallel\right|^2 + \left|\vec{\boldsymbol{\Omega}}_\perp\right|^2 = |\hat{\boldsymbol{n}}|^2\left|\vec{\boldsymbol{\Omega}}\right|^2 \cos^2\theta + |\hat{\boldsymbol{n}}|^2\left|\vec{\boldsymbol{\Omega}}\right|^2 \sin^2\theta = \left|\vec{\boldsymbol{\Omega}}\right|^2$. Since $\vec{\boldsymbol{\Omega}}_\top + \vec{\boldsymbol{\Omega}}_\parallel = \vec{\boldsymbol{\Omega}}$, we know that $\left|\vec{\boldsymbol{\Omega}}_\parallel\right|^2 + \left|\vec{\boldsymbol{\Omega}}_\top\right|^2 = \left|\vec{\boldsymbol{\Omega}}_\parallel\right|^2 + \left|\vec{\boldsymbol{\Omega}}_\perp\right|^2$ and therefore $|\vec{\boldsymbol{\Omega}}_\top| = |\vec{\boldsymbol{\Omega}}_\perp|$.

For neutrally buoyant particles in isotropic turbulence, we expect no globally preferential particle orientation, and therefore $\langle\vec{\boldsymbol{\Omega}}\rangle = \langle\hat{\boldsymbol{n}}\rangle = 0$ (where the angle brackets denote the expectation value). It follows that the variance of particle angular velocity is simply $\langle\vec{\boldsymbol{\Omega}}\cdot\vec{\boldsymbol{\Omega}}\rangle = \langle|\vec{\boldsymbol{\Omega}}|^2\rangle$. Any given value of $\hat{\boldsymbol{n}}$ represents the same geometric configuration as $-\hat{\boldsymbol{n}}$; therefore, the distribution of $\vec{\boldsymbol{\Omega}}_\perp = \vec{\boldsymbol{\Omega}}\times\hat{\boldsymbol{n}}$ should be symmetric about zero with $\langle\vec{\boldsymbol{\Omega}}_\perp\rangle = 0$. From these observations, we can deduce that the average spinning rate $\langle\vec{\boldsymbol{\Omega}}_\parallel\rangle$ is also zero, and that the following relations apply:

$$\left|\vec{\boldsymbol{\Omega}}\right|^2 = \left|\vec{\boldsymbol{\Omega}}_\parallel\right|^2 + \left|\vec{\boldsymbol{\Omega}}_\perp\right|^2 \tag{4.3}$$

$$\mathrm{var}(\vec{\boldsymbol{\Omega}}) = \mathrm{var}(\vec{\boldsymbol{\Omega}}_\parallel) + \mathrm{var}(\vec{\boldsymbol{\Omega}}_\perp) = \langle|\vec{\boldsymbol{\Omega}}|^2\rangle = \langle|\vec{\boldsymbol{\Omega}}_\parallel|^2\rangle + \langle|\vec{\boldsymbol{\Omega}}_\perp|^2\rangle \tag{4.4}$$

For neutrally-buoyant particles in isotropic turbulence, as we have stated, there should be no *global* preferential orientation of $\hat{\boldsymbol{n}}$—that is, particles should not have a tendency to align with a particular direction in the −xyz or laboratory frame. However, there is a great deal of research on the preferential alignment of $\hat{\boldsymbol{n}}$ with turbulent vorticity structures, and how this alignment depends on particle shape (Parsa et al. 2012; Chevillard and Meneveau 2013; Gustavsson, Einarsson, and Mehlig 2014). A brief discussion of this topic is provided at the end of Section 4.1.1.

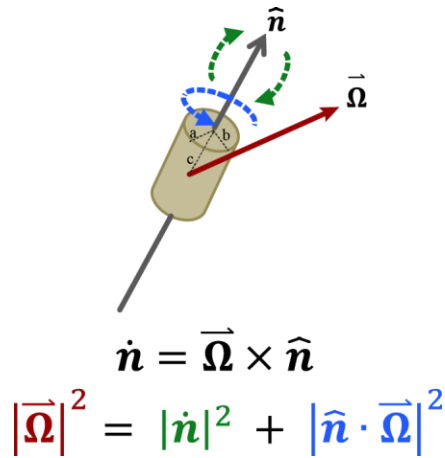

$$\dot{\boldsymbol{n}} = \vec{\boldsymbol{\Omega}}\times\hat{\boldsymbol{n}}$$

$$\left|\vec{\boldsymbol{\Omega}}\right|^2 = |\dot{\boldsymbol{n}}|^2 + \left|\hat{\boldsymbol{n}}\cdot\vec{\boldsymbol{\Omega}}\right|^2$$

**Figure 4.1:** Schematic showing the quantities $\hat{\boldsymbol{n}}$ and $\vec{\boldsymbol{\Omega}}$, and their relationship to one another.

Throughout this chapter, we will use the following symbolic convention:

| | |
|---|---|
| $\vec{\boldsymbol{\Omega}}$ | Particle angular velocity |
| $\hat{\boldsymbol{n}}$ | Unit vector parallel to the particle's axis of symmetry |
| $\dot{\boldsymbol{n}} = \vec{\boldsymbol{\Omega}} \times \hat{\boldsymbol{n}} = \vec{\boldsymbol{\Omega}}_{\perp}$ | Time rate of change of $\hat{\boldsymbol{n}}$, or "tumbling rate" of the particle |
| $\hat{\boldsymbol{n}} \cdot \vec{\boldsymbol{\Omega}} = \vec{\boldsymbol{\Omega}}_{\parallel}$ | Component of $\vec{\boldsymbol{\Omega}}$ parallel to the particle's axis of symmetry, or "spinning rate" of the particle |
| $a, b, c$ | Particle axes as illustrated in Figure 4.1. |
| $\alpha \equiv c/a$ | Aspect ratio of the particle |
| $\Lambda$ | Dimensionless particle shape factor as defined below |
| $\vec{\boldsymbol{\omega}}$ | Fluid vorticity |
| $\vec{\boldsymbol{\zeta}} \equiv \dfrac{1}{2} \vec{\boldsymbol{\omega}}$ | Fluid rotation |
| $\mathbb{A}$ | Fluid velocity gradient tensor |
| $\mathbb{S} \equiv \dfrac{(\mathbb{A} + \mathbb{A}^T)}{2}$ | Strain-rate tensor, or the symmetric part of $\mathbb{A}$ |
| $\mathbb{R} \equiv \dfrac{(\mathbb{A} - \mathbb{A}^T)}{2}$ | Rotation tensor, or the antisymmetric part of $\mathbb{A}$ |

**Table 4.1: Definitions of terms relating to particle and fluid rotation.**

## 4.1.1   Simulation of small particles

The particle equation of motion, as outlined in Chapters 2 and 3, is known only for several special cases. For small rigid spheres in unsteady flow, the Maxey-Riley equation may be used (Maxey and Riley 1983). However, this equation does not account for shape. If particles are assumed to have no inertia, Jeffery's equations can be used to compute particle rotation (Jeffery 1922). This assumption is valid for neutrally-buoyant particles that are much smaller than the Kolmogorov lengthscale, which may be considered as point-particles that take up no space in the flow. Jeffery's equations account for shape via the introduction of a "shape parameter" $\Lambda$, where

$$\Lambda \equiv \frac{\alpha^2 - 1}{\alpha^2 + 1} \tag{4.5}$$

and $\alpha$ is the particle aspect ratio; therefore, $\Lambda = 1$ for an infinitely thin rod, and $\Lambda = -1$ for an infinitely flat disk. This model does not allow for non-axisymmetric shapes that require more than one parameter for their characterization. However, it is a useful starting point for our analysis.

For very small particles, Jeffery's equations are (Jeffery 1922):

$$\dot{\boldsymbol{x}} = \vec{\boldsymbol{u}}_{\boldsymbol{p}} = \vec{\boldsymbol{u}}_f(\boldsymbol{x}_t, t) \tag{4.6}$$

$$\vec{\boldsymbol{\Omega}} = \vec{\boldsymbol{\zeta}}(\boldsymbol{x}_t, t) + \Lambda \hat{\boldsymbol{n}} \times \mathbb{S}(\boldsymbol{x}_t, t)\hat{\boldsymbol{n}} \tag{4.7}$$

where $\dot{x} = \vec{u}_p$ is the particle's translational velocity; $\vec{\zeta}(x_t, t) = \frac{1}{2}\vec{\omega}(x_t, t)$ is the fluid rotation rate, which is equal to half the fluid vorticity; and $\vec{\Omega}$ is the particle's angular velocity. In this case, the particles' center-of-masses are advected exactly with the Lagrangian fluid trajectories defined by $\vec{u}_f(x_t, t)$.

Using these equations, we conduct one-way coupled simulations in which nonspherical particles are passively advected and rotated by the flow. The presence of the particles does not affect the surrounding turbulence, which is prescribed. Both the fluid vorticity and the fluid strain contribute to particle rotation, so that $\vec{\Omega}$ does not simply equal the fluid angular velocity $\vec{\zeta} = \frac{1}{2}\vec{\omega}$. We note that the fluid strain $\mathbb{S}$ contributes to $\vec{\Omega}$ only in the case of nonspherical particles; when the shape factor $\Lambda$ is zero, the contribution from the strain is also zero. In this case, $\vec{\Omega} = \vec{\zeta}$ and the (spherical) particle rotates exactly according to one-half the fluid vorticity.

Because the particles are neutrally-buoyant and very small, the particle center-of-mass trajectory is prescribed by the Lagrangian fluid velocity. For the ambient fluid flow, we use the direct numerical simulation (DNS) data provided by the Johns Hopkins University turbulence database (Li et al. 2008), containing time-series data for $\vec{u}_f(x_t, t)$ and $\mathbb{S}(x_t, t)$ within a 1024 x 1024 x 1024 grid at a Reynolds number (based on the Taylor microscale) of $\text{Re}_\lambda = 433$. This turbulence is continually forced and isotropic, as is our laboratory case. Particle position and orientation are initialized randomly, and particles are permitted to advect and rotate according to equations (4.6) and (4.7) for approximately 45 Kolmogorov timescales $\tau_k$. Further details of the simulation parameters and results can be found in (Byron et al 2015).

Averaging over a large number of trajectories, we find that regardless of aspect ratio, particles inherit roughly the same amount of total rotation (i.e., the expected value of the rotation magnitude squared, $\langle\vec{\Omega}\cdot\vec{\Omega}\rangle$) from the flow (red circles in Figure 4.2a, which show a value of $\langle\vec{\Omega}\cdot\vec{\Omega}\rangle \approx 2.5$ for all aspect ratios). This equivalence holds over a wide range of aspect ratios. However, the rotation components $\vec{\Omega}_\perp$ and $\vec{\Omega}_\parallel$ are *not* equivalent across particle aspect ratio. High-aspect-ratio particles (rods) bias towards spinning, $\vec{\Omega}_\parallel$ (blue squares in Figure 4.2a), and low-aspect-ratio particles (disks) bias towards tumbling, $\vec{\Omega}_\perp$ (green triangles in Figure 4.2a). This bias saturates relatively quickly as the aspect ratio departs from unity, where we observe the predicted two-to-one ratio of tumbling and spinning. Within one order of magnitude, where $\alpha > 10$ or $\alpha < 0.1$, further changes in aspect ratio do not affect the tumbling-to-spinning ratio. For spherical particles, no meaningfully different principal axes can be defined. At this point ($\alpha = 1$), the tumble-to-spin ratio is two-to-one, as discussed previously.

This preferential distribution of rotation, parceled out between $\vec{\Omega}_\perp$ and $\vec{\Omega}_\parallel$, arises from preferential particle orientation in turbulence (see figures 1, 7, and 8 in Byron et al 2015; reprinted in Appendix C, with permission). In long-lived turbulent vortex structures, (non-inertial) particles' longest lengths tend to align with the vorticity. For rods, this effect is known and well-studied (Pumir and Wilkinson 2011; Ni, Ouellette, and Voth 2014; Parsa et al. 2011). Very thin rods behave in turbulence like material lines, and align well with the vorticity. This is due to the stretching term that is present in the equation of motion for both material lines and vorticity (Wilkinson and Kennard 2012; Batchelor 1952). The vorticity itself is often considered in the context of the fluid strain, and has been found to align with the intermediate eigenvector of the Eulerian fluid strain-rate tensor (Pumir and Wilkinson 2011; Ashurst et al. 1987).

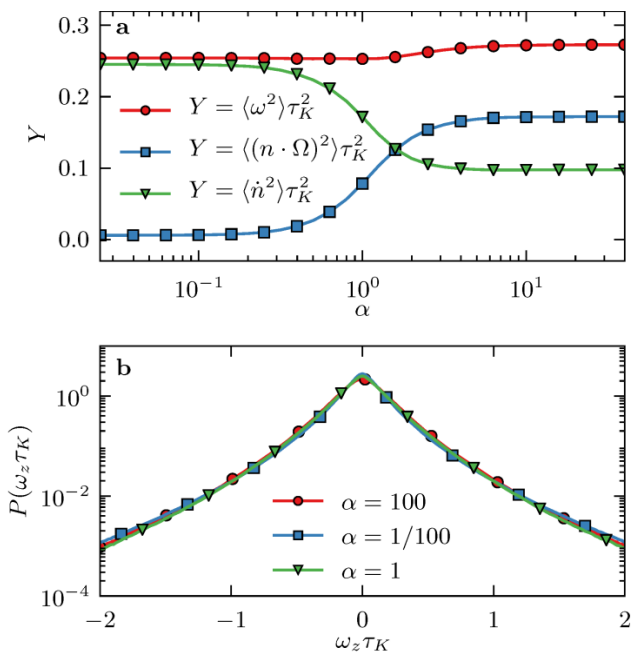

**Figure 4.2: Variance of Spinning and tumbling rates of small particles in turbulence. (a) tumbling (green triangles), spinning (blue squares), and total rotation (red circles) for axisymmetric particles of varying aspect ratio α. (b) pdf of particle rotation component $\omega_z$ for high-aspect ratio particles (α=100, red circles), low-aspect ratio particles (α=1/100, blue squares), and unity aspect ratio particles (α=1, green triangles). Reprinted from Byron et al 2015, with permission.**

Another framework for analysis is the Cauchy-Green strain tensor, which describes the Lagrangian stretching and deformation of a fluid element. The eigenvectors of the left Cauchy-Green tensor describe the orientation of the principal axes of an originally-spherical fluid element after a period of deformation (Malvern 1969; Ni, Ouellette, and Voth 2014). When considered in the context of the Cauchy-Green tensor (rather than the Eulerian strain-rate tensor), vorticity has been shown to preferentially align with the strongest Lagrangian stretching direction (Ni, Ouellette, and Voth 2014). This is somewhat more physically appealing than the previous framework, since the strongest stretching should intuitively correspond with alignment. The Cauchy-Green view of turbulent vortex stretching is somewhat related to the use of Lyapunov exponents as they are used in studies of Lagrangian coherent structures (Shadden, Dabiri, and Marsden 2006; Green, Rowley, and Haller 2007). We note that the subject of vorticity alignment with fluid strain is a subject of extremely interesting ongoing research, but is not the primary focus of this thesis.

The preferential distribution of particle rotation which we observe when the rotation is projected onto the particle axes (Figure 4.2a) is not present in the global (-xyz) coordinate system (Figure 4.2b). The distribution of particle angular velocity (here shown by a single component $\Omega_z$, denoted in the referenced paper as $\omega_z$) is virtually identical for particles of vastly different aspect ratio ($\alpha = 1$, $\alpha = 1/100$, and $\alpha = 100$). Though this is perhaps to be expected due to the homogeneity and isotropy of the ambient turbulence, it is remarkable that particles of such dramatically different shape experience not only the same angular velocity variance, but also the same general distribution in the global coordinate system.

## 4.1.2  Measurements of large particles

For particles which are smaller than the Kolmogorov scale, such as those discussed in the previous section, rotation is determined only by the locally-linear fluid velocity gradients. By contrast, large particles must integrate fluid velocity gradients which are nonlinear on the scale of the particle. The dependence of large-particle rotation on these nonlinear gradients is analytically unknown, and computationally difficult to determine. A further difficulty posed by finite-size is the effect of inertia on particle motion. This effect—safely ignored in the infinitesimally-small, neutrally-buoyant case—must be accounted for when dealing with large particles. Because they are inertial, large particles will not follow Lagrangian fluid trajectories. This establishes a slip velocity, discussed extensively in Chapter 3. Additionally, one-way coupling is not adequate for modeling inertial particle motion. The presence of the particle alters the fluid flow, and particle inertia allows for two-way momentum transfer between the particle and fluid phases. Two-way coupling simulations have mostly been performed on spherical particles (Eaton 2009; L. Zhao and Andersson 2011), though the one-way coupling case is much easier to simulate (Balachandar and Eaton 2010). It is possible to compute a two-way coupled simulation for nonspherical particles (usually done with axisymmetric spheroids), but these studies are not as common (Andersson, Zhao, and Barri 2012; F. Zhao and Wachem 2013). As discussed in Section 1.2.1, an effective numerical simulation must compute the turbulence on a grid whose spacing is much smaller than the particle size. Grid points are not likely to fall directly on the particle surface, and so the surface-fluid interaction must be interpolated. If the particle is freely suspended, boundary conditions must be re-enforced at every timestep. These limitations combine to make this approach computationally expensive and difficult to implement.

In the face of these complications, we turn to experiment in order to investigate the rotation of large nonspherical particles. By "large particles", we here mean particles whose size scales are within the inertial subrange of the ambient turbulence—more specifically, particles whose size scales roughly correspond to the Taylor microscale, $\lambda$.

Data were collected according to the protocol described in Chapter 2, which yields simultaneous vector fields of a suspended solid particle and the fluid flowfield around it. Particle angular velocity was computed from interior (solid-phase) velocity fields via the solid-body rotation equation:

$$\vec{u}_m = \vec{u}_n + \vec{\Omega} \times (r_m - r_n) \tag{4.8}$$

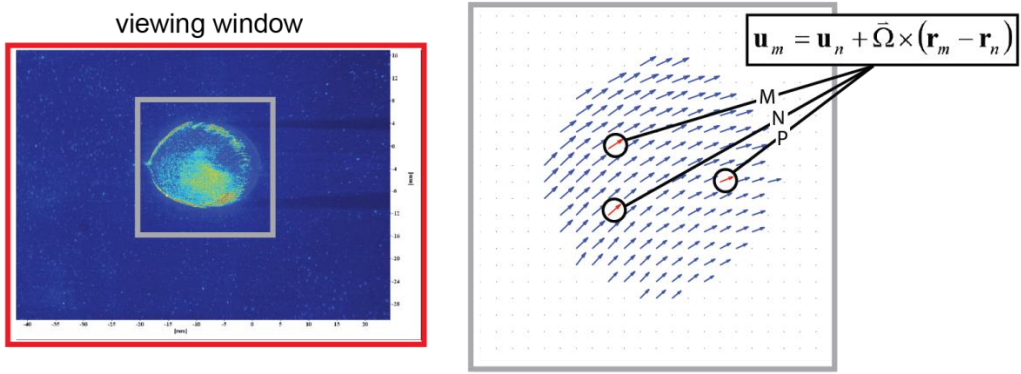

**Figure 4.3: Angular velocity is computed from vector triplets via the solid-body rotation equation. Because all measurements are coplanar in z, three points are needed to solve the equation for $\vec{\Omega}$.**

In theory, only two interior data points are needed to solve for $\vec{\Omega}$. However, in our reference frame, all data are coplanar in z, ensuring that the equation is overdetermined in z and underdetermined in x and y:

$$\begin{bmatrix} u_m \\ v_m \\ w_m \end{bmatrix} = \begin{bmatrix} u_n \\ v_n \\ w_n \end{bmatrix} + \begin{pmatrix} 0 & 0 & -(y_m - y_n) \\ 0 & 0 & -(x_m - x_n) \\ y_m - y_n & x_m - x_n & 0 \end{pmatrix} \cdot \begin{pmatrix} \Omega_x \\ \Omega_y \\ \Omega_z \end{pmatrix} \qquad (4.9)$$

$$\Omega_z = \frac{(u_m - u_n)}{(y_n - y_m)}$$
$$\Omega_z = \frac{(v_m - v_n)}{(x_n - x_m)} \qquad (4.10)$$
$$\Omega_x(y_m - y_n) + \Omega_y(x_m - x_n) = w_m - w_n$$

Therefore, three interior data points are needed to solve the solid-body rotation equation (as illustrated in Figure 4.3). For each particle, we apply this equation across all possible vector triplets within the particle and take the median value for robustness (Bellani et al. 2012). We collect a large number of instantaneous measurements of particles in flow (N = 101 - 413), with each measurement including both the interior (particle) vector field and exterior (fluid) vector field. Image thresholding and vector field separation are performed as described in Section 2.2.

Because the ambient turbulence is homogeneous and isotropic, $\vec{\Omega}$ averages to zero in every (global) component $\Omega_x$, $\Omega_y$, and $\Omega_z$. This is also true for the numerically simulated small particles (Figure 4.2; Figure 4.6). To quantify rotation, we therefore investigate the square of the angular velocity, $\langle \vec{\Omega} \cdot \vec{\Omega} \rangle = |\vec{\Omega}|^2$, across all particle shapes and densities.

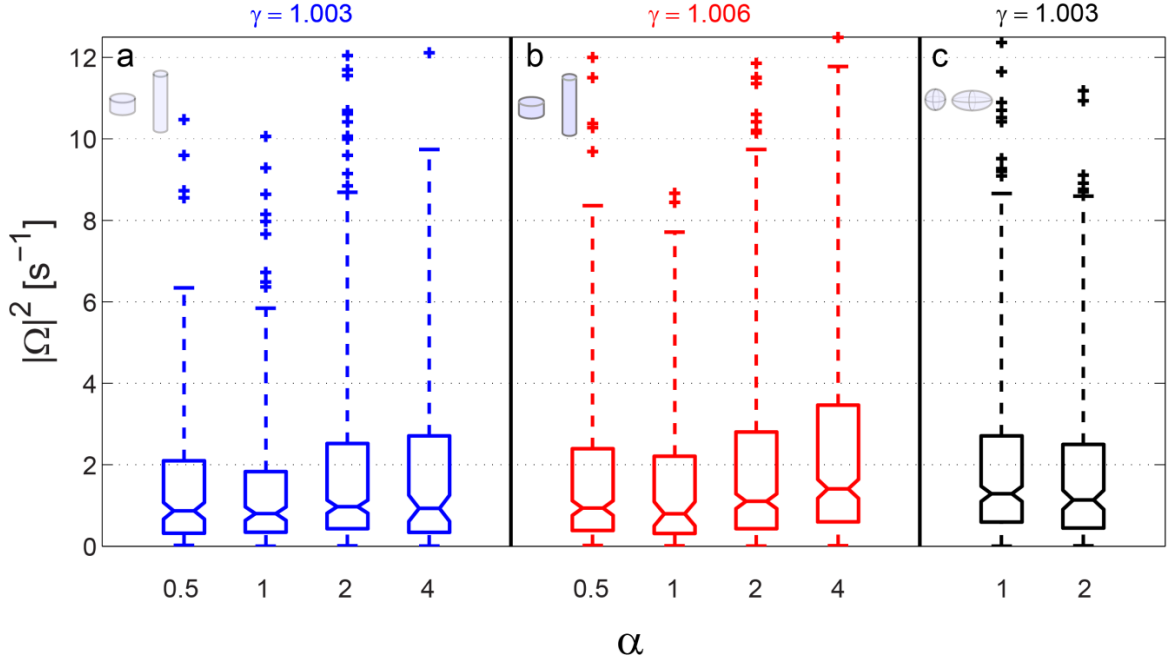

**Figure 4.4:** Angular velocity magnitude across particle shape and density. Left panel (a) shows data from cylinders at $\gamma = 1.003$; center panel (b) shows cylinders at $\gamma = 1.006$; right panel (c) shows spheroids at $\gamma = 1.003$. Solid horizontal lines within each box are the median value of $\langle \vec{\Omega} \cdot \vec{\Omega} \rangle$ for each particle type. Notches represent 95% confidence interval on the median value; boxes represent interquartile range; whiskers represent approximate full data range (thrice interquartile range). Outliers are shown as crosses, and those above $|\mathbf{\Omega}|^2 = \mathbf{12.5 \ s^{-2}}$ (comprising < 3% of the dataset in all cases) are not shown.

Figure 4.4 shows the measured rotation rates for all ten particle types (eight cylinders and two spheroids), including the median, confidence intervals, interquartile range, and full data range. Figure 4.5 shows the mean rotation magnitude squared, $\langle |\vec{\Omega}|^2 \rangle$, and confidence intervals for each particle type. We can see from Figure 4.5 that for the $\gamma_1$ particles ($\gamma$=1.003), there is no significant difference in the rotation magnitude between particle types. This mirrors our findings for very small, non-inertial, neutrally buoyant particles (Figure 4.2): both large and small neutrally-buoyant particles experience the same amount of total rotation across all values of $\alpha$.

A slightly different picture is presented for the $\gamma_2$ particles ($\gamma$=1.006). Though the median values of rotation are still very similar to one another, a subtle trend is beginning to emerge, where $\alpha$=1 forms a local minimum of rotation, and departures from $\alpha$=1 cause particles to experience more rotation. This trend may be present in the $\gamma_1$ data, but is not statistically significant and/or cannot be precisely illustrated with our method. However, the data range (as denoted by the whiskers in Figure 4.4) for both $\gamma_1$ and $\gamma_2$ hint at the same pattern, in which particles with longer lengthscales show more extreme values of rotation.

For comparison, we also show the rotation of two spheroids: a sphere at α=1 and a prolate ellipsoid at α=2 (see Table 2.7). Because the spheroids are not volume, length, or surface-area matched to the cylinders, they are not directly comparable. However, they provide an interesting counterpoint to the data from the cylinders. The spheroids are fabricated at density $\gamma_1$=1.003, and show no significant difference in rotation between α=1 and α=2. This is also the case for the cylinders at $\gamma_1$=1.003.

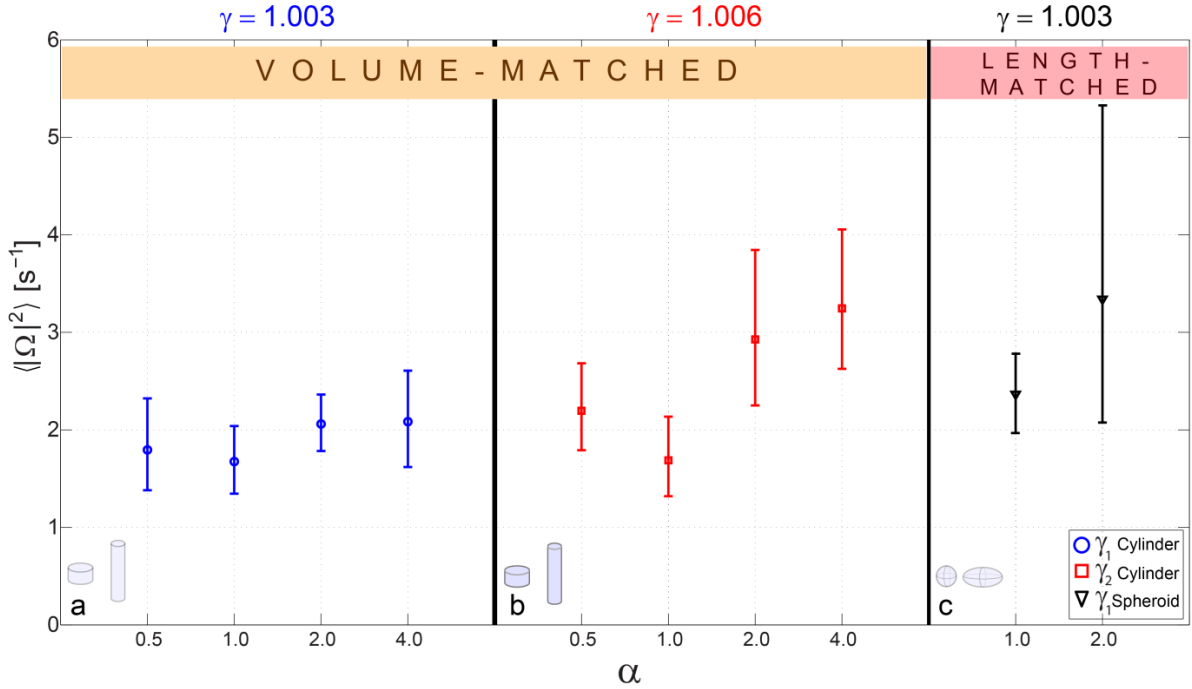

Figure 4.5: Mean angular velocity magnitude across particle shape for both tested densities. Errorbars represent 95% confidence intervals, calculated via bootstrapping with 1000 replicates.

| | variance of $|\Omega|$ for $\gamma_1$ | variance of $|\Omega|$ for $\gamma_2$ |
|---|---|---|
| **C Y L I N D E R S** | | |
| **α = 0.5** | 0.5699 [0.361, 0.8652] | 0.7442 [0.5249, 0.9793] |
| **α = 1** | 0.5103 [0.3434, 0.6875] | 0.5097 [0.3425, 0.6813] |
| **α = 2** | 0.6145 [0.4673, 0.7668] | 1.1650 [0.6744, 1.8568] |
| **α = 4** | 0.6568 [0.6568, 0.4353] | 1.1332 [0.7484, 1.6377] |
| **S P H E R O I D S** | | |
| **α = 1** | 0.6608 [0.4578, 0.9063] | |
| **α = 2** | 1.5495 [0.6266, 3.0575] | |

Table 4.2: Variance of $\overline{|\Omega|}$ for cylindrical and spheroidal particles at two specific gravities, $\gamma_1$=1.003 and $\gamma_2$ = 1.006. Bracketed values represent 95% confidence intervals, calculated via bootstrapping with 1000 replicates.

The variance of $|\Omega|$ is another useful metric for illustrating the differences between the cylinders at $\gamma_1$ and the cylinders at $\gamma_2$. For both densities, $\alpha=1$ displays the lowest variance of $|\Omega|$ (statistically significant only for $\gamma_2$, between $\alpha=1$ and $\alpha=4$). This is also reflected in the spheres, where the $\alpha=1$ spheres display a lower variance of $|\Omega|$ than the prolate ellipsoids at $\alpha=2$. This indicates that particles with longer lengthscales may be experiencing a broader range of rotation. This makes intuitive sense: particles inherit rotation from the surrounding turbulent flow, which is multi-scaled. Particles with longer lengthscales experience a broader distribution of turbulent eddy sizes, and should therefore display a broader range of rotation. However, the passing-down of rotation from the fluid to the particle phase is a complex problem, whose functional dependence is still uncertain. This topic is discussed in depth in Section 4.3.

In general, our results suggest that the physics governing the rotation of large particles is inextricably tied to the presence of inertia. Particles which are very close to neutral buoyancy, like our $\gamma_1$ particles, to a degree behave like small, non-inertial point particles. Particles of differing aspect ratio may experience the same average rotation. However, even a slight departure from neutral buoyancy may break this trend. Our $\gamma_2$ particles, which have slightly more inertia than our $\gamma_1$ particles, experience higher rotation for $\alpha\neq1$. This indicates that for inertial particles, rotation may be governed in part by the longest lengthscale of the particle. This is consistent with the theories proposed by some other researchers (Parsa and Voth 2014). Additionally, the longest lengthscale of a particle is a good predictor of the variance of the total rotation magnitude (though this effect is more pronounced for denser particles).

## 4.2   Rotation distribution

In the previous section, we saw that differences in particle shape did not lead to differences in the magnitude of the overall rotation for either small non-inertial particles or large neutrally-buoyant particles, though a slight increase in density results in an interesting trend across particle shape (Figure 4.5). We also saw that the variance of the angular velocity magnitude $|\vec{\Omega}|$ tended to be larger for particles with longer lengthscales (Table 4.2). Here, we continue or examination of the higher-order moments of $\vec{\Omega}$, and examine the complete distribution of a representative component $\Omega_z$. For very small particles, there was no difference in the angular velocity distribution $\Omega_z$ between particle shapes (Figure 4.2b). It will be useful to see whether this is also true for large particles, and may shed further light on the emerging shape-dependent trends that we observed in Table 4.2 and Figure 4.5.

Before taking $\Omega_z$ as a representative rotation component for large particles, as we did in Figure 4.2 for small particles, we must verify that the three components are comparable. Because the turbulence is homogeneous and isotropic, we should not expect to see major differences between the distributions of $\Omega_x$, $\Omega_y$, and $\Omega_z$. This was our argument in Section 4.1.1, in which the numerically-simulated particles were very small and neutrally buoyant. However, our large particles are slightly negatively buoyant, which may break the symmetry between the gravity-coupled and non-gravity-coupled component of $\vec{\boldsymbol{\Omega}}$.

From our discussion in Section 3.3.2, we know that buoyancy forces are small compared to the unsteady particle drag that arises in turbulent flow. To ground this in our previous analysis of the turbulent slip velocity, we can compare the vertical slip $v_q$ to the overall slip magnitude $|\vec{\boldsymbol{u}}_s|$. In still water, the settling velocity of the particles is on the order of 1-3 cm s$^{-1}$. In turbulence, the vertical slip velocity $v_q$ is much less than this, and is small compared to the overall slip magnitude $|\vec{\boldsymbol{u}}_s|$ (Table 3.3). The ratio of $v_q$ to $|\vec{\boldsymbol{u}}_s|$ (approximately one to three, in our case) is qualitatively representative of the ratio between buoyancy forces and fluid drag forces. We therefore expect to see no systematic differences between $\Omega_y$ (the gravity-coupled component of the angular velocity) vs. $\Omega_x$ and $\Omega_z$ (the lateral components of angular velocity). Indeed, this is what we observe in Figure 4.6: there are no significant differences between the distributions of the three global (i.e., laboratory reference frame) angular velocity components.

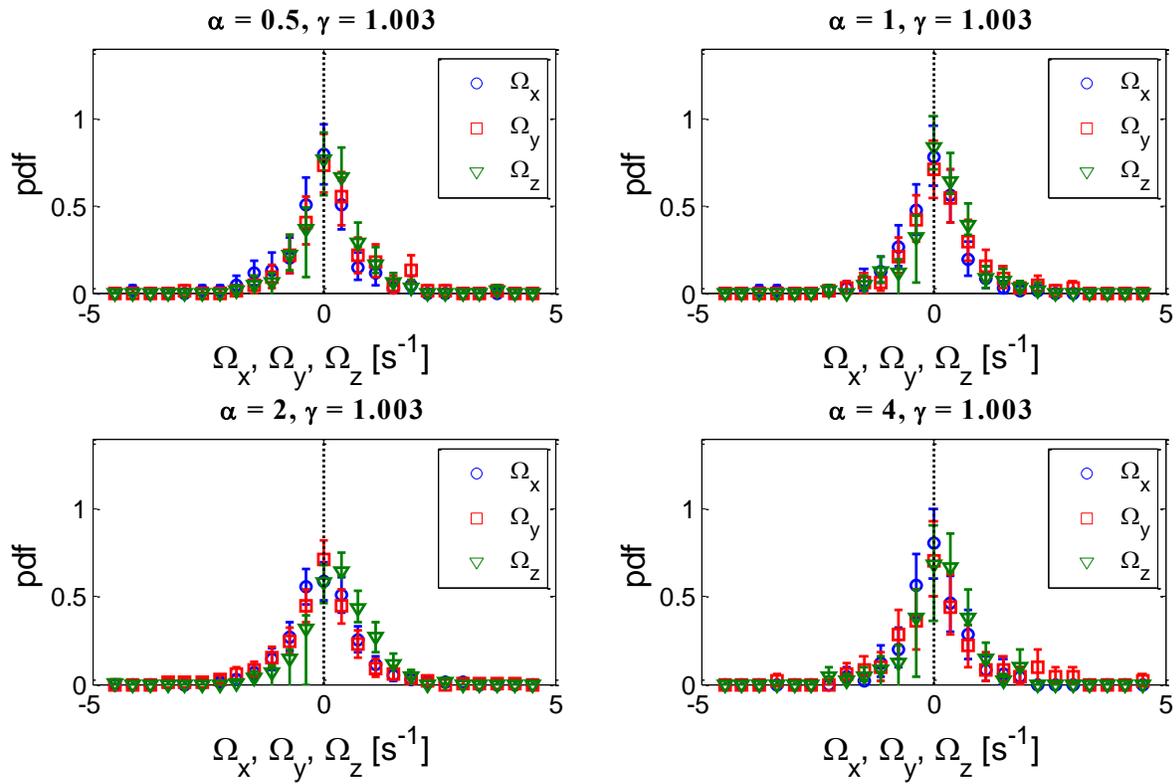

Figure 4.6: Probability density function of particle angular velocity components for each aspect ratio α, for the case γ = 1.003 (shown as a representative case for both particle densities). Errorbars represent 95% confidence interval, calculated via bootstrapping with 1000 replicates.

Though the data displayed in Figure 4.6 are taken from the case where $\gamma=1.003$, these data are also representative of the other density case. No significant differences in the distributions of $\Omega_x$, $\Omega_y$, or $\Omega_z$ were found at either of the two examined particle densities.

Even if the particles were not neutrally buoyant, the homogeneity and isotropy of the turbulence would lead us to expect identical results for the distributions of $\Omega_x$ and $\Omega_z$, the lateral (non-gravity-coupled) components of angular velocity. This is reinforced by the experimental results shown in Figure 4.6. We therefore take $\Omega_z$ as a representative of the total rotation distribution for these near-neutrally-buoyant particles. An added benefit of this approach is that $\Omega_z$ is also the lowest-noise component of the particle rotation (see Equation (4.9)), since it is overdetermined in the solid-body rotation equation. $\Omega_z$ is therefore the most robust measurement of overall particle angular velocity.

In Figure 4.7, we see that there are no significant differences between the angular velocity distributions of the four tested shapes ($\alpha = 0.5, 1, 2,$ and $4$) for particles with specific gravity $\gamma_1=1.003$. This result is particularly remarkable for its similarity to the result for small simulated particles (Figure 4.2b). This is also consistent with our findings from Table 4.2, which found no significant difference in variance for cylinders or spheroids at $\gamma_1=1.003$.

Previous studies (Parsa et al. 2012; Parsa et al. 2011) have held that rotation of neutrally-buoyant particles is controlled by the particle's longest lengthscale. This may be true for particles which have very high aspect ratio (i.e. long rods), but does not necessarily hold for our large (Taylor-scale) particles, whose aspect ratios are close to unity. In our particles, the longest lengthscale of the $\alpha = 4$ rods is more than twice that of the $\alpha = 1$ cylinders, for which H = D. If the longest lengthscale were the sole predictor of rotation, we would expect to see a dramatic difference in the overall magnitude of rotation experienced for these two particle shapes; however, we see no difference in the rotation magnitude nor in its underlying distribution, at least in the $\gamma_1=1.003$ case. This suggests that the longest lengthscale is not the only controlling factor for particle rotation, and that large neutrally-buoyant particles of similar volume and surface area experience very similar rotation in the global (laboratory) coordinate system.

For slightly denser cylinders, a different trend emerges. The interesting behavior we observed in the mean rotation rate for $\gamma_2=1.006$ particles (Figure 4.5) is also present when we examine the angular velocity distributions. In Figure 4.7, we see the discrete probability density function of each particle shape at both of the tested densities. At $\gamma_1=1.003$, there is no significant difference between any of the pdfs. They are all centered around $\Omega_z = 0$, as we expected (also reflected in Figure 4.6). However, at $\gamma_2 = 1.006$, there is a slight difference between the most extreme shapes. To illustrate this, we have drawn a continuous line between the points that make up the discrete pdfs of the $\alpha=1$ and $\alpha=4$ particles—recall that these two particle types had significantly different variances of $|\vec{\boldsymbol{\Omega}}|$ and significantly different average values of $|\vec{\boldsymbol{\Omega}}|$.

In Figure 4.7b, we see that the distribution for $\alpha=4$ is slightly broader, indicating that the $\alpha=4$ particles experience a broader range of rotation than the $\alpha=1$ particles. Computing the overall variance of $|\vec{\Omega}|$ confirms this, as shown in Table 4.2. This suggests that particle density matters: the cylinders at $\gamma_1 = 1.003$ display different behavior than the cylinders at $\gamma_2 = 1.006$.

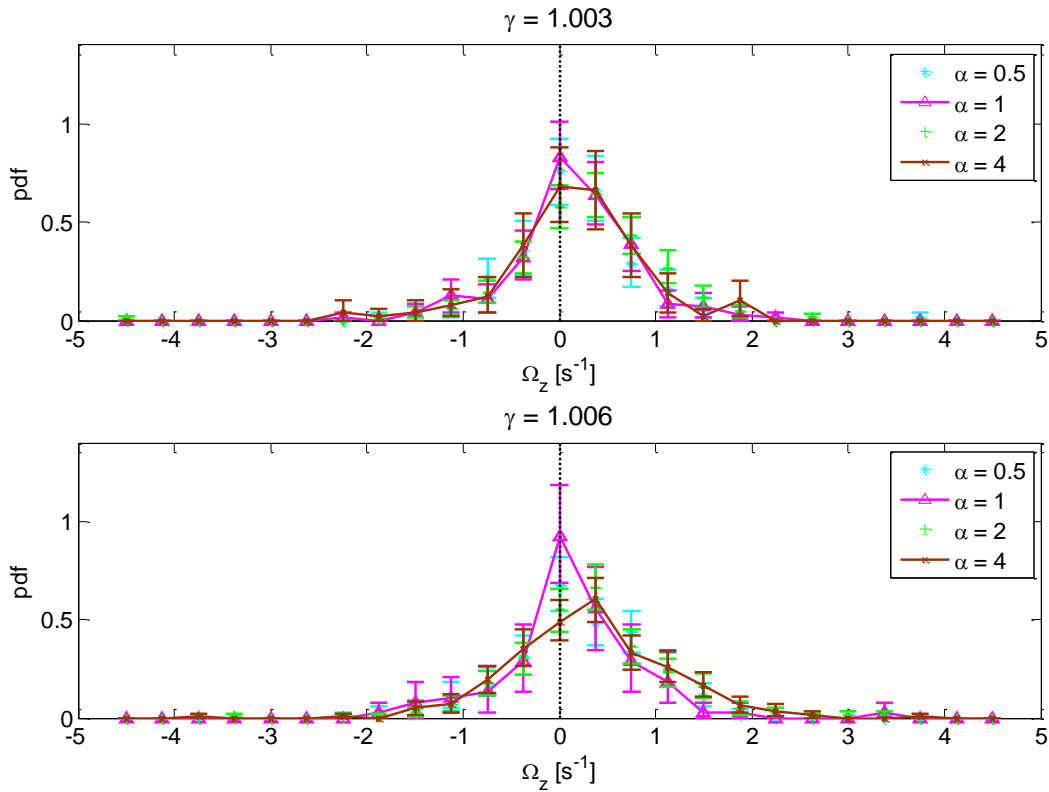

**Figure 4.7:** Comparison of rotation distribution across shape, for two different particle densities. Errorbars represent 95% confidence interval, calculated via bootstrapping with 1000 replicates. Connecting lines have been drawn for the $\alpha = 1$ and $\alpha = 4$ cases to illustrate the differences in variance between different particle shapes.

We also recall that small particles showed a preferential distribution of rotation about the particle axes—high-aspect-ratio rods showed a higher proportion of spinning, and low-aspect-ratio disks showed a higher proportion of tumbling. It is an open question whether or not large particles display similar behavior, as our experimental method is not able to decompose the angular velocity vector along the particle axes.

## 4.3 Inheritance of vorticity from flow

Because finite-size particles are inertial, their relationship with the surrounding fluid is complex. Unlike infinitesimal particles, finite-size particles alter the flow around themselves simply by being present in the fluid. Infinitesimal particles are assumed to be one-way-coupled to the flow: their motion is dependent only on the local fluid velocity, which is not affected by their presence. In contrast, the motion of large particles feeds back into the flow, altering the local gradients. Particle inertia can act as filter for fluid motion, changing the way that energy is passed from larger to smaller scales in turbulence. After a particle inherits momentum from the flow, it will eventually pass it back to the fluid in a complex way. One aspect of this process can be observed in the modulation of the turbulent cascade that is seen in particle laden flows (Bellani et al. 2012; Balachandar and Eaton 2010). We seek to investigate another aspect of this filtering process: how particles inherit vorticity from the flow.

The transfer of fluid-phase vorticity to particle rotation is a multifactorial process, and is affected by particle inertia, shape, and lengthscale. This relationship has been studied in sub-Kolmogorov-scale particles (L. Zhao and Andersson 2011; Andersson, Zhao, and Barri 2012), but larger particles can be expected to exhibit more complex dynamics. As a first-order approximation, we assume that particles sample the vorticity field at their own lengthscale, then return vorticity to the flow at smaller scales. In effect, we hypothesize that large particles rotate according to a spatially-filtered vorticity: particles average the fluid vorticity along their own length, and rotate according to this filtered vorticity.

We collect a series of independent and identically-distributed (IID) fluid-phase vector fields in which there are no particles in the image window, but particles are still present in the tank. This ensures that we are not including locally-altered flow fields, but only the background vorticity. We use N=800 IID vector fields, using 100 vector fields from each of the eight experiments for agarose cylinders.

Because our measurement window is two-dimensional, we cannot calculate fluid velocity gradients in the out-of-plane direction (i.e., $du_i/dz$). Our method therefore yields only the out-of-plane component of the vorticity, $\omega_z$. We compute $\omega_z$ based on the eight-point circulation method (Luff et al. 1999; Abrahamson and Lonnes 1995). Both the velocity and vorticity fields are resolved at 0.67 mm, roughly twice the Kolmogorov scale. Vector fields are computed via multipass PIV processing with decreasing subwindow sizes (128 x 128 and 64 x 64 pixels), overlapping by 75%. Our vorticity-calculation method yields a vorticity field that is resolved at the same scale as the velocity, but does not give measurements at the outer edges of the grid.

We hypothesize that large particles rotate according to a spatially-filtered fluid vorticity, with a filter length close to the particle's longest lengthscale. We calculate a filtered vorticity $\mathcal{F}(\omega_z)$ by applying a two-dimensional Gaussian spatial filter to the vorticity field $\omega_z$. The rotationally-symmetric Gaussian window has a diameter of 9.2mm, the diameter of a sphere with equivalent volume to our cylinders, and a standard deviation of 3.07mm so that the filter decays to nearly zero at the edges. We compare the distribution of particle angular velocity $\Omega_z$ with the out-of-plane fluid rotation rate $\zeta_z = \frac{1}{2}\omega_z$, along with a filtered rotation rate $\mathcal{F}(\zeta_z)$. Recall that for the particle rotation rate, $\Omega_z$ is the lowest-noise component and is representative of overall particle rotation.

Figure 4.8 shows the distributions of the particle rotation rate $\Omega_z$, the fluid rotation rate $\zeta_z$, and the filtered fluid rotation rate $\mathcal{F}(\zeta_z)$. In general, we see that $\mathcal{F}(\zeta_z)$ is a much better predictor of $\Omega_z$ than $\zeta_z$. However, the agreement between $\mathcal{F}(\zeta_z)$ and $\Omega_z$ is better at rotation rates which are close to zero. At larger rotation rates, this fit is not as pronounced, with the pdf of $\mathcal{F}(\zeta_z)$ exhibiting longer tails than the pdf of $\Omega_z$.

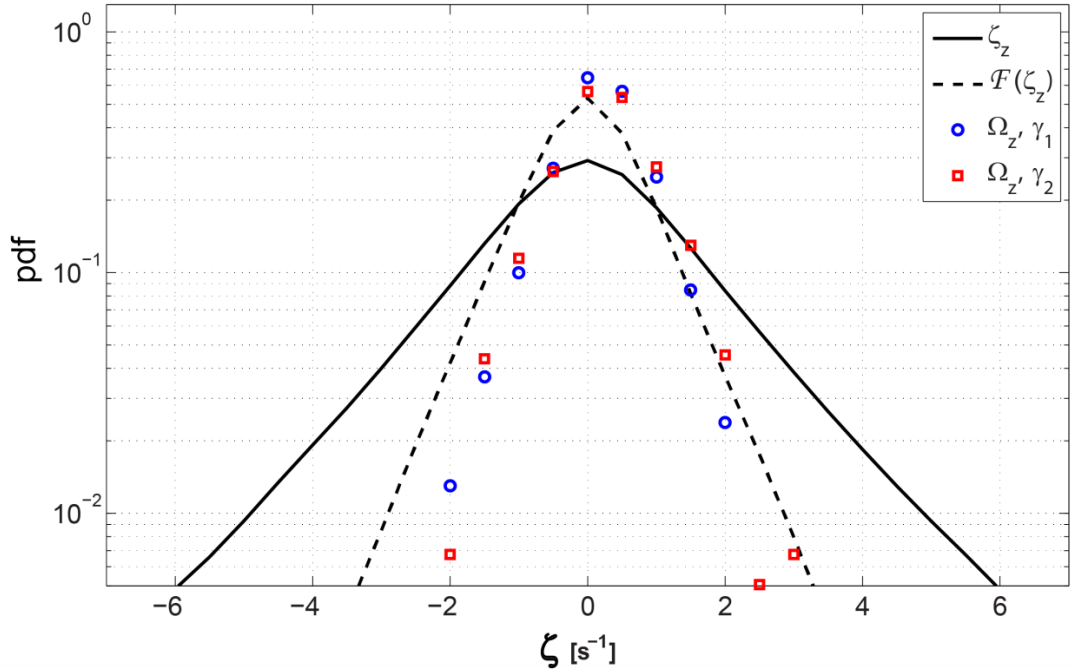

**Figure 4.8:** Fluid-phase rotation $\zeta_z$, filtered fluid-phase rotation $F(\zeta_z)$, and particle rotation at two densities, averaging across all four aspect ratios.

| Fluid vs. particle rotation/enstrophy | Variance of rotation [s⁻²] |
|---|---|
| Fluid rotation: $\langle \zeta_z^2 \rangle$ | 4.024 *[4.013, 4.035]* |
| Filtered fluid rotation: $\langle \mathrm{filt}(\zeta_z)^2 \rangle$ | 0.931 *[0.929, 0.934]* |
| Cylinders, $\gamma_1$: $\langle \Omega_z^2 \rangle$ | 0.495 *[0.424, 0.583]* |
| Cylinders, $\gamma_2$: $\langle \Omega_z^2 \rangle$ | 0.597 *[0.534, 0.663]* |

**Table 4.3:** variance of fluid rotation, filtered fluid rotation, and particle rotation at two different densities. Bracketed intervals represent 95% confidence intervals, calculated via bootstrapping with 1000 replicates.

At extreme rotation rates ($|\Omega_z| \geq 2$), the distribution of particle rotation rates begin to diverge from that of $\mathcal{F}(\zeta_z)$—in other words, particles experience extreme rotation less often than would be predicted by the filtered vorticity field. This may be due to the interplay between turbulent intermittency and particle inertia. High-rotation events in the fluid phase are short-lived and infrequent. When a particle encounters this kind of event, it may take some time to be "spun up" by the turbulence, and a corresponding amount of time to "spin down" once the intermittent event is over. The inertia of the particles keeps them from ever reaching the rapid rotation rates that are present in the fluid. Therefore, high-rotation events in the particle phase will not occur very often, if at all. This is reflected in the pdfs shown in Figure 4.8, and shown in the shorter tails of the pdf of $\Omega_z$ as compared to $\mathcal{F}(\zeta_z)$. It is also reflected by the variances of each distribution: the variance of $\Omega_z{}^2$ is significantly smaller than the variance of $\mathcal{F}(\zeta_z)^2$ (Table 4.3).

We note that this simple model—a single-scale filter, which filters only in space—is likely to be incomplete, and will not perfectly predict the particle rotation. The spatial filter approach can roughly predict the particle angular velocity variance, but not the higher-order moments. It is useful as a first guess. We also note that by increasing the filter size, we could achieve a better fit to the distribution of $\Omega_z$. This suggests a possible definition for a "rotation-equivalent lengthscale", similar to the commonly-used volume-equivalent lengthscale. We used the volume-equivalent lengthscale (i.e., the diameter of the sphere of equivalent volume) as the filter lengthscale. By adjusting the filter size so that it more exactly matched the rotation distribution, we could find another equivalent lengthscale that may be useful in modeling and predictions of particle behavior.

## 4.4    Rotation: Conclusions

In the preceding chapter, we discussed the rotation of both small and large non-spherical particles in turbulence, as well as the mechanisms by which they inherit that rotation from the surrounding flow. We explored the physics of small (sub-Kolmogorov scale) particles via numerical simulation, "seeding" particles into already-computed high-Reynolds number turbulence. By following these particles along their trajectories, we can examine their rotation in two reference frames: the particles' principal or body axes (tumble vs. spin), and the fluid strain eigen-system. In the first reference frame, we see that across a wide range of aspect ratio $\alpha$, particles inherit roughly the same amount of rotation $\langle \vec{\boldsymbol{\Omega}} \cdot \vec{\boldsymbol{\Omega}} \rangle$ from the flow. However, this rotation is not isotropically distributed about the particle's body axes, but can be divided into two components, $\vec{\boldsymbol{\Omega}}_{\perp}$ (tumble) and $\vec{\boldsymbol{\Omega}}_{\parallel}$ (spin). Particles with a very high aspect ratio (rods) tend to spin more than they tumble, whereas particles with a very low aspect ratio (disks) tend to tumble more than they spin. When we examine particle rotation in the context of the fluid strain eigen-system, we see that the longest length of the particle tends to align with fluid vorticity—which is itself primarily aligned with $\hat{e}_2$, the eigenvector associated with the intermediate eigenvalue of the fluid strain-rate tensor $\mathbb{S}(\vec{x}, t)$ (see Appendix C).

Using the experimental methods outlined in Chapter 2, we are able to investigate the rotation rate of large particles, whose lengthscales are comparable to the Taylor microscale and therefore within the inertial subrange of the ambient turbulence. Interestingly, we find that for particles very close to neutral buoyancy ($\gamma$=1.003), there is no significant difference in the expected value of $\langle \vec{\boldsymbol{\Omega}} \cdot \vec{\boldsymbol{\Omega}} \rangle$ for particles with aspect ratios varying from $0.5 < \alpha < 4$. This mirrors our result for small particles, despite the shift from locally-linear fluid shear (for sub-Kolmogorov scale particles) to locally-nonlinear fluid shear. However, as particle specific gravity increases ($\gamma$=1.006), we observe a local minimum of $\langle \vec{\boldsymbol{\Omega}} \cdot \vec{\boldsymbol{\Omega}} \rangle$ at $\alpha = 1$, with rotation increasing as aspect ratio departs from unity. We also observe differences in the variance of $|\vec{\Omega}|$: particles with longer lengthscales (i.e., non-unity aspect ratios) show higher variances. Higher-order moments also show that when $\vec{\Omega}$ is considered component-wise, with $\Omega_z$ as a representative component, longer-lengthscale particles show a greater variance of $\vec{\Omega}$. This shape-dependent difference in the variance of $\vec{\Omega}$ is very slightly present for particles at $\gamma=1.003$, but is not significant; differences become more pronounced for the denser particles, with $\gamma=1.006$.

Lastly, we begin to investigate the complex transfer of momentum between the fluid turbulence and the suspended particles, and how this transfer may impact particle rotation. We construct a spatially-filtered fluid vorticity field, using a rotationally-symmetric, Gaussian-weighted averaging window. This filtered vorticity, $\mathcal{F}(\omega_z)$, is our approximation of what the particle "feels". As a filter lengthscale, we use the diameter of the sphere which has equivalent volume to our cylindrical particles. The distribution of $\mathcal{F}(\zeta_z)$, the filtered fluid rotation rate, closely matches the distribution of $\Omega_z$, the particle rotation. This signifies that for large particles, rotation is inherited from the surrounding flow according to the spatially-filtered vorticity field: particles integrate nonlinear velocity gradients over their length, and rotate according to the averaged fluid vorticity.

This view—that particles spatially filter the vorticity, and it is this filtered vorticity that determines particle rotation—is consistent with our observations of large-particle rotation. In general, particles with longer lengthscales experience more rotation (in the form of higher values of $\langle \vec{\Omega} \cdot \vec{\Omega} \rangle$), since larger flow structures correspond with higher velocities. Particles with longer lengthscales must also perform this averaging over a wider wavenumber spectrum, and thus the variance of $|\vec{\Omega}|$ is also higher. The longest lengthscale of a particle is therefore an important driving factor in the way the particle inherits vorticity from the flow.

We note that the shape effects that we observed for cylindrical particles were much more pronounced in the more negatively-buoyant particles, even though there was only a 0.3% difference between the two specific gravities. This signifies that particle inertia plays an extremely important role in the transfer of momentum from fluid turbulence to suspended particles, and that studies of particle rotation in turbulence must take inertia into account. This is discussed further in Chapter 5.

# V.    Future Horizons

In the preceding chapters, we have discussed several pertinent questions relating to the motion of large particles in homogeneous, isotropic turbulence, as well as a novel measurement method for our investigation.    In this chapter, we will discuss the major conclusions, implications for the field, and potential future applications of our results.

## 5.1    Discussion and summary of conclusions

### 5.1.1    Refractive-index-matched particle image velocimetry

We have described a new method for the study of suspended objects in flow, using refractive-index-matched materials along with traditional velocimetry techniques.    We fabricate large particles out of hydrogels, which are refractive-index-matched to water, and suspend them in our desired flow (in this case, homogeneous and isotropic turbulence).    This gives us optical access to all parts of the flow, since the hydrogel particles don't block or bend the laser sheet.    Since we embed glass microspheres into the gel, as well as in the fluid, we can use the same PIV algorithm to calculate both the in-particle vector field and the surrounding fluid flowfield.    This simultaneous velocity measurement is information-rich, enabling the study of the turbulent slip velocity (Chapter 3), the study of particle rotation and angular momentum transfer from the fluid to the particle (Chapter 4), and the future exploration of many other topics (Section 5.2).

Though refractive-index matching has been used in conjunction with PIV before (Budwig 1994; Butscher et al. 2012)—and has even been used to study freely-suspended particles in turbulence (Wiederseiner et al. 2010; Klein et al. 2013)—our method has several advantages over those currently in widespread use.    First, our method is cheap.    Many previous studies, e.g. (Hassan and Dominguez-Ontiveros 2008), have required the use of aqueous sodium iodide or other reagents in order to shift the refractive index of the working fluid to match the particle. Sodium iodide is expensive, as are many other materials that make up common refractive-index-matched solid-fluid pairs.    Our method allows the solid-phase objects (particles, structures, etc) to be made in large quantities for little cost, and the working fluid can be provided for no additional cost.    Second, our method is clean. Since the working fluid in our approach is water, there is no need to add salt (Daviero, Roberts, and Maile 2001), sugar (Budwig 1994) sodium iodide, or any other agent to the working fluid.    This allows the experimenter to use the equipment that is already available, and obviates the need for a special dedicated facility for refractive-matched experiments.    Third, our method is adaptable.    Since hydrogels are easily injection-molded into virtually any shape, one can simply make a mold for the desired shapes (via 3D printing, as in our approach, or some other fabrication method).    There are some limits to the injection-molding method; features smaller than ~1mm were not able to be captured by this method.    However, hydrogels may also be cut using laser sintering or stereolithography if finer features are desired (Arcaute, Mann, and Wicker 2010). We find that the 1mm resolution is quite sufficient for our purposes (Figure 5.1).

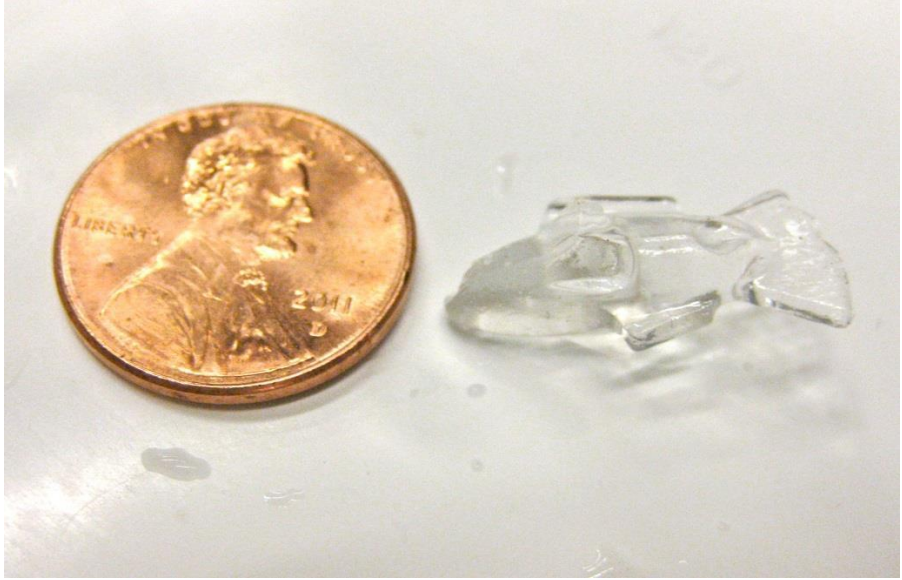

**Figure 5.1: Polyacrylamide fish model, approximately 1cm in length—penny (U.S.) is shown for scale. This demonstrates the resolution limit (~1mm) of features using the injection-molding method.**

This method may be used for future studies which require any of the discussed three attributes (frugality, cleanliness, and adaptability). In fact, our method has already been adapted and used by other researchers seeking to study flow through complex terrain (Weitzman et al. 2014). Furthermore, unlike most materials in widespread use for RIM-PIV, hydrogels are flexible. This flexibility may allow experimenters to use hydrogels as a stand-in for biological tissue. Hydrogels are used widely in the medical field to mimic tissue (Steffen, Turner, and Vanderlaan 2005; Pallua and Wolter 2010; L. H. Christensen et al. 2003), but have not yet been used as models for biological fluid flows (e.g. arterial or respiratory flows). The flexibility of hydrogels, combined with the optical access granted by their matched refractive index, may shed new light on longstanding problems in biology and medicine.

### 5.1.2 Particle translation: settling and slip

Using the method described above and in Chapter 2, we have fabricated hydrogel particles of varying shape, with aspect ratios from 0.5 to 4. All particles are very close to neutral buoyancy, with two slightly different specific gravities $\gamma_1$=1.003 and $\gamma_2$=1.006 (0.3% and 0.6% denser than water, respectively). Despite this very small departure from neutral buoyancy, particles have nontrivial settling velocities in both quiescent and turbulent flows.

To measure the quiescent settling velocity $v_q$, we released particles at the top of a column of still water and precisely measured the time $dt$ taken to fall a prescribed distance $dz$. All particles displayed a preference for the drag-maximizing orientation, with large aspect ratios (rods) falling broad-side and small aspect ratios (disks) falling with their axis of symmetry parallel to gravity. At the lower density $\gamma_1$, settling velocity was relatively constant across all four tested shapes. At the higher density $\gamma_2$, particles with an aspect ratio of unity displayed a much higher settling velocity.

This may be accounted for in part by the smaller cross-sectional area of the falling particles at this aspect ratio, which arises due to the particles' preferential orientation.

The shape-dependent trends in $v_q$ are somewhat predicted by empirical models, but not entirely. We propose three reasons for the lack of fit in this model. The first is that the empirical models are extremely sensitive to density; though we have estimated the particle density very precisely, it is possible that our estimates are not quite precise enough to predict the settling velocity *a priori*. The model curves in Figure 3.2 have overestimated the settling velocity in almost every case, which may signify that we have slightly overestimated particle density. This illustrates the need for an extremely precise density measurement method, and validates our choice of using the settling-velocity method rather than Archimedes' principle (see Section Particle density measurement method3.1.3).

The second reason for the lack of fit is that the empirical/analytical models (drawn from (Loth 2008)) were originally developed for spheres and spheroidal particles, and then extended to cylinders. The model does not do well in describing cylinders whose aspect ratios are close to one, as ours are; the coefficients and correction factors were intended for thin rods or flat disks, with correspondingly large or small aspect ratios. Because of these assumptions, the empirical model will not perfectly describe "fat" cylinders such as the ones we examine.

The third and most interesting reason for the discrepancy between the model and our measurements is the potential onset of wake instabilities. There is a large body of literature on falling spheres which describe experimental observations of the breaking of the axisymmetric wake at an approximate Reynolds number of 200. Classical fluid mechanics predicts that the von Kármán vortex street behind a fixed sphere will shift from periodic and laminar to aperiodic and turbulent at approximately the same Reynolds number threshold. We should therefore expect to see some wake instabilities behind the denser cylinders at $\gamma_2 = 1.006$, which have Reynolds numbers right around this threshold. Along with the issues described above (lack of precision in density measurement and flawed correction coefficients), this accounts for the departure from the empirical model.

In addition to the quiescent settling velocity $v_q$, we also calculate the turbulent slip velocity $\overline{\boldsymbol{u}}_s \equiv \overline{\boldsymbol{u}}_p - \overline{\boldsymbol{u}}_f$, where $\overline{\boldsymbol{u}}_p$ is the average in-plane particle velocity and $\overline{\boldsymbol{u}}_f$ is the average fluid velocity in the neighborhood of the particle. We calculate $\overline{\boldsymbol{u}}_f$ by drawing an annulus around the in-plane particle slice, whose inner bound is set at the outer boundary layer (to exclude fluid influenced by the no-slip boundary condition) and whose outer bound is set at a distance $D_{sph}$ from the surface of the particle (to include all fluid that is within one equivalent-volume-sphere diameter from the particle). The dependence of the slip velocity upon the annulus bounds is calculated and tabulated in Appendix B.

We find that the relationship between $\overline{\boldsymbol{u}}_f$ and $\overline{\boldsymbol{u}}_p$ is linear, according to the following approximation:

$$\overline{\boldsymbol{u}}_p = \overline{\boldsymbol{u}}_f + \langle v_s \rangle \hat{\boldsymbol{j}} + \boldsymbol{\varepsilon} \qquad (5.1)$$

where $v_s$ is the ensemble-averaged vertical slip velocity—that is, a constant offset in the vertical velocity due to the slight negative buoyancy of the particles. This relationship is most easily illustrated as a one-to-one line, with some scatter about that line. This scatter, denoted as $\varepsilon$ in Equation (5.1), is a complex function of the particle history. The instantaneous slip velocity $\vec{u}_s \equiv \vec{u}_p - \vec{u}_f$ is as much a function of the particle history as it is a function of the particle geometry or Stokes number. However, the ensemble-averaged vertical slip $\langle v_s \rangle$ is not a function of $\vec{u}_f$, the instantaneous fluid velocity.

This important result—the independence of $\langle v_s \rangle$ with respect to $\vec{u}_f$—is *not* seen in simple flows, such as the one discussed in Section 3.3.2. A large particle which is embedded in a one-dimensional, oscillatory flow does not follow Equation (5.1). Many naturally-occurring flows of interest are cellular or oscillatory, e.g. convective flows in the atmosphere; in such flows, one must be aware that the slip velocity is likely to be different between updrafts downdrafts. Additionally, our experiments were conducted in moderate-intensity turbulence, in which $v_q$ was comparable to $u_T$, the turbulent velocity scale. Our results do not prove the independence of $\langle v_s \rangle$ for the cases of weak turbulence our strong turbulence. We therefore caution those who would seek to study the slip velocity or turbulence-altered settling velocity to take care to measure $\vec{u}_s$ across the full spectrum of velocities $\vec{u}_f$ which are present in the desired flow.

In our experiments—large, near-neutrally-buoyant particles in moderate-intensity turbulence—we find a very strong correlation between the instantaneous particle velocity $\vec{u}_p$ and the instantaneous surrounding-fluid velocity $\vec{u}_f$. This indicates that though the influence of the particle history may be important, the influence of the instantaneous fluid velocity $\vec{u}_f$ is also very important in determining $\vec{u}_p$. Given the large discrepancy between the particle response time (5.3 s) and the fluid timescale at the same size (0.77 s), it is not apparent that the particle would "care" at all about the surrounding fluid—$\vec{u}_f$ is changing so fast that $\vec{u}_p$ should not have time to adjust to it. The fact that we see such a strong dependence of $\vec{u}_p$ on $\vec{u}_f$ is remarkable and deserves further study.

Lastly, in our experiments, $\langle v_s \rangle$ is strongly reduced relative to $v_q$—we observed a reduction of 40-60%, much more of a reduction than has been previously observed (Nielsen 2007). This is another difference between our experiment and our one-dimensional simulation. In a simple oscillatory flow, settling was reduced by only 15%. This discrepancy points to phenomenological effects within the turbulence which may cause a much more dramatic reduction in comparison to simpler flows. Many studies of settling velocity reduction are performed in simple vortex or cellular flows, and their results are extrapolated to turbulence. Given the dissimilarity between the reduction in the oscillatory flow and the reduction in turbulence, we caution against this sort of extrapolation. Furthermore, because the slip velocity $v_s$ is strongly dependent on $v_f$ in these cases, it is possible to unknowingly measure a biased value of $\langle v_s \rangle$. Therefore, when measuring $\vec{u}_s$, it is doubly necessary to conduct experiments exploring the full range of $\vec{u}_f$ in the desired flow.

Our study of turbulent slip velocity and settling velocity reduction has many implications for both environmental engineering and biological oceanography. We have established that the slip velocity $\langle v_s \rangle$ of large particles is dependent on the flow regime, and that it is the exception and not the norm for $\langle v_s \rangle$ to be independent of $\overline{\boldsymbol{u}}_f$. For scientists studying sediment transport, this may inform the placement of field sites or data collection methods, since we know that it is important to measure $\langle v_s \rangle$ across the full spectrum of $\overline{\boldsymbol{u}}_f$. This may also help us to further develop and advance the state-of-the-art in computational modeling of sediment transport; current models (e.g. Delft3D) use a constant settling velocity which is not dependent on local flow conditions

The question of particle settling through a turbulent water column is also highly relevant to studies of plankton and marine snow, as outlined in Section 1.2.2. Previous work on phytoplankton has shown that "fast-tracking" may occur for very small organisms, enhancing their overall settling velocity in turbulence (Ruiz, Macías, and Peters 2004). However, many species of plankton and nektoplankton are large enough to be a good size match for the particles we study. It is likely that the larger plankton experience *reduced* turbulent settling velocities, as our particles do. This should be taken into account into studies of plankton suspension and energetics. In still water, a large plankter must expend more energy to stay in suspension than the same animal in turbulence. If plankton energetics are measured in still water tanks, rather than *in situ* or in tanks that approximate marine turbulence, the result will not be representative of the animals' natural interactions with their environment.

The settling velocity of marine snow particles is a key parameter in numerical models of the biological carbon pump. Our study provides further support for the curve shown in Figure 1.1, in which the turbulence-altered settling velocity depends heavily on $v_q / u_T$. Marine snow particles show considerable variance in size and shape; this suggests that a single value for the settling velocity may not be appropriate, especially since turbulence will alter the settling velocity in different ways for differently-sized particles. Very small marine snow particles will likely display an enhanced settling velocity in turbulence, whereas very large marine snow particles will have a reduced settling velocity in turbulence. This variation, dependent as it is on shape, size, and turbulence level, will require a nuanced approach for any models which seek to parametrize marine snow settling.

### 5.1.3 Particle rotation

Through experiments and numerical simulation, we investigate the rotation of both small and large nonspherical particles in homogeneous isotropic turbulence. In both small and large particles at neutral buoyancy, we find that total rotation is conserved: for both rod-like and disk-like particles, the expected value of the angular velocity magnitude remains constant. In the small (numerically-simulated) particles, we can break this down further into the components of rotation which are parallel and perpendicular to the particle's axis of symmetry. Though there is no difference in total rotation, we find that this rotation is anisotropically distributed about the particle axes: rods experience more spinning than tumbling, and disks experience more tumbling with almost no spinning. This shape-dependent difference saturates quickly in that rods with an aspect ratio of greater than ten do not experience proportionally more spinning, and disks with an aspect ratio of less than 0.1 do not experience proportionally more tumbling.

This anisotropic distribution of rotation does not occur in the global (-xyz) reference frame. This is due to preferential particle alignment with both the vorticity and the fluid strain. Rods may be considered similarly to fluid material lines, whose equation of motion contains the same stretching term as the vorticity transport equation; therefore, it is expected that small rods align with the flow. Disks, unlike rods, experience a very strong contribution from the fluid strain (rather than merely the fluid rotation). Questions of small-particle alignment with fluid strain and vorticity are discussed in detail in Appendix C. For large particles, response time is very slow compared to fluid timescales, and so we do not expect the same type of alignment with vorticity as in small particles.

Large particles in turbulence experience a very different environment than small particles. For one thing, the fluid velocity gradients at the scale of the particle are no longer linear. For another, the drag force on the particle is no longer linearly proportional to the velocity. The presence of both particle and fluid inertia complicates things as well—particles can no longer be considered as infinitesimally small, their presence displaces fluid as they move through the flow. It is therefore surprising that total rotation should be conserved, as it is in small particles. We find that this is true for particles with $\gamma_1$=1.003, but not true for particles with $\gamma_2$=1.006. This further underscores the role of inertia in particle rotation; a very small increase in density has a large effect on particle motion.

For these denser particles, shape effects arise. Particles with longer lengthscales experience more total rotation, along with a higher rotation variance. This is because particles with longer lengthscales integrate across a larger range of wavenumbers. Smaller wavenumbers, corresponding to larger flow structures, are coupled with higher velocities. Particles with longer lengthscales are more likely to be spun by these larger flow structures, and therefore experience more rotation in general. The higher rotation variance in these particles may be attributed to the same phenomenon: rotation variance is determined by the range of wavenumbers that the particle encounters in, or inherits from, the flow.

This framework of vorticity inheritance is supported by our exploration of the distribution of rotation. We see that large particles rotate according to a spatially-filtered fluid vorticity field, where the filter scale is comparable to the diameter of an equivalent-volume-sphere. The vorticity variance of the unfiltered out-of-plane vorticity field (that is, the field which is resolved at the PIV grid scale) is much larger than the variance of the out-of-plane particle rotation. When the vorticity field is filtered at the scale of the particles, the variance is much more comparable to the particles' rotation variance. The pdf of particle rotation displays noticeably shorter tails than that of the filtered fluid vorticity. This may be attributed to particle inertia. In the fluid, extreme rotation events are infrequent and short-lived. Since the particle response time is very slow, any high-rotation event would be over before particles could "spin up" to the level of the fluid.

In our analysis, we have used $D_{sph}$, the diameter of the equivalent-volume-sphere, as our filter lengthscale. As discussed in the previous paragraph, this creates almost-overlapping pdfs between the filtered fluid rotation rate $\mathcal{F}(\zeta_z)$ and the particle rotation rate $\Omega_z$. However, due to the presence of particle inertia, the pdf of $\Omega_z$ displays shorter tails than the pdf of $\mathcal{F}(\zeta_z)$. An interesting exercise would be to compute a rotation-equivalent lengthscale, $L_\Omega$: the filter length at which the pdf of $\mathcal{F}(\zeta_z)$ best matched the pdf of $\Omega_z$. This lengthscale could be added to the pantheon of equivalent-lengthscales commonly used in studies of nonspherical particle-turbulence interaction, and may prove useful in studies of momentum transfer and vorticity inheritance.

Of course, this concept of vorticity inheritance is just one aspect of the two-way momentum transfer between the fluid and the particles suspended therein. We know that the presence of particles in turbulence interrupts and alters the turbulent cascade, but the exact mechanisms of how this occurs are largely unexplored. We discuss this further in Section 5.2.3.

The shape- and size-dependent differences we observe have implications for both biological navigation and underwater robotics. Firstly, total rotation is not dependent on body shape for neutrally-buoyant particles, animals, or robots. An animal (or robot) cannot minimize the total rotation it experiences by changing its body shape. However, it can select (via its shape) how that rotation may be distributed about the body axes. This may impact the formation and local thickness of boundary layers, and therefore affect nutrient uptake in planktonic organisms. This tendency may also represent an evolutionary pressure on the propulsive mechanisms for a given organism: a disk-shaped organism, which experiences a great deal of tumbling and not much spinning, may employ different propulsive mechanisms than a rod-shaped organism which is given to spinning and not much tumbling. The greatest marginal return for rotation alteration due to shape-change occurs at aspect ratios which are close to unity. Many planktonic animals, such as cydippid ctenophores, display roughly spheroidal body plans with aspect ratios of about $\alpha=1$. This places them in the range for which small changes in shape will greatly influence rotation, and suggests that shape may play a major role in their behavior and locomotion. Continued cross-disciplinary study is needed to elucidate the full impact of body shape on plankton biology in complex flows.

## 5.2   Future work: particle-laden turbulence

In the previous section, we have summarized and discussed the results of our experiments, as well as their broader implications. However, the results of our experiments may also be used to explore the physics of particle-interaction in greater depth. The following sections contain suggestions for future work using the approach outlined in this thesis (stereoscopic PIV of hydrogel particles in homogeneous isotropic turbulence).

### 5.2.1   Particle boundary layer

The boundary layer on a large freely-suspended particle is not well-defined, even for a sphere. Our method gives us the ability to investigate the finely-resolved boundary layer on a large number of particles. In particular, we are interested in the variation of velocity, shear, and turbulent kinetic energy within the particle boundary layer. In this investigation, we propose classifying two regions: the "fore" and "aft" wake of suspended particles (illustrated in Figure 5.2).

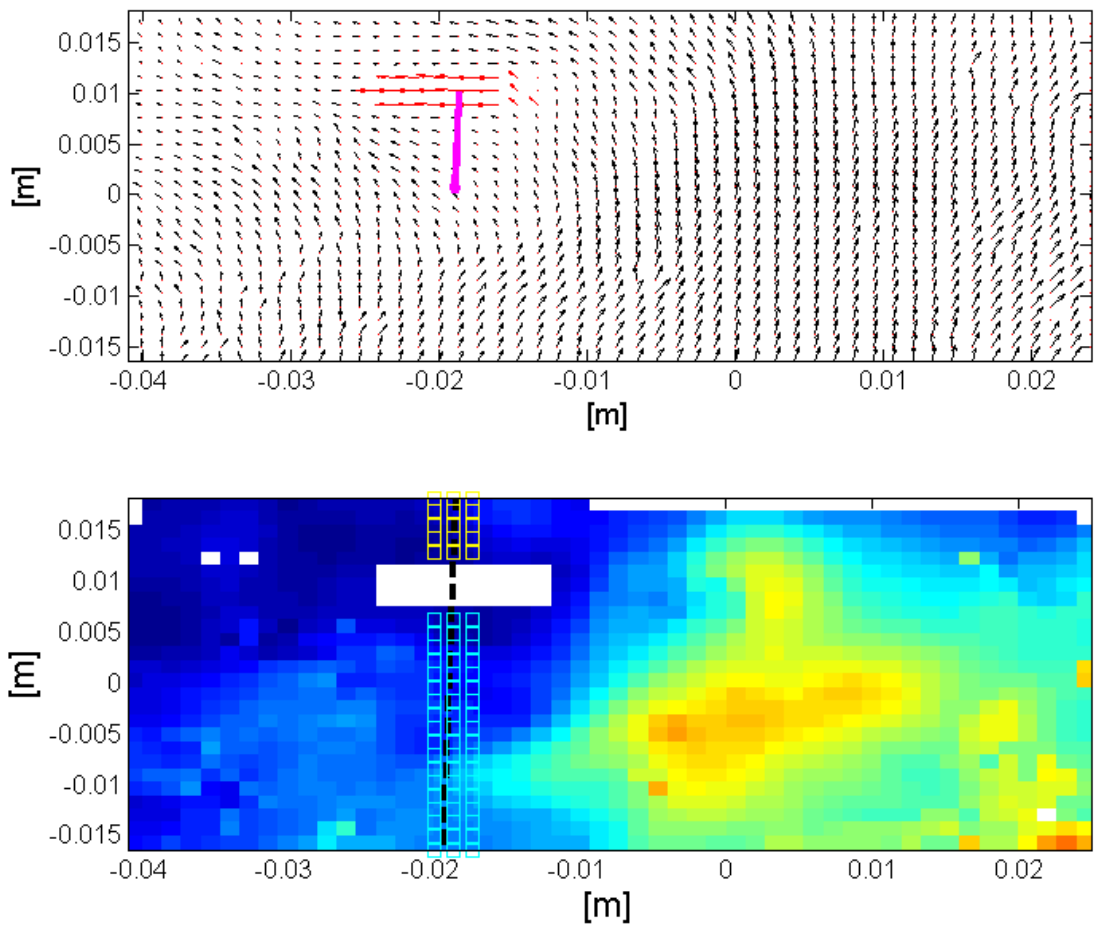

Figure 5.2: Example data for a single particle ($\alpha$=4, $\gamma_1$=1.003), showing (a) fluid vector field (black arrows) with in-particle velocities (red arrows), along with the calculated slip velocity (magenta arrow, not to scale). (b) shows the turbulent kinetic energy field (background colors), overlaid with a line drawn through the particle center and parallel to the particle slip. Square markings represent points considered to be in the "fore" wake (cyan) and the "aft" wake (yellow). In the example shown, the two wake regions are calculated based on any fluid that is within 2mm of the slip-direction line; future analysis will change the size of this region to see if turbulence properties (e.g. TKE, fluid shear, et al) are dependent on their proximity to the particle and/or their location fore or aft of the particle.

In Figure 5.2, we have shown an example of this kind of calculation. A particle is embedded in a turbulent eddy, with some slip velocity and slip direction (Figure 5.2a). The particle is moving relative to the fluid; we divide this fluid into an "aft" region, containing the fluid immediately behind the particle with respect to the slip direction, and a "fore" region, containing the fluid immediately in front of the particle with respect to the slip direction. We expect that these two regions will have significantly different single-point statistics. By analyzing differences between the fore-wake, aft-wake, and the background turbulence (e.g. in turbulent kinetic energy or shear), we may be able to provide some insights into boundary layer formation in large nonspherical particles, along with vortex shedding and wake dynamics on freely-suspended particles. This analysis will also help us to elucidate potential mechanisms of turbulence modulation in particle-laden flows (see Section 5.2.3).

## 5.2.2 Alignment and orientation

As we discussed in Section 4.1, our measurement method does not allow us to break down the particle rotation $\vec{\mathbf{\Omega}}$ into its tumbling ($\vec{\mathbf{\Omega}}_\perp$) and spinning ($\vec{\mathbf{\Omega}}_\parallel$) components. This limits our ability to draw comparisons between the behavior of large and small particles. Additionally, we have not explored the alignment of our large particles with the local strain eigensystem, as we have in the numerically-simulated small particles. This is in part due to the ease of information access in the numerical simulation: $\mathbb{S}(\vec{x}, t)$ is already computed and available for every point in the flow, and is unaffected by the presence of particles. This information is much harder to come by in the experimental case, where particles take up space and affect the flow around them.

To study either flow alignment or tumble/spin distribution, we must obtain the particles' instantaneous orientations with respect to the global (-xyz) coordinate system. The particle orientation angles (often referred to as "Euler angles" in studies of solid-body dynamics), in conjunction with our measurements of the rotation vector $\vec{\mathbf{\Omega}}$, will allow us to calculate $\vec{\mathbf{\Omega}}_\perp$ and $\vec{\mathbf{\Omega}}_\parallel$ for each particle. The Euler angles will also provide a basis for comparison with $\mathbb{S}(\vec{x}, t)$, analogous to our use of $\hat{\mathbf{n}}$ in the small particle case.

In theory, all that is needed to determine an object's Euler angles are two independent views of that object, using two cameras whose fields-of-view have been calibrated in 3D space. However, in practice, it is very difficult to use only two cameras. Often three or more cameras are used in order to minimize error. Most 3D-position-finding systems operate as follows: first, focus two or more cameras on the same volume of space. Second, place a three-dimensional object into the space, with clearly marked points or features distributed throughout the volume. Third, mark the position of all the features in each of the camera views. Lastly, use an algebraic reconstruction algorithm, e.g. (Herman and Lent 1976), to reconcile the views of all cameras, so that each local camera coordinate system is mapped to 3D space. Now, when an object passes into the calibrated volume, the experimenter needs only to locate the object (or features on that object) in each camera view, and feed the local-view coordinates into the calibrated algorithm to calculate the global, 3D coordinates.

This seemingly simple approach becomes very difficult in our experimental context. Due to the near-transparency of the hydrogels, it is difficult to locate the borders of the suspended particles (since most of a particle's surface actually lies outside the laser sheet). Further complications arise due to the brightness of the laser sheet. The laser sheet is necessary to perform PIV and therefore to compute particle and fluid velocity fields. However, due to the difference in background light scattering between the hydrogel and the surrounding fluid (discussed in Section 2.2.2), the in-plane particle slice becomes very bright compared to the rest of the particle. The dim outlines of the out-of-plane sections of the particle are still visible, but in many cases are indistinct (Figure 5.3). Image contrast must be highly elevated to even see the particles or particle edges that lie outside the plane. Additionally, algebraic reconstruction algorithms (and the various pieces of software that make use of them) require the user to find the exact same point in all camera views. Our hydrogel particles are not marked with specific features or tracking points, which makes this requirement difficult. We can approximate the endpoints of the particle's primary axis, but this approach may lead to substantial error. A third camera, as well as fluorescent trackers embedded in the particles (such as those used by (Klein et al. 2013)) could potentially enable us to find 3D particle orientation, and unlock the various analyses mentioned above.

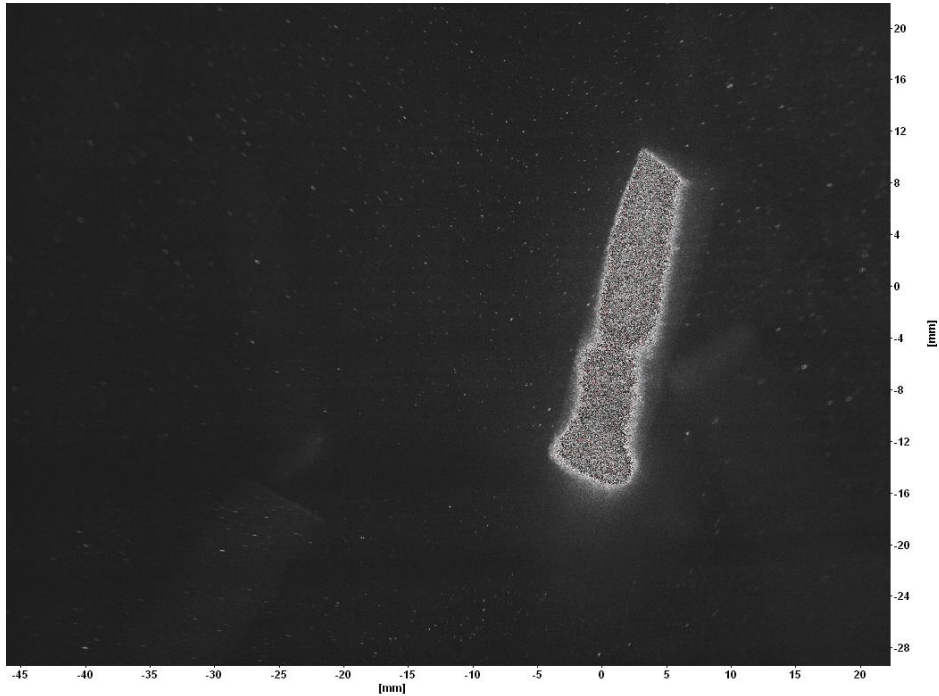

**Figure 5.3: Example raw image of hydrogel particles (α=4) in turbulence. The particles and edges that lie outside the laser plane are indistinct, and it is difficult to approximate the location of the particle's primary axis. (Note: for this image, contrast was heightened to view particle outlines, which makes the in-particle slice appear saturated).**

## 5.2.3 Turbulence modulation and momentum transfer

The mere presence of suspended particles necessarily alters the characteristics of the surrounding turbulence (Balachandar and Eaton 2010). In single-phase turbulent flow, classical Kolmogorov theory states that energy is passed losslessly from larger to smaller eddies, and that viscosity begins to dissipate energy only when eddy sizes are below a certain scale. However, suspended solid particles change this model in many ways. When particles are added to the flow, the surface area of each particle must adhere to the no-slip boundary condition. If the particles are inertial, particle slip will create sharp velocity gradients and corresponding shear stresses, creating numerous opportunities for viscosity to dissipate energy. Therefore, the presence of large inertial particles will always lead to a decrease in the total kinetic energy of a turbulent flow (Lucci, Ferrante, and Elghobashi 2010).

Though the total energy of the turbulent flow decreases when particles are added, this attenuation is not constant across wavenumber space. Large particles subtract TKE at small wavenumbers but add it at large wavenumbers, due to the vortex shedding and shear production at and below the scale of the particle. This leads to a "spectral pivot", observed and recorded by many researchers (Schreck and Kleis 1993; Poelma, Westerweel, and Ooms 2007). Energy is damped at the large scales and overall, but enhanced at the small scales, and the resultant spectra flatten out in comparison to the characteristic power law in which $E(\kappa) \propto \kappa^{-5/3}$ (where $E(\kappa)$ is the kinetic energy and $\kappa$ is the wavenumber). In particle-laden flows, the power-law exponent will be greater than $-5/3$.

It is an open question whether nonspherical particles attenuate turbulence in the same way that spherical particles do. In some previous work, we have begun investigating this question using hydrogel RIM-PIV (Bellani et al. 2012). This study used spheroidal particles (a sphere at α=1 and a prolate ellipsoid at α=2; the minor axis of the ellipsoid was matched to the diameter of the sphere). Our initial experiments showed that at the same volume fraction, spherical particles removed more energy in total than ellipsoidal particles. Though both shapes injected energy at high wavenumbers (compared to the single-phase case), the ellipsoidal particles injected more energy into these small scales. This resulted in a total attenuation of 3% for ellipsoids, but 15% for spheres. However, in this study, particles were not volume-matched. Therefore, though the volume fraction of the two experiments was held constant, the number concentration of the spheres was much higher (almost twice that of the ellipsoids). This suggests that it may be inappropriate to compare these two particle shapes, since the number concentration was radically different in this experiment.

In the work presented here, particles are volume-matched to within 20% variation and surface-area matched to within 5% variation, but vary in aspect ratio from $0.5 < \alpha < 4$. Our data are therefore ideal for studying the effects of particle shape on turbulence modulation, building on the work of (Bellani et al. 2012). However, in order to gain a high resolution of the fluid flow field, our window size is small (6 by 8 cm). Because of this, we are limited in the range of wavenumbers we can examine. If shorter focal-length lenses were used on the PIV cameras, a larger window size (such as the one described in (Bellani et al. 2012)) could be obtained. This would allow us to study turbulence modulation over the full inertial subrange, including any differences in the spectral pivot which arise due to particle shape.

The transfer of momentum between a turbulent flow and the particles suspended within it is complex. We have hypothesized that particles behave, in a way, as a high-pass wavenumber filter: particles rotation is controlled by eddies at their own size scale, and the spinning particles shed small eddies in their wake. We can observe this at the particle scale (see Figure 4.8 and the accompanying discussion). However, we can also observe it in the aforementioned macro-scale investigations of turbulence modulation, in the spectral pivot and small-scale energy injection. Further investigation of the particle boundary layer (see Section 5.2.1) will allow us to explore the mechanisms by which turbulence modulation occurs, as we investigate shear production and vortex shedding at the scale of the particle.

## 5.3    Future work: marine animal navigation and robotics[1]

Our insights on particle rotation and translation will provide a base from which to explore the biomechanics of meso-scale aquatic organisms. As discussed in Sections 1.1.2 and 1.1.3, animals and robots whose lengthscales are within the inertial subrange of turbulence face unique navigational challenges; they must navigate in, around, and through flow structures that are both larger and smaller than themselves. They will have intermittent control over their position and orientation in the water column; when passing through strong gusts (i.e., large flow structures), they will behave like passive particles, but in relatively calm spatiotemporal regions, they may regain control. Therefore, the passive behavior of particles in turbulence and the investigation contained within this thesis will be helpful to those seeking to study marine animal locomotion. Even when animals are relatively in control, there will still be a passive component to their observed behavior. Indeed, many animals which fall into the "nektoplankton" category spend much of their lifecycle as passive drifters.

Building on the research contained in this dissertation, we will investigate the swimming and navigation of cydippid ctenophores (Figure 1.2). This order of nektoplankton can be from one to several centimeters in length—within the inertial subrange of oceanic turbulence. They are approximately neutrally buoyant and are roughly spheroidal in shape, with body aspect ratios from one to three. Additionally, their unique ciliated propulsion system has not yet been studied in the context of possible adaptation for underwater vehicles, despite ctenophore's agility and maneuverability in complex environments. The study of ctenophores is therefore a logical next step for our investigation.

### 5.3.1    Background and motivation

Ctenophores, or comb jellyfish, are the largest known ciliated organisms. When active, their motion is facilitated by the use of cilia, despite ctenophores' relatively large size (1-15cm) compared to other cilia-reliant organisms (typically tens to hundreds of micrometers (Lynn 2008)). In ctenophores, specialized long cilia are packed into paddle-like plates called ctenes, which are organized into rows or "combs" along the side of the body. Many ctenophores combine other propulsive mechanisms, such as jetting (*Thalassocalyce inconstans*) or undulating (*Cestum veneris*), with fine-level position control by ctene rows (Hutchins and Olendorf 2004). Others (e.g. *Pleurobrachia*) rely exclusively on their eight ctene rows for both propulsion and control (Barlow and Sleigh 1993).

---

[1] The work described in section 5.3 will be conducted by the author during a postdoctoral appointment from 2015 - 2017, under the joint supervision of Dr. Matthew McHenry (Ecology and Evolutionary Biology, University of California Irvine) and Dr. John Dabiri (Mechanical and Civil Engineering, Stanford University).

Virtually all ctenophore species are predators that feed on smaller plankton, such as copepods, larvae, and other ctenophores. They are typically voracious, often indiscriminate in prey choice, and can eat over 10 times their body weight per day (Suthers 2009). This contributes to the great success of some ctenophores as invasive species (Purcell et al. 2001; Zaika and Sergeeva 1990). Ctenophores also serve as a food source for jellyfish, turtles, and marine fishes, with recent research showing that they form a large portion of the diet of commercially important fish such as chum salmon and spiny dogfish (Ford and Link 2014; Purcell and Arai 2001). They are a crucial link in the marine food chain, limiting the populations of planktonic organisms that would otherwise graze on phytoplankton, the ocean's major carbon-capturers. Despite the ecological importance of ctenophores, little is known about their distribution within the water column, their methods of navigation and control, or their predation effectiveness in different flow conditions. Our research will strive to close this gap by investigating ctenophore locomotion in complex flows, including the effects of turbulence on feeding and swimming behavior.

As we have outlined in Section 1.2.3, the utility of small underwater robots is evident for a number of applications. Most AUVs in widespread usage today are larger than 5 meters in length; even the smallest AUVs are usually around half a meter in length Table 1.2. The development of a centimeter-scale robot that is both efficient and maneuverable would enable the exploration of many sensitive habitats or complex terrains that are inaccessible with current technology. Additionally, centimeter-scale robots could potentially be deployed in networks, allowing researchers to study larger areas with more facility. Ctenophores' unique ciliated propulsion system and small size can provide inspiration for vehicles that are much smaller than those in common usage today. Our study, by investigating biological propulsion and control mechanisms in ctenophores, will enable the development of new technologies to meet the rising environmental challenges of climate change, ocean acidification, and pollution.

### 5.3.2 Research objectives

Our first objective will be to quantify ctenophore agility in turbulent and nonturbulent conditions, including adaptive behaviors and predator-prey interactions. We will measure, both in the laboratory and in situ, a suite of standard performance metrics including maximum observed acceleration, length-specific turning radius and stopping distance. This analysis will include the effects of turbulence and shear on ctenophore navigation and locomotion, and identify any special behaviors associated with flow condition (e.g. turbulence avoidance, changing gait patterns, leveraging periodic vortex structures, etc). We will also examine changes in predation effectiveness with increasing flow complexity, which will provide insight into ctenophores' preferred locations in the water column and help explain their success as non-native invaders in ecosystems around the world. Our ultimate goal is to situate ecological observations of these animals within a physical context.

Our second objective will be to extract translational principles of ctenophore locomotion for use in the advancing field of autonomous underwater vehicle design. As stated in Section 1.2.3, two of the major current limitations on AUV design are navigation and power ((Z. Wang, Thiébaut, and Dauvin 1995; Heidemann, Stojanovic, and Zorzi 2012)). The second of these is quickly advancing: today's vehicles are equipped with the latest, lightest, longest-lasting onboard power supplies. Our study will illuminate the physical mechanisms behind efficient ctenophore swimming, which can be applied to the design of AUVs for longer mission durations, more complex maneuvers, or a higher payload. Our exploration of the outer limits of ciliate agility and control will also provide insight into AUV design and potentially allow roboticists to decrease path deviation due to turbulent gusts, improving the accuracy of navigational methods. The development of small, agile AUVs will also be a significant benefit for underwater ecologists and biological oceanographers, as they can be used to measure ocean properties and organism-habitat interactions in areas that are currently inaccessible.

### 5.3.3 Proposed experiments

We will use the latest technology to investigate ctenophore swimming and mechanisms of locomotion control, including stereoscopic imaging, PIV, and high-speed video. Using these tools, we will test the following hypotheses:

- **H$_1$:** Ctenophore agility, as well as predation effectiveness, decreases with increasing turbulence/shear.
- **H$_2$:** Navigational control of ctenophores is governed exclusively by asymmetric beating of comb rows.
- **H$_3$:** Comb rows are able to respond to turbulent velocity fluctuations and maintain animal orientation and heading in complex flows.
- **H$_4$:** Cilia are an effective mechanism for the control of meso-scale AUVs in variable flow conditions.

Our first experiment will measure the dependence of ctenophore agility on fluid shear. We will collaborate with the Monterey Bay Aquarium (MBA) and aquarist Thomas Knowles to stereoscopically film *Pleurobrachia bachei* and *Hormiphora californiensis*, two species of cydippid ctenophore. MBA will provide animals and planktonkreisel tanks in which we will produce simple shear (Rakow and Graham 2006). We will quantify this shear flow using PIV. To induce active behavior, we will maintain a high ambient density of typical prey items (copepods or live *Artemia*) and record successful vs. unsuccessful captures. Animals will be digitally tracked to examine translational and angular velocities and accelerations, as well as length-specific turning radii (Figure 5.4). These agility metrics will be quantified in varying levels of shear (H$_1$). We will also simultaneously image animals using high-speed, high-resolution videography, allowing us to correlate the movement of individual ctene rows with large-scale movement (H$_2$).

Our second experiment will measure the dependence of ctenophore agility on fluid turbulence. We will measure ctenophores in turbulence, using DPIV to simultaneously image animals together with the surrounding fluid flow. Animals will be studied in a custom-built tank, using grid structures to generate turbulence at the appropriate scales. Similar tanks have been used with great success to study turbulence-organism interactions ((Webster, Braithwaite, and Yen 2004; Hondzo and Lynn 1999)). We will study ctenophore responses to turbulent velocity structures (including predator-prey interactions), quantifying both macro-scale agility and individual ctene-row responses to turbulence ($H_3$, $H_2$).

Lastly, we will use the results from the first two experiments to create a numerical model of a ctenophore-inspired ciliated swimmer. We will subject our model swimmer to typical environmental flow conditions to further test the effectiveness of cilia as a control system for navigation and orientation ($H_2$, $H_3$) and to explore the possibility of a similar system for use in the design of underwater vehicles ($H_4$).

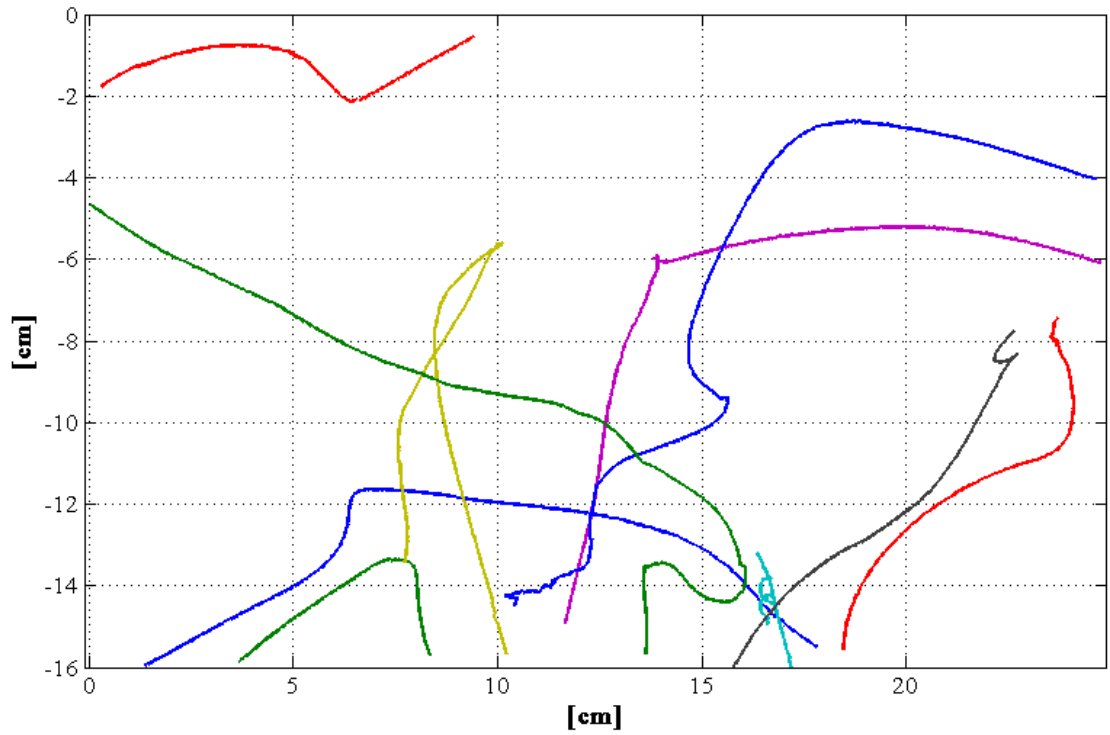

Figure 5.4: Preliminary data, collected at the Monterey Bay Aquarium, showing center-of-mass trajectories of *Pleurobrachia bachei* in low-shear flow, displaying small turning radii and rapid directional reversal. Each color is a different animal, and flow is right-to-left.

### 5.3.4 Impact

The impact of this study is both immediate and far-reaching. The quantification of ctenophore navigation methods and agility in turbulence will provide biologists and oceanographers with much-needed data on these important and ubiquitous animals; our investigation of macro-scale ciliate locomotion will pave the way for work in biomimetic robotics, and may also shed light on issues of human health, ocean mixing and circulation, and animal phylogeny.

As the field of underwater robotics continues to advance, and vehicles continue to shrink, new needs will arise in vehicle design. Smaller vehicles occupy a different physical flow regime where the impact of turbulence becomes nontrivial, which necessitates new control systems. Our investigation of ctenophore navigation and locomotion will progress in parallel with the development of robotics technology, so that when small robots become readily deployable, a bio-inspired control system will be ready to implement. This will address the challenge of fine-scale control for AUVs while providing ecological and behavioral data on predation effectiveness, turbulence avoidance, and agility in different flow conditions. Though ctenophore locomotion has been studied in the past, much of the prior work was completed before the advent of PIV, high-speed/high-resolution cameras, or advanced computer models. We now have the technology for a much more quantitative analysis of performance characteristics like power output (Barlow and Sleigh 1993), velocity fields and feeding currents, and 3D orientation and position.

Ctenophores are currently of great interest to the genetics and neurological communities, as recent research (Moroz et al. 2014) has indicated that ctenophores' nervous systems may have evolved independently from those of other animals. They are also widespread and important contributors to the global ecosystem (Harbison, Madin, and Swanberg 1978), and they exist within complex, turbulent fluid environments. Knowledge of ctenophore-environment interactions could prove to be a crucial missing link in questions of conservation and sustainability, such as the removal of invasive species, oceanic carbon sinks, and restoring balance to marine food chains.

Recent research has also shown that small organisms may make major contributions to ocean mixing (Noss and Lorke 2014; Thiffeault and Childress 2010; Dabiri 2010; Kunze et al. 2006). Most ctenophores range in size from approximately 1-15cm (Matsumoto 1991), which makes them ideal candidates for fluid transport via drift (Katija and Dabiri 2009). They are also exceedingly numerous and reach high population densities (e.g. 17-33 individuals/m$^3$ (Miller 1974; van der Veer and Sadée 1984), (Z. Wang, Thiébaut, and Dauvin 1995)) and are distributed throughout the world's oceans (Harbison, Madin, and Swanberg 1978). Further analysis of the near- and far-field flows surrounding ctenophores would elucidate their potential role in ocean mixing. A thorough understanding of the biomechanics of cilia is also necessary for advances in treatment for ciliopathic and respiratory diseases (Waters and Beales 2011; Pan, Wang, and Snell 2005; S. T. Christensen et al. 2007).

Lastly, though ciliated propulsion has been proposed and somewhat developed as a potential mechanism for underwater microrobots (Ghanbari and Bahrami 2011; Palagi et al. 2013; Sareh et al. 2012) (1-200µm), it has not yet been proposed as a method of control for meso-scale underwater vehicles (1-10 cm). Additionally, no one has explored the mechanics of ciliated navigation in conjunction with other "primary" propulsive modes such as jetting or undulation, which are common among ctenophores. A detailed study of ctenophore biomechanics will inform robotics design and open new avenues of research in underwater propulsion.

## 5.4   Concluding thoughts

Through our use of novel techniques in refractive-index-matching, we obtain simultaneous measurements of large, freely-suspended, nonspherical particles and the surrounding homogeneous isotropic turbulence. We find that in turbulence, the instantaneous particle velocity and the instantaneous fluid velocity follow a one-to-one relationship, and that the slip velocity is not a function of the fluid environment. This is not true for numerically-modeled similar particles in a simple oscillatory flow. We also find that slip velocity is drastically reduced in comparison to quiescent settling velocity. Future work must consider the complicated relationship between particle slip, turbulence-altered settling velocities, and the surrounding flows. This is especially important for numerical models which use a single value to parametrize particle settling, such as models for sediment transport or marine snow settling in the biological pump.

We explore particle rotation and find that for neutrally-buoyant particles, the expected angular velocity magnitude in turbulence does not depend on particle shape. With increasing inertia, particles' longest lengthscales begin to play a role. We present a qualitative framework for particle rotation in which particles are rotated by the range of wavenumbers corresponding to their longest lengthscale. Particles with longer lengthscales therefore experience a greater variance in angular velocity, as well as a higher overall angular velocity magnitude (since larger eddies are coupled with higher velocities). This qualitative framework is further supported by the similarity in the distribution of rotation for both large particles and the fluid vorticity field which is spatially filtered at the lengthscale of the particles.

Though the dynamics of non-spherical particles remain enigmatic in many ways, we have begun to step towards a better understanding. There is much still to be explored, and we look forward to new computational and experimental work which will illustrate a clearer picture of the complex physics behind particle translation and rotation in turbulence.

# List of works cited

Abrahamson, S., and S. Lonnes. 1995. "Uncertainty in Calculating Vorticity from 2D Velocity Fields Using Circulation and Least-Squares Approaches." *Experiments in Fluids* 20 (1): 10–20. doi:10.1007/BF00190593.

Aidun, C.K., and J.R. Clausen. 2010. "Lattice-Boltzmann Method for Complex Flows." *Annual Review of Fluid Mechanics* 42: 439–72. doi:10.1146/annurev-fluid-121108-145519.

Aleev, IŪriĭ Glebovich. 1977. *Nekton*. Kluwer Academic.

Alldredge, A. L., and C. C. Gotschalk. 1989. "Direct Observations of the Mass Flocculation of Diatom Blooms: Characteristics, Settling Velocities and Formation of Diatom Aggregates." *Deep Sea Research Part A. Oceanographic Research Papers* 36 (2): 159–71. doi:10.1016/0198-0149(89)90131-3.

Alldredge, A.L., U. Passow, and B.E. Logan. 1993. "The Abundance and Significance of a Class of Large, Transparent Organic Particles in the Ocean." *Deep Sea Research Part I: Oceanographic Research Papers* 40 (6): 1131–40. doi:10.1016/0967-0637(93)90129-Q.

Alldredge, A.L., and M.W. Silver. 1988. "Characteristics, Dynamics and Significance of Marine Snow." *Progress in Oceanography* 20 (1): 41–82. doi:10.1016/0079-6611(88)90053-5.

Andersson, H.I., and A. Soldati. 2013. "Anisotropic Particles in Turbulence: Status and Outlook." *Acta Mechanica* 224 (10): 2219–23. doi:10.1007/s00707-013-0926-y.

Andersson, H.I., L. Zhao, and M. Barri. 2012. "Torque-Coupling and Particle–turbulence Interactions." *Journal of Fluid Mechanics* 696: 319–29. doi:10.1017/jfm.2012.44.

Andrews, John C. 1983. "Deformation of the Active Space in the Low Reynolds Number Feeding Current of Calanoid Copepods." *Canadian Journal of Fisheries and Aquatic Sciences* 40 (8): 1293–1302. doi:10.1139/f83-147.

Arcaute, K., B.K. Mann, and R.B. Wicker. 2010. "Fabrication of Off-the-Shelf Multilumen Poly(Ethylene Glycol) Nerve Guidance Conduits Using Stereolithography." *Tissue Engineering Part C: Methods* 17 (1): 27–38. doi:10.1089/ten.tec.2010.0011.

Armenio, V., and V. Fiorotto. 2001. "The Importance of the Forces Acting on Particles in Turbulent Flows." *Physics of Fluids (1994-Present)* 13 (8): 2437–40. doi:10.1063/1.1385390.

Ashurst, W.T., A R. Kerstein, R. M. Kerr, and C. H. Gibson. 1987. "Alignment of Vorticity and Scalar Gradient with Strain Rate in Simulated Navier–Stokes Turbulence." *Physics of Fluids (1958-1988)* 30 (8): 2343–53. doi:10.1063/1.866513.

Asper, V.L. 1987. "Measuring the Flux and Sinking Speed of Marine Snow Aggregates." *Deep Sea Research Part A. Oceanographic Research Papers* 34 (1): 1–17. doi:10.1016/0198-0149(87)90117-8.

Bagchi, P., and S. Balachandar. 2003. "Effect of Turbulence on the Drag and Lift of a Particle." *Physics of Fluids (1994-Present)* 15 (11): 3496–3513. doi:10.1063/1.1616031.

Balachandar, S., and J.K. Eaton. 2010. "Turbulent Dispersed Multiphase Flow." *Annual Review of Fluid Mechanics* 42 (1): 111–33. doi:10.1146/annurev.fluid.010908.165243.

Bandyopadhyay, P.R. 2005. "Trends in Biorobotic Autonomous Undersea Vehicles." *IEEE Journal of Oceanic Engineering* 30 (1): 109–39. doi:10.1109/JOE.2005.843748.

Barlow, D., and M. A. Sleigh. 1993. "Water Propulsion Speeds and Power Output by Comb Plates of the Ctenophore Pleurobrachia Pileus Under Different Conditions." *Journal of Experimental Biology* 183 (1): 149–64.

Barrett, D., M. Grosenbaugh, and M. Triantafyllou. 1996. "The Optimal Control of a Flexible Hull Robotic Undersea Vehicle Propelled by an Oscillating Foil." In *, Proceedings of the 1996 Symposium on Autonomous Underwater Vehicle Technology, 1996. AUV '96*, 1–9. doi:10.1109/AUV.1996.532833.

Basset, A.B. 1888. *A Treatise on Hydrodynamics: With Numerous Examples*. Deighton, Bell and Company.

Batchelor, G. K. 1952. "The Effect of Homogeneous Turbulence on Material Lines and Surfaces." *Proceedings of the Royal Society of London. Series A, Mathematical and Physical Sciences* 213 (1114): 349–66. doi:10.1098/rspa.1952.0130.

Bellani, G., M.L. Byron, A.G. Collignon, C.R. Meyer, and E.A. Variano. 2012. "Shape Effects on Turbulent Modulation by Large Nearly Neutrally Buoyant Particles." *Journal of Fluid Mechanics* 712: 41–60. doi:10.1017/jfm.2012.393.

Bellani, G., M.A. Nole, and E.A. Variano. 2013. "Turbulence Modulation by Large Ellipsoidal Particles: Concentration Effects." *Acta Mechanica* 224 (10): 2291–99. doi:10.1007/s00707-013-0925-z.

Bellani, G., and E. A. Variano. 2012. "Slip Velocity of Large Neutrally Buoyant Particles in Turbulent Flows." *New Journal of Physics* 14 (12): 125009. doi:10.1088/1367-2630/14/12/125009.

Bellani, G., and E.A. Variano. 2014. "Homogeneity and Isotropy in a Laboratory Turbulent Flow." *Experiments in Fluids* 55 (1): 1–12. doi:10.1007/s00348-013-1646-8.

Bernstein, O., and M. Shapiro. 1994. "Direct Determination of the Orientation Distribution Function of Cylindrical Particles Immersed in Laminar and Turbulent Shear Flows." *Journal of Aerosol Science* 25 (1): 113–36. doi:10.1016/0021-8502(94)90185-6.

Bingham, D., T. Drake, A. Hill, and R. Lott. 2002. "The Application of Autonomous Underwater Vehicle (AUV) Technology in the Oil Industry–vision and Experiences." In *FIG XXII International Congress Washington, DC USA*, 1–13.

Birouk, M., B. Sarh, and I. Gökalp. 2003. "An Attempt to Realize Experimental Isotropic Turbulence at Low Reynolds Number." *Flow, Turbulence and Combustion* 70 (1-4): 325–48. doi:10.1023/B:APPL.0000004974.74706.6d.

Blidberg, R.D. 2001. "The Development of Autonomous Underwater Vehicles (AUV); A Brief Summary." In *International Conference on Robotics and Automation--ICRA*. AUSI.

Boatright, J.L., D.E. Balint, W.A. Mackay, and J.M. Zajicek. 1997. "Incorporation of a Hydrophilic Polymer into Annual Landscape Beds." *Journal of Environmental Horticulture* 15 (1): 37–40.

Borzacchiello, A., and L. Ambrosio. 2009. "Structure-Property Relationships in Hydrogels." In *Hydrogels*, 9–20. Springer Milan. http://link.springer.com/chapter/10.1007/978-88-470-1104-5_2.

Boussinesq, Joseph. 1903. *Théorie analytique de la chaleur mise en harmonic avec la thermodynamique et avec la théorie mécanique de la lumière*. Gauthier-Villars.

Brush, L. M., H.-W. Ho, and S.R. Singamsetti. 1962. "A Study of Sediment in Suspension." *Bulletin of the International Association of Scientific Hydrology* 59.

Budwig, R. 1994. "Refractive Index Matching Methods for Liquid Flow Investigations." *Experiments in Fluids* 17 (5): 350–55. doi:10.1007/BF01874416.

Buevich, Y.A. 1966. "Motion Resistance of a Particle Suspended in a Turbulent Medium." *Fluid Dynamics* 1 (6): 119–119. doi:10.1007/BF01022298.

Burgers, J. M., and M. Mitchner. 1953. "On Homogeneous Non-Isotropic Turbulence Connected with a Mean Motion Having a Constant Velocity Gradient. I." Edited by F. T. M. Nieuwstadt and J. A. Steketee. *Proceedings of the Koninklijke Nederlandse Akademie van Wetenschappen* 56: 228–35. doi:10.1007/978-94-011-0195-0_25.

Butscher, D., C. Hutter, S. Kuhn, and P.R. von Rohr. 2012. "Particle Image Velocimetry in a Foam-like Porous Structure Using Refractive Index Matching: A Method to Characterize the Hydrodynamic Performance of Porous Structures." *Experiments in Fluids* 53 (4): 1123–32. doi:10.1007/s00348-012-1346-9.

Byron, M.L., J. Einarsson, K. Gustavsson, G. Voth, B. Mehlig, and E. Variano. 2015. "Shape-Dependence of Particle Rotation in Isotropic Turbulence." *Physics of Fluids* 27 (3): 035101. doi:10.1063/1.4913501.

Byron, M.L., and E.A. Variano. 2013. "Refractive-Index-Matched Hydrogel Materials for Measuring Flow-Structure Interactions." *Experiments in Fluids* 54 (2). doi:10.1007/s00348-013-1456-z.

Cada, G. F. 1997. "Shaken, Not Stirred: The Recipe for a Fish-Friendly Turbine." In *Proceedings of an International Conference & Exposition on Hydropower*, 374–82. New York, NY: American Society of Civil Engineers.

Calzavarini, Enrico, Romain Volk, Emmanuel Lévêque, Jean-François Pinton, and Federico Toschi. 2012. "Impact of Trailing Wake Drag on the Statistical Properties and Dynamics of Finite-Sized Particle in Turbulence." *Physica D: Nonlinear Phenomena*, Special Issue on Small Scale Turbulence, 241 (3): 237–44. doi:10.1016/j.physd.2011.06.004.

Challabotla, Niranjan Reddy, Lihao Zhao, and Helge I. Andersson. 2015. "Orientation and Rotation of Inertial Disk Particles in Wall Turbulence." *Journal of Fluid Mechanics* 766 (March). doi:10.1017/jfm.2015.38.

Cheng, N. 1997. "Simplified Settling Velocity Formula for Sediment Particle." *Journal of Hydraulic Engineering* 123 (2): 149–52. doi:10.1061/(ASCE)0733-9429(1997)123:2(149).

Chevillard, L., and C. Meneveau. 2013. "Orientation Dynamics of Small, Triaxial–ellipsoidal Particles in Isotropic Turbulence." *Journal of Fluid Mechanics* 737: 571–96. doi:10.1017/jfm.2013.580.

Chhabra, R. P., L. Agarwal, and N. K. Sinha. 1999. "Drag on Non-Spherical Particles: An Evaluation of Available Methods." *Powder Technology* 101 (3): 288–95. doi:10.1016/S0032-5910(98)00178-8.

Christensen, L.H., V.B. Breiting, A. Aasted, A. Jørgensen, and I. Kebuladze. 2003. "Long-Term Effects of Polyacrylamide Hydrogel on Human Breast Tissue." *Plastic and Reconstructive Surgery* 111 (6): 1883–90. doi:10.1097/01.PRS.0000056873.87165.5A.

Christensen, S.T., L.B. Pedersen, L. Schneider, and P. Satir. 2007. "Sensory Cilia and Integration of Signal Transduction in Human Health and Disease." *Traffic* 8 (2): 97–109. doi:10.1111/j.1600-0854.2006.00516.x.

Christiansen, E. B., and D.H. Barker. 1965. "The Effect of Shape and Density on the Free Settling of Particles at High Reynolds Numbers." *AIChE Journal* 11 (1): 145–51. doi:10.1002/aic.690110130.

Cisse, Mamadou, Holger Homann, and Jérémie Bec. 2013. "Slipping Motion of Large Neutrally Buoyant Particles in Turbulence." *Journal of Fluid Mechanics* 735: null – null. doi:10.1017/jfm.2013.490.

Clark, N. N., P. Gabriele, S. Shuker, and R. Turton. 1989. "Drag Coefficient of Irregular Particles in Newton's Settling Regime." *Powder Technology* 59 (1): 69–72.

Clift, R., and W.H. Gauvin. 1971. "Motion of Entrained Particles in Gas Streams." *Canadian Journal of Chemical Engineering* 49 (4): 439 – &. doi:10.1002/cjce.5450490403.

Clift, R., J. R. Grace, and M. E. Weber. 2013. *Bubbles, Drops, and Particles*. Courier Corporation.

Corrsin, S., and J. Lumley. 1956. "On the Equation of Motion for a Particle in Turbulent Fluid." *Applied Scientific Research, Section A* 6 (2-3): 114–16. doi:10.1007/BF03185030.

Corrsin, Stanley. 1942. "Decay of Turbulence behind Three Similar Grids." Engd, California Institute of Technology. http://resolver.caltech.edu/CaltechETD:etd-04062006-141720.

Csanady, G. T. 1963. "Turbulent Diffusion of Heavy Particles in the Atmosphere." *Journal of the Atmospheric Sciences* 20 (3): 201–8. doi:10.1175/1520-0469(1963)020<0201:TDOHPI>2.0.CO;2.

Dabiri, J.O. 2010. "Role of Vertical Migration in Biogenic Ocean Mixing." *Geophysical Research Letters* 37 (11). http://onlinelibrary.wiley.com/doi/10.1029/2010GL043556/full.

Daviero, G. J., P. J. W. Roberts, and K. Maile. 2001. "Refractive Index Matching in Large-Scale Stratified Experiments." *Experiments in Fluids* 31 (2): 119–26. doi:10.1007/s003480000260.

Davis, R.E., C.C. Eriksen, and C.P. Jones. 2002. "Autonomous Buoyancy-Driven Underwater Gliders." In *The Technology and Applications of Autonomous Underwater Vehicles*, 37–58. Taylor and Francis, London. http://www.ifremer.fr/lpo/gliders/donnees_tt/references/techno/4Gliders.pdf.

De La Rocha, C.L., and U. Passow. 2007. "Factors Influencing the Sinking of POC and the Efficiency of the Biological Carbon Pump." *Deep Sea Research Part II: Topical Studies in Oceanography*, The Role of Marine Organic Carbon and Calcite Fluxes in Driving Global Climate Change, Past and Future, 54 (5–7): 639–58. doi:10.1016/j.dsr2.2007.01.004.

Dietrich, W.E. 1982. "Settling Velocity of Natural Particles." *Water Resources Research* 18 (6): 1615–26. doi:10.1029/WR018i006p01615.

Dijksman, J.A., F. Rietz, K.A. Lőrincz, M. van Hecke, and W. Losert. 2012. "Refractive Index Matched Scanning of Dense Granular Materials." *Review of Scientific Instruments* 83 (1): 011301. doi:10.1063/1.3674173.

Domenici, P. 2001. "The Scaling of Locomotor Performance in Predator–prey Encounters: From Fish to Killer Whales." *Comparative Biochemistry and Physiology Part A: Molecular & Integrative Physiology* 131 (1): 169–82. doi:10.1016/S1095-6433(01)00465-2.

Durham, W.M., and R. Stocker. 2012. "Thin Phytoplankton Layers: Characteristics, Mechanisms, and Consequences." *Annual Review of Marine Science* 4 (1): 177–207. doi:10.1146/annurev-marine-120710-100957.

Dusenbery, D.B. 1998. "Fitness Landscapes for Effects of Shape on Chemotaxis and Other Behaviors of Bacteria." *Journal of Bacteriology* 180 (22): 5978–83.

———. 2009. *Living at Micro Scale: The Unexpected Physics of Being Small*. Harvard University Press.

Eaton, J.K. 2009. "Two-Way Coupled Turbulence Simulations of Gas-Particle Flows Using Point-Particle Tracking." *International Journal of Multiphase Flow*, Special Issue: Point-Particle Model for Disperse Turbulent Flows, 35 (9): 792–800. doi:10.1016/j.ijmultiphaseflow.2009.02.009.

Eaton, J.K., and J. Fessler. 1994. "Preferential Concentration of Particles by Turbulence." *International Journal of Multiphase Flow* 20 (August): 169–209. doi:10.1016/0301-9322(94)90072-8.

Elghobashi, S. 1994. "On Predicting Particle-Laden Turbulent Flows." *Applied Scientific Research* 52 (4): 309–29. doi:10.1007/BF00936835.

Elimelech, M., J. Gregory, and X. Jia. 2013. *Particle Deposition and Aggregation: Measurement, Modelling and Simulation*. Butterworth-Heinemann.

Eriksen, C.C., T.J. Osse, R.D. Light, T. Wen, T.W. Lehman, P.L. Sabin, J.W. Ballard, and A.M. Chiodi. 2001. "Seaglider: A Long-Range Autonomous Underwater Vehicle for Oceanographic Research." *IEEE Journal of Oceanic Engineering* 26 (4): 424–36. doi:10.1109/48.972073.

Ezzein, F.M., and R.J. Bathurst. 2011. "A Transparent Sand for Geotechnical Laboratory Modeling." *Geotechnical Testing Journal* 34 (6): 103808. doi:10.1520/GTJ103808.

Faxén, H. 1922. "Der Widerstand Gegen Die Bewegung Einer Starren Kugel in Einer Zähen Flüssigkeit, Die Zwischen Zwei Parallelen Ebenen Wänden Eingeschlossen Ist." *Annalen Der Physik* 373 (10): 89–119. doi:10.1002/andp.19223731003.

Ferruzzi, G. G., N. Pan, and W.H. Casey. 2000. "Mechanical Properties of Gellan and Polyacrylamide Gels with Implications for Soil Stabilization." *Soil Science* 165: 778–92. doi:10.1097/00010694-200010000-00003.

Fields, D.M., and J. Yen. 1997. "The Escape Behavior of Marine Copepods in Response to a Quantifiable Fluid Mechanical Disturbance." *Journal of Plankton Research* 19 (9): 1289–1304. doi:10.1093/plankt/19.9.1289.

Field, W.G. 1968. "Effects of Density Ratio on Sedimentary Similitude." *Journal of the Hydraulics Division* 94 (3): 705–20.

Ford, M.D., and J.S. Link. 2014. "Bounds on Biomass Estimates and Energetic Consequences of Ctenophora in the Northeast U.S. Shelf Ecosystem." *International Journal of Oceanography* 2014: e851809. doi:10.1155/2014/851809.

Franklin, J., and Z.Y. Wang. 2003. "Additive-induced enhancement of optical clarity of polyacrylamide hydrogel." 高分子科学 *(Chinese Journal of Polymer Science)* 21 (5): 533–39.

Friedman, P. D., and J. Katz. 2002. "Mean Rise Rate of Droplets in Isotropic Turbulence." *Physics of Fluids (1994-Present)* 14 (9): 3059–73. doi:10.1063/1.1497377.

Ganser, G.H. 1993. "A Rational Approach to Drag Prediction of Spherical and Nonspherical Particles." *Powder Technology* 77 (2): 143–52. doi:10.1016/0032-5910(93)80051-B.

Garau, B., A. Alvarez, and G. Oliver. 2006. "AUV Navigation through Turbulent Ocean Environments Supported by Onboard H-ADCP." In *Proceedings 2006 IEEE International Conference on Robotics and Automation, 2006. ICRA 2006*, 3556–61. doi:10.1109/ROBOT.2006.1642245.

Geurts, B., H. Clercx, and W. Uijttewaal. 2007. *Particle-Laden Flow: From Geophysical to Kolmogorov Scales*. Springer Science & Business Media.

Ghanbari, A., and M. Bahrami. 2011. "A Novel Swimming Microrobot Based on Artificial Cilia for Biomedical Applications." *Journal of Intelligent & Robotic Systems* 63 (3-4): 399–416. doi:10.1007/s10846-010-9516-6.

Gilbert, O.M., and E.J. Buskey. 2005. "Turbulence Decreases the Hydrodynamic Predator Sensing Ability of the Calanoid Copepod Acartia Tonsa." *Journal of Plankton Research* 27 (10): 1067–71. doi:10.1093/plankt/fbi066.

Gill, Adrian E. 1982. *Atmosphere-Ocean Dynamics*. Academic Press.

Gitterman, M., and V. Steinberg. 1980. "Memory Effects in the Motion of a Suspended Particle in a Turbulent Fluid." *Physics of Fluids (1958-1988)* 23 (11): 2154–60. doi:10.1063/1.862909.

Gögüs, M., N. Ipekçi, and M.A. Kökpinar. 2001. "Effect of Particle Shape on Fall Velocity of Angular Particles." *Journal of Hydraulic Engineering* 127 (10): 860–69. doi:10.1061/(ASCE)0733-9429(2001)127:10(860).

Good, G. H., P. J. Ireland, G. P. Bewley, E. Bodenschatz, L. R. Collins, and Z. Warhaft. 2014. "Settling Regimes of Inertial Particles in Isotropic Turbulence." *Journal of Fluid Mechanics* 759. doi:10.1017/jfm.2014.602.

Grattoni, C.A., H.H. Al-Sharji, C. Yang, A.H. Muggeridge, and R.W. Zimmerman. 2001. "Rheology and Permeability of Crosslinked Polyacrylamide Gel." *Journal of Colloid and Interface Science* 240 (2): 601–7. doi:10.1006/jcis.2001.7633.

Green, M. A., C. W. Rowley, and G. Haller. 2007. "Detection of Lagrangian Coherent Structures in Three-Dimensional Turbulence." *Journal of Fluid Mechanics* 572: 111–20. doi:10.1017/S0022112006003648.

Griffiths, G. 2002. *Technology and Applications of Autonomous Underwater Vehicles*. CRC Press.

Grünbaum, D. 1998. "Schooling as a Strategy for Taxis in a Noisy Environment." *Evolutionary Ecology* 12 (5): 503–22. doi:10.1023/A:1006574607845.

Grünbaum, D., and R.R. Strathmann. 2003. "Form, Performance and Trade-Offs in Swimming and Stability of Armed Larvae." *Journal of Marine Research* 61 (5): 659–91. doi:10.1357/002224003771815990.

Gustavsson, K., J. Einarsson, and B. Mehlig. 2014. "Tumbling of Small Axisymmetric Particles in Random and Turbulent Flows." *Physical Review Letters* 112 (1): 014501. doi:10.1103/PhysRevLett.112.014501.

Haider, A., and O. Levenspiel. 1989. "Drag Coefficient and Terminal Velocity of Spherical and Nonspherical Particles." *Powder Technology* 58 (1): 63–70. doi:10.1016/0032-5910(89)80008-7.

Happel, J., and H. Brenner. 1983. *Low Reynolds Number Hydrodynamics: With Special Applications to Particulate Media*. Springer Science & Business Media.

Harbison, G. R., L. P. Madin, and N. R. Swanberg. 1978. "On the Natural History and Distribution of Oceanic Ctenophores." *Deep Sea Research* 25 (3): 233–56. doi:10.1016/0146-6291(78)90590-8.

Hassan, Y.A., and E. E. Dominguez-Ontiveros. 2008. "Flow Visualization in a Pebble Bed Reactor Experiment Using PIV and Refractive Index Matching Techniques." *Nuclear Engineering and Design*, HTR-2006: 3rd International Topical Meeting on High Temperature Reactor Technology, 238 (11): 3080–85. doi:10.1016/j.nucengdes.2008.01.027.

Heidemann, John, Milica Stojanovic, and Michele Zorzi. 2012. "Underwater Sensor Networks: Applications, Advances and Challenges." *Philosophical Transactions of the Royal Society A: Mathematical, Physical and Engineering Sciences* 370 (1958): 158–75. doi:10.1098/rsta.2011.0214.

Herman, G.T., and A. Lent. 1976. "Iterative Reconstruction Algorithms." *Computers in Biology and Medicine*, Advances in Picture Reconstruction—Theory and Applications, 6 (4): 273–94. doi:10.1016/0010-4825(76)90066-4.

Hines, S. 2005. "Pairs of Seagliders Set Endurance Records." *UW News and Information*. April 5. http://www.washington.edu/news/2005/04/05/pairs-of-seagliders-set-endurance-records/.

Hirche, H. -J, M. E. M Baumann, G Kattner, and R Gradinger. 1991. "Plankton Distribution and the Impact of Copepod Grazing on Primary Production in Fram Strait, Greenland Sea." *Journal of Marine Systems* 2 (3–4): 477–94. doi:10.1016/0924-7963(91)90048-Y.

Hobson, B., B. Schulz, J. Janet, M. Kemp, R. Moody, C. Pell, and H. Pinnix. 2001. "Development of a Micro Autonomous Underwater Vehicle for Complex 3-D Sensing." In *MTS/IEEE Conference and Exhibition OCEANS, 2001*, 4:2043–45 vol.4. doi:10.1109/OCEANS.2001.968311.

Hondzo, M., and D. Lynn. 1999. "Quantified Small-Scale Turbulence Inhibits the Growth of a Green Alga." *Freshwater Biology* 41 (1): 51–61. doi:10.1046/j.1365-2427.1999.00389.x.

Hudson, I. R., D. Jones, and B. D. Wigham. 2005. "A Review of the Uses of Work-Class ROVs for the Benefits of Science: Lessons Learned from the SERPENT Project." *Underwater Technology* 26 (3): 83–88. doi:10.3723/175605405784426637.

Hutchins, M., and D. Olendorf. 2004. *Grzimek's Animal Life Encyclopedia: Lower Metazoans and Lesser Deuterosomes*. Vol. 1. Gale/Cengage Learning.

Hwang, W., and J. K. Eaton. 2004. "Creating Homogeneous and Isotropic Turbulence without a Mean Flow." *Experiments in Fluids* 36 (3): 444–54. doi:10.1007/s00348-003-0742-6.

Jeffery, G. B. 1922. "The Motion of Ellipsoidal Particles Immersed in a Viscous Fluid." *Proceedings of the Royal Society of London. Series A* 102 (715): 161–79. doi:10.1098/rspa.1922.0078.

Jenkins, S.A., D.E. Humphreys, J. Sherman, J. Osse, C. Jones, N. Leonard, J. Graver, et al. 2003. *Underwater Glider System Study*. 53. Scripps Institution of Oceanography Technical Report. San Diego, CA: UC San Diego. http://escholarship.org/uc/item/1c28t6bb.

Jenny, M., G. Bouchet, and J. Dušek. 2003. "Nonvertical Ascension or Fall of a Free Sphere in a Newtonian Fluid." *Physics of Fluids* 15 (1): L9–12. doi:10.1063/1.1529179.

Jimenez, J. 1997. "Oceanic Turbulence at Millimeter Scales." *Scientia Marina* 61: 47–56.

Jong, J. de, L. Cao, S. H. Woodward, J. P. L. C. Salazar, L. R. Collins, and H. Meng. 2009. "Dissipation Rate Estimation from PIV in Zero-Mean Isotropic Turbulence." *Experiments in Fluids* 46 (3): 499–515. doi:10.1007/s00348-008-0576-3.

Karp-Boss, L., E. Boss, and P. A. Jumars. 1996. "Nutrient Fluxes to Planktonic Osmotrophs in the Presence of Fluid Motion." *Oceanography and Marine Biology, Vol 34* 34: 71–107.

Katija, Kakani, and John O. Dabiri. 2009. "A Viscosity-Enhanced Mechanism for Biogenic Ocean Mixing." *Nature* 460 (7255): 624–26. doi:10.1038/nature08207.

Kiørboe, T. 1997. "Small-Scale Turbulence, Marine Snow Formation, and Planktivorous Feeding." *Scientia Marina* 61 (1): 141–58.

Kiørboe, T., and A. Visser. 1999. "Predator and Prey Perception in Copepods due to Hydromechanical Signals." *Marine Ecology - Progress Series* 179: 81–95. doi:10.3354/meps179081.

Kjellerup, S., and T. Kiørboe. 2012. "Prey Detection in a Cruising Copepod." *Biology Letters* 8 (3): 438–41. doi:10.1098/rsbl.2011.1073.

Klein, S., M. Gibert, A. Bérut, and E. Bodenschatz. 2013. "Simultaneous 3D Measurement of the Translation and Rotation of Finite-Size Particles and the Flow Field in a Fully Developed Turbulent Water Flow." *Measurement Science and Technology* 24 (2): 024006. doi:10.1088/0957-0233/24/2/024006.

Kodati, P., J. Hinkle, A. Winn, and X. Deng. 2008. "Microautonomous Robotic Ostraciiform (MARCO): Hydrodynamics, Design, and Fabrication." *Ieee Transactions on Robotics* 24 (1): 105–17. doi:10.1109/TRO.2008.915446.

Krawczynski, J. F., B. Renou, and L. Danaila. 2010. "The Structure of the Velocity Field in a Confined Flow Driven by an Array of Opposed Jets." *Physics of Fluids* 22 (4): 045104. doi:10.1063/1.3371820.

Kundu, P.K., and I.M. Cohen. 2007. *Fluid Mechanics*. 4th ed. Amsterdam ; Boston: Academic Press.

Kunze, E., J.F. Dower, I. Beveridge, R. Dewey, and K.P. Bartlett. 2006. "Observations of Biologically Generated Turbulence in a Coastal Inlet." *Science* 313 (5794): 1768–70. doi:10.1126/science.1129378.

Kuo, K.K., and R. Acharya. 2012. *Fundamentals of Turbulent and Multi-Phase Combustion*. Wiley.

Lachhab, A., Y.-K. Zhang, and M.V.I. Muste. 2008. "Particle Tracking Experiments in Match-Index-Refraction Porous Media." *Ground Water* 46 (6): 865–72. doi:10.1111/j.1745-6584.2008.00479.x.

Largier, J. L., C. A. Lawrence, M. Roughan, D. M. Kaplan, E. P. Dever, C. E. Dorman, R. M. Kudela, et al. 2006. "WEST: A Northern California Study of the Role of Wind-Driven Transport in the Productivity of Coastal Plankton Communities." *Deep Sea Research Part II: Topical Studies in Oceanography*, The Role of Wind-Driven Flow in Shelf Productivity Results from the Wind Events and Shelf Transport (CoOP WEST) Program, 53 (25–26): 2833–49. doi:10.1016/j.dsr2.2006.08.018.

LaRue, J.C., and P.A. Libby. 1981. "Thermal Mixing Layer Downstream of Half-heated Turbulence Grid." *Physics of Fluids (1958-1988)* 24 (4): 597–603. doi:10.1063/1.863426.

Leonard, J.J., A.A. Bennett, C.M. Smith, and H. Feder. 1998. "Autonomous Underwater Vehicle Navigation." In *IEEE ICRA Workshop on Navigation of Outdoor Autonomous Vehicles*. Citeseer.

Liao, James C. 2007. "A Review of Fish Swimming Mechanics and Behaviour in Altered Flows." *Philosophical Transactions of the Royal Society B: Biological Sciences* 362 (1487): 1973–93. doi:10.1098/rstb.2007.2082.

Li, Y., E. Perlman, Y. Yang, R. Burns, C. Meneveau, S. Chen, A. Szalay, and G. Eyink. 2008. "A Public Turbulence Database Cluster and Applications to Study Lagrangian Evolution of Velocity Increments in Turbulence." *Journal of Turbulence* 9 (31).

Logan, B.E., and D.B. Wilkinson. 1990. "Fractal Geometry of Marine Snow and Other Biological Aggregates." *Limnology and Oceanography* 35 (1): 130–36. doi:10.4319/lo.1990.35.1.0130.

Loth, E. 2000. "Numerical Approaches for Motion of Dispersed Particles, Droplets and Bubbles." *Progress in Energy and Combustion Science* 26 (3): 161–223. doi:10.1016/S0360-1285(99)00013-1.

———. 2008. "Drag of Non-Spherical Solid Particles of Regular and Irregular Shape." *Powder Technology* 182 (3): 342–53. doi:10.1016/j.powtec.2007.06.001.

Lucci, F., A. Ferrante, and S. Elghobashi. 2010. "Modulation of Isotropic Turbulence by Particles of Taylor Length-Scale Size." *Journal of Fluid Mechanics* 650: 5–55. doi:10.1017/S0022112009994022.

Luff, J. D., T. Drouillard, A. M. Rompage, M. A. Linne, and J. R. Hertzberg. 1999. "Experimental Uncertainties Associated with Particle Image Velocimetry (PIV) Based Vorticity Algorithms." *Experiments in Fluids* 26 (1-2): 36–54. doi:10.1007/s003480050263.

Lundell, F., L.D. Söderberg, and P.H. Alfredsson. 2011. "Fluid Mechanics of Papermaking." *Annual Review of Fluid Mechanics* 43 (1): 195–217. doi:10.1146/annurev-fluid-122109-160700.

Lynn, D.H. 2008. *The Ciliated Protozoa: Characterization, Classification, and Guide to the Literature*. Springer Science & Business Media.

Machemer, H, and R. Braucker. 1991. "Gravireception and Graviresponses in Ciliates." *Acta Protozoologica* 31 (4): 185–214.

MacKenzie, B.R., T.J. Miller, S. Cyr, and W.C. Leggett. 1994. "Evidence for a Dome-Shaped Relationship between Turbulence and Larval Fish Ingestion Rates." *Limnol. Oceanogr* 39 (8): 1790–99. doi:10.4319/lo.1994.39.8.1790.

Magarvey, R. H., and Roy L. Bishop. 1961. "Transition Ranges for Three-Dimensional Wakes." *Canadian Journal of Physics* 39 (10): 1418–22. doi:10.1139/p61-169.

Makita, H. 1991. "Realization of a Large-Scale Turbulence Field in a Small Wind Tunnel." *Fluid Dynamics Research* 8 (1-4): 53. doi:10.1016/0169-5983(91)90030-M.

Malvern, L. E. 1969. *INTRODUCTION TO THE MECHANICS OF A CONTINUOUS MEDIUM*. http://trid.trb.org/view.aspx?id=199874.

Mandø, M., and L. Rosendahl. 2010. "On the Motion of Non-Spherical Particles at High Reynolds Number." *Powder Technology* 202 (1–3): 1–13. doi:10.1016/j.powtec.2010.05.001.

Marchioli, Cristian, and A. Soldati. 2013. "Rotation Statistics of Fibers in Wall Shear Turbulence." *Acta Mechanica* 224 (10): 2311–29. doi:10.1007/s00707-013-0933-z.

Marcos, H.C. Fu, T.R. Powers, and R. Stocker. 2012. "Bacterial Rheotaxis." *Proceedings of the National Academy of Sciences* 109 (13): 4780–85. doi:10.1073/pnas.1120955109.

Matsumoto, G. I. 1991. "Swimming Movements of Ctenophores, and the Mechanics of Propulsion by Ctene Rows." In *Coelenterate Biology: Recent Research on Cnidaria and Ctenophora*, edited by R. B. Williams, P. F. S. Cornelius, R. G. Hughes, and E. A. Robson, 319–25. Developments in Hydrobiology 66. Springer Netherlands. http://link.springer.com/chapter/10.1007/978-94-011-3240-4_46.

Maxey, M.R. 1987. "The Gravitational Settling of Aerosol Particles in Homogeneous Turbulence and Random Flow Fields." *Journal of Fluid Mechanics* 174 (January): 441–65. doi:10.1017/S0022112087000193.

Maxey, M. R. 1990. "On the Advection of Spherical and Non-Spherical Particles in a Non-Uniform Flow." *Philosophical Transactions of the Royal Society of London. Series A: Physical and Engineering Sciences* 333 (1631): 289–307. doi:10.1098/rsta.1990.0162.

Maxey, M.R., and S. Corrsin. 1986. "Gravitational Settling of Aerosol-Particles in Randomly Oriented Cellular-Flow Fields." *Journal of the Atmospheric Sciences* 43 (11): 1112–34. doi:10.1175/1520-0469(1986)043<1112:GSOAPI>2.0.CO;2.

Maxey, M.R., and J.J. Riley. 1983. "Equation of Motion for a Small Rigid Sphere in a Nonuniform Flow." *Physics of Fluids (1958-1988)* 26 (4): 883–89. doi:10.1063/1.864230.

Meek, C.C., and B.G. Jones. 1973. "Studies of the Behavior of Heavy Particles in a Turbulent Fluid Flow." *Journal of the Atmospheric Sciences* 30 (2): 239–44. doi:10.1175/1520-0469(1973)030<0239:SOTBOH>2.0.CO;2.

Miller, R. J. 1974. "Distribution and Biomass of an Estuarine Ctenophore population,Mnemiopis Leidyi (A. Agassiz)." *Chesapeake Science* 15 (1): 1–8. doi:10.2307/1350952.

Monaghan, J. J. 2012. "Smoothed Particle Hydrodynamics and Its Diverse Applications." Edited by S. H. Davis and P. Moin. *Annual Review of Fluid Mechanics, Vol 44* 44: 323–46. doi:10.1146/annurev-fluid-120710-101220.

Moored, K.W., P.A. Dewey, M.C. Leftwich, H. Bart-Smith, and A.J. Smits. 2011. "Bioinspired Propulsion Mechanisms Based on Manta Ray Locomotion." *Marine Technology Society Journal* 45 (4): 110–18. doi:10.4031/MTSJ.45.4.3.

Moroz, L.L., K.M. Kocot, M.R. Citarella, S. Dosung, T.P. Norekian, I.S. Povolotskaya, A.P. Grigorenko, et al. 2014. "The Ctenophore Genome and the Evolutionary Origins of Neural Systems." *Nature* 510 (7503): 109–14. doi:10.1038/nature13400.

Mukhopadhyay, S., and J. Peixinho. 2011. "Packings of Deformable Spheres." *Physical Review E* 84 (1): 011302. doi:10.1103/PhysRevE.84.011302.

Murray, S.P. 1968. "Simulation of Horizontal Turbulent Diffusion of Particles under Waves." *Coastal Engineering Proceedings* 1 (11). https://icce-ojs-tamu.tdl.org/icce/index.php/icce/article/view/2531.

Murray, S. P. 1970. "Settling Velocities and Vertical Diffusion of Particles in Turbulent Water." *Journal of Geophysical Research* 75 (9): 1693–96.

Natarajan, R., and A. Acrivos. 1993. "The Instability of the Steady Flow Past Spheres and Disks." *Journal of Fluid Mechanics* 254: 323–44. doi:10.1017/S0022112093002150.

Nguyen, H., L. Karp-Boss, P.A. Jumars, and L. Fauci. 2011. "Hydrodynamic Effects of Spines: A Different Spin." *Limnology and Oceanography Fluids & Environments* 1: 110–19. doi:10.1215/21573698-1303444.

Nielsen, P. 2007. "Mean and Variance of the Velocity of Solid Particles in Turbulence." In *Particle-Laden Flow*, 385–91. Springer. http://link.springer.com/chapter/10.1007/978-1-4020-6218-6_30.

Ni, R., S. Kramel, N.T. Ouellette, and G.A. Voth. 2015. "Measurements of the Coupling between the Tumbling of Rods and the Velocity Gradient Tensor in Turbulence." *Journal of Fluid Mechanics* 766. doi:10.1017/jfm.2015.16.

Ni, R., N.T. Ouellette, and G.A. Voth. 2014. "Alignment of Vorticity and Rods with Lagrangian Fluid Stretching in Turbulence." *Journal of Fluid Mechanics* 743. doi:10.1017/jfm.2014.32.

Noss, C., and A. Lorke. 2014. "Direct Observation of Biomixing by Vertically Migrating Zooplankton." *Limnol Oceanogr*. doi:10.4319/lo.2014.59.3.0724.

Ohlberger, J., G. Staaks, and F. Holker. 2006. "Swimming Efficiency and the Influence of Morphology on Swimming Costs in Fishes." *Journal of Comparative Physiology B-Biochemical Systemic and Environmental Physiology* 176 (1): 17–25. doi:10.1007/s00360-005-0024-0.

Ormières, D., and M. Provansal. 1999. "Transition to Turbulence in the Wake of a Sphere." *Physical Review Letters* 83 (1): 80–83. doi:10.1103/PhysRevLett.83.80.

Oseen, C.W. 1927. *Hydrodynamik*. Akad. Verl.-Ges.

Pahlow, M., U. Riebesell, and D.A. Wolf-Gladrow. 1997. "Impact of Cell Shape and Chain Formation on Nutrient Acquisition by Marine Diatoms." *Limnology and Oceanography* 42 (8): 1660–72. doi:10.4319/lo.1997.42.8.1660.

Palagi, S., E.W.H. Jager, B. Mazzolai, and L. Beccai. 2013. "Propulsion of Swimming Microrobots Inspired by Metachronal Waves in Ciliates: From Biology to Material Specifications." *Bioinspiration & Biomimetics* 8 (4): 046004. doi:10.1088/1748-3182/8/4/046004.

Pallua, N., and T.P. Wolter. 2010. "A 5-Year Assessment of Safety and Aesthetic Results after Facial Soft-Tissue Augmentation with Polyacrylamide Hydrogel (Aquamid): A Prospective Multicenter Study of 251 Patients:" *Plastic and Reconstructive Surgery* 125 (6): 1797–1804. doi:10.1097/PRS.0b013e3181d18158.

Pan, J. M., Q. Wang, and W. J. Snell. 2005. "Cilium-Generated Signaling and Cilia-Related Disorders." *Laboratory Investigation* 85 (4): 452–63. doi:10.1038/labinvest.3700253.

Parry, D. 2013. "Navy 'Mine-Hunter' AUV Sets Mission Endurance Record - U.S. Naval Research Laboratory." *NRL Public Affairs & Media*. November 20. http://www.nrl.navy.mil/media/news-releases/2013/navy-mine-hunter-auv-sets-mission-endurance-record.

Parsa, S., E. Calzavarini, F. Toschi, and G.A. Voth. 2012. "Rotation Rate of Rods in Turbulent Fluid Flow." *Physical Review Letters* 109 (13): 134501. doi:10.1103/PhysRevLett.109.134501.

Parsa, S., J.S. Guasto, M. Kishore, N.T. Ouellette, J. P. Gollub, and G.A. Voth. 2011. "Rotation and Alignment of Rods in Two-Dimensional Chaotic Flow." *Physics of Fluids* 23 (4): 043302. doi:10.1063/1.3570526.

Parsa, S., and G.A. Voth. 2014. "Inertial Range Scaling in Rotations of Long Rods in Turbulence." *Physical Review Letters* 112 (2): 024501. doi:10.1103/PhysRevLett.112.024501.

Paull, L., S. Saeedi, M. Seto, and H. Li. 2014. "AUV Navigation and Localization: A Review." *Ieee Journal of Oceanic Engineering* 39 (1): 131–49. doi:10.1109/JOE.2013.2278891.

Peng, J., and S. Alben. 2012. "Effects of Shape and Stroke Parameters on the Propulsion Performance of an Axisymmetric Swimmer." *Bioinspiration & Biomimetics* 7 (1): 016012. doi:10.1088/1748-3182/7/1/016012.

Peng, J., and J. O. Dabiri. 2009. "Transport of Inertial Particles by Lagrangian Coherent Structures: Application to Predator–prey Interaction in Jellyfish Feeding." *Journal of Fluid Mechanics* 623 (March): 75. doi:10.1017/S0022112008005089.

Pettyjohn, E.S., and E. B. Christiansen. 1948. "Effect of Particle Shape on Free-Settling Rates of Isometric Particles." *Chemical Engineering Progress* 44 (2): 157–72. doi:10.1002/aic.690110130.

Piazza, R. 2014. "Settled and Unsettled Issues in Particle Settling." *Reports on Progress in Physics* 77 (5): 056602. doi:10.1088/0034-4885/77/5/056602.

Poelma, C., J. Westerweel, and G. Ooms. 2007. "Particle-Fluid Interactions in Grid-Generated Turbulence." *Journal of Fluid Mechanics* 589 (1): 315–51.

Poorte, R. E. G., and A. Biesheuvel. 2002. "Experiments on the Motion of Gas Bubbles in Turbulence Generated by an Active Grid." *Journal of Fluid Mechanics* 461 (June): 127–54. doi:10.1017/S0022112002008273.

Pope, S. B. 2000. *Turbulent Flows*. Cambridge University Press.

Prairie, J.C., K.R. Sutherland, K.J. Nickols, and A.M. Kaltenberg. 2012. "Biophysical Interactions in the Plankton: A Cross-Scale Review." *Limnology and Oceanography Fluids & Environments* 2: 121–45. doi:10.1215/21573689-1964713.

Pringle, J.M. 2007. "Turbulence Avoidance and the Wind-Driven Transport of Plankton in the Surface Ekman Layer." *Continental Shelf Research* 27 (5): 670–78. doi:10.1016/j.csr.2006.11.011.

Pumir, Alain, and Michael Wilkinson. 2011. "Orientation Statistics of Small Particles in Turbulence." *New Journal of Physics* 13 (9): 093030. doi:10.1088/1367-2630/13/9/093030.

Purcell, J.E., and M.N. Arai. 2001. "Interactions of Pelagic Cnidarians and Ctenophores with Fish: A Review." *Hydrobiologia* 451 (1-3): 27–44. doi:10.1023/A:1011883905394.

Purcell, J.E., T.A. Shiganova, M.B. Decker, and E.D. Houde. 2001. "The Ctenophore Mnemiopsis in Native and Exotic Habitats: US Estuaries versus the Black Sea Basin." In *Jellyfish Blooms: Ecological and Societal Importance*, 145–76. Springer. http://link.springer.com/chapter/10.1007/978-94-010-0722-1_13.

Qureshi, N. M., U. Arrieta, C. Baudet, A. Cartellier, Y. Gagne, and M. Bourgoin. 2008. "Acceleration Statistics of Inertial Particles in Turbulent Flow." *The European Physical Journal B* 66 (4): 531–36. doi:10.1140/epjb/e2008-00460-x.

Qureshi, N.M., M. Bourgoin, C. Baudet, A. Cartellier, and Y. Gagne. 2007. "Turbulent Transport of Material Particles: An Experimental Study of Finite Size Effects." *Physical Review Letters* 99 (18): 184502. doi:10.1103/PhysRevLett.99.184502.

Raffel, M., C.E. Willert, S.T. Wereley, and J. Kompenhans. 2007. *Particle Image Velocimetry: A Practical Guide*. Experimental Fluid Mechanics. Berlin, Heidelberg: Springer Berlin Heidelberg. http://link.springer.com/10.1007/978-3-540-72308-0.

Rakow, K. C., and W. M. Graham. 2006. "Orientation and Swimming Mechanics by the Scyphomedusa Aurelia Sp in Shear Flow." *Limnology and Oceanography* 51 (2): 1097–1106. doi:10.4319/lo.2006.51.2.1097.

Reddy, R.K., J.B. Joshi, K. Nandakumar, and P.D. Minev. 2010. "Direct Numerical Simulations of a Freely Falling Sphere Using Fictitious Domain Method: Breaking of Axisymmetric Wake." *Chemical Engineering Science* 65 (6): 2159–71. doi:10.1016/j.ces.2009.12.009.

Reeks, M. W., and S. McKee. 1984. "The Dispersive Effects of Basset History Forces on Particle Motion in a Turbulent Flow." *Physics of Fluids (1958-1988)* 27 (7): 1573–82. doi:10.1063/1.864812.

Reidenbach, M.A., N. George, and M.A.R. Koehl. 2008. "Antennule Morphology and Flicking Kinematics Facilitate Odor Sampling by the Spiny Lobster, Panulirus Argus." *Journal of Experimental Biology* 211 (17): 2849–58. doi:10.1242/jeb.016394.

Reidenbach, M.A., J.R. Koseff, and M.A.R. Koehl. 2009. "Hydrodynamic Forces on Larvae Affect Their Settlement on Coral Reefs in Turbulent, Wave-Driven Flow." *Limnology and Oceanography* 54 (1): 318–30. doi:10.4319/lo.2009.54.1.0318.

Riley, J.J. 1971. "Computer Simulations of Turbulent Dispersion." Baltimore. https://catalyst.library.jhu.edu/catalog/bib_1826426.

Rizon, M., H. Yazid, P. Saad, A.Y.M. Shakaff, A.R. Saad, M. Sugisaka, S. Yaacob, M.R. Mamat, and M. Karthigayan. 2005. "Object Detection Using Circular Hough Transform." *American Journal of Applied Sciences* 2 (12): 1606–9.

Roberts, A. M., and F. M. Deacon. 2002. "Gravitaxis in Motile Micro-Organisms: The Role of Fore-Aft Body Asymmetry." *Journal of Fluid Mechanics* 452 (February): 405–23. doi:10.1017/S0022112001006772.

Robinson, J. 2006. *The Oxford Companion to Wine*. OUP Oxford.

Roper, D. T., S. Sharma, R. Sutton, and P. Culverhouse. 2011. "A Review of Developments towards Biologically Inspired Propulsion Systems for Autonomous Underwater Vehicles." *Proceedings of the Institution of Mechanical Engineers Part M-Journal of Engineering for the Maritime Environment* 225 (M2): 77–96. doi:10.1177/1475090210397438.

Rothschild, B. J., and T. R. Osborn. 1988. "Small-Scale Turbulence and Plankton Contact Rates." *Journal of Plankton Research* 10 (3): 465–74. doi:10.1093/plankt/10.3.465.

Ruiz, J., D. Macías, and Francesc Peters. 2004. "Turbulence Increases the Average Settling Velocity of Phytoplankton Cells." *Proceedings of the National Academy of Sciences of the United States of America* 101 (51): 17720–24. doi:10.1073/pnas.0401539101.

Sareh, S., J. Rossiter, A. Conn, K. Drescher, and R.E. Goldstein. 2012. "Swimming like Algae: Biomimetic Soft Artificial Cilia." *Journal of The Royal Society Interface*, October, rsif20120666. doi:10.1098/rsif.2012.0666.

Schiller, L., and A. Naumann. 1933. "Fundamental Calculations Relating to Gravitational Preparation of Materials." *VDI Zeitschrift* 77: 318–20.

Schreck, S., and S.J. Kleis. 1993. "Modification of Grid-Generated Turbulence by Solid Particles." *Journal of Fluid Mechanics* 249: 665–88. doi:10.1017/S0022112093001326.

Sciacchitano, A., R.P. Dwight, and F. Scarano. 2012. "Navier–Stokes Simulations in Gappy PIV Data." *Experiments in Fluids* 53 (5): 1421–35. doi:10.1007/s00348-012-1366-5.

Seo, K., S.-J. Chung, and J.-J.E. Slotine. 2010. "CPG-Based Control of a Turtle-like Underwater Vehicle." *Autonomous Robots* 28 (3): 247–69. doi:10.1007/s10514-009-9169-0.

Shadden, S.C., J.O. Dabiri, and J.E. Marsden. 2006. "Lagrangian Analysis of Fluid Transport in Empirical Vortex Ring Flows." *Physics of Fluids* 18 (4): 047105. doi:10.1063/1.2189885.

Shanks, A.L., and J>D. Trent. 1980. "Marine Snow: Sinking Rates and Potential Role in Vertical Flux." *Deep Sea Research Part A. Oceanographic Research Papers* 27 (2): 137–43. doi:10.1016/0198-0149(80)90092-8.

Shapiro, A.L., E. Viñuela, and J.V. Maizel Jr. 1967. "Molecular Weight Estimation of Polypeptide Chains by Electrophoresis in SDS-Polyacrylamide Gels." *Biochemical and Biophysical Research Communications* 28 (5): 815–20. doi:10.1016/0006-291X(67)90391-9.

Simmons, L. F. G., and C. Salter. 1934. "Experimental Investigation and Analysis of the Velocity Variations in Turbulent Flow." *Proceedings of the Royal Society of London. Series A, Containing Papers of a Mathematical and Physical Character* 145 (854): 212–34.

Sojka, R.E., R.D. Lentz, and D.T. Westerman. 1998. "Water and Erosion Management with Multiple Applications of Polyacrylamide in Furrow Irrigation." *Soil Science Society of America Journal* 62 (6): 1672–80.

Soo, S. L. 1975. "Equation of Motion of a Solid Particle Suspended in a Fluid." *Physics of Fluids* 18 (2): 263–64. doi:10.1063/1.861113.

Sreenivasan, K. R., A. Prabhu, and R. Narasimha. 1983. "Zero-Crossings in Turbulent Signals." *Journal of Fluid Mechanics* 137 (1): 251. doi:10.1017/S0022112083002396.

Steffen, R.B., D.C. Turner, and D. Vanderlaan. 2005. "Contact Lenses."

Stein, D.B. 2010. *Handbook of Hydrogels: Properties, Preparation & Applications*. New York: Nova Science Publishers, Inc.

Stokes, G.G. 1851. "On the Effect of the Internal Friction of Fluids on the Motion of Pendulums." *Transactions of the Cambridge Philosophical Society* 9: 8–106.

Stoots, C., S. Becker, K. Condie, F. Durst, and D. McEligot. 2001. "A Large-Scale Matched Index of Refraction Flow Facility for LDA Studies around Complex Geometries." *Experiments in Fluids* 30 (4): 391–98. doi:10.1007/s003480000216.

Stout, J. E., S. P. Arya, and E. L. Genikhovich. 1995. "The Effect of Nonlinear Drag on the Motion and Settling Velocity of Heavy Particles." *Journal of the Atmospheric Sciences* 52 (22): 3836–48. doi:10.1175/1520-0469(1995)052<3836:TEONDO>2.0.CO;2.

Stringham, G. E., D.B. Simons, and H.P. Guy. 1969. *The Behavior of Large Particles Falling in Quiescent Liquids*. US Government Printing Office. http://pubs.usgs.gov/pp/0562c/report.pdf.

Subramaniam, S. 2013. "Lagrangian-Eulerian Methods for Multiphase Flows." *Progress in Energy and Combustion Science* 39 (2-3): 215–45. doi:10.1016/j.pecs.2012.10.003.

Sutherland, K.R., J.H. Costello, S.P. Colin, and J.O. Dabiri. 2014. "Ambient Fluid Motions Influence Swimming and Feeding by the Ctenophore Mnemiopsis Leidyi." *Journal of Plankton Research*, June, fbu051. doi:10.1093/plankt/fbu051.

Suthers, I.M. 2009. *Plankton: A Guide to Their Ecology and Monitoring for Water Quality*. Csiro Publishing.

Taylor, G. I. 1935. "Statistical Theory of Turbulence." *Proceedings of the Royal Society of London. Series A, Mathematical and Physical Sciences* 151 (873): 421–44.

Tchen, C.-M. 1947. "Mean Value and Correlation Problems Connected with the Motion of Small Particles Suspended in a Turbulent Fluid." TU Delft, Delft University of Technology. http://repository.tudelft.nl/view/ir/uuid:65d1454c-e070-44f3-a28a-9c03cc9868ef/.

Thiffeault, J.-L., and S. Childress. 2010. "Stirring by Swimming Bodies." *Physics Letters A* 374 (34): 3487–90. doi:10.1016/j.physleta.2010.06.043.

Thompson, B.E., C. Vafidis, and J.H. Whitelaw. 1987. *Velocimetry with Refractive Index Matching for Complex Flow Configurations*. NAS8-37320. Glastonbury, CT: Scientific Research Associates, Inc. http:// ntrs.nasa.gov/archive/nasa/casi.ntrs.nasa.gov/19870017719_198 7017719.pdf.

Thompson, T.L., and N. N. Clark. 1991. "A Holistic Approach to Particle Drag Prediction." *Powder Technology* 67 (1): 57–66. doi:10.1016/0032-5910(91)80026-F.

Thornton, D. 2002. "Diatom Aggregation in the Sea: Mechanisms and Ecological Implications." *European Journal of Phycology* 37 (2): 149–61. doi:10.1017/S0967026202003657.

Thorpe, S. A. 2005. *The Turbulent Ocean*. Cambridge University Press.

Tooby, P.F., G.L. Wick, and J.D. Isaacs. 1977. "The Motion of a Small Sphere in a Rotating Velocity Field: A Possible Mechanism for Suspending Particles in Turbulence." *Journal of Geophysical Research* 82 (15): 2096–2100. doi:10.1029/JC082i015p02096.

Toschi, F., and E. Bodenschatz. 2009. "Lagrangian Properties of Particles in Turbulence." *Annual Review of Fluid Mechanics* 41 (1): 375–404. doi:10.1146/annurev.fluid.010908.165210.

Towle, W. L., T. K. Sherwood, and L. A. Seder. 1939. "Effect of a Screen Grid on the Turbulence of an Air Stream." *Industrial & Engineering Chemistry* 31 (4): 462–63. doi:10.1021/ie50352a014.

Tritico, H. M., and A. J. Cotel. 2010. "The Effects of Turbulent Eddies on the Stability and Critical Swimming Speed of Creek Chub (Semotilus Atromaculatus)." *The Journal of Experimental Biology* 213 (13): 2284–93. doi:10.1242/jeb.041806.

Turian, R.M., F.-L. Hsu, K.S. Avramidis, D.-J. Sung, and R.K. Allendorfer. 1992. "Settling and Rheology of Suspensions of Narrow-Sized Coal Particles." *AIChE Journal* 38 (7): 969–87. doi:10.1002/aic.690380702.

Tytell, E.D., I. Borazjani, F. Sotiropoulos, T.V. Baker, E.J. Anderson, and G.V. Lauder. 2010. "Disentangling the Functional Roles of Morphology and Motion in the Swimming of Fish." *Integrative and Comparative Biology* 50 (6): 1140–54. doi:10.1093/icb/icq057.

Uzol, O., Y.-C. Chow, J. Katz, and C. Meneveau. 2002. "Unobstructed Particle Image Velocimetry Measurements within an Axial Turbo-Pump Using Liquid and Blades with Matched Refractive Indices." *Experiments in Fluids* 33 (6): 909–19. doi:10.1007/s00348-002-0494-8.

Van der Veer, H. W., and C. F. M. Sadée. 1984. "Seasonal Occurrence of the Ctenophore Pleurobrachia Pileus in the Western Dutch Wadden Sea." *Marine Biology* 79 (3): 219–27. doi:10.1007/BF00393253.

Variano, E.A., E. Bodenschatz, and E.A. Cowen. 2004. "A Random Synthetic Jet Array Driven Turbulence Tank." *Experiments in Fluids* 37 (4): 613–15. doi:10.1007/s00348-004-0833-z.

Variano, E.A., and E.A. Cowen. 2008. "A Random-Jet-Stirred Turbulence Tank." *Journal of Fluid Mechanics* 604 (June): 1–32. doi:10.1017/S0022112008000645.

Veldhuis, C., A. Biesheuvel, L. van Wijngaarden, and D. Lohse. 2005. "Motion and Wake Structure of Spherical Particles." *Nonlinearity* 18 (1): C1. doi:10.1088/0951-7715/18/1/000.

Villanueva, A., C. Smith, and S. Priya. 2011. "A Biomimetic Robotic Jellyfish (Robojelly) Actuated by Shape Memory Alloy Composite Actuators." *Bioinspiration & Biomimetics* 6 (3): 036004. doi:10.1088/1748-3182/6/3/036004.

Villermaux, E., B. Sixou, and Y. Gagne. 1995. "Intense Vortical Structures in Grid-generated Turbulence." *Physics of Fluids (1994-Present)* 7 (8): 2008–13. doi:10.1063/1.868512.

Visser, A. 2001. "Hydromechanical Signals in the Plankton." *Marine Ecology - Progress Series* 222: 1–24. doi:10.3354/meps222001.

Von Mises, R. 1959. *Theory of Flight*. Courier Corporation.

Walker, D. 2006. "Micro Autonomous Underwater Vehicle Concept for Distributed Data Collection." In *OCEANS 2006*, 1–4. doi:10.1109/OCEANS.2006.307050.

Wang, L.-P., and M.R. Maxey. 1993. "Settling Velocity and Concentration Distribution of Heavy Particles in Homogeneous Isotropic Turbulence." *Journal of Fluid Mechanics* 256 (November): 27–68. doi:10.1017/S0022112093002708.

Wang, W. H., R. C. Engelaar, X. Q. Chen, and J. G. Chase. 2009. "The State-of-Art of Underwater Vehicles - Theories and Applications." In *Mobile Robots - State of the Art in Land, Sea, Air, and COllaborative Missions*, 129–52. I-Tech Education and Publishing. http://ir.canterbury.ac.nz/handle/10092/4130.

Wang, Z., E. Thiébaut, and J. C. Dauvin. 1995. "Spring Abundance and Distribution of the Ctenophore Pleurobrachia Pileus in the Seine Estuary: Advective Transport and Diel Vertical Migration." *Marine Biology* 124 (2): 313–24. doi:10.1007/BF00347135.

Waters, A.M., and P.L. Beales. 2011. "Ciliopathies: An Expanding Disease Spectrum." *Pediatric Nephrology* 26 (7): 1039–56. doi:10.1007/s00467-010-1731-7.

Watson, S.A, D.J.P. Crutchley, and P.N. Green. 2011. "The Design and Technical Challenges of a Micro-Autonomous Underwater Vehicle." In *2011 International Conference on Mechatronics and Automation (ICMA)*, 567–72. doi:10.1109/ICMA.2011.5985723.

Watson, S.A., D.J.P. Crutchley, and P.N. Green. 2012. "The Mechatronic Design of a Micro–autonomous Underwater Vehicle (μAUV)." *International Journal of Mechatronics and Automation* 2 (3): 157–68. doi:10.1504/IJMA.2012.049397.

Watson, S.A, and P.N. Green. 2010. "Propulsion Systems for Micro-Autonomous Underwater Vehicles." In *2010 IEEE Conference on Robotics Automation and Mechatronics (RAM)*, 435–40. doi:10.1109/RAMECH.2010.5513155.

Webb, P. W. 1984a. "Form and Function in Fish Swimming." *Scientific American* 251 (1): 58–68.

———. 1984b. "Body Form, Locomotion and Foraging in Aquatic Vertebrates." *American Zoologist* 24 (1): 107–20. doi:10.1093/icb/24.1.107.

Webb, P. W., and A. J. Cotel. 2010. "Turbulence: Does Vorticity Affect the Structure and Shape of Body and Fin Propulsors?" *Integrative and Comparative Biology* 50 (6): 1155–66. doi:10.1093/icb/icq020.

Webster, D.R., A. Braithwaite, and J. Yen. 2004. "A Novel Apparatus for Simulating Isotropic Oceanic Turbulence at Low Reynolds Number." *Limnology and Oceanography: Methods* 2: 1–12.

Weihs, D. 2002. "Stability Versus Maneuverability in Aquatic Locomotion." *Integrative and Comparative Biology* 42 (1): 127–34. doi:10.1093/icb/42.1.127.

Weissburg, M., L. Atkins, K. Berkenkamp, and D. Mankin. 2012. "Dine or Dash? Turbulence Inhibits Blue Crab Navigation in Attractive–aversive Odor Plumes by Altering Signal Structure Encoded by the Olfactory Pathway." *The Journal of Experimental Biology* 215 (23): 4175–82. doi:10.1242/jeb.077255.

Weitzman, J.S., L.C. Samuel, A.E. Craig, R.B. Zeller, S.G. Monismith, and J.R. Koseff. 2014. "On the Use of Refractive-Index-Matched Hydrogel for Fluid Velocity Measurement within and around Geometrically Complex Solid Obstructions." *Experiments in Fluids* 55 (12): 1–12. doi:10.1007/s00348-014-1862-x.

Wiederseiner, S., N. Andreini, G. Epely-Chauvin, and C. Ancey. 2010. "Refractive-Index and Density Matching in Concentrated Particle Suspensions: A Review." *Experiments in Fluids* 50 (5): 1183–1206. doi:10.1007/s00348-010-0996-8.

Wilkinson, M., and H. R. Kennard. 2012. "A Model for Alignment between Microscopic Rods and Vorticity." *Journal of Physics A: Mathematical and Theoretical* 45 (45): 455502. doi:10.1088/1751-8113/45/45/455502.

Xu, H., and E. Bodenschatz. 2008. "Motion of Inertial Particles with Size Larger than Kolmogorov Scale in Turbulent Flows." *Physica D: Nonlinear Phenomena*, Euler Equations: 250 Years On Proceedings of an international conference, 237 (14–17): 2095–2100. doi:10.1016/j.physd.2008.04.022.

Yamamoto, I., T. Aoki, S. Tsukioka, H. Yoshida, T. Hyakudome, T. Sawa, S. Ishibashi, et al. 2004. "Fuel Cell System of AUV 'Urashima.'" In *OCEANS '04. MTTS/IEEE TECHNO-OCEAN '04*, 3:1732–37 Vol.3. doi:10.1109/OCEANS.2004.1406386.

Yamazaki, H., and K.D. Squires. 1996. "Comparison of Ocean Turbulence and Copepod Swimming." *Marine Ecology Progress Series* 144: 299–301.

Yang, G., and R. Zhang. 2009. "Path Planning of AUV in Turbulent Ocean Environments Used Adapted Inertia-Weight PSO." In *Fifth International Conference on Natural Computation, 2009. ICNC '09*, 3:299–302. doi:10.1109/ICNC.2009.355.

Yoerger, D.R., and J.-J.E. Slotine. 1985. "Robust Trajectory Control of Underwater Vehicles." *IEEE Journal of Oceanic Engineering* 10 (4): 462–70. doi:10.1109/JOE.1985.1145131.

Yudine, M. I. 1959. "Physical Considerations on Heavy-Particle Diffusion." In *Advances in Geophysics*, edited by H. E. Landsberg and J. Van Mieghem, 6:185–91. Elsevier. http://www.sciencedirect.com/science/article/pii/S0065268708601065.

Yuh, J., G. Marani, and D.R. Blidberg. 2011. "Applications of Marine Robotic Vehicles." *Intelligent Service Robotics* 4 (4): 221–31. doi:10.1007/s11370-011-0096-5.

Yu, J., R. Ding, Q. Yang, M. Tan, W. Wang, and J. Zhang. 2012. "On a Bio-Inspired Amphibious Robot Capable of Multimodal Motion." *IEEE/ASME Transactions on Mechatronics* 17 (5): 847–56. doi:10.1109/TMECH.2011.2132732.

Zaika, V., and N. Sergeeva. 1990. "Morphology and Development of Mnemiopsis-Mccradyi (ctenophora, Lobata)." *Zoologichesky Zhurnal* 69 (2): 5–11.

Zeng, Q. 2001. "Motion of Particles and Bubbles in Turbulent Flows." Brisbane: University of Queensland.

Zhang, H., G. Ahmadi, F.-G. Fan, and J.B. McLaughlin. 2001. "Ellipsoidal Particles Transport and Deposition in Turbulent Channel Flows." *International Journal of Multiphase Flow* 27 (6): 971–1009. doi:10.1016/S0301-9322(00)00064-1.

Zhao, F., and B. G. M. van Wachem. 2013. "Direct Numerical Simulation of Ellipsoidal Particles in Turbulent Channel Flow." *Acta Mechanica* 224 (10): 2331–58. doi:10.1007/s00707-013-0921-3.

Zhao, L., and H.I. Andersson. 2011. "On Particle Spin in Two-Way Coupled Turbulent Channel Flow Simulations." *Physics of Fluids* 23 (9): 093302. doi:10.1063/1.3626583.

Zhao, L., N. Challabotla, H. Anderssen, and E. Variano. 2015. "Rotation of Non-Spherical Particles in Turbulent Channel Flow." *Under Review.*

Zhao, L., C. Marchioli, and H. I. Andersson. 2014. "Slip Velocity of Rigid Fibers in Turbulent Channel Flow." *Physics of Fluids* 26 (6): 063302. doi:10.1063/1.4881942.

Zimmermann, Robert, Yoann Gasteuil, Mickael Bourgoin, Romain Volk, Alain Pumir, and Jean-François Pinton. 2011. "Rotational Intermittency and Turbulence Induced Lift Experienced by Large Particles in a Turbulent Flow." *Physical Review Letters* 106 (15): 154501. doi:10.1103/PhysRevLett.106.154501.

# Appendix A: Shape-Dependent Particle Drag

In our experiments, we encounter two main difficulties in finding analytic models for the drag on our particles. The first is that the particles are relatively large, meaning that Stokesian (creeping) flow is not appropriate; the particle Reynolds numbers are on the order of $10^2$. The second is that they are nonspherical, so that even Stokes's expressions (derived for a sphere) are not directly applicable. A number of corrections to the Stokes equations have been proposed and somewhat validated by experiment (Pettyjohn and Christiansen 1948; Stringham, Simons, and Guy 1969; Haider and Levenspiel 1989; Clark et al. 1989; Göğüş, Ipekçi, and Köpkinar 2001). We find a convenient collection of these in (Loth 2008), and follow his derivations below. From the diverse models presented in this excellent review, we select the ones that are most appropriate to our case (cylinders and spheroids at intermediate Reynolds number). We therefore obtain the curves plotted in Figure 3.2 and the values listed in Table 3..

## A1. Drag force

We begin with a simple expression for the drag force on an object:

$$\vec{\mathbf{F}}_{\mathbf{D}} = -\frac{1}{2}\rho_f A C_D |\vec{\mathbf{w}}|\vec{\mathbf{w}} \tag{A.2}$$

where $\rho_f$ is the fluid density, $A$ is the object's cross-sectional area, $C_D$ is the drag coefficient, and $\vec{\mathbf{w}}$ is the particle velocity with respect to the surrounding fluid. In the case of Stokes flow ($Re_p < 1$), the drag coefficient is $C_D = 24/Re_p$, which makes the drag linear in $\vec{\mathbf{w}}$. For large Reynolds number ($Re_p > 2000$), the drag coefficient is approximately constant. However, at intermediate Reynolds number, the drag coefficient varies nonlinearly with $Re_p$.

In this nonlinear regime, it is necessary to use a corrected version of the Stokesian drag coefficient. The most common correction for spheres at intermediate $Re_p$, developed by (Clift and Gauvin 1971), is:

$$C_D = \left[\frac{24}{Re_p}\left(1 + 0.15 Re_p^{0.687}\right)\right] + \frac{0.42}{1 + \dfrac{42{,}500}{Re_p^{1.16}}} \tag{A.3}$$

where the quantity in brackets is known as the Schiller-Naumann correction (Schiller and Naumann 1933). It is also useful to define a "Stokes correction factor" $f$:

$$f = \frac{C_D(Re_p)}{24/Re_p} \qquad (A.4)$$

such that in Equation (A.3), $f = \left(1 + 0.15 Re_p{}^{0.687}\right)$ (this definition neglects the final correcting term from Clift and Gauvin's expression, which is small for $Re_p < 10^3$, but we will carry this term through our calculations).

## A2. Stokes correction factors for nonspherical particles

The equations above are derived for spherical particles, but we seek a model for nonspherical (but regularly-shaped) particles. This necessitates the use of a different Stokes correction factor $f$, which depends on the particle aspect ratio, $\alpha$. These correction factors are well-defined for spheroids, and may be used for cylinders as well. The formulation of the correction factor is dependent on the orientation of the particle with respect to the surrounding flow: the drag parallel to the axis of symmetry is determined by $f_{\parallel}$, and the drag perpendicular to the axis of symmetry is determined by $f_{\perp}$. For oblate particles, these factors are:

$$f_{\parallel,\text{ob}} = \frac{(4/3)\alpha^{-1/3}(1 - \alpha^2)}{\alpha + \dfrac{(1 - 2\alpha^2)\cos^{-1}\alpha}{\sqrt{1 - \alpha^2}}} \; ; \; f_{\perp,\text{ob}} = \frac{(8/3)\alpha^{-1/3}(\alpha^2 - 1)}{\alpha - \dfrac{(3 - 2\alpha^2)\cos^{-1}\alpha}{\sqrt{1 - \alpha^2}}} \qquad (A.5)$$

For prolate particles, these factors are:

$$f_{\parallel,\text{pr}} = \frac{(4/3)\alpha^{-1/3}(1 - \alpha^2)}{\alpha - \dfrac{(2\alpha^2 - 1)\ln\left(\alpha + \sqrt{\alpha^2 - 1}\right)}{\sqrt{\alpha^2 - 1}}} \; ; \; f_{\perp,\text{pr}} = \frac{(8/3)\alpha^{-1/3}(\alpha^2 - 1)}{\alpha + \dfrac{(2\alpha^2 - 3)\ln\left(\alpha + \sqrt{\alpha^2 - 1}\right)}{\sqrt{\alpha^2 - 1}}} \qquad (A.6)$$

In the case of quiescent flow, we find that oblate particles ($\alpha \leq 1$) always fall with a vertical axis of symmetry, so in this case we would use $f_{\parallel,\text{ob}}$. Prolate particles ($\alpha > 1$) always fall broad-side, with the axis of symmetry parallel to the ground, so in this case we would use $f_{\perp,\text{pr}}$. In turbulent flow, however, we expect no preferential orientation. We can therefore use an averaged Stokes correction factor:

$$\frac{3}{\langle f \rangle} = \frac{1}{f_{\parallel}} + \frac{1}{f_{\perp}} + \frac{1}{f_{\perp}} \qquad (A.7)$$

(since cylinders and ellipsoids have one axis of symmetry, and two axes which are perpendicular to the axis of symmetry). For oblate particles, this yields:

$$\langle f_{\mathrm{ob}} \rangle = \frac{\alpha^{-1/3}\sqrt{1-\alpha^2}}{\cos^{-1}\alpha} \tag{A.8}$$

For prolate particles, this yields:

$$\langle f_{\mathrm{pr}} \rangle = \frac{\alpha^{-1/3}\sqrt{\alpha^2-1}}{\ln(\alpha+\sqrt{\alpha^2-1})} \tag{A.9}$$

These Stokes correction factors help us to determine the adjustments that must be made to the drag coefficient of our particles. We also note, however, that drag is also proportional to the particle's area. In the Newtonian (large Reynolds number) regime, the dominant part of the drag force is proportional to the cross-sectional area. However, at lower Reynolds numbers, viscosity is important and therefore skin-friction drag (which is proportional to total surface-area) is also important.

## A3. Normalized surface and projected areas

We can normalize the particle's surface area by the surface area of a sphere of equivalent volume:

$$A_{\mathrm{surf}}^* \equiv \frac{A_{\mathrm{surf}}}{\pi d_{\mathrm{eq}}{}^2} = \frac{2\alpha+1}{(18\alpha^2)^{1/3}} \text{ (cylinders)} \tag{A.10}$$

where $d_{\mathrm{eq}}$ is the diameter of the sphere of equivalent volume and $A_{\mathrm{surf}}$ is the cylinder's surface area.

In the previous section, we treated spheroids and cylinders as equivalent. However, due to the substantially different surface area between these two shape classes, we will calculate more specific values for $A_{\mathrm{surf}}^*$ and $A_{\mathrm{proj}}^*$. For prolate spheroids, $A_{\mathrm{surf}}^*$ is given by:

$$A_{\mathrm{surf}}^* = \frac{\frac{2\pi c^2}{\alpha^2}\left(1+\frac{\alpha}{\sqrt{1-\alpha^{-2}}}\sin^{-1}\left(\sqrt{1-\alpha^{-2}}\right)\right)}{\pi d_{\mathrm{eq}}{}^2}$$
$$= \frac{1}{2\alpha^{2/3}}\left(1+\frac{\alpha}{\sqrt{1-\alpha^{-2}}}\sin^{-1}\left(\sqrt{1-\alpha^{-2}}\right)\right) \tag{A.11}$$

where $c$ is the ellipsoid's major axis. For oblate spheroids, $A_{\mathrm{surf}}^*$ is given by:

$$A_{\mathrm{surf}}^* = \frac{\frac{2\pi c^2}{\alpha^2}\left(1+\frac{\alpha^2}{\sqrt{1-\alpha^2}}\tanh^{-1}\left(\sqrt{1-\alpha^2}\right)\right)}{\pi d_{\mathrm{eq}}{}^2}$$
$$= \frac{1}{2\alpha^{2/3}}\left(1+\frac{\alpha^2}{\sqrt{1-\alpha^2}}\tanh^{-1}\left(\sqrt{1-\alpha^2}\right)\right) \tag{A.12}$$

We can similarly normalize the particle's projected area. This quantity, like the Stokes correction factors, is orientation-dependent. In quiescent flow, oblate particles fall with the principal axis vertical:

$$A_{\text{proj,ob}}^* \equiv \frac{A_{\text{proj}}}{\pi r_{\text{eq}}^2} = \frac{\pi R_{\text{cyl}}^2}{\pi r_{\text{eq}}^2} = \left(\frac{3\alpha}{2}\right)^{-2/3} \text{ (oblate cylinder)} \tag{A.13}$$

Prolate cylinders fall with the principal axis parallel to the ground:

$$A_{\text{proj,pr}}^* = \frac{2R_{\text{cyl}} \cdot H_{\text{cyl}}}{\pi r_{\text{eq}}^2} = \frac{4R_{\text{cyl}}^2 \cdot \alpha}{\pi r_{\text{eq}}^2} = \frac{4\alpha}{\pi}\left(\frac{3\alpha}{2}\right)^{-2/3} \text{ (prolate cylinder)} \tag{A.14}$$

For spheroids, the derivation is similar:

$$A_{\text{proj,ob}}^* = \frac{\pi \frac{c^2}{\alpha^2}}{\pi r_{\text{eq}}^2} = \frac{1}{\alpha^2}\frac{c^2}{r_{\text{eq}}^2} = \alpha^{-1/3} \text{ (oblate spheroid)} \tag{A.15}$$

$$A_{\text{proj,pr}}^* = \frac{\pi \frac{c^2}{\alpha}}{\pi r_{\text{eq}}^2} = \frac{1}{\alpha}\frac{c^2}{r_{\text{eq}}^2} = \alpha^{1/3} \text{ (prolate spheroid)} \tag{A.16}$$

In turbulent flow, where no preferential orientation is expected, $A_{\text{surf}}^*$ will be the same, but $A_{\text{proj}}^*$ will change. We can treat $A_{\text{proj}}^*$ as we treated the Stokes correction factors and find an orientation-averaged version of $A_{\text{proj}}^*$:

$$3\langle A_{\text{proj}}^* \rangle = 2A_{\text{proj,pr}}^* + A_{\text{proj,ob}}^* = \left(\frac{8\alpha}{\pi} + 1\right)\left(\frac{3\alpha}{2}\right)^{-2/3} \text{ (cylinders)} \tag{A.17}$$

$$3\langle A_{\text{proj}}^* \rangle = 2A_{\text{proj,pr}}^* + A_{\text{proj,ob}}^* = 2\alpha^{1/3} + \alpha^{-1/3} \text{ (spheroids)} \tag{A.18}$$

Note that for a spherical particle, $A_{\text{surf}}^* = A_{\text{proj}}^* = \langle A_{\text{proj}}^* \rangle = 1$. We have now derived the Stokes correction factor, the normalized surface area, and the normalized projected area for all of the particle shapes discussed in this and other chapters.

## A4. Combining forces: Stokes vs. Newton regime

At intermediate Reynolds numbers, where we perform our experiments, a robust drag model is elusive (Chhabra, Agarwal, and Sinha 1999). Previous work has focused on combining drag models from the Stokes and Newton regimes (Ganser 1993; Cheng 1997). In the Newtonian regime, the drag coefficient is expected to be constant; this is often called the "critical drag". Just as we defined the Stokes drag correction factor as the ratio between the particle's actual drag coefficient and the drag coefficient of a volume-equivalent sphere (Equation (A.4)), so may we define a Newton drag correction factor that is the ratio of the particle's critical drag coefficient to a volume-equivalent sphere's critical drag coefficient:

$$C_{\text{shape}} = \frac{C_{\text{D,shape,crit}}}{C_{\text{D,sphere,crit}}}\bigg|_{\substack{\text{constant} \\ \text{volume}}}$$

We know that the critical drag coefficient of a given particle is dependent on its shape in complex ways, making this problem difficult to model exactly. We therefore approximate $C_{\text{shape}}$ based only on particle sphericity, which is the ratio of the particle's surface area to the surface area of the sphere of equivalent volume (that is, $A_{\text{surf}}^*$):

$$C_{\text{shape}} = 1 + 1.5(A_{\text{surf}}^* - 1)^{1/2} + 6.7(A_{\text{surf}}^* - 1) \text{ (oblate particles)} \tag{A.19}$$

$$C_{\text{shape}} = 1 + 0.7(A_{\text{surf}}^* - 1)^{1/2} + 2.4(A_{\text{surf}}^* - 1) \text{ (prolate particles)} \tag{A.20}$$

These models are based on the theoretical work of (Clift, Grace, and Weber 2013; T. L. Thompson and Clark 1991; Ganser 1993) and the experimental work of (Pettyjohn and Christiansen 1948; Stringham, Simons, and Guy 1969; Haider and Levenspiel 1989; Clark et al. 1989; Gögüs, Ipekçi, and Kökpinar 2001). As expected, $C_{\text{shape}} = 0$ for spheres ($A_{\text{surf}}^* = 1$) for both the oblate and prolate expressions above. However, because the sphericity of a cylinder is never exactly equal to one (for $\alpha = 1$, $A_{\text{surf}}^* = 1.14$ for a cylinder), the value of $C_{\text{shape}}$ will be discontinuous at $\alpha = 1$ for cylinders. So, this is not an exact model for cylinders, but it has shown reasonable accuracy in experiments involving disks and rods (Haider and Levenspiel 1989; Pettyjohn and Christiansen 1948).

Using a dimensional-collapse approach, (Ganser 1993) defines a normalized drag coefficient and a normalized Reynolds number:

$$C_{\text{D}}^* = \frac{C_{\text{D}}}{C_{\text{shape}}} \tag{A.21}$$

$$Re_{\text{p}}^* = \frac{C_{\text{shape}} Re_{\text{p}}}{f_{\text{shape}}} \tag{A.22}$$

These two parameters, $C_D^*$ and $Re_p^*$, encompass shape variation in the Newton-drag regime (as quantified by $C_{shape}$) and shape variation in the Stokes-drag regime (as quantified by $f_{shape}$).

By substituting these quantities into Equation (A.3), we achieve this expression:

$$C_D^* = \left[\frac{24}{Re_p^*}\left(1 + 0.15Re_p^{*\,0.687}\right)\right] + \frac{0.42}{1 + \dfrac{42,500}{Re_p^{*\,1.16}}} \tag{A.23}$$

We now have a motley collection of normalized and non-normalized parameters, all of which are non-dimensional and some of which depend on particle orientation (but not all). As a refresher for the reader, we qualitatively summarize these parameters in Table A4.

| $C_D$ | Particle drag coefficient; unknown. |
|---|---|
| $C_{shape}$ | Newton-drag correction factor, based entirely on particle sphericity $A_{surf}^*$ (which is a function of particle aspect ratio $\alpha$). This is a property of particle geometry only, and does not vary with orientation. |
| $f_{shape}$ | Stokes-drag correction factor, based entirely on particle aspect ratio $\alpha$. This is a property of geometry, but it varies with particle orientation. When there is no preferential particle orientation (e.g. Brownian motion, turbulence), an averaged version $\langle f \rangle$ may be used. |
| $Re_p$ | Particle Reynolds number based on the particle's velocity, the particle's diameter, and the density and viscosity of the surrounding fluid. |
| $Re_p^*$ | Normalized particle Reynolds number, which depends on $C_{shape}$, $f_{shape}$, and $Re_p$. |
| $C_D^*$ | Normalized particle drag coefficient for the intermediate Reynolds-number regime, found via equation XX. |
| $A_{surf}^*$ | The surface area of the particle divided by the surface area of a sphere of equivalent volume, also called sphericity. This is a property of particle geometry only, and does not vary with orientation. |
| $A_{proj}^*$ | The projected area of the particle divided by the projected area of a sphere of equivalent volume. This is a property of geometry, but it varies with particle orientation. When there is no preferential particle orientation (e.g. Brownian motion, turbulence), an averaged version $\langle A_{proj}^* \rangle$ may be used. |

Table A4: Parameters involved in the calculation of drag on non-spherical particles in the intermediate Reynolds number regime. It is the particle drag coefficient, $C_D$, which must be found in order to calculate the true drag on the particle and therefore the still-water settling velocity for various shapes.

## A5. Solving for the particle drag: iterative approach

We return to Equation (A.2), the drag on a particle:

$$\vec{F}_D = -\frac{1}{2}\rho_f A C_D |\vec{w}|\vec{w} \tag{A.24}$$

This expression holds for both the Newtonian and Stokesian drag regimes; it is only the drag coefficient $C_D$ that must be changed. If we substitute $C_D = 24/Re_p$, we recover the familiar expression for a sphere in creeping flow: $\vec{F}_D = 6\pi\mu R\vec{w}$. For our problem—nonspherical particles at intermediate Reynolds number—the drag coefficient is nonlinearly dependent on the particle Reynolds number. Specifically, we seek to discover the still-water settling velocity of our particles, when the buoyant force is equal to the drag force. In equilibrium, the particle settling velocity is governed by the following equation:

$$\vec{\mathbf{F}}_{\mathbf{B}} = \frac{1}{2}\rho_f A C_D |\vec{\mathbf{w}}| \vec{\mathbf{w}} \qquad (A.25)$$

The buoyant force $\vec{\mathbf{F}}_{\mathbf{B}}$ is dependent only on particle volume, particle density, and fluid density—not particle velocity. We can therefore take an iterative approach, using the following steps:

1. Guess a value for the particle velocity $w$.
2. Determine the resulting particle Reynolds number $Re_{\mathrm{p}}$.
3. Using $C_{\mathrm{shape}}$ and $f_{\mathrm{shape}}$, which are properties of geometry alone (Equations (A.5)(A.6) and (A.19)(A.20)), calculate the normalized particle Reynolds number $Re_{\mathrm{p}}^*$ (Equation (A.22)).
4. Using $Re_{\mathrm{p}}^*$ and Equation (A.23), determine the normalized drag coefficient $C_{\mathrm{D}}^*$.
5. Using $C_{\mathrm{D}}^*$ and $C_{\mathrm{shape}}$, determine the actual particle drag coefficient $C_D$.
6. Solve for $w$ using Equation (A.25) and the known value of $\vec{\mathbf{F}}_{\mathbf{B}}$, which is a property of geometry alone. Compare this value with the initial guess from Step 1.
7. If the difference between the two values of $w$ is too large, begin again at Step 1, this time using the value obtained in step 6.

The particle's fall orientation is taken into account in both step 3 and step 6, via $f_{\mathrm{shape}}$ and $A$ (the particle's cross-sectional area).

With this approach, we can determine the expected still-water settling velocity of prolate and oblate cylinders. For our size range ($r_{\mathrm{eq}} = 4.6 \, \mathrm{mm}$), the results are plotted in Figure A1. The settling velocity rises to a peak at $\alpha = 1$, which is to be expected as this aspect ratio has the lowest projected surface area. We also note the small discontinuity at $\alpha = 1$. This discontinuity arises from two parameters: first, the difference between the definition of $C_{\mathrm{shape}}$ for prolate and oblate particles; second, the difference between the preferred fall orientations (and therefore the projected cross-sectional areas) of oblate vs. prolate particles. Our observations indicate that cylinders with $\alpha = 1$ fall with their symmetry axis vertical, whereas cylinders with $\alpha = 2$ fall with their symmetry axis parallel to the ground ("broadside"). We have no observational data for $1 < \alpha < 2$; we have therefore left the discontinuity as-is. The discontinuity does not appear for spheroids, as both $A_{\mathrm{proj}}^*$ and $C_{\mathrm{shape}}$ are continuous for this case.

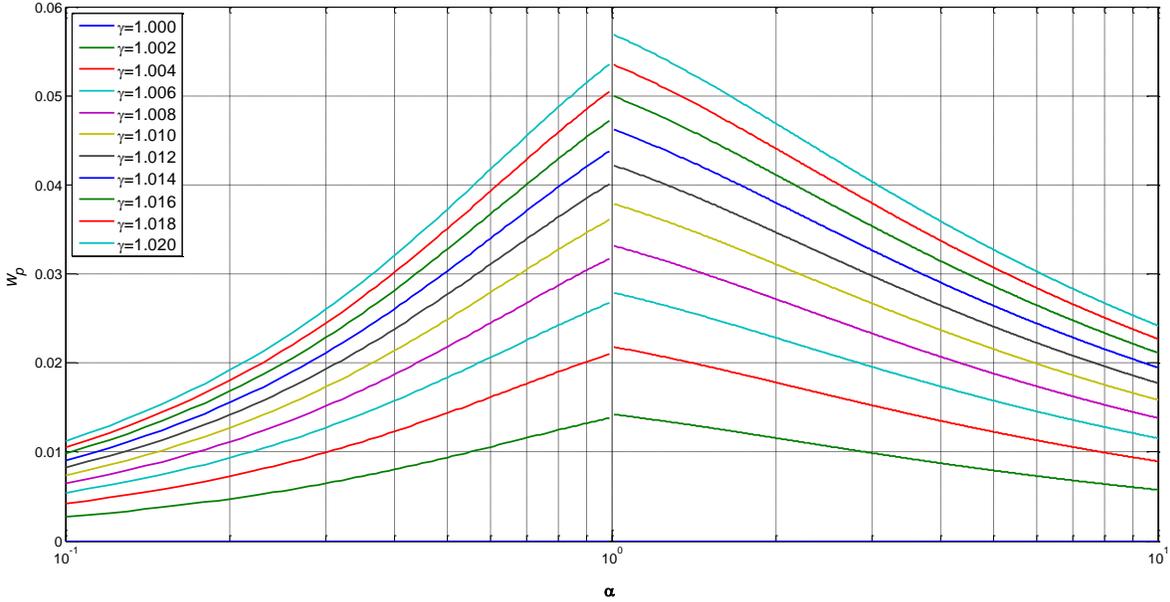

**Figure A1:** Still-water settling velocities of prolate and oblate cylinders with $r_{eq} = 4.6\ \mathrm{mm}$ for varying particle specific gravity. The discontinuity at $\alpha=1$ arises because of the model's abrupt shift in preferential settling orientation.

## A6. A note on still-water settling velocity vs. turbulent slip velocity

Though we may predict the still-water settling velocity using a combination of analytical and empirical methods as described above, the turbulent slip velocity is altogether a different animal. An iterative approach which assumes equilibrium between buoyancy and drag is not applicable. In this situation, there are many other forces on the particle such as the acceleration reaction force and the history force (as described earlier in this chapter). Additionally, particles are not in equilibrium, but are continuously accelerating and decelerating.

Despite these limitations, it is illustrative to consider the effects of preferential orientation on particle drag. We can use the same model and iterative approach as used in section A5, but remove the effects of preferential orientation in Steps 3 and 6. We use an averaged shape factor $\langle f \rangle$ as defined by equations (A.8) and (A.9), and instead of using a preferential cross-sectional area, we can use the average projected area $\langle A^*_{proj} \rangle$. The results for both cylinders and spheroids are shown in Figure A2, for particles with $r_{eq} = 4.6\ \mathrm{mm}$ and $\gamma = 1.01$. The discontinuity at $\alpha = 1$ is still present, due to the definition of $C_{shape}$. It is interesting that the discontinuity is actually somewhat larger in the random case, showing that perhaps a more continuous model of $C_{shape}$ is needed for cylinders. Outside this small contested area, however, we see that in general, randomly-oriented-particle fall velocity is *increased* with respect to the preferred-orientation case. This is logical—in the preferred-orientation case, cylinders always appear in a drag-maximizing orientation due to the pressure distribution and recirculation in the particle wake. This scenario is what we empirically observe.

This observation gives us further confidence that the sharply reduced settling velocity we observe in turbulent flow is most likely due to complex interactions with turbulent vortex structures, and cannot be explained by the erasure of preferential orientation. In fact, if particles have no preferred orientation, the opposite is seen—settling velocity should increase. Since this is not what we observe, we may conclude that the settling velocity reduction may be attributed to phenomenological effects related to turbulence.

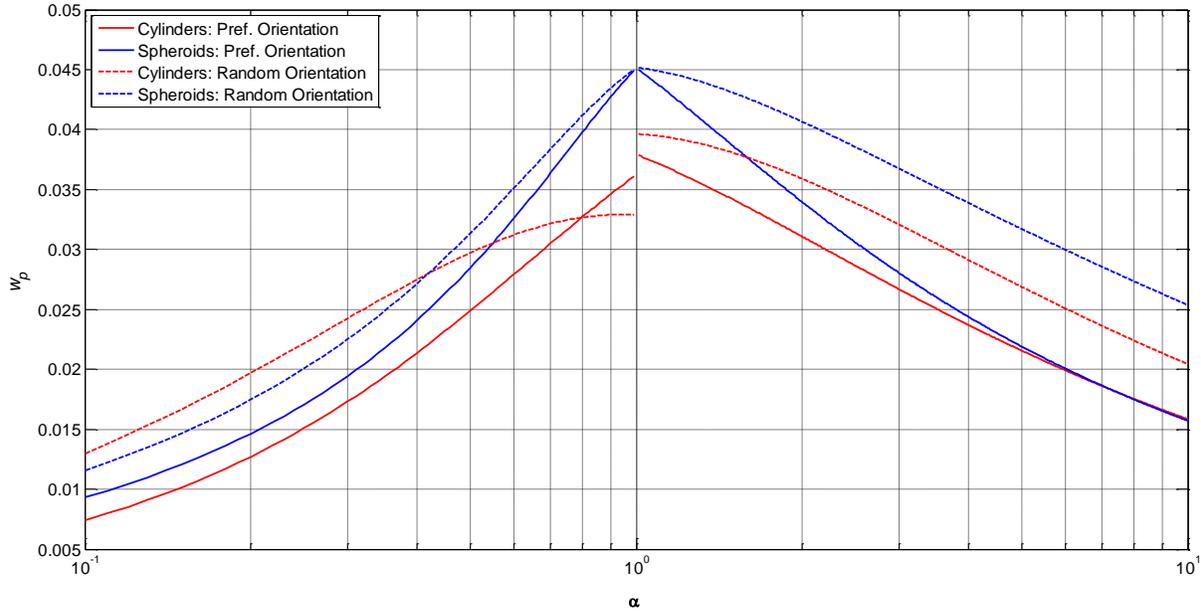

**Figure A2: :** Settling velocity vs. aspect ratio for cylinders (red) and spheroids (blue) falling at a preferred orientation (broadside for $\alpha > 1$, vertical for $\alpha \leq 1$), shown as solid lines, and at random orientations, shown as dotted lines. Data are for particles with $r_{eq} = 4.6\text{ mm}$ and $\gamma = 1.01$. Discontinuity at $\alpha = 1$ for cylinders arises from differences in the definition of $C_{shape}$ and $A_{proj}$.

# Appendix B: Finding $\langle \vec{u}_f \rangle$

We define the turbulent slip velocity as $\vec{u}_s \equiv \langle \vec{u}_p \rangle - \langle \vec{u}_f \rangle$, where $\langle \vec{u}_p \rangle$ is the average particle velocity and $\langle \vec{u}_f \rangle$ is the average fluid velocity in an annulus around the particle. In this annulus, we would like to exclude the fluid which is immediately adjacent to the particle and include all the fluid that is likely to influence the particle. We set the inner bound of the annulus at approximately 2.5 mm from the particle border, which is the start of the "outer layer" of the logarithmic boundary layer in wall turbulence. We set the outer bound of the annulus at approximately 9.2mm from the particle border, which is equal to the diameter of the sphere of equivalent volume to the cylindrical particles we examine. However, it is worthwhile to shift the boundaries and examine the impact on our results. Since changing the bounds on the fluid annulus does not change $\langle \vec{u}_p \rangle$, we can look at the effects on $\langle \vec{u}_f \rangle$ and the subsequent effects on $\vec{u}_s$.

As stated previously, we choose 2.5mm as our inner bound. If we set the inner bound to be smaller than 2.5mm (which in our resolution corresponds to 2·dx), the value of $\langle \vec{u}_f \rangle$ will grow closer to the value of $\langle \vec{u}_p \rangle$ due to the no-slip boundary condition at the particle surface. This will have the effect of decreasing the magnitude of $\vec{u}_s$. If we set the inner bound to be larger than 2.5mm, we will include less of the fluid being dragged along with the particle, and $\langle \vec{u}_f \rangle$ will take on values that are (presumably) not similar to $\langle \vec{u}_p \rangle$. This will increase the magnitude of $\vec{u}_s$. This is borne out in Figure B1; as the inner bound varies from dx to 4dx, the slip velocity increases.

For our outer bound, we choose the diameter of the sphere of equivalent volume to the cylinders, which is 9.2mm. A larger outer bound, including more of the surrounding fluid, should bias $\langle \vec{u}_f \rangle$ toward zero due to the very low mean flow of the tank—if enough of the image area is averaged, $\langle \vec{u}_f \rangle$ will approach the tank-scale mean flow. A lower $\langle \vec{u}_f \rangle$ will result in a higher magnitude of $\vec{u}_s$, which we see in Figure B2.

For both cases, drastic changes of the annulus bounds result in only small changes to the calculated slip velocity. We are therefore comfortable in our choice to set the annulus bounds of 2.5mm (approximately 2dx) and 9.2mm (approximately 7dx).

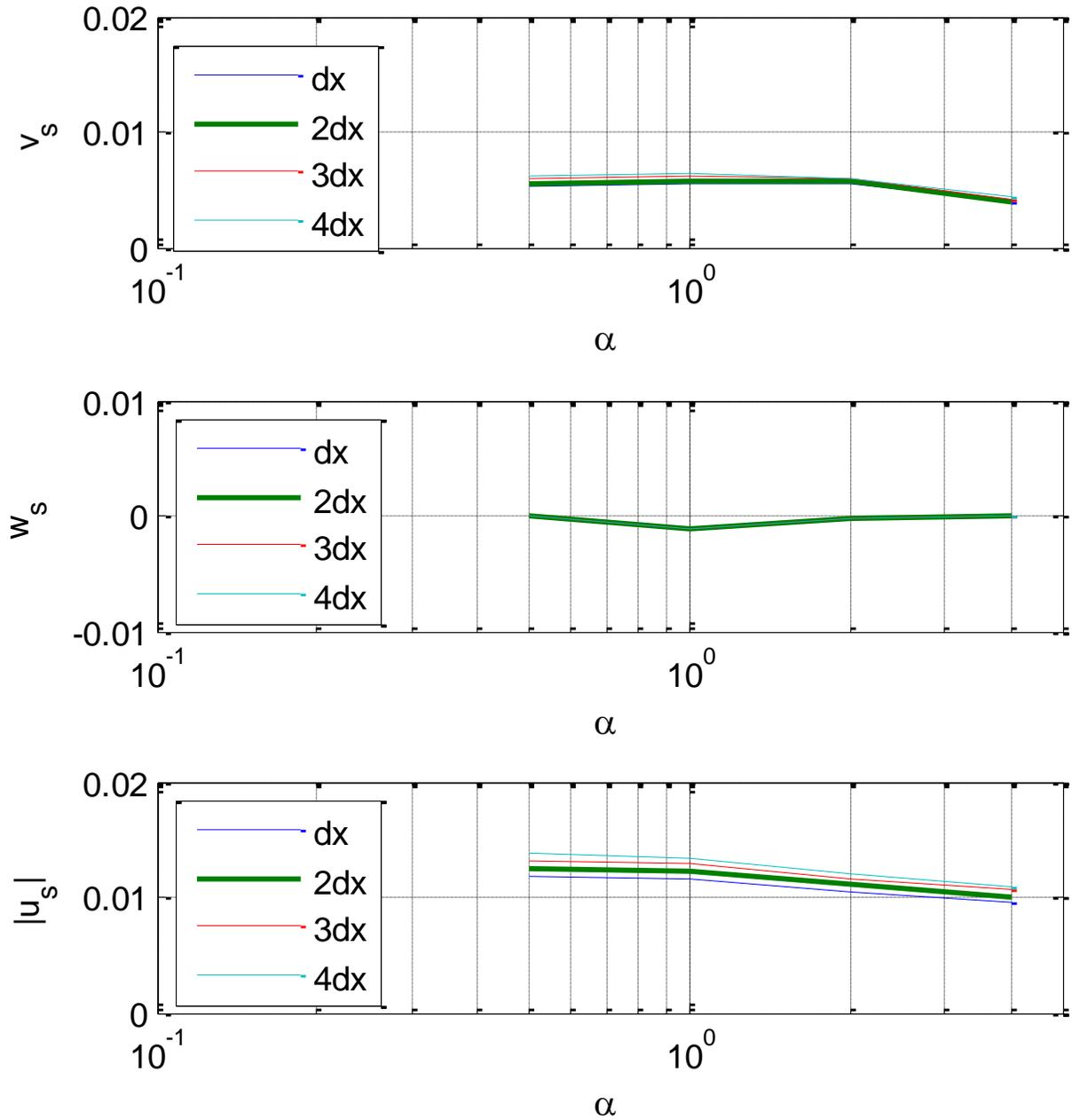

**Figure B1:** Effects of changing the inner bound of the fluid-averaging annulus on $v_s$, $w_s$, and $|\vec{u}_s|$. In these plots, the outer bound is held fixed at $d_{eq}$ = 9.2 mm while the inner bound varies from dx=1.3mm to 4dx=5.2mm. These data are from cylinders at $\gamma$ = 1.003, $\alpha$ = 0.5, 1, 2, and 4. Bolded line represents the bounds that were used for all calculations in Chapter 3.

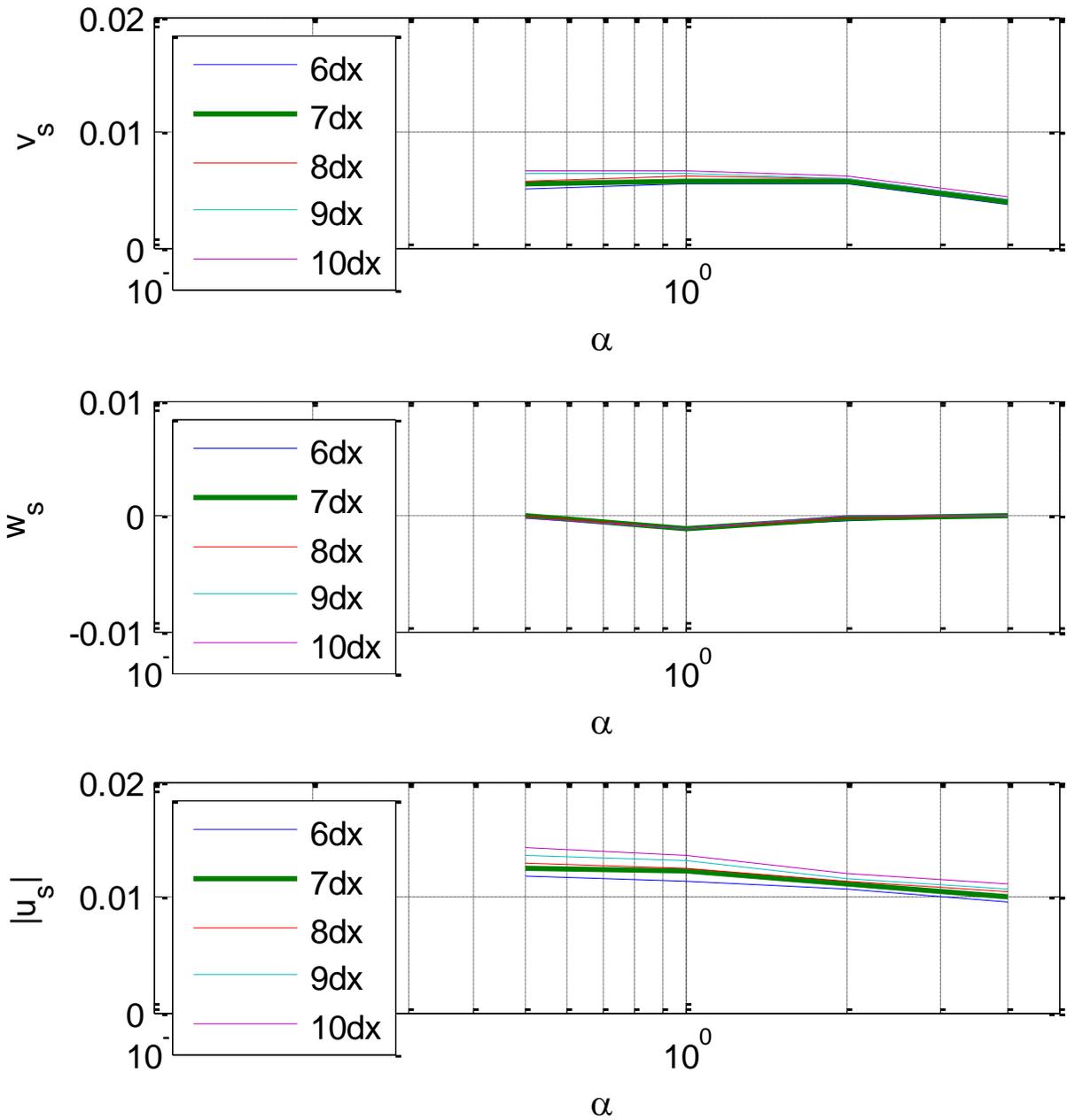

**Figure B2:** Effects of changing the outer bound of the fluid-averaging annulus on $v_s$, $w_s$, and $|\vec{u}_s|$. In these plots, the inner bound is held fixed at 3 mm while the inner bound varies from 6dx=7.8mm to 10dx=13mm. These data are from cylinders at $\gamma$ = 1.003, $\alpha$ = 0.5, 1, 2, and 4. Bolded line represents the bounds that were used for all calculations in Chapter 3.

# Appendix C: Small-particle simulation

Infinitesimally-small, neutrally-buoyant, nonspherical particles are suspended in numerically-simulated turbulence (Li et al. 2008) at a Reynolds number of $Re_\lambda$=433. The particles are initialized at random positions $\vec{x}$ and orientations $\hat{n}$, and advected exactly with the fluid such that $D\vec{x}/Dt = \vec{u}_p(\vec{x}, t) = \vec{u}_f(\vec{x}, t)$. Particles rotate according to Jeffery's approximations (Equations (4.6) and (4.7)). From the turbulence data, downloaded from the Johns Hopkins University turbulence database, we obtain both the fluid velocity $\vec{u}_f(\vec{x}, t)$ and the fluid-velocity-gradient tensor $\mathbb{A}(\vec{x}, t)$, from which we may compute the strain-rate tensor $\mathbb{S} = (\mathbb{A} + \mathbb{A}^T)/2$. As the particles' positions and orientations evolve over time, we may follow them along their trajectory. Particle shape factor, $\Lambda$ (defined in Equation (4.5)), is continuously varied from $\Lambda = -1$ to $\Lambda = 1$ to explore the effects of particle shape on rotation and its components (Figure 4.2). We may also investigate the interaction of the particle and its orientation vector, $\hat{n}$, with the surrounding fluid.

In particular, we are interested in the alignment between the particle's axis of symmetry $\hat{n}$, particle rotation (here denoted as $\boldsymbol{\omega}$), fluid rotation rate (equal to half the fluid vorticity, here denoted as $\boldsymbol{\Omega}$), and the eigenvalues of the strain-rate tensor $\mathbb{S}$ (here denoted as $\boldsymbol{e}_i$)[1]. To explore the effects of shape on alignment with the surrounding flow, we consider particles with a shape factor of either $\Lambda = 1$ (an infinitely-thin rod, shown as red circles in the figures below) or $\Lambda = -1$ (an infinitely-thin disk, shown as blue squares in the figures below). Three simulated trajectories are shown in Figure C1, Figure C2, and Figure C3.

The first panel of each figure shows the magnitude of the total rotation experienced by both the fluid ($\Omega^2$) and the rod and disk ($\omega^2$), normalized by the Kolmogorov timescale $\tau_K{}^2$. We see that in general, the magnitude of the rotation experienced by the rod is basically equivalent to the fluid rotation, while the disk exhibits large fluctuations. This is clarified in the second panel of each figure, which shows the alignment of the particles' principal axis $\hat{n}$ with the fluid vorticity. We see that the rod is generally well-aligned with the vorticity, with a normalized alignment coefficient close to one throughout the trajectory. By contrast, the disk is not well-aligned with the vorticity vector: in fact, the disk tends to align in such a way that its symmetry axis $\hat{n}$ is perpendicular to the fluid vorticity (with a normalized alignment coefficient close to zero). It is this misalignment which causes the rapid fluctuations seen in the first panel of each figure. The alignment of $\hat{n}_{\text{disk}}$ in the plane perpendicular to $\vec{\Omega}$ leads the disk to have a weak spinning rate and a strong tumbling rate (Figure 4.2).

Because $\vec{\Omega}$ tends to align with $\boldsymbol{e}_2$ in regions of high vorticity, the intermediate eigenvalue of the strain-rate tensor $\mathbb{S}$, we expect $\hat{n}_{\text{disk}}$ to fall into the plane spanned by $\boldsymbol{e}_1$ and $\boldsymbol{e}_3$, the other two eigenvalues of $\mathbb{S}$. This may be observed in the fourth panel of each figure: during large rotation events (high vorticity), $\hat{\boldsymbol{n}}_{\mathbf{disk}}$ oscillates rapidly between alignment with $\boldsymbol{e}_1$ and alignment with $\boldsymbol{e}_3$.

---

[1] We caution that this notation differs somewhat from the notation used throughout the rest of the thesis. We also note that the cross-product, which tends to be denoted as $\vec{a} \times \vec{b}$ in engineering, is denoted as $\vec{a} \wedge \vec{b}$ in physics.

Recall that the rotation rates of these tiny particles are determined by both the fluid rotation and the fluid strain (see Equations (4.6) and (4.7)). Because $\hat{\boldsymbol{n}}_{\mathbf{disk}}$ does not tend to be aligned with $\vec{\boldsymbol{\Omega}}$, the strain-contribution term $\hat{\boldsymbol{n}} \times \mathbb{S}\hat{\boldsymbol{n}}$ is larger than the same term for a rod. Since $\hat{\boldsymbol{n}}_{\mathbf{disk}}$ tends to lie within the plane spanned by $\boldsymbol{e}_1$ and $\boldsymbol{e}_3$, the strain contribution term $\hat{\boldsymbol{n}} \times \mathbb{S}\hat{\boldsymbol{n}}$ is likely to be parallel to $\boldsymbol{e}_2$, but rapidly changing signs. Therefore, the strain contribution is alternately enhancing and opposing the rotation of $\hat{\boldsymbol{n}}_{\mathbf{disk}}$ about the fluid vorticity vector. This is shown in the third panel of each figure, which shows the alignment between $\vec{\boldsymbol{\Omega}}$ and $\hat{\boldsymbol{n}} \times \mathbb{S}\hat{\boldsymbol{n}}$ for both a rod and a disk. Note that for the disk, the value alternates rapidly between approximately negative one (opposing the rotation of $\hat{\boldsymbol{n}}_{\mathbf{disk}}$ about $\vec{\boldsymbol{\Omega}}$) and approximately one (enhancing the rotation of $\hat{\boldsymbol{n}}_{\mathbf{disk}}$ about $\vec{\boldsymbol{\Omega}}$). For rods, the strain contribution is nearly always perpendicular to $\vec{\boldsymbol{\Omega}}$.

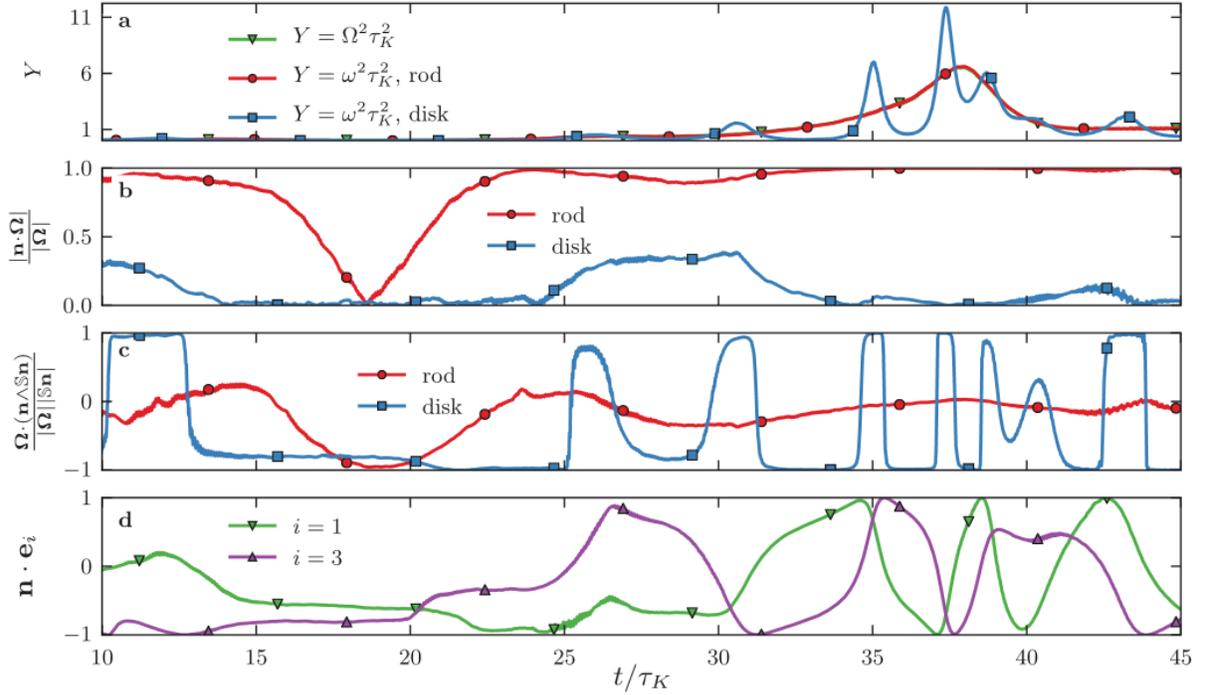

Figure C1: A simulated infinitely thin rod (red circles) and an infinitely thin disk (blue squares) following the same trajectory, reprinted directly from (Byron et al. 2015) with permission. Panel (a) shows the magnitude of the fluid rotation (green triangles), along with the magnitude of the rotation of the rod and the disk. Panel (b) shows the alignment of the rod and the disk orientation (which is also the direction of the "spinning" rate) with the fluid vorticity. Panel (c) shows the alignment of the rod and disk "tumbling" rate with the vorticity. Panel (d) shows data only for disks, and shows the alignment of the disk's principal axes with the two extreme eigenvalues of the strain-rate tensor $\mathbb{S}$, denoted as $\boldsymbol{e}_1$ (green point-down triangles) and $\boldsymbol{e}_3$ (purple point-up triangles). Note the region of intense vorticity at $35\tau_K < t < 40\tau_K$.

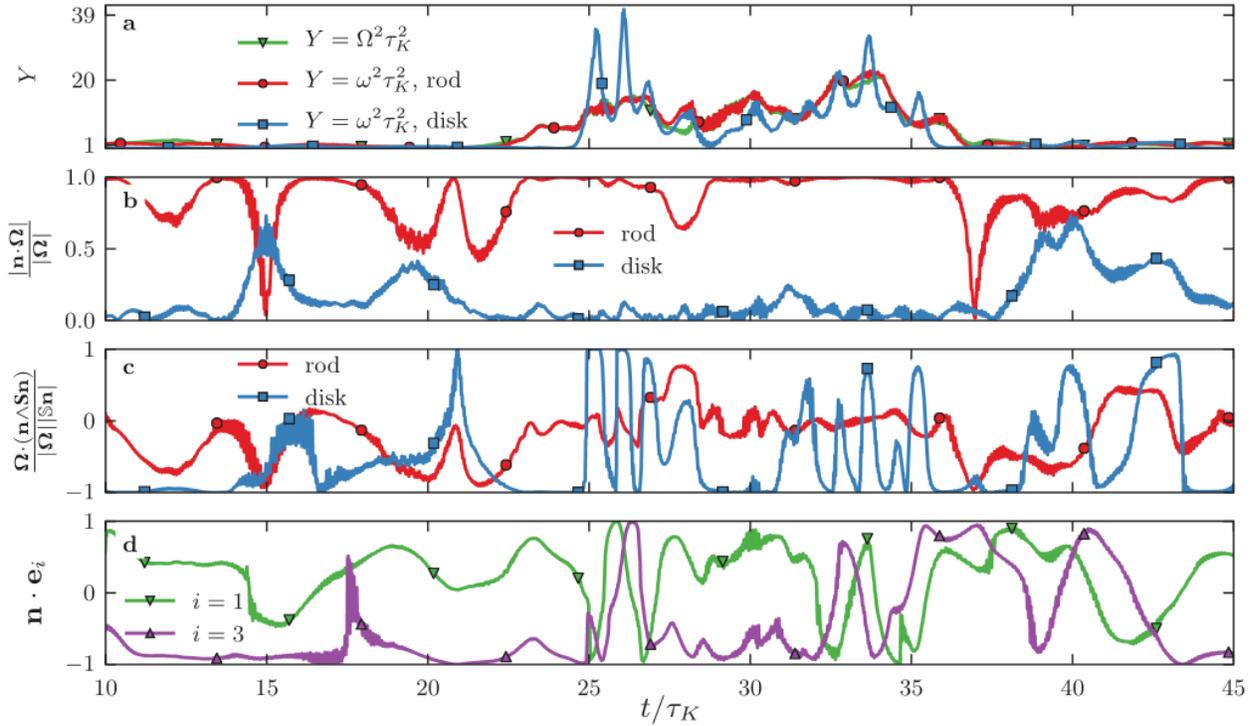

**Figure C2:** A simulated infinitely thin rod (red circles) and an infinitely thin disk (blue squares) following the same trajectory, reprinted directly from (Byron et al. 2015) with permission. Panel (a) shows the magnitude of the fluid rotation (green triangles), along with the magnitude of the rotation of the rod and the disk. Panel (b) shows the alignment of the rod and the disk orientation (which is also the direction of the "spinning" rate) with the fluid vorticity. Panel (c) shows the alignment of the rod and disk "tumbling" rate with the vorticity. Panel (d) shows data only for disks, and shows the alignment of the disk's principal axes with the two extreme eigenvalues of the strain-rate tensor $\mathbb{S}$, denoted as $e_1$ (green point-down triangles) and $e_3$ (purple point-up triangles). The high-frequency components of the data are due to numerical artifacts.

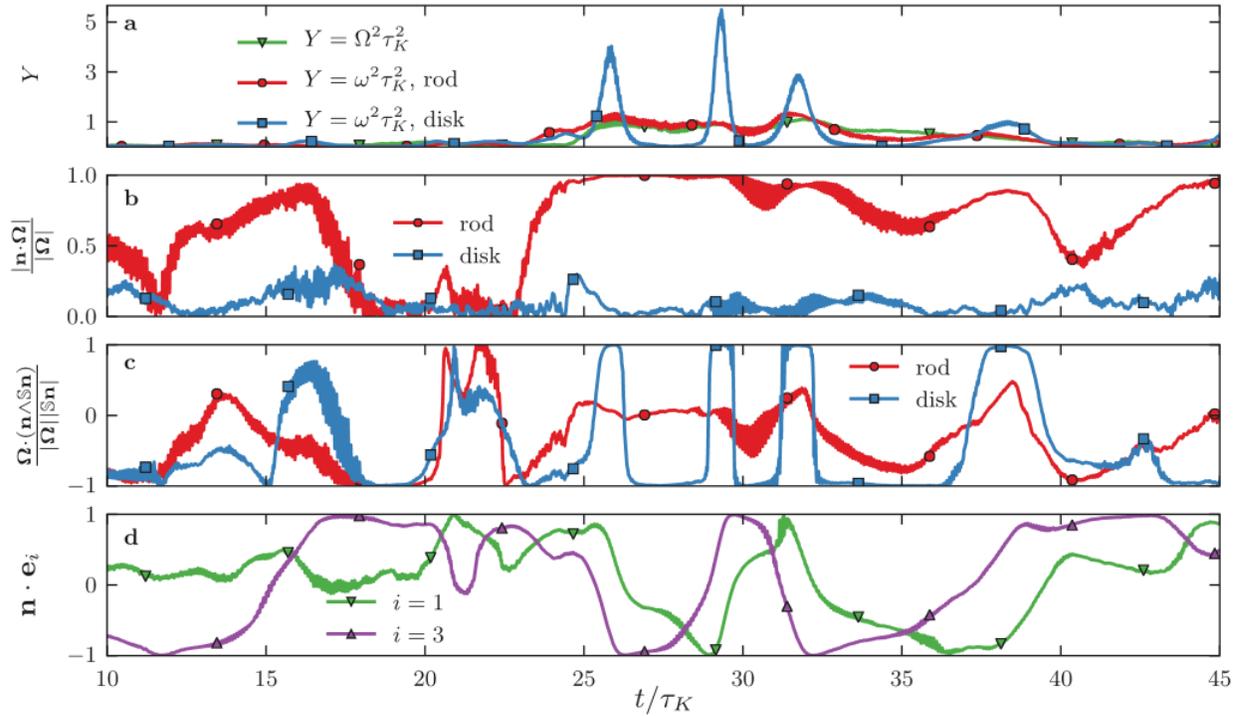

Figure C3: A simulated infinitely thin rod (red circles) and an infinitely thin disk (blue squares) following the same trajectory, reprinted directly from (Byron et al. 2015) with permission. Panel (a) shows the magnitude of the fluid rotation (green triangles), along with the magnitude of the rotation of the rod and the disk. Panel (b) shows the alignment of the rod and the disk orientation (which is also the direction of the "spinning" rate) with the fluid vorticity. Panel (c) shows the alignment of the rod and disk "tumbling" rate with the vorticity. Panel (d) shows data only for disks, and shows the alignment of the disk's principal axes with the two extreme eigenvalues of the strain-rate tensor $\mathbb{S}$, denoted as $e_1$ (green point-down triangles) and $e_3$ (purple point-up triangles). The high-frequency components of the data are due to numerical artifacts.

This work contained in this dissertation was funded by the National Science Foundation through NSF IGERT #0903711 and NSF DGE #1106400.